\documentclass[10pt, a4paper, twoside]{Thesis} 

\graphicspath{{Pictures/}} 

\usepackage{float}
\usepackage{minitoc}
\usepackage{color}
\usepackage[final]{pdfpages} 

\usepackage[comma, sort&compress]{natbib} 
\hypersetup{urlcolor=blue, colorlinks=true} 
\title{\ttitle} 
\newcommand{\sa}{\mathrm{S}}
\newcommand{\te}{\mathrm{T}}
\newcommand{\pk}{\mathcal{P}}
\newcommand{\he}{\mathrm{H}}
\newcommand{\wjjj}[6]
{{
\left( 
\begin{array}{lcr} #1 & #2 & #3 \\#4 & #5 & #6 \end{array}
\right) 
}}

\begin{document}

\frontmatter 

\setstretch{1.3} 

\fancyhead{} 
\rhead{\thepage} 
\lhead{} 

\pagestyle{fancy} 

\newcommand{\HRule}{\rule{\linewidth}{0.5mm}} 
\newcommand{\sz}{{\sc sz}}
\newcommand{\zb}{{\sc zb}}
\newcommand{\kn}{{\sc kn}}

\newcommand{\etal}{{et~al.}}

\def\aj{AJ}%
\def\actaa{Acta Astron.}%
\def\araa{ARA\&A}%
\def\apj{Astrophys. J.}%
\def\apjl{ApJ}%
\def\apjs{ApJS}%
\def\ao{Appl.~Opt.}%
\def\apss{Ap\&SS}%
\def\aap{Astron. \& Astrophys.}%
\def\aapr{A\&A~Rev.}%
\def\aaps{A\&AS}%
\def\azh{AZh}%
\def\baas{BAAS}%
\def\bac{Bull. astr. Inst. Czechosl.}%
\def\caa{Chinese Astron. Astrophys.}%
\def\cjaa{Chinese J. Astron. Astrophys.}%
\def\icarus{Icarus}%
\def\jcap{J. Cosmology Astropart. Phys.}%
\def\jrasc{JRASC}%
\def\mnras{MNRAS}%
\def\memras{MmRAS}%
\def\na{New A}%
\def\nar{New A Rev.}%
\def\pasa{PASA}%
\def\pra{Phys.~Rev.~A}%
\def\prb{Phys.~Rev.~B}%
\def\prc{Phys.~Rev.~C}%
\def\prd{Phys.~Rev.~D}%
\def\pre{Phys.~Rev.~E}%
\def\prl{Phys.~Rev.~Lett.}%
\def\pasp{PASP}%
\def\pasj{PASJ}%
\def\qjras{QJRAS}%
\def\rmxaa{Rev. Mexicana Astron. Astrofis.}%
\def\skytel{S\&T}%
\def\solphys{Sol.~Phys.}%
\def\sovast{Soviet~Ast.}%
\def\ssr{Space~Sci.~Rev.}%
\def\zap{ZAp}%
\def\nat{Nature}%
\def\iaucirc{IAU~Circ.}%
\def\aplett{Astrophys.~Lett.}%
\def\apspr{Astrophys.~Space~Phys.~Res.}%
\def\bain{Bull.~Astron.~Inst.~Netherlands}%
\def\fcp{Fund.~Cosmic~Phys.}%
\def\gca{Geochim.~Cosmochim.~Acta}%
\def\grl{Geophys.~Res.~Lett.}%
\def\jcp{J.~Chem.~Phys.}%
\def\jgr{J.~Geophys.~Res.}%
\def\jqsrt{J.~Quant.~Spec.~Radiat.~Transf.}%
\def\memsai{Mem.~Soc.~Astron.~Italiana}%
\def\nphysa{Nucl.~Phys.~A}%
\def\physrep{Phys.~Rep.}%
\def\physscr{Phys.~Scr}%
\def\planss{Planet.~Space~Sci.}%
\def\procspie{Proc.~SPIE}%
\let\astap=\aap
\let\apjlett=\apjl
\let\apjsupp=\apjs
\let\applopt=\ao

\hypersetup{pdftitle={\ttitle}}
\hypersetup{pdfsubject=\subjectname}
\hypersetup{pdfauthor=\authornames}
\hypersetup{pdfkeywords=\keywordnames}


\thispagestyle{empty}

\vspace*{-3cm}
\begin{figure}[htb!]
        \hspace{-0.4cm}\begin{minipage}[t]{5cm}
        \begin{flushleft}
					\includegraphics[height=2cm]{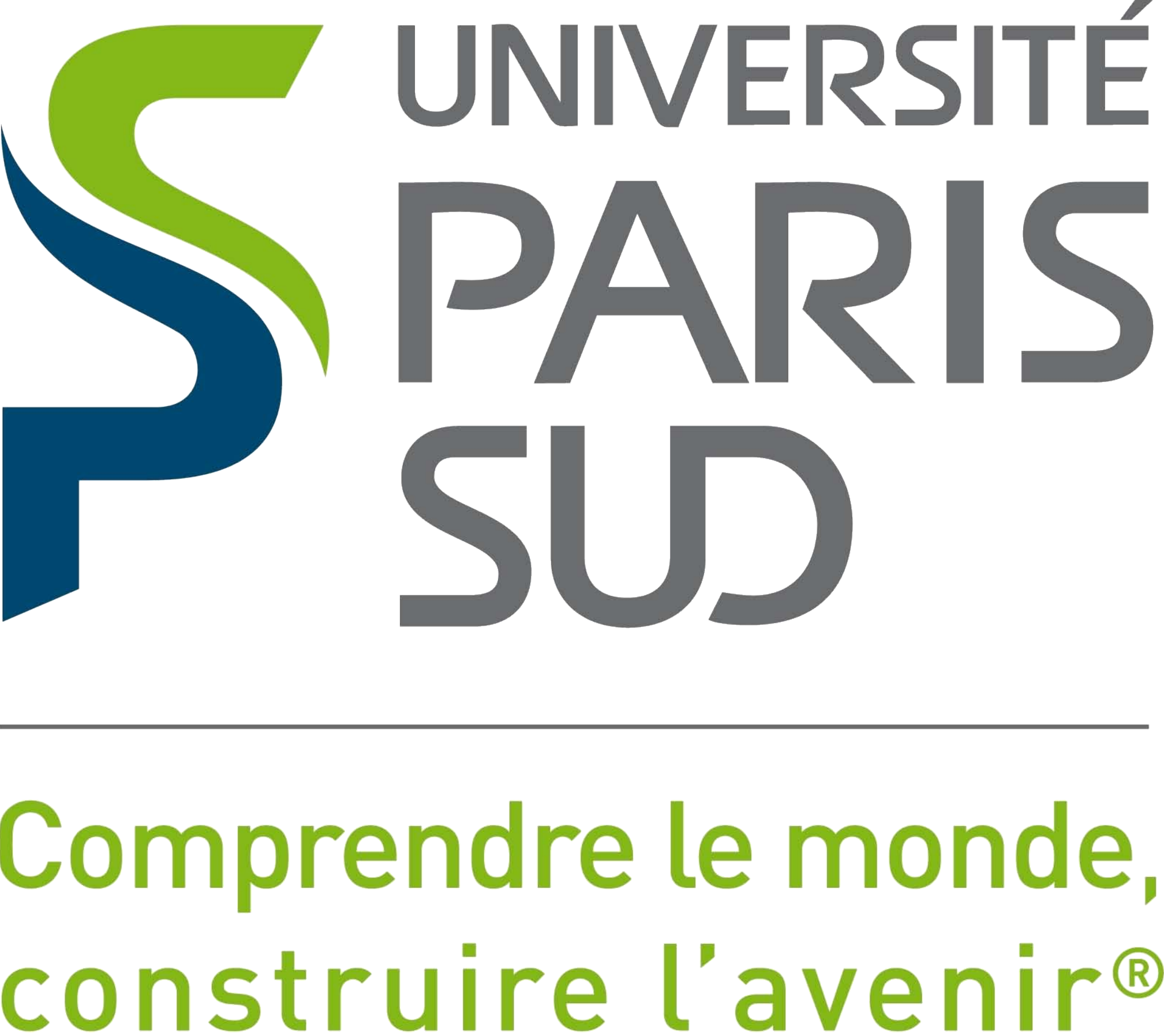}
				\end{flushleft}
				\end{minipage}
        \hspace{-1cm}\begin{minipage}[t]{5cm}
        \begin{flushright}
          \includegraphics[height=2.15cm]{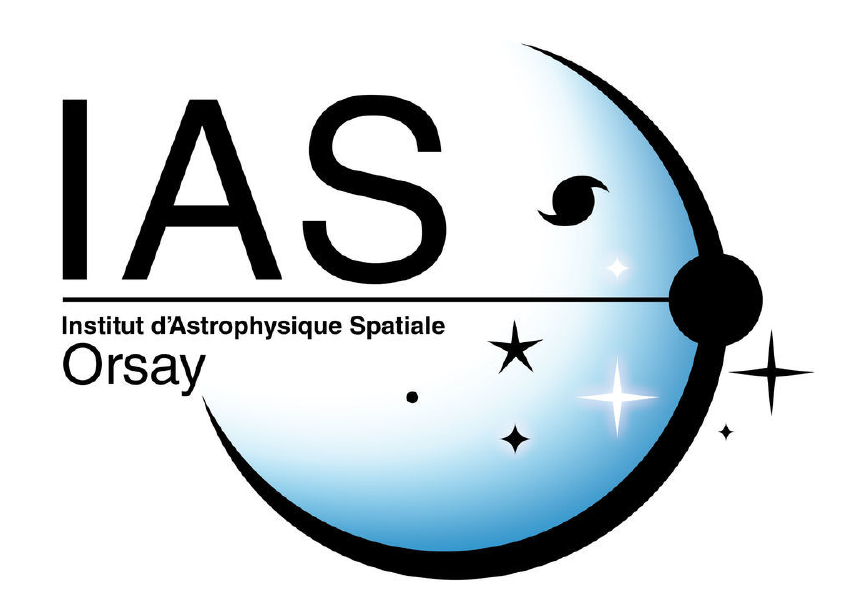} 
        \end{flushright}
        \end{minipage}
        \hspace{1cm}
            \begin{minipage}[t]{5cm}
        \begin{flushright}
          \includegraphics[height=2cm]{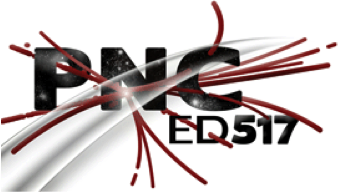} 
        \end{flushright}
        \end{minipage}
\end{figure}

\vspace{1cm}
\begin{center} 
\begin{tabular}{c}
		\\
    {\LARGE \textsc{Universit\'e Paris-Sud}} \\
		\\
    {\LARGE \textsc{\'Ecole doctorale 517 :}} \\
    {\LARGE \textsc{particules, noyaux et cosmos}}\\
    \\
		{\LARGE \textsc{Institut d'Astrophysique Spatiale}} \\
		\\
		{\LARGE \textsc{Discipline : Physique}} \\
\\
\\
    \huge \textsc{Th\`ese de doctorat}\\
\\
{\Large Soutenue le 26 septembre 2014 par}\\[0.8cm]
    \Huge{{{Agn\`es Fert\'e}}}\\
\\
\\
\\
\\
\hspace{-1cm} 
\begin{minipage}{1.1\textwidth}
\begin{center}
  \hspace{1cm} \Huge\bf{Statistics of the CMB} \\ 
     \hspace{1cm}\Huge\bf{Polarised Anisotropies}\\
      \hspace{1cm}\LARGE\bf{Unveiling the Primordial Universe}
\end{center}
\end{minipage}
\end{tabular}
\end{center}

\vspace{1.8cm}

\begin{tabular}{p{0cm} p{3.5cm} p{5.4cm} l }
& \bf\underline{Composition du jury :}& &\\ \\
	& Directeur de th\`ese : & Julien Grain & Charg\'e de recherche (IAS)\\
	& Pr\'esident du jury : & Ken Ganga  & Directeur de recherche (APC) \\
	& Rapporteurs : & Jean-Christophe Hamilton & Directeur de recherche (APC) \\
  &	& Julien Lesgourgues & Chargé de recherche (CERN) \\
  & Examinateurs : & Martin Kunz & Professeur (ITP) \\
  &	& Jean-Loup Puget  & Directeur de recherche (IAS)  \\
\end{tabular}

\newpage

\addtotoc{Abstract} 

\abstract{\addtocontents{toc}{\vspace{1em}} 

A deep understanding of the first instants of the Universe would not only complete our description of the cosmic history but also enable an exploration of new fundamental phsyics at energy scales unexplored on Earth laboratories and colliders. The most favoured scenario which describes these first instants is the cosmic inflation, an ephemeral period of accelerated expansion shortly after the big bang. Some hints are in favour of this scenario which is however still waiting for a smoking-gun observational signature. The cosmic microwave background (CMB) $B$ modes would be generated at large angular scales by primordial gravitational waves produced during the cosmic inflation. In this frame, the primordial CMB $B$-modes are the aim of various ongoing or being-deployed experiments, as well as being-planned satellite mission. However, unavoidable instrumental and astrophysical features makes its detection difficult. More specifically, a partial sky coverage of the CMB polarisation (inherent to any CMB measurements) leads to the $E$-to-$B$ leakage, a major issue on the estimation of the CMB $B$ modes power spectrum. This effect can prevent from a detection of the primordial $B$ modes even if the polarisation maps are perfectly cleaned, since the (much more intense) leaked E-modes mask the B-modes. Various methods have been proposed in the literature offering a $B$ modes estimation theoretically free from any leakage. However, when applied to real data, they are no longer completely leakage-free and remove part of the information on B-modes. These methods consequently need to be validate in the frame of real data analysis. In this purpose, I have worked on the implementation and numerical developments of three typical pseudospectrum methods. Afterwards, I have tested each of them in the case of two fiducial experimental set ups, typical of current balloon-borne or ground based experiments and of potential satellite mission. I have therefore stated on the efficiency and necessity of one of them: the so-called pure method. I have also shown that the case of nearly full sky coverage is not trivial because of the intricate shape of the contours of the point-sources and galactic mask. As a result this method is also required for an optimal $B$ modes pseudospectrum estimation in the context of a satellite mission. 

With this powerful method, I performed realistic forecasts on the constraints that a CMB polarisation detection could set on the physics of the primordial universe. First of all, I have studied the detectability of the tensor-to-scalar ratio $r$, amounting the amplitude of primordial gravity waves and directly related to the energy scale of inflation, in the case of current suborbital experiments, a potential array of telescopes and a potential satellite mission. I have shown that a satellite-like experiment dedicated to the CMB polarisation detection will enable us to measure a tensor-to-scalar ratio of about 0.001, thus allowing for distinguishing between large and small field models of inflation. Moreover, in extension of the standard model of cosmology, the CMB $EB$ and $TB$ correlations can be generated. In particular, I have forecast the constraints that one could set on a parity violation in the gravitational waves during the primordial universe from observations on a small and a large part of the sky. Our results have shown that a satellite-like experiment is mandatory to set constraints on a range of parity violation models. I finally address the problematic of the detectability of observational signature of a primordial magnetic field.

\addtotoc{R\'esum\'e}

\begin{center}
    {\huge{\textit{R\'esum\'e}} \par}
  \end{center}

La compréhension des premiers instants de notre Univers complèterait notre description de son histoire et permettrait également une exploration de la physique fondamentale à des échelles d'énergie jusque là inatteignables. L'inflation cosmique est le scénario privilégié pour décrire ces premiers instants car il s'intègre très bien dans le modèle standard de la cosmologie. Selon ce scénario l'Univers aurait connu une courte période d'expansion accélérée peu après le Big Bang. Quelques indices favorisent ce modèle cependant toujours en attente d'une signature observationnelle décisive. Les modes $B$ du fond diffus comologique (FDC) aux grandes échelles angulaires sont générés par les ondes gravitationnelles primordiales, produites durant l'inflation cosmique. Dans ce cadre, la détection des modes $B$ primordiaux est le but de nombreuses expériences, actuelles ou à venir. Cependant, des effets astrophysiques et instrumentaux rendent sa détection difficile. Plus précisément, une couverture incomplète de la polarisation du FDC (inhérente à toute observation du FDC) entraine la fuite des modes $E$ dans $B$, un problème majeur dans l'estimation des modes $B$. Cet effet peut empêcher une détection des modes $B$ même à partir de cartes parfaitement nettoyées, car les modes $E$ fuyant (beaucoup plus intenses) masquent les modes $B$. Diverses méthodes offrant une estimation de modes $B$ théoriquement non affectés par cette fuite, ont été récemment proposées dans la littérature. Cependant, lorsqu'elles sont appliquées à des expériences réalistes, elles ne corrigent plus exactement cette fuite. Ces méthodes doivent donc être validées dans le cadre d'expériences réalistes. Dans ce but, j'ai travaillé sur l'implémentation et le développement numérique de trois méthodes typiques de pseudospectres. Ensuite, je les ai testé dans le cas de deux expériences fiducielles, typiques d'une expérience suborbitale et d'une potentielle mission satellite. J'ai alors montré l'efficacité et la nécessité d'une méthode en particulier: la méthode dite pure. J'ai également montré que le cas d'une couverture quasi complète du ciel n'est pas trivial, à cause des contours compliqués du masque galactique et des points sources. Par conséquent, une estimation optimale de pseudospectre des modes B exige l'utilisation d'une telle méthode également dans le contexte d'une mission satellite. 

Grâce à cette méthode, j'ai fait des prévisions réalistes sur les contraintes qu'une détection de la polarisation du FDC pourra apporter sur la physique de l'Univers primordial. J'ai tout d'abord étudié la détectabilité du rapport tenseur-sur-scalaire $r$ qui quantifie l'amplitude des ondes gravitationnelles primordiales, directement relié à l'échelle d'énergie de l'inflation, dans le cas de différentes expériences dédiées à la détection de la polarisation du FDC. J'ai montré qu'une mission satellite nous permettrait de mesurer un rapport tenseur-sur-scalaire de l'ordre de 0.001, autorisant une distinction entre les modèles d'inflation à champ fort et faible. De plus, dans le cas d'une extension du modèle standard de la cosmologie, des corrélations $EB$ et $TB$ du FDC peuvent être générées. En particulier, j'ai prévu les contraintes que nous pourrons mettre sur une violation de parité durant l'univers primordial à partir d'observations sur une grande ou une petite partie du ciel. Mes résultats ont montré qu'une expérience satellite est nécessaire pour mettre des contraintes sur une gamme de modèles de violation de parité. J'ai finalement abordé la problématique de la détectabilité d'une signature observationnelle d'un champ magnétique primordial.

}

\clearpage 

%
\setstretch{1.3} 
\acknowledgements{\addtocontents{toc}{\vspace{1em}} 

Elle est un peu difficile à écrire cette partie. D'abord car c'est important pour moi de remercier les personnes qui ont compté, ensuite parce qu'il parait qu'on ne lit généralement que cette partie, ça fout la pression. Mais bon, je suis encore à la bourre pour rendre le manuscrit alors voilà : 

Tout d'abord, \textbf{nonante mercis} à Jean-Christophe Hamilton et Julien Lesgourgues d'avoir accepter de lire, commenter, questionner et reporter ce manuscrit. \textbf{Merci beaucoup} aussi à Ken Ganga, Martin Kunz et Jean-Loup Puget (bien qu'absent) pour vos commentaires et questions intéressantes. Merci à vous d'être sympathiques et de bonne humeur, rendant la soutenance légèrement moins stressante. 

Travailler avec des personnes scientifiquement excellentes m'a énormément stimulé, \textbf{merci} à Julien Peloton (je te souhaite une \emph{pure} fin de thèse!), Radek Stompor (entre autres pour avoir pointé des choses à étudier qui ont doublé mon temps de travail sur le sujet) et Matthieu Tristram.

Ca fait quelques temps que je trainasse à Orsay. \textbf{Merci} à mes profs motivants et aux encadrants de mes premiers stages de recherche. \textbf{Merci} à Hervé pour (G)ALCOR, de m'avoir dit 'tiens, tu devrais aller avoir Julien Grain \& co à l'IAS, ils ont peut être une proposition de thèse' un jour de mars 2011, et pour ton entrain. \textbf{D\`iky moc} à Mathieu, son bureau (lieu de craquage, s'il en est) et JB pour les geek nights, les mots fléchés, votre curiosité et vision différente.
Orsay, c'est le magistère et NPAC où j'ai rencontré des personnes sans qui les études et la thèse auraient été différentes (et moins bien pour sûr): \textbf{tellement merci} à Gé$^2$ et ses pestougnèses défiant la gravité, Schmi et son petit macaquon (bon courage à vous pour la fin de thèse!), et puis Matthieu, Benjamin, Pierre, (I am not a...) Tico, Guillaume, Julian, Flavien qui ont trop participé à l'élévation de mon alcoolémie, et enfin Vincent, Guigui, Asénath, Estelle, Samuel Franco, Jérémy, Marie, Marco mon binome, ses pastas et les repas NPAC. On s'est bien marré quand même, bon vent pour vos chemins respectifs. 

Je suis contente d'avoir fait mes débuts en recherche à l'Insitut d'Astrophysique Spatiale, un laboratoire à la fois travailleur et vivant. \textbf{Enorme merci} à tout le laboratoire, du personnel administratif jusqu'aux doctorants, pour tout dont les petit mots, votre générosité et cette soirée. \textbf{Merci} à l'équipe MIC et en particulier au groupe cosmologie dont Nabila et Marian, pour vos explications, votre curiosité et garder votre sympathie malgré la pression. \textbf{Héxa merci} à {\color{magenta}Véronique}, Sébitouf \& Heddy (heureusement que vous êtes partis sinon j'aurais fait une thèse en QPUC/TLMVPSP/blagues salaces (d'ailleurs...)), 2Fab, Aurélie, cobureau Cédricounet, Guillaume H. pour les pauses scientifiques (et les stupides aussi), Antoine, c'était cool d'être ta suppléante: merci pour ta volonté de changer les choses (j'aime), et puis à tous les thésard-es et post-docs d'hier et d'aujourd'hui, pour les marrades et réflexions. 

Aussi, j'ai fait des rencontres enrichissantes (Marta pour n'en citer qu'une) au gré des nouvelles expériences, des voyages (comme les écoles d'été, les conférences, les interventions auprès du grand public). En particulier, \textbf{merci} aux personnes avec qui nous avons organisé la conférence Elbereth, \textbf{merci} @AstroLR Loïc pour m'avoir convaincue de twitter la science: ça m'a permis quelques actions et rencontres intéressantes (radio thésards, conscience dont je remercie le soutien, ...)

\textbf{Merci infiniment} à toute ma famille (dont mes parents, mes soeurettes d'amour, et puis mon frère pour avoir déclenché ma passion pour la science et avoir rendu mon parcours si normal) qui s'en fichent que je veuille faire de la recherche scientifique du moment que ça me plait. \textbf{Merci} à Clémence GB, d'être toujours là par delà les kilomètres.

\textbf{Cimer à donf} à Faustine, d'avoir été là pendant l'accouchement de ce manuscrit (merci tellement de m'avoir relue). \textbf{Merci} aussi à ta maman pour le mignon mouton birefracté ! Belle thèse à toi, je suis sure que tu iras très loin (mais pas trop de moi j'espère). Du fond du battant, \textbf{le grand merci} à Cécile, mon indispensable bouteille d'oxygène. \textbf{Au delà du merci} et GG à Stéphane pour avoir tout supporté. Vous êtiez là pendant mes doutes, mes échecs et pour les chouettes choses: j'espère vous rendre la pareille. Tout ça, c'est aussi pour vous. 

Et enfin, \textbf{énorme merci inflationnaire} à Julien pour cette thèse. D'abord pour ce sujet si passionnant, ensuite pour m'avoir encadré et passé tant de temps à m'expliquer l'Univers, puis pour m'avoir laissé libre de faire ce que je voulais et enfin pour m'avoir soutenue. Si je devais te remercier en pintes de stout, tu en aurais pour la vie. C'est vraiment chouette de bosser avec toi, j'espère que ça continuera et qu'on trouvera plein de choses cools. 
\\
\\
}
\clearpage 


\pagestyle{fancy} 

\lhead{\emph{Contents}} 
\dominitoc
\tableofcontents 

\lhead{\emph{List of Figures}} 
\listoffigures 

\lhead{\emph{List of Tables}} 
\listoftables 
%
 \frontmatter

\chapter{Introduction} 
\label{Chapter0} 
\lhead{\textit{Introduction}} 

Among all electromagnetic radiations surrounding us, there exists one particular light that originates from the first moments of the Universe, today constituting the \textit{cosmic microwave background} (CMB). This light is composed of the oldest photons emitted in the Universe and is a valuable source of knowledge since it holds information about the formation and evolution of the Universe from the Big Bang up to now. Shortly after its discovery in 1964, the CMB was at the origin of many questions for instance, the origin of its statistical isotropy along with its tiny temperature fluctuations remain a mystery. Several theories have been proposed to solve these enigma but one is favoured for its simple explanation of most observed phenomena: the \textit{cosmic inflation}. According to this paradigm, the Universe would have known a violent and ephemeral accelerated expansion about $10^{-30} s$ after the Big Bang. Its existence would assert the consistency of the standard model of cosmology and is only waiting for a firm experimental verification. It might seem impossible at first to observe such a distant epoch but cosmic inflation would have left perturbations of the space-time curvature, first under the shape of variations of the gravitational potential, but also under the shape of \textit{primordial gravitational waves}. These are ripples of space-time which would have left imprints in the fluctuations of the CMB temperature and polarisation. By investigating its temperature and polarisation anisotropies, the CMB may hold the answers to the issues it raised.

\begin{figure}[!h]
\begin{center}
	\includegraphics[scale=0.25]{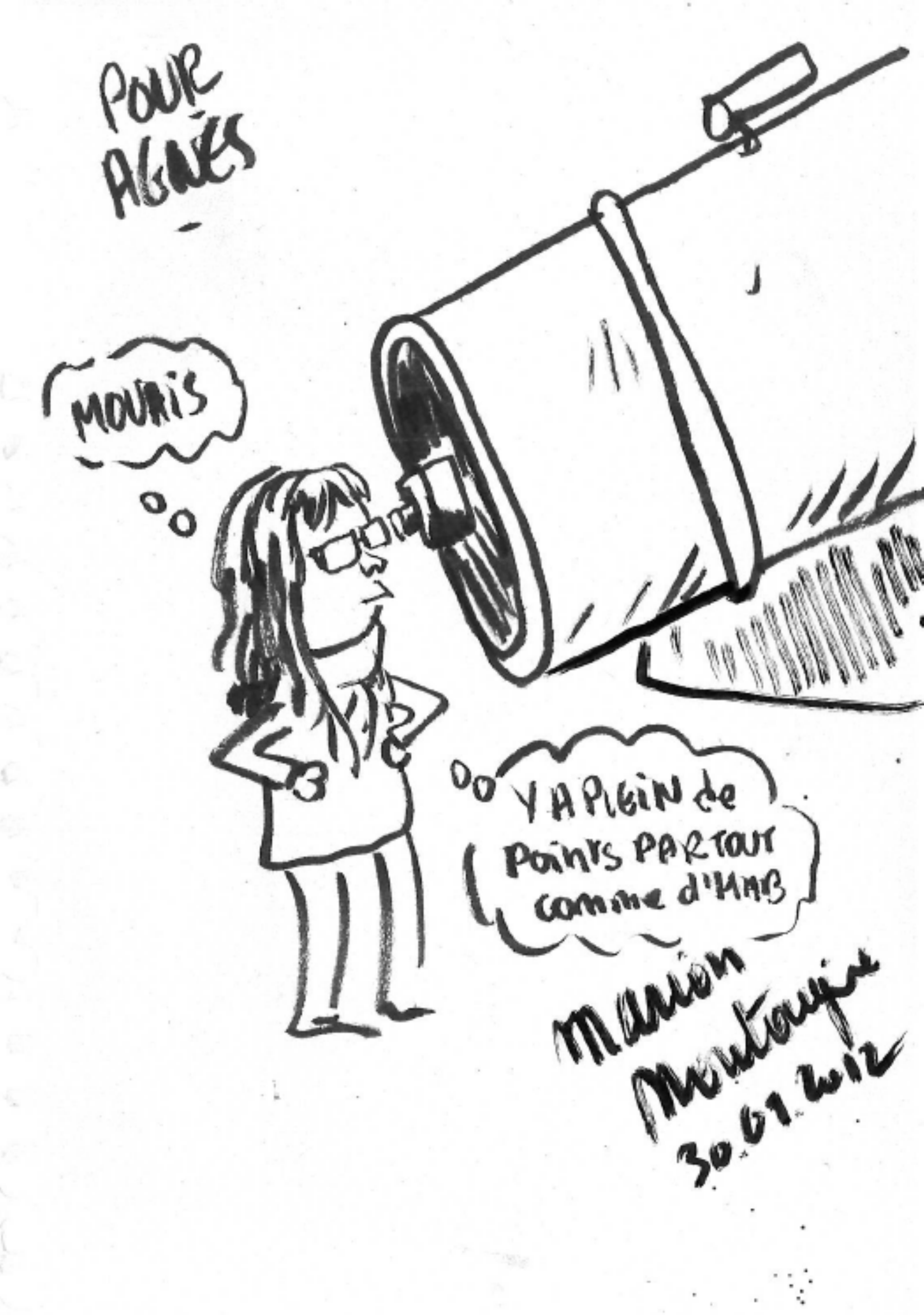}
	\caption{During my PhD, I have focused my research on the CMB, a light much older than the light of any stars in the sky (picture by Marion Montaigne\protect\footnotemark[1])}
	\label{fig:cltbeb}
\end{center}
\end{figure}
\setcounter{footnote}{1}
\footnotetext{\url{tumourrasmoinsbete.blogspot.com}}

However, the evidence of the primordial gravitational waves is so tiny that it can be hidden by the other sources of fluctuations in the CMB, except for one of the CMB polarisation modes: the $B$ modes. The gravitational waves are indeed the only source of CMB $B$ modes at large angular scales. The detection of the CMB $B$ modes would therefore be a smoking gun for cosmic inflation. Their amplitude is however expected to be extremely low making their detection an instrumental and data processing challenge. 

During my PhD thesis, my endeavour has consisted in developing, testing and validating new statistical methods to perform an accurate estimation of the CMB $B$ modes in order to set thigh constraints on the cosmological parameters describing the very early universe. Due to its cosmological origin, the CMB is a peculiar observable since the signal itself contributes to the statistical uncertainties on its reconstruction, this is the so-called cosmic variance. This effect becomes more puzzling in the case of a low signal such as the CMB polarisation because the $E$ modes can contribute to the $B$ modes signal. This effect is known as the \textit{$E$-to-$B$ leakage} and leads to a laborious, if not impossible, reconstruction of the $B$ modes. The leakage can strongly prevent from detecting the $B$-modes, even in an ideal \textit{noiseless} case, because of the partial sky coverage. These uncertainties can however be statistically reduced thanks to clever methods correcting for this contamination. In that case, the CMB polarisation unveils clues about the first instants of the Universe, in particular about the inflation period.

As an introduction, I will first detail in part~\ref{part1} the formalism of polarisation, a key observable in astrophysics. I will then portray the CMB in the frame of the current standard model of cosmology, the $\Lambda$CDM model. The CMB polarisation and its current detection status will be developed next. Afterwards, the part~\ref{part2} is dedicated to the reconstruction of the CMB polarisation power spectrum from the CMB maps. After introducing the formalism of the CMB statistics, I will demonstrate the efficiency of different pseudospectrum methods aiming at estimating accurately the uncertainties on the CMB power spectra. As a result, one of them proves to be the most efficient, so I use it as a tool to properly estimate the CMB $B$ modes and derive the constraints on the primordial Universe (part~\ref{part3}). In particular, I investigated the potential detection of the primordial Universe physics: the energy scale of the cosmic inflation, a parity violation and finally the existence of a primordial magnetic field, each of these issues being dealt with in separated chapters. I eventually conclude in part~\ref{part4} by summarizing the results of my PhD research and outlining my future projects.

\adjustmtc[+1]

 \mainmatter 

\pagestyle{fancy} 


\part{Introduction: Light Polarisation and the CMB}
\label{part1}

\chapter{Light Polarisation} 

\label{Chapter1} 

\lhead{Chapter 1. \textit{Light Polarisation}} 


\noindent \hrulefill \\
\textit{During the $19^{th}$ century, Malus and Arago have highlighted the light polarisation, a curious behaviour of light that they observed in peculiar crystals such as calcite. Ever since, the light polarisation is widely used for technological purposes such as 3D vision or remote sensing. It also explains numerous natural processes like the sky polarisation or the birefringence.}
\noindent \hrulefill \\


\section{Polarisation Formalism in Optics}\label{polFor}

The light is classically described as a propagating electromagnetic wave. It is often reduced to its electric field component, from which the magnetic field is deduced thanks to the Maxwell equations. I present here the light description in the ideal but easier case of a monochromatic plane wave with a frequency $\nu$ and a direction of propagation along its wave number $\vec{k}$. In a $(\vec{e_x},\vec{e_y},\vec{e_z})$ orthonormal system coordinate, the electric field representing such a wave propagating in the $z$ direction is: 
\begin{equation}
\vec{E}(\vec{r},t) = [E_x\vec{e_x}+E_y\vec{e_y}]e^{i(kz - 2\pi\nu t)},
\label{eq:light}
\end{equation}
with $E_i$ the complex amplitude of the electric field in the $i$-direction. In order to characterise this electromagnetic wave, the two main quantities to be measured are its intensity and its polarisation. 

$\bullet$ \underline{\textbf{Intensity}}

Measuring the electric field components at any moment would give a complete description of the electromagnetic wave. Nonetheless, in the case of high frequency light (in the order of hundreds of GHz) that are of interest to this thesis, the current detectors do not have high enough sampling frequency to have access to the electric field every fraction of microseconds. Nonetheless, the energy on a given surface during a given time lapse gives enough information to describe the wave. This measurable energy is the light \textit{intensity} and is therefore defined as:  
\begin{equation}
I(t) = \left<|E_x|^2\right>_t + \left<|E_y|^2\right>_t,
\label{eq:1.2}
\end{equation}
where the brackets $\left<.\right>_t$ stand for time average. 

The light intensity $I(t)$ is an essential quantity that permits to characterise most optical phenomenons such as photometry, interferometry or polarimetry. It however eludes the light vectorial behaviour. 

$\bullet$ \underline{\textbf{Polarisation}} 

\begin{figure}[h]
\begin{center}
	\includegraphics[scale=0.05]{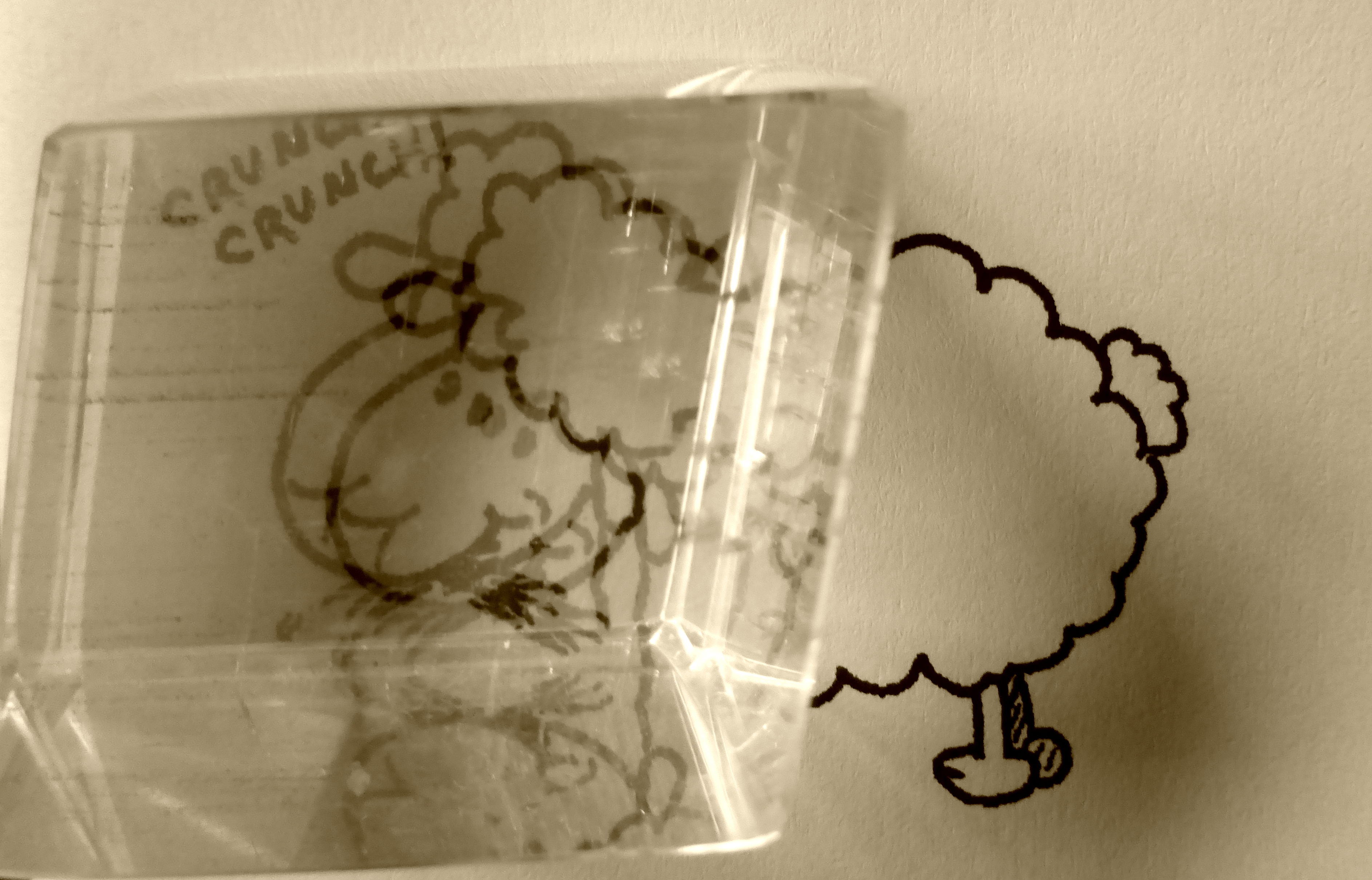}
	\caption{An example of birefringence: the drawing of the sheep is split into two parts when looked through the birefringent crystal.}
	\label{fig:mouton}
\end{center}
\end{figure}

The polarisation of the light holds the information on the vectorial structure of the electromagnetic field. More specifically, it defines the direction of its oscillations. Indeed, regarding the process at the origin of the light emission or in the way of the light path, the electromagnetic field can adopt a preferential direction with respect to the direction of propagation. A non-polarized light, such as the natural light, has its electric and magnetic fields varying too fast with respect to the sampling frequency of our best detectors and in an unpredictable direction. Nonetheless, there are processes that favour a peculiar direction of these electromagnetic oscillations. For instance, any kind of light outgoing from a linear polariser acquires one specific direction of the electric field oscillations given by the device which acts as a filter. The light polarisation may therefore give consequential information on the phenomenon from which it originates or on the medium in which the light is propagating. An interesting example which is only explained thanks to the polarisation is the \textit{birefringence}. This peculiar effect is due to anisotropies in the atomic distribution of some medium such as crystal quartz which has two favoured directions. This irregular arrangement implies local variations of the optical index. Consequently the different polarisation directions of the propagating light see different optical index of the medium. The different polarisations of the light thus refract in different directions. The light therefore emerges from the medium in two separated light beams perpendicularly polarised to each other as shown in Fig.~\ref{fig:mouton}. The birefringence is therefore tightly related to the light polarisation -- and was at the origin of polarisation discovery.

The light polarisation characterises the direction of the electric field oscillations, a general case of which is displayed in Fig.~\ref{fig:ellipse}. It is thus an intrinsic property of the light ensuring an access to additional information on the nature of the light and the medium the light is going through. A specific formalism is therefore required to describe it.

\begin{figure}[h]
\begin{center}
	\includegraphics[scale=0.3]{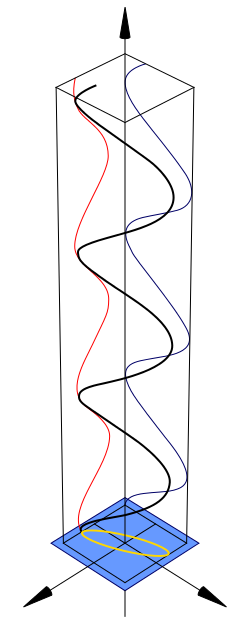}
	\caption{A diagram of a polarised light. The black line stands for the direction of the electric field in time and its projection onto a plane is shown in yellow, drawing an ellipse: the light is elliptically polarised (from \cite{polar}).}
	\label{fig:ellipse}
\end{center}
\end{figure}

\subsection{Fully polarized light}

\subsubsection{Jones formalism}

A specific case of polarisation is the case of a totally polarized light: the electromagnetic field direction evolution is deterministic. Usually, in such a case, the Jones formalism is used to describe the light polarisation. The light can be characterized by a Jones vector while the properties of the medium it propagates in is embodied in the Jones matrix. 

In this case, the equation~(\ref{eq:1.1}) describing the behaviour of the electric field can be written as a vector $\vec{E}$ verifying:
\begin{eqnarray}
\mathfrak{Re}(\vec{E}(\vec{r},t)) =  \mathfrak{Re}\left(\left[ \begin{array}{l} E_{ox} \\ E_{oy} e^{i\phi} \\ \end{array} \right] e^{i(kz - 2\pi\nu t)}\right)  \nonumber \\
= \left[ \begin{array}{l} E_{ox}cos(kz - 2\pi \nu t) \\ E_{oy}cos(\phi + kz - 2\pi\nu t) \\ \end{array} \right],
\label{eq:1.3}
\end{eqnarray}
with $\phi$ the phase between the two transverse components of the electric field and $E_{oi}$ is the real amplitude of the $i$ electric field component.

Following this expression, the Jones vector is defined as:
\begin{equation}
\vec{J} = \left[ \begin{array}{l} J_x \\ J_y \\ \end{array} \right] = \left[ \begin{array}{l} E_{ox} \\ E_{oy} e^{i\phi} \\ \end{array} \right].
\label{eq:1.4}
\end{equation}
 
By reformulating Eq.~(\ref{eq:1.3}) and combinating the two components of the electric field, an equation giving insight on the evolution of each component $E_i$ of the electric field is obtained: 
\begin{equation}
\frac{E_x^2}{E_{ox}^2} + \frac{E_y^2}{E_{oy}^2} - 2\frac{E_x}{E_{ox}}\frac{E_y}{E_{oy}}\cos(\phi) = \sin(\phi)^2.
\label{eq:1.6}
\end{equation}

This equation is the equation of an ellipse inscribed in a rectangle of side width $2E_{ox}$ and $2E_{oy}$, as shown in Fig.\ref{fig:ellipse}. The cross term in $E_xE_y$ indicates that the ellipse is rotated. The polarisation is said to be elliptic: the electric field vector draws an ellipse over the time. The polarisation is either left- or right-handed depending on the sign of the phase shift $\phi$. If $\phi > 0$, the polarisation is then left-handed (trigonometric orientation) and vice versa.   

\begin{figure}[h]
\begin{center}
	\includegraphics[scale=0.5]{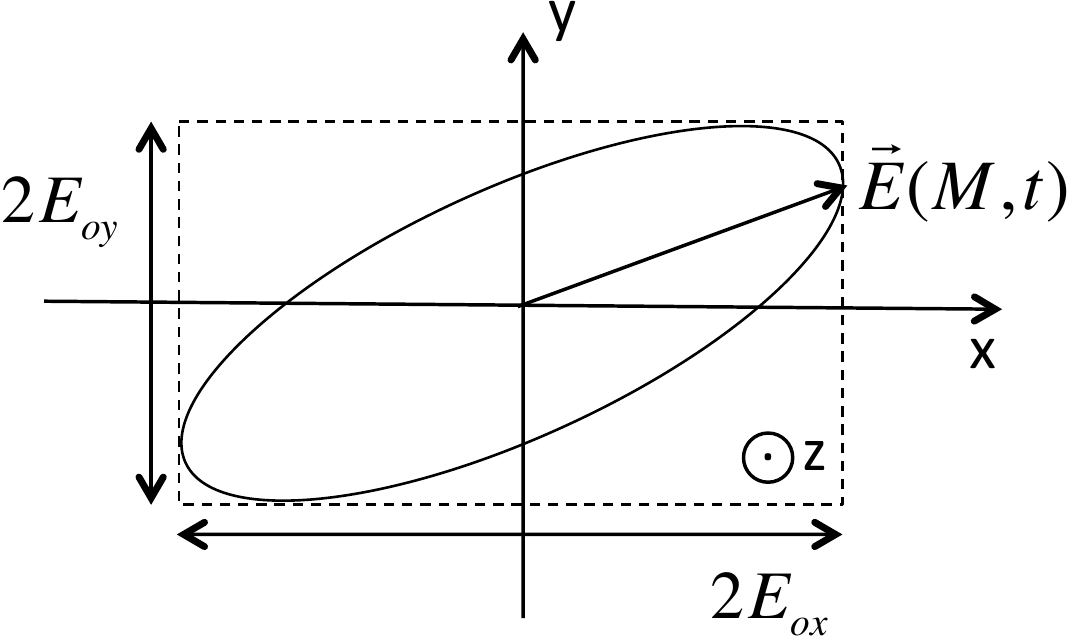}
	\caption{The ellipse drawn by the evolution of the electric field $\vec{E}(M,t)$ in time when the light is elliptically polarised (see Eq.~(\ref{eq:1.6})).}
	\label{fig:ellipse}
\end{center}
\end{figure}

From the general case of an elliptic polarisation, two peculiar cases come out: the linear and circular polarisations. In the former case, the electric field oscillations are contained in a plane and its components verify $\frac{E_x}{E_y} = cste$. The Jones vector of a light linearly polarised along $(Ox)$ is therefore written: $\vec{J} = \left[ \begin{array}{l} 1 \\ 0 \\ \end{array} \right] $. The electric field direction of a circularly polarised light follows a circle in time. The $E_x$ and $E_y$ components of the electric field are therefore equal and the phase is equal to $\pm\frac{\pi}{2}$. Thus, the Jones vector of right-handed circular polarisation is: $\vec{J} =  \frac{1}{\sqrt{2}}\left[ \begin{array}{l} 1 \\ -i \\ \end{array} \right]$ because here $\phi = -\frac{\pi}{2}$. 


\subsubsection{Poincar\'e Sphere}

The Poincar\'e sphere is a tool introduced in \cite{Poincare_1892} to easily describe the polarisation state of the light and its modification. It is very useful since it enables to quickly derive the resulting polarisation state of light after going through a polarising device such as a waveplate.

\begin{figure}[h]
\begin{center}
	\includegraphics[scale=0.5]{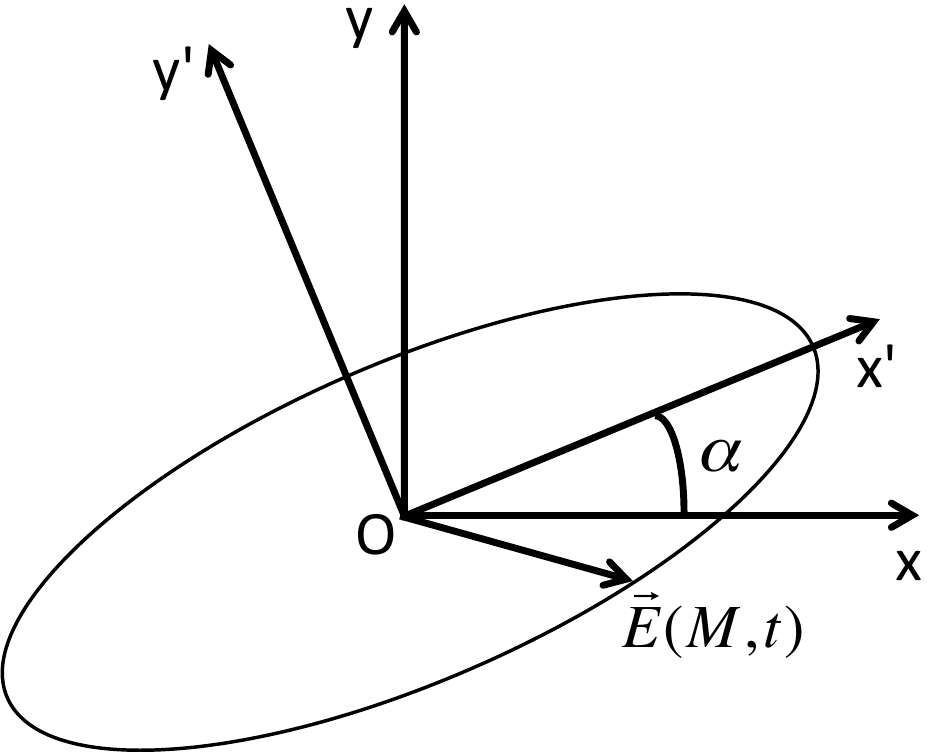}
	\caption{The coordinate change between $(x,y)$ and $(x',y')$, the proper axes of the polarisation ellipse drawn by the electric field $\vec{E}(M,t)$ evolution.}
	\label{fig:coord}
\end{center}
\end{figure}

As shown in figure \ref{fig:coord}, the coordinate system $(Ox'y'z')$ in which the electric field is measured is rotated by an angle $\alpha$ with respect to the proper coordinate system $(Oxyz)$ of the polarisation ellipse, with $z = z'$. Therefore, the electric field transforms as: 
\begin{equation}
\left \{ \begin{array}{l} E_{x'} = E_x\cos(\alpha) + E_y\sin(\alpha), \\ E_{y'} = E_y\cos(\alpha) - E_x\sin(\alpha), \end{array}
\right .
\label{eq:1.7}
\end{equation}
where $E_{x',y'}$ denotes the electric field components defined in the coordinate system $(Ox'y'z')$. 

Thus, the electric field in the $(Oxyz)$ system is:
\begin{equation}
\left \{ \begin{array}{l} E_x = E_{x'}\cos(\alpha) - E_{y'}\sin(\alpha), \\ E_y = E_{x'}\sin(\alpha) + E_{y'}\cos(\alpha). \end{array}
\right .
\label{eq:1.7}
\end{equation}

In $(Ox'y'z')$ the equation~(\ref{eq:1.6}) of the polarisation ellipse consequently writes:
\begin{eqnarray}
E_x'^2 \left( \frac{\cos(\alpha)^2}{E_{ox}^2} + \frac{\sin(\alpha)^2}{E_{oy}^2} - \frac{2\cos(\alpha)\sin(\alpha)\cos(\phi)}{E_{ox}E_{oy}} \right) \nonumber \\
+ E_y'^2 \left( \frac{\sin(\alpha)^2}{E_{ox}^2} + \frac{\cos(\alpha)^2}{E_{oy}^2} - \frac{2\cos(\alpha)\sin(\alpha)\cos(\phi)}{E_{ox}E_{oy}} \right) \nonumber \\
+ 2 \bold{E_x'E_y'}\left(\cos(\alpha)\sin(\alpha)\left(-\frac{1}{E_{ox}^2} + \frac{1}{E_{oy}^2}\right) - \frac{\cos(\alpha)}{E_{oy}E_{ox}}\right)\cos(\phi) = \sin(\alpha).
\label{eq:1.8}
\end{eqnarray}

In this coordinate system $(O'x'y'z')$, the ellipse is not rotated as its proper axes are along $(O'x')$ and $(O'y')$. The coefficient of the cross term in bold letters of Eq.~(\ref{eq:1.8}) must therefore be set equal to zero. We therefore obtain a relation between $\alpha$ and the characteristics of the polarisation ellipse $E_{oi}$ and $\phi$: 
\begin{equation}
\tan(2\alpha) = 2\frac{E_{ox}E_{oy}}{E_{ox}^2-E_{oy}^2}\cos(\phi).
\label{eq:1.9}
\end{equation}

The semi-major axis $a$ and semi-minor axis $b$ of the ellipse write:
\begin{equation}
\left\{ \begin{array}{l}  a^2 = E_{ox}^2\cos{\alpha}^2 + E_{oy}^2\sin{\alpha}^2 + 2E_{ox}E_{oy}\cos{\alpha}\sin{\alpha}\cos{\phi}, \\  
b^2 = E_{ox}^2\sin{\alpha}^2 + E_{oy}^2\cos{\alpha}^2 - 2E_{ox}E_{oy}\cos{\alpha}\sin{\alpha}\cos{\phi}. \end{array} \right.
\label{eq:1.10}
\end{equation}

\begin{figure}[!h]
\begin{center}
	\includegraphics[scale=0.5]{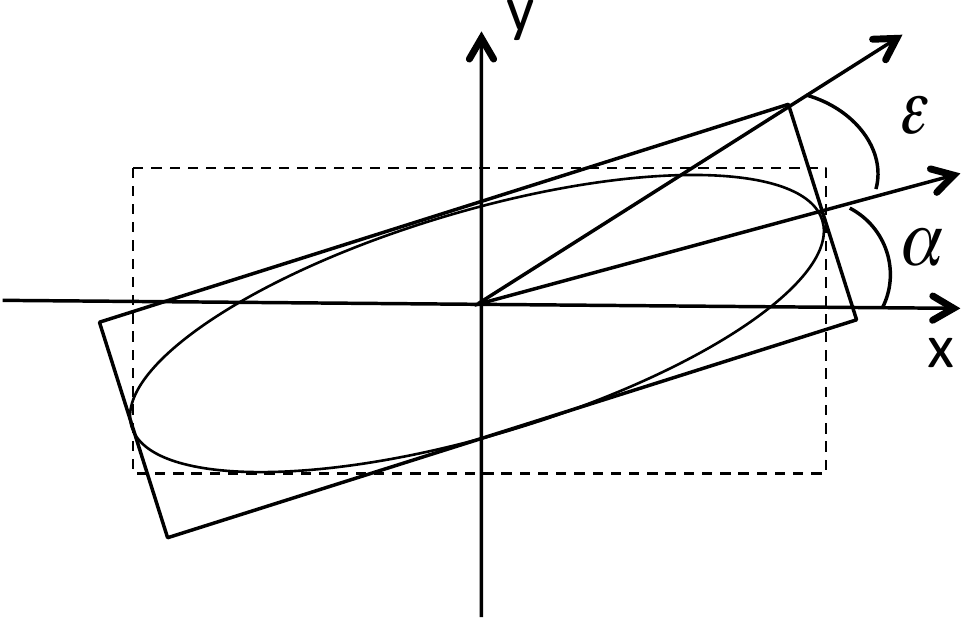}
	\caption{The ellipticity, $\epsilon$, and azimuth, $\alpha$, angles of the polarisation ellipse.}
	\label{fig:ellipse2}
\end{center}
\end{figure}

Instead of using the three previously derived parameters $a$, $b$ and $\alpha$, the ellipse can be described by the three following parameters: \\
- the light beam intensity $I = E_{ox}^2 + E_{oy}^2 = a^2 + b^2$; \\
- the azimuth angle $\alpha$ that characterises the tilt of the polarisation ellipse and varies from 0 to $\pi$; \\
- the ellipticity $\epsilon$ given by: $\tan{\epsilon} = \pm \frac{b}{a}$, representing the width of the ellipse. $\epsilon$ varies from $-\frac{\pi}{4}$ to $\frac{\pi}{4}$. 
Fig.~\ref{fig:ellipse2} illustrate a generic elliptic polarisation showing the $\alpha$ and $\epsilon$ angles. 

These three parameters $I$, $\alpha$ and $\epsilon$ therefore utterly describe the light polarisation. The Poincar\'e sphere is a sphere of centre $O$, unity radius and axis $OX$, $OY$ and $OZ$, where the polarisation state is represented as a point which coordinates depend on the three parameters.

\begin{figure}[!h]
\begin{center}
	\includegraphics[scale=0.6]{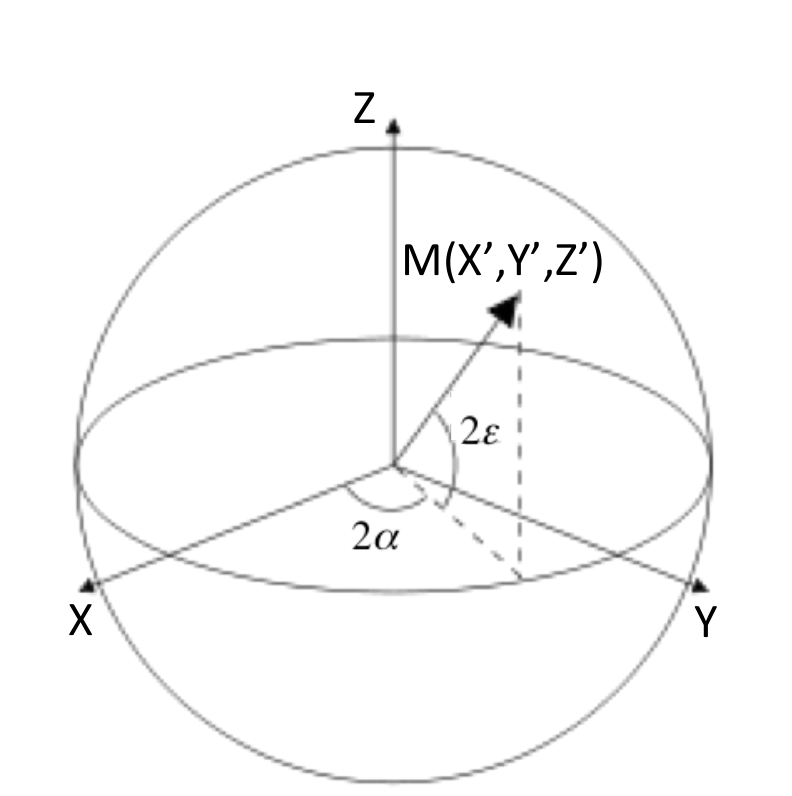}
	\caption{An elliptical polarisation of azimuth $\alpha$ and ellipticity $\epsilon$ placed on the Poincar\'e sphere.}
	\label{fig:ellipse3}
\end{center}
\end{figure}

By convention, the $(OXY)$ plane is the plane with $\epsilon = 0$ and the $(OXZ)$ the one with $\alpha = 0$ as illustrated in Fig.~\ref{fig:ellipse3}. The polarisation state of a totally polarised light with an azimuth $\alpha$ and an ellipticity $\epsilon$ is thus represented by a point M(X',Y',Z') at the surface of the Poincar\'e sphere whose coordinates are defined as:
\begin{equation}
\left\{ \begin{array}{l} X' = \cos{2\alpha} \cos{2\epsilon},\\  
 Y' = \sin{2\alpha}\cos{2\epsilon},\\
 Z' = \sin{2\epsilon}. \end{array} \right.
\label{eq:1.12}
\end{equation}

By convention, the upper hemisphere is the location of the left-handed polarisation, while the bottom is the one of the right-handed polarisation. The two particular cases of circular and linear polarisation evoked above can thus be placed on the Poincar\'e sphere. For a fully polarised beam light with intensity $I$: \\
- a right-handed circularly polarised light is represented by $M(0,0,-1)$; \\
- a light linearly polarised with an angle $\alpha$ is represented by $M(\cos(2\alpha),\sin(2\alpha),0)$. \\

The Jones formalism is a useful formalism to describe a fully polarized light and it can be easily pictured thanks to the Poincar\'e sphere. Nonetheless, it cannot be used to describe a partially polarised light which is however the case of most polarised electromagnetic waves.


\subsection{Partially polarized light}

\subsubsection{Stokes parameters}

Between the two cases of a fully polarised and an unpolarised light lies the case of the partially polarised light that concerns most radiations. In this more realistic case, the Jones vector no longer has a deterministic behaviour and is therefore not appropriate anymore. Proposed by Stokes in 1852, the Stokes parameters are very handy quantities which contain all the needed information about the light to accurately analyse its polarisation. 

The Stokes parameters can be deduced from the so-called \textit{coherence matrix} which encodes for the covariance of the randomly evolving Jones vector. 

However, as done in \cite{Collett_1992}, the Stokes parameters can be easily derived from an empirical perspective. 
The equation~(\ref{eq:1.6}) describing an elliptical polarisation is valid in the case of fully polarised light. The typical time scale involved here -- such as the time for the electric field to draw the ellipse -- is nonetheless in the order of $10^{-15}s$. Thus the polarisation ellipse remains elusive to our detectors. Therefore, in the same way we have introduced the intensity, we may want to transcribe the polarisation ellipse characterised by the $\epsilon$, $\alpha$ variables into observable quantities. 

Using current detectors gives access to the electric field integrated over a time period. We therefore need to integrate in time the equation~(\ref{eq:1.6}) of a monochromatic wave propagating in the $z$-direction with an elliptical polarisation. As the electric field is periodic, it is equivalent to average it in time over one oscillation:
 \begin{equation}
\frac{\left<E_x(t)^2\right>_t}{E_{ox}^2} + \frac{\left<E_y(t)^2\right>_t}{E_{oy}^2} - 2\frac{\left<E_x(t)E_y(t)\right>_t}{E_{ox}E_{oy}}cos(\phi) = \sin(\phi)^2,
\label{eq:1.11}
\end{equation}
with $\left<E_i\right>_t = \lim\limits_{T\rightarrow \infty} \frac{1}{T}\int_0^{\infty} E_i(t) dt$, $i = x,y$. For a monochromatic wave propagating in the $z$-direction, whose expression is given by Eq.~(\ref{eq:1.3}), we obtain: $\left<E_i\right>_t = \frac{1}{2}E_{oi}^2$ and $\left<E_i E_j\right>_t = \frac{1}{2}E_{ox}E_{oy}\cos(\phi)$.

Multiplying by $4E_{ox}E_{oy}$ and inserting the expression of $\left<E_i\right>_t$ in the above equation gives: 
 \begin{equation}
4E_{ox}^2E_{oy}^2 - 4E_{ox}^2E_{oy}^2\cos^2(\phi) = 4E_{ox}^2E_{oy}^2\sin^2(\phi).
\label{eq:1.12}
\end{equation}
We thus obtain the canonical expression:
\begin{equation}
(E_{ox}^2 + E_{oy}^2)^2 - (E_{ox}^2 - E_{oy}^2)^2 - (2E_{ox}E_{oy}\cos(\phi))^2 = (2E_{ox}E_{oy}\sin(\phi))^2.
\label{eq:1.13}
\end{equation}

By rewriting the previous equation~(\ref{eq:1.13}) in the form: $S_0^2 - S_1^2 - S_2^2 = S_3^2$, we deduce:
\begin{equation}
\left \{ \begin{array}{l} S_0 = E_{ox}^2 + E_{oy}^2, \\ S_1 =  E_{ox}^2 - E_{oy}^2, \\ S_2 =  2E_{ox}E_{oy}\cos(\phi), \\ S_3 = 2E_{ox}E_{oy}\sin(\phi). \end{array} \right .
\label{eq:1.14}
\end{equation}

These four quantities are the so-called \textit{Stokes parameters} and fully describe any type of light beam. 
The first Stokes parameter $S_0$ is, compared to Eq.~(\ref{eq:1.2}), the total intensity of the light and is linked to the other parameters via the relation: $S_0^2 \geqslant S_1^2 + S_2^2 + S_3^2$, the equality being the case of a fully polarised light. Also, from these parameters, the polarisation degree $P$ is defined. It quantifies the ratio between the polarised contribution $I_{pol}$ with respect to the total intensity: 
\begin{equation}
P = \frac{I_{pol}}{I_{tot}} = \frac{\sqrt{S_1^2+S_2^2+S_3^2}}{S_0}.
\label{eq:1.15}
\end{equation}
$P$ is consequently ranging from 0 for an unpolarised light to 1 for a totally polarised light. A partially polarised light has its polarisation degree such as $0 < P < 1$.

In the general case of a light beam described with a complex amplitude as in Eq.~(\ref{eq:1.1}), the same reasoning gives: 
\begin{equation}
\left \{ \begin{array}{l} S_0 = E_xE_x^* + E_yE_y^*, \\ S_1 =  E_xE_x^* - E_yE_y^*, \\ S_2 =  E_xE_y^* + E_yE_x^*,  \\ S_3 = i(E_xE_y^* - E_yE_x^*).\end{array} \right .
\label{eq:1.16}
\end{equation}

From these parameters, the so-called \textit{Stokes vector} are built: 
\begin{equation}
S = \left( \begin{array}{l} S_0 \\ S_1 \\ S_2 \\ S_3 \end{array} \right).
\label{eq:1.17}
\end{equation}

The Stokes parameters are also denoted $Q$, $U$ and $V$, corresponding to $S_1$, $S_2$ and $S_3$ respectively.

On the one hand, the $(Q,U)$ Stokes parameters describe the linear polarisation and are tightly coupled because $U$ is equivalent to the $Q$ quantity but defined in a coordinate system rotated by $45^o$. On the other hand, the $V$ Stokes parameter characterises the circular polarisation. Thus, for three peculiar examples of light with intensity $I$ and polarisation degree $P$, the Stokes vectors give: \\
- $S = I\left( 1 ~0 ~0  ~0 \right)^T$  if unpolarised; \\
- $S = I\left( 1 ~0 ~0  ~-P \right)^T$ if right-handed circularly polarised; \\
- $S = I\left( 1 ~P ~0  ~0 \right)^T$ if linearly polarised along $(Ox)$; \\
where ${}^{T}$ denotes the transpose operation.

Any polarisation state can be described by the linear combination of a circularly polarised light and a linerly polarised light. The $(Q,U,V)$ Stokes parameters thus ensure an utter description of the light polarisation. The Jones formalism is used in the peculiar case where $P = 1$.   

\subsubsection{Poincar\'e sphere} 

The reduced Stokes vector is defined as: 
\begin{equation}
\vec{s} = \left( \begin{array}{l} S_1 \\ S_2 \\ S_3 \end{array} \right).
\label{eq:1.17}
\end{equation}

The norm of this vector $\vec{s}$ is: 
\begin{equation}
|\vec{s}| = \sqrt{S_1^2 + S_2^2 + S_3^2}  = PI_{tot},
\label{eq:1.18}
\end{equation}
by definition of the polarisation degree $P$.

The expression of the azimuth parameter $\alpha$ in equation~(\ref{eq:1.9}) and the properties of the semi-axis $a$ and $b$ give a relation between the Stokes parameters and the characteristics of the polarisation ellipse: 
\begin{equation}
 \left\{ \begin{array}{lll} \tan(2\alpha) & = & \frac{S_2}{S_1}, \\ \sin(2\epsilon) & = & \frac{S_3}{S_0}. \end{array} \right . 
\label{eq:1.19}
\end{equation}

From these relations, we deduce the following expressions for the reduced Stokes vector: 
\begin{equation}
 s = \left( \begin{array}{c} Q \\ U \\ V \end{array} \right ) =  PI_{tot} \left( \begin{array}{c} \cos(2\alpha)\cos(2\epsilon) \\ \sin(2\alpha)\cos(2\epsilon) \\ \sin(2\epsilon) \end{array} \right ).
\label{eq:1.20}
\end{equation}

From this reduced Stokes vector, it is noticeable that $\vec{s}$ sets the location of the polarisation state inside the Poincar\'e sphere at a distance $P$ from the centre. Moreover, it is the generalisation of Eq.~(\ref{eq:1.12}) for any polarisation degree.    

For instance a partially linearly polarised light with an azimuth angle $\alpha$ and a polarisation degree $P$ will be a point $M$ inside the Poincar\'e sphere whose coordinates are: $M(P\cos{2\alpha},P\sin(\alpha),0)$. In the specific case of an unpolarised light, $M$ is located at the center of the sphere ($P = 0$).

 
\subsubsection{Stokes parameters properties}

By measuring a polarised light beam, we implicitly choose a reference coordinate system $(O\vec{e_x}\vec{e_y}\vec{e_z})$ where the intensity parallel or orthogonal to $(Ox)$ is measured, in order to measure the Stokes parameters. Now we define a coordinate system $(O\vec{e_{x'}}\vec{e_{y'}}\vec{e_{z'}})$ such as: 
\begin{equation}
 \left\{ \begin{array}{rcl} \vec{e_{x'}} & = & \cos(\Psi)\vec{e_x} + \sin(\Psi)\vec{e_y}, \\ \vec{e_{y'}} & = & - \sin(\Psi)\vec{e_x} + \cos(\Psi)\vec{e_y}, \\ \vec{e_{z'}} & = & \vec{e_z}, \end{array} \right .
\label{eq:1.21}
\end{equation}
with $\Psi$ the rotation angle between the old coordinate system and the new one. 

The electric field in the rotated frame is then:
\begin{equation}
 \left\{ \begin{array}{rcl} E_{0'x} & = & \cos(\Psi)E_{0x} + \sin(\Psi)E_{0y}, \\ E_{0y'} &  = & - \sin(\Psi)E_{0x} + \cos(\Psi)E_{0y}, \\ E_{0z'} & = &E_{0z}.  \end{array} \right .
\label{eq:1.22}
\end{equation}

And the intensity measured in the system $(O\vec{e_{x'}}\vec{e_{y'}}\vec{e_{z'}})$ is: 
\begin{equation}
I' = E_{0x'}^2 + E_{0y'}^2 = E_{0x}^2 + E_{0y}^2 = I.
\end{equation}
As expected, the intensity being an intrinsic scalar property of the light, it does not depend on the choice of the coordinate system. 

Moreover, because it characterises the circularly polarised part of the light, the Stoke parameter $V$ does not vary when rotating the reference frame:
\begin{equation}
V' = V.
\end{equation}

From their definition, the other Stokes parameters $Q$ and $U$ are transformed under this rotation following:
\begin{equation}
 \left\{ \begin{array}{l}  Q' = \cos(2\Psi)Q + \sin(2\Psi)U, \\ U' = -\sin(2\alpha)Q + \cos(2\alpha)U. \end{array} \right .
\label{eq:stokesrot}
\end{equation}

This shows the peculiar feature of the $(Q,U)$ Stokes parameters: they are dependent upon the coordinate system in which they are defined, contrary to the intensity or $V$. We therefore have to be very careful to the way we compute the Stokes parameters and we will see in Chapter~\ref{Chapter3} how this problem is circumvent in cosmology. 

To conclude, the Stokes parameters are very useful as they are linear combinations of measurable quantities. The next section is dedicated to the method of detection of these polarisation parameters. Some examples of application of the polarisation features in the field of astrophysics will be presented.


\section{Polarisation Detection}

\subsection{Instruments}

\begin{figure}[!h]
\begin{center}
	\includegraphics[scale=0.35]{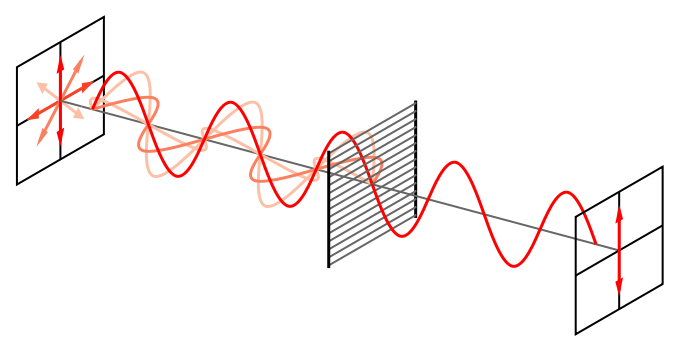}
	\caption{ The effect of a linear polariser pictured as grid on an incident unpolarised light: the outgoing light is linearly polarised (from \cite{polariser}).}
	\label{fig:polariser}
\end{center}
\end{figure}

Various materials, either natural or artificial, modify the polarisation state of the light. The most common is the linear polariser: it transforms any light into a light linearly polarised in the direction of its axis. The figure~\ref{fig:polariser} displays this effect in the case of a polariser, pictured as a grid, with its axis along the vertical. Its action on the Poincar\'e sphere defined in the coordinate system $(OXYZ)$ is the projection of any point $M$ of the sphere in a point $M'$ in the $(OXY)$ plane. More precisely, $M'$ has the coordinate $(\cos(2\alpha),\sin(2\alpha),0)$ with $\alpha$ the angle made by the polariser axis with respect to $(OX)$.  

Two other kinds of useful analysers are the half-wave and the quarter-wave plates. Their principle lies in introducing a phase shift of $\pi$ and $\frac{\pi}{2}$ (for the half-wave and quarter-wave plates respectively) between the two orthogonal components of the incident light. On the Poincar\'e sphere, it induces a rotation of an angle equal to the phase shift around the axis of the wave plate. These wave plates are widely used for the detection of the light polarisation.

Finally, the devices dedicated to the detection of light polarisation are well known. We can therefore have access to the information held in the light polarisation which are complementary to the usual observed quantity $I(t)$, the light intensity. The polarisation indeed probes various physical processes up to the microscopic scale, such as the ones I will explain in the following section.

\subsection{Polarisation in astrophysics}

In the field of astrophysics, the key property involved in polarisation studies is that any light reflecting on a surface gives an outgoing polarised light. The induced polarisation can therefore be a crucial source of information on distant medium.      

\begin{figure}[!h]
\begin{center}
	\includegraphics[scale=0.2]{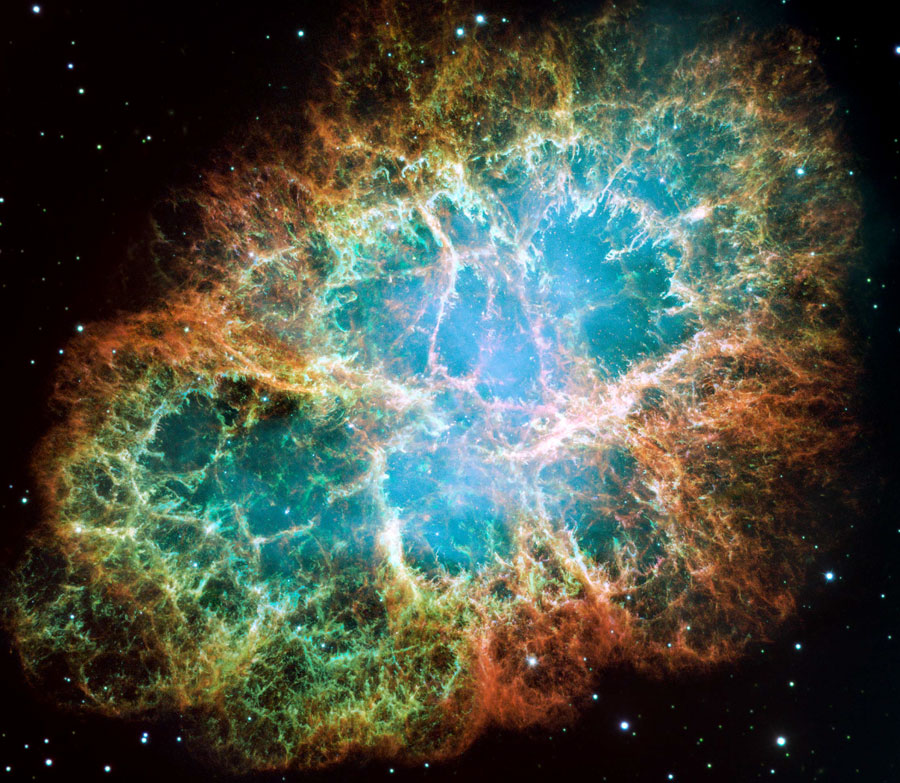}
	\caption{ The Crab Nebula as seen from the Hubble telescope. Its polarisation is well known and is therefore a calibration source for CMB experiments (from \cite{apod}).}
	\label{fig:Crab}
\end{center}
\end{figure}

The light polarisation is an observable of choice for planetary science. First, in the same way our sky is polarised due to light scattering, Venus atmosphere is polarised. As this planet surface is otherwise unobservable by current instruments, its atmosphere emission is thus crucial for the study of this planet. Its polarimetry is consequently of interest as shown in~\cite{Hansen_1974}. The same reasoning can be applied to Jupiter-like exoplanets as in~\cite{Stam_2004}. Their polarised emission can indeed help to detect and characteris them. This statement has driven the construction of the promising {\sc zimpol} instrument (which specifications are found in~\cite{Zimpol_2008}) included in the {\sc sphere} exoplanet imager, recently installed at the Very Large Telescope. Moreover, other astrophysical objects such as nebula have also a polarised emission. It originates from the reflection of the light from the inner object on the surrounding matter. For instance, the Crab nebula, pictured in Fig.~\ref{fig:Crab}, has been intensively observed and its polarisation is now very well known. It is thus a convenient astrophysical object for the calibration of CMB experiments such as the {\sc polarbear} experiment.

\begin{figure}[!h]
\begin{center}
	\includegraphics[scale=0.2]{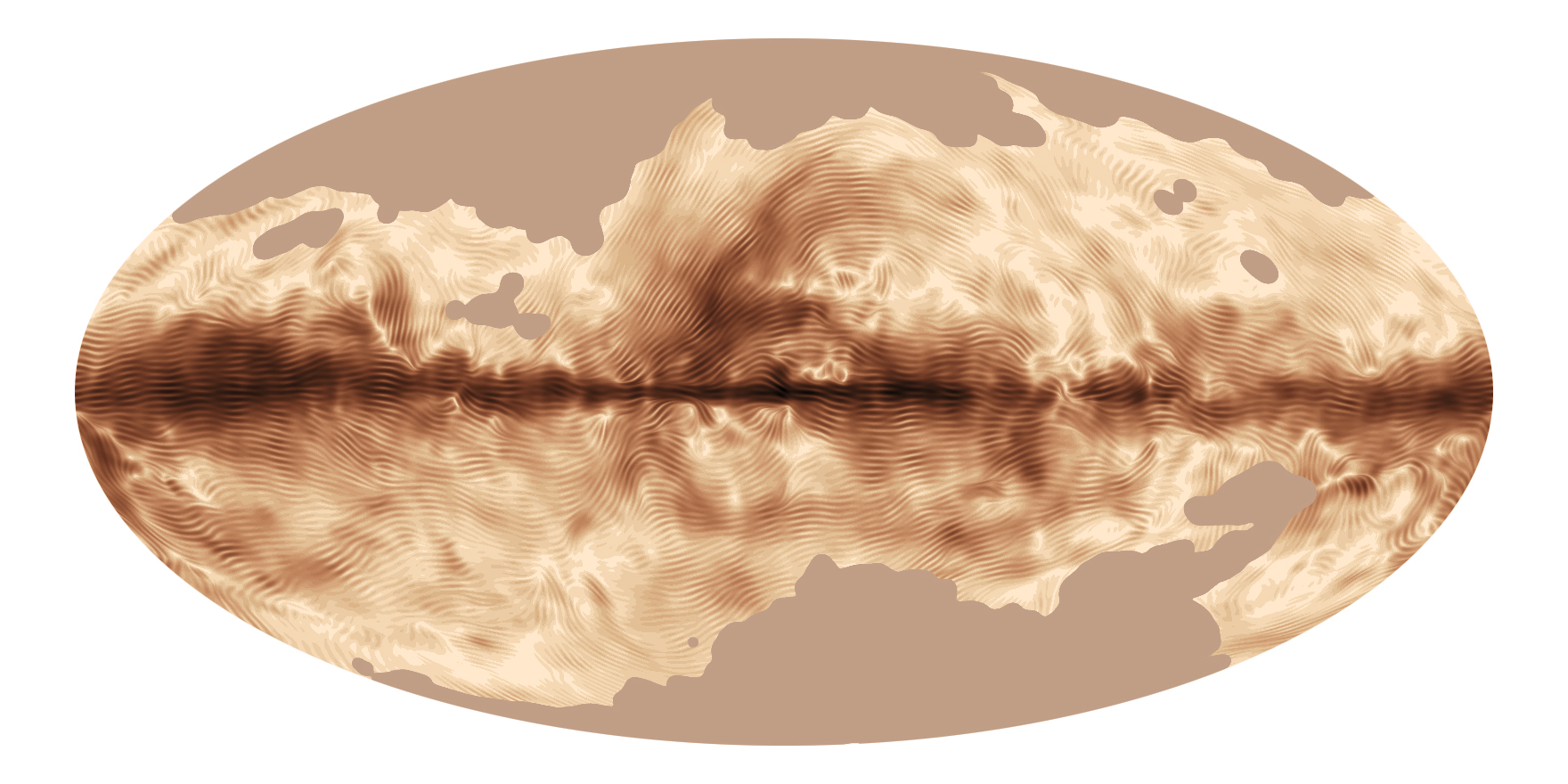}
	\caption{The magnetic field in the Milky Way deduced from the dust polarisation as seen from the Planck satellite (from \cite{Planck_2014_Pol}).}
	\label{fig:planck_mwmag}
\end{center}
\end{figure}

Furthermore, the dust is an appropriate example of the microphysics information revealed by the polarised emission. The dust is indeed prominent in the interstellar medium of our galaxy and give access to the process responsible for stellar formation. As a consequence, its constitution and evolution are currently under scrutiny. The interstellar dust is polarised under the effect of the galactic magnetic field. It is thus a key observable to better understand the magnetic field at galactic scales. In particular, in \cite{Planck_2014_Pol}, the Planck team has released a map of the magnetic field in the Milky Way shown at Fig.~\ref{fig:planck_mwmag}. However, the dust polarisation is also a contaminant for the study of cosmological observable as developed in Chapter~\ref{Chapter4}. Therefore, for the understanding of the interstellar medium and also for cleanliness of the cosmological surveys, a perfect knowledge of the dust polarisation is crucial. Also, the ionised medium surrounding stars can be polarised by scattering.

To conclude, several physical phenomenons including scattering, reflexion or magnetic field, can induce light polarisation. Polarimetry is thus a powerful observational tool in astrophysics as it can probe the microscopic scales physics of astrophysical objects. In order to share the knowledge of the different communities using this key observable, an e-COST (European Cooperation in Science and Technology) has been dedicated to the polarisation~\footnote{\url{http://www.polarisation.eu}} in astrophysics and cosmology showing that the polarisation detection is decisive in our understanding of the Universe.

\section*{Conclusion} 
The light polarisation is of great importance for astrophysics and cosmology as it reveals a multitude of information on the underlying processes such as the interstellar dust properties for instance. The cosmic microwave background being one of the main cosmological probes, its polarisation might tell us a lot about our Universe. In the next chapter, I will present the current standard model of cosmology followed by a close look at the cosmic microwave background.


\chapter{The Cosmic Microwave Background in the Frame of $\Lambda$CDM Model} 

\label{Chapter2} 

\lhead{Chapter 2. \textit{The Cosmic Microwave Background}} 


\noindent \hrulefill \\
\textit{The existence of a remaining light from the first instants of the Universe was predicted by Gamow, Alpher and Bethe in the late 40s. An indirect insight of this relic light has been found by \cite{McKellar_1941} who studied the interestellar molecules. However, a real asset would have been a direct detection of this cosmological background. After Dicke's attempts for its observation, it was unexpectedly discovered by \cite{Penzias_1965}. The existence of a cosmic microwave backgroung radiation was therefore firmly confirmed, leading the way to numerous observations for its utter characterisation. Meanwhile, this detection has permitted to establish the model of the Hot Big Bang on observational ground.}
\noindent \hrulefill \\

\section{The $\Lambda$ CDM Model}

Our current description of the Universe succeeds in explaining the origin of the large scale structures, the presence of relics contents such as photons or atoms, the dynamics of the Universe and its evolution. The observations are well outlined by the standard model but the underlying microphysics remains under scrutiny.

\subsection{A dynamic Universe}

\underline{Universe in expansion} 

By virtue of his observation, \cite{hubble_1929} has asserted that the furthest are the galaxies the faster they recede. He has indeed measured the distance of galaxies thanks to the Cepheids emission and their velocity via the Doppler effect. This observation has been interpreted as a universe in expansion. A scale factor $a(t)$ embodying the expansion of the universe is therefore introduced and defined as: 
\begin{equation}
r(t) = a(t)r(t=t_{today}), 
\label{eq:scale}
\end{equation}
with $r(t)$ the physical distance between two objects and $a(t=t_{today}) = 1$. The scale factor is then varying from 0 to 1 today.

From this, we introduce the Hubble rate such as: 
\begin{equation}
H(t) = \frac{\dot{a}(t)}{a(t)},
\label{eq:scale}
\end{equation}
with $\dot{a}$ the time derivative of the scale factor.

This effect is translated in the Hubble law which quantifies the velocity of a galaxy as its own peculiar velocity and the receding speed due to the stretch of the Universe.   

The redshift, denoted $z$, is introduced to quantify the elongation of any light wavelength due to the expansion of the universe. It is thus related to the scale factor as: $1 + z = \frac{a_0}{a(t)}$ and defines the distance to an object.

\begin{figure}[h!]
\begin{center}
	\includegraphics[scale=0.51]{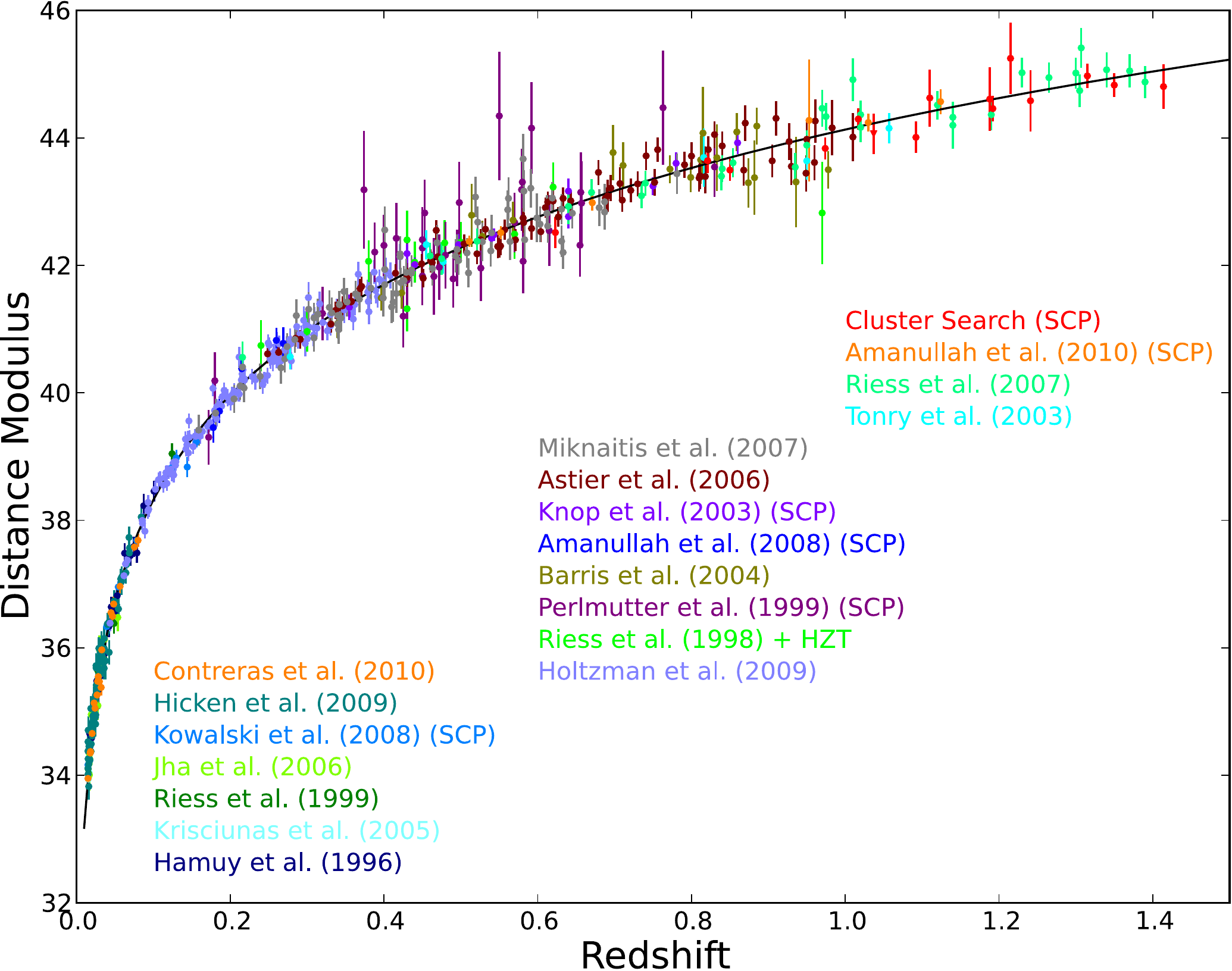}
	\caption{Hubble diagram from past and modern supernovae measurements (from \cite{suzuki_2012}). The black line is the expected profile for a flat $\Lambda$CDM universe while the coloured dots are observations.}
	\label{fig:hubblediagram}
\end{center}
\end{figure}

Since the first evidence of Hubble law, many observations have been dedicated to the refinement of the Hubble diagram which displays the velocity of an object with respect to its distance. A modern version of this Hubble diagram is shown in Fig.~\ref{fig:hubblediagram}. Nonetheless, this theory is not sufficient to explain the behaviour of the furthest galaxies. In 1998, observations of supernovae indicated that the receding velocity is growing with the distance, as shown in \cite{Riess_1998} and \cite{Perlmutter_1999}. This is commonly understood as a recent accelerating expansion of the universe which has entered an era dominated by the so-called \textit{dark energy}. The numerous latest accurate observations have confirmed that the Universe seems to be indeed in an accelerating expansion state as displayed in Fig.~\ref{fig:hubblediagram}. 

\underline{General relativity and Friedmann equations} 

The newtonian formalism of the gravitation is the main ingredient required to roughly describe our local environment. However, to understand the \textit{small} trajectories deviations or the \textit{large} scale structures such as the galaxies, one needs a more general theory of gravitation: general relativity. This theory first developed by Einstein tells us that the gravitation is a geometric effect: the space-time is curved by the presence of a massive object. The Einstein equations then drive the relation between the space-time curvature and the contained mass. These equations are very difficult to solve in the general case. Thus the description of the Universe dynamics requires to make a powerful assumption: the \textit{cosmological principle}. It can be formulated as: \textit{the Universe is homogeneous and isotropic at large scales}. In this frame, the Einstein equations can be solved and the metric describing the Universe is the one of Friedmann-Lema\^itre-Robertson-Walker (FLRW). The space-time interval $ds^2$ between two objects therefore writes: 
\begin{equation}
ds^2 = c^2dt^2 - a(t)^2\left[\frac{dr^2}{1-kr^2} + r^2(d\theta^2 + \sin^2(\theta)d\phi^2)\right],
\label{eq:flrw}
\end{equation}
$k$ being the spatial curvature with $k = -1,0,1$ for an open, flat and closed universe respectively and $dr$, $d\theta$ and $d\phi$ the spherical coordinates. 

The Einstein equations are therefore simplified in the so-called Friedmann equations:
\begin{equation}
\begin{array}{rcl} H^2(t) & = & \frac{8\pi G\rho}{3} - \frac{\kappa}{a^2(t)} + \frac{\Lambda}{3}, \\ 
\frac{\ddot{a}}{a} & = & - \frac{4\pi G}{3}(\rho+3p) + \frac{\Lambda}{3}, \\ 
\dot{\rho} + 3H(\rho+p) & = & 0, \end{array}
\label{eq:friedmann}
\end{equation}
with $\rho$ the energy density of the universe contents, $p$ the pressure, $\Lambda$ the cosmological constant and $G$ the gravitation constant. For perfect fluids, the pressure is related to the energy density by the equation of state: $p = w\rho$. From Eqs.~(\ref{eq:friedmann}), the equation of state is such that $w < -1/3$ for an accelerated expansion of the Universe ($\ddot{a}>0$) while $w > -1/3$ for a decelerated expansion ($\ddot{a} < 0$).

Besides, the first equation of Eqs.~(\ref{eq:friedmann}) gives:
\begin{equation}  
\Omega_k = \Omega_m + \Omega_{\Lambda} - 1,
\end{equation}
with:
\begin{eqnarray}
\Omega_m & = & \frac{\rho}{\rho_c}, \\ \nonumber
\Omega_{k} & = & \frac{k}{a^2H^2}, \\ \nonumber
\Omega_{\Lambda} & = & \frac{\Lambda}{3H^2},
\end{eqnarray}
with $\rho_c = \frac{3H^2}{8\pi G}$ the critical density for which the Universe is flat.

As the Universe is expanding, it is expected to be denser and hotter in the past. In particular, the Universe is assumed to be once in such an extreme state that general relativity cannot be applied: it is the Big Bang singularity. This prediction from general relativity is part of the standard model of cosmology. The model of the hot Big Bang indeed well explain the observations such as the origin of the lightest elements in the universe as first explored by \cite{alpher_1948}, or the existence of the cosmic microwave background (CMB) for instance.


\subsection{Components of the Universe and thermal history}

From the observations and according to the standard model, the main components of our Universe today are, by decreasing abundance: 
\textit{dark energy}, responsible for the recent accelerated expansion of the universe, \textit{dark matter} which weakly interacts with known matter, \textit{baryonic matter}, \textit{photons} and finally the \textit{curvature}. 

The Friedmann equations~(\ref{eq:friedmann}) applied to each component give different kind of dilution of their respective density. The figure~\ref{fig:density} shows the behaviour of the different components with time: we directly see that the Universe go through epochs dominated by different components. The temperature of the Universe is given by the photon temperature which scale as $\sim \frac{1}{a(t)}$. Within this frame, the main events that occured in our universe are recalled below along with the Universe temperature at the time of the event, following a chronological order. This permits to give an overview if our Universe history which is illustrated by Fig.~\ref{fig:density}. 

\begin{figure}[h!]
\begin{center}
	\includegraphics[scale=0.5]{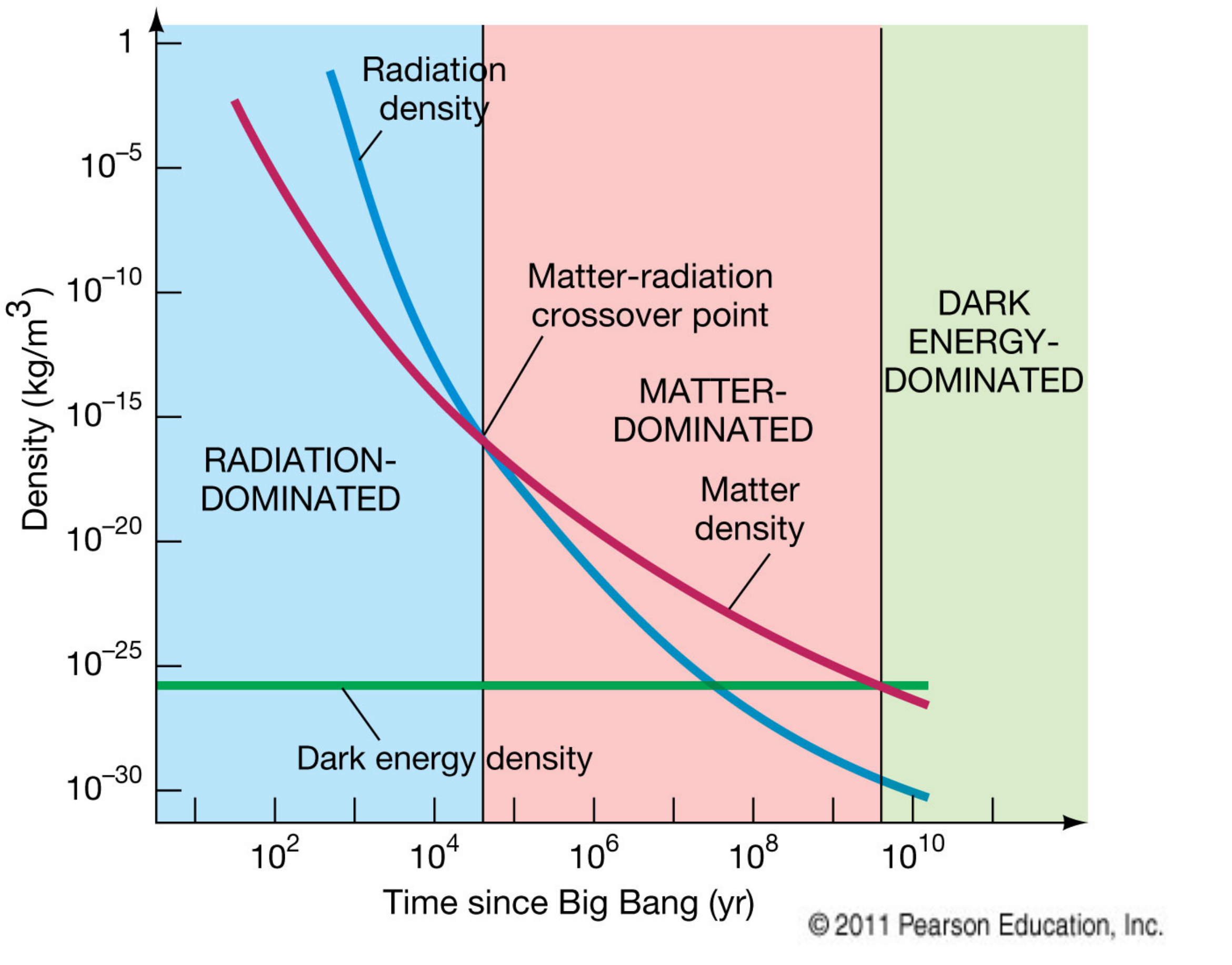}
	\caption{The dependence of each components density with respect to the time from the Big Bang singularity.}
	\label{fig:density}
\end{center}
\end{figure}

\textbf{GUT Scale:} all the interactions except gravitation are coupled (GUT standing for Grand Unification Theory). $T_{GUT} \sim 10^{26} K$. \\
\textbf{Electroweak Phase Transition:} the electromagnetism and the weak interaction are decoupled. The Higgs mechanism breaks $SU(2)\times U(1)$\footnote{SU(2) (U(1)) is the group representing the weak (electromagnetic) interaction which has therefore 3 (1) gauge bosons.}. Quarks and gluons are in a plasma state. $T \sim 10^{16} K$. \\
\textbf{QCD Phase Transition:} quarks are confined in nucleons (protons $p$ and neutrons $n$) by the strong interaction. $T \sim 10^{13} K$. \\

\textbf{Big Bang Nucleosynthesis:} the previously created protons $p$ and neutrons $n$ interact via: $n + p \rightarrow D + \gamma$ releasing a photon $\gamma$ and a Deuterium nucleus $D$. The lightest nuclei (Deuterium, Tritium) are formed, followed by heavier ones (${}^{3}H$,${}^{4}H$), up to the lithium (${}^{6}Li$,${}^{7}Li$). $T \sim 10^{11}-10^{9} K$. During this time, the neutrino decouple and propagate freely, forming the cosmic neutrino background with a temperature today of 1.96K.  \\
\textbf{Matter-Radiation equality:} the Universe ceases being dominated by radiation thus starting the matter dominated era.  $T \sim 65 000 K$. \\
\textbf{Recombination:} the electrons are bound to the nuclei forming the first atoms. In absence of scattering particles, the photons decouple and then propagate freely. The temperature corresponding to the photon decoupling is $T_{dec} = 3700 K$ corresponding to an energy $E_{dec} = 0.3 eV$. The photons are more abundant than the electrons (of a factor $\sim 10^{10}$), in consequence there is high energy photons remaining in the Universe when the energy is equal to the one of hydrogen ionisation ($E_{ion} = 13.6 eV$). These photons freely propagating in the expanding universe constitute the cosmic microwave background at a temperature $T \sim 3000 K$ at the time of their release.\\

\textbf{Structures Formation:} the galaxies are gathered in clusters and superclusters along dense matter filaments. The structure formation process is not yet fully understood although the `bottom-up' scenario\footnote{Small galaxies are formed first followed later by the large scale structures.} is favoured nowadays. The galaxy surveys such as SDSS (in \cite{Tegmark_2004}) done so far seem to match the N-body simulations which tends to lean towards this model. The cold dark matter is however required in this scenario. $T \sim 15 K$ \\    
\textbf{Today:} the Universe today is structured in galaxies aggregated in clusters and superclusters. The photon bath released during recombination has a temperature today of $T \sim 2.725 K$.


\section{Cosmic Inflation Paradigm}

In the previous section, the consistency of the hot Big Bang model has been show to succeed very well in outlining the evolution and the properties of the Universe. However, a closer look reveals some flaws in the model, three of which will be enumerated along with a qualitative explanation.  

\subsection{Three examples of the $\Lambda$CDM model inconsistency} 

\underline{Flatness problem}

A nearly null spatial curvature, $\Omega_k << 1$, of the Universe is observed today: the Universe is nearly flat. From the Friedmann equation~(\ref{eq:friedmann}), the density of the curvature $\Omega_k$ evolves as $\dot{a}^{-2}$. In a decelerated universe, $\dot{a}$ is decreasing, which means that the curvature density is growing with time. Assuming a decelerated universe expansion from GUT scale to today, $\Omega_k(t_0) \sim 10^{-5}$ 
today would require $\Omega_k(t_{GUT}) \sim 10^{-60}$. Therefore, back in the past, the curvature must have been incredibly closed to $0$. It consequently suggests that the initial conditions ought to be extremely precise: this is known as the \textit{flatness problem}.

\underline{Horizon problem} 

\begin{figure}[h!]
\begin{center}
	\includegraphics[scale=0.2]{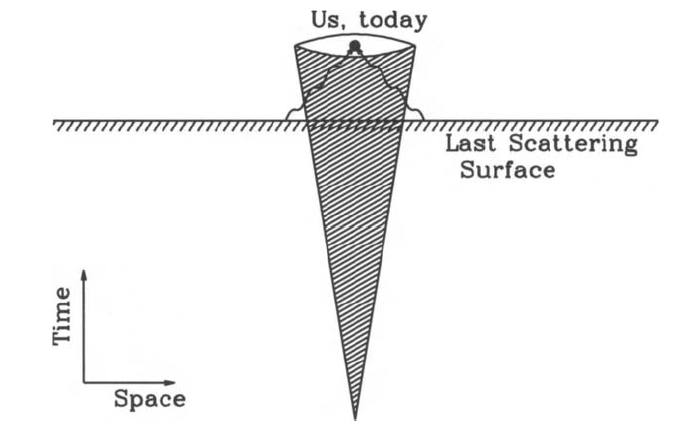}
	\caption{A sketch of the horizon problem from \cite{dodelson_2003}. Two photons emitted from the last scattering surface outside the horizon - outside of the grey shaded conical region - were causally disconnected. We however observe that they have roughly the same temperature.}
	\label{fig:horizon}
\end{center}
\end{figure}

A region of causally connected events (inside the horizon) is growing with the scale factor as pictured in Fig.~\ref{fig:horizon}. The CMB photons we receive today seem to roughly have the same temperature regardless of the line of sight. Every photon is therefore expected to be causally connected to the other in the past. However as the horizon was much smaller in the past, most photons were necessarily causally disconnected and then not thermalised. This disagreement between the observed CMB isotropy and the size of the horizon when photons were released is called the \textit{horizon problem}.

\underline{Inhomogeneities problem}

The so-called \textit{inhomogeneities problem} lies in the origin of the CMB fluctuations and consequently in today's structures. Indeed, in the frame of a FLWR Universe, the Universe is expected to be homogeneous and isotropic contrary to our highly structured Universe. The evolution of the primordial fluctuations is well understood as it simply consists in a competition between the gravitation and the expansion of the universe.Nonetheless, the origin of these primordial perturbations is not explained in the standard model.

These three problems are examples illustrating some discrepancies that exist between the observations and the hot Big Bang model. An elegant theory have been therefore established to solve these significant issues: the cosmic inflation.

\subsection{Inflation as an answer}

\underline{Accelerated expansion}

In \cite{Starobinsky_1979} and \cite{guth_1981}, the inflation scenario was proposed to solve the flatness and horizon controversies over the standard model. The inflation epoch is defined as a period of \textit{accelerating} expansion of the Universe, the second derivative of the scale factor is therefore positive. 
On the one hand, the curvature density decreases with time during the inflation period, thus the curvature is diluted in the accelerating expansion since $\dot{a}$ is increasing: the issue of the curvature fine tuning is solved.
On the other hand, the inflationary epoch is such that the Universe is tremendously stretched while the comoving Hubble radius, quantifying the size of the causally connected regions, is decreasing. At the beginning of the inflation, a causally connected region of the Universe is expected to expand and it can become a region with size greater than the Hubble radius at the end of inflation. Consequently, the same region observed today will \textit{seem} causally disconnected. Therefore, accounting for cosmic inflation, the CMB photons, which appear to come from causally disconnected regions, were in causally connected region before inflation. It explains the CMB properties such as its isotropy and homogeneity. 

Hence, the inflation brings a solution to the flatness and horizon problems. It is an appealing answer which however needs a source that accelerates the expansion of the Universe. To describe this process, the economical theory of a scalar field may be used.

\subsection{Scalar field inflation}

\begin{figure}[h!]
\begin{center}
	\includegraphics[scale=3]{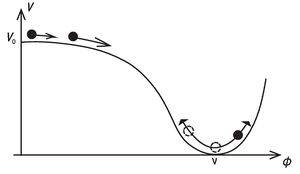}
	\caption{A scalar field slowly rolling down towards the potential minimum.}
	\label{fig:slowroll}
\end{center}
\end{figure}

The simplest way of having an ephemeral accelerating expansion of the Universe is to introduce the presence of a scalar field at the beginning of the Universe. The main current model is a scalar field whose potential is slowly falling towards its minima as shown on Fig.~\ref{fig:slowroll}.

At the time of inflation, the Universe was in a fluid state with a density $\rho_{\phi}$ and pression $P_{\phi}$ driven by the scalar field $\phi$ following:
\begin{equation}
\left \{ \begin{array}{l} \rho_{\phi} = \frac{1}{2}(\dot{\phi}^2 +V(\phi)), \\ 
p_{\phi} = \frac{1}{2}(\dot{\phi}^2 - V(\phi)).\end{array} \right . 
\label{eq:scalfield}
\end{equation} 

To induce an acceleration of the Universe expansion, the scalar field has to satisfy: $\dot{\phi}^2 < V(\phi)$, making $p_{\phi} \sim \rho_{\phi}$ (\textit{i.e.} $w < -1/3$).  Profiling the scalar field potential by a simple well is thus favoured. When this condition is no longer true, the inflation period stops. 

Furthermore, the slowing rolling conditions on the potential and the scalar field are: 
\begin{equation}
\left \{ \begin{array}{l} \dot{\phi}^2 << V(\phi), \\ 
\ddot{\phi} << 3H\dot{\phi}. \end{array} \right . 
\label{eq:slowroll}
\end{equation} 
The first equation warrants	an accelerated expansion while the second one ensures the slow evolution of the scalar field, thus guaranteeing a long enough acceleration period. 

The slow roll parameters $\epsilon_V$ and $\eta_V$ are introduced and defined as a function of the scalar field potential and its derivatives: 
\begin{eqnarray}
\epsilon_V & = & \frac{1}{16\pi G } \left(\frac{V_{\phi}}{V}\right)^2,\\
\eta_V & = & \frac{1}{8\pi G } \left(\frac{V_{\phi \phi}}{V}\right),
\end{eqnarray} 
where the subscript ${}_{\phi}$ stands for the derivative with respect to $\phi$. The first parameter quantifies the slope of the potential shape while the second one measures its curvature.
Therefore, the slow-rolling conditions in Eq.~(\ref{eq:slowroll}) boils down to $\epsilon_V << 1$ and $\eta_V << 1$. As shown in the next section, the slow-roll parameters will be relevant for they are a pivot between the properties of the scalar field and the primordial perturbations. 

Therefore, the conditions for the inflationary period are such that the scalar field is expected to be in a slow rolling potential. Such an inflationary epoch also ensures the generation of primordial perturbations, crucial for our description of the universe.

\subsection{From micro- to macro-fluctuations}
\label{ssec:primpert}

Besides answering the flatness and horizon problem, the inflation also enables to enlarge the microscopic quantum fluctuations to macroscopic scales. The acceleration of the universe expansion indeed produces growing modes of the perturbations. The quantum fluctuations of the scalar field yield scalar perturbations described by a primordial power spectrum. These scalar fluctuations are translated into density inhomogeneities which will be able to gravitationally collapse to form great structures such as galaxies or clusters later in the Universe history. Thus, the scalar perturbations generated during the inflationary epoch are the seeds of today structures. In addition, tensor perturbations, the so-called primordial gravitational waves, emerge from the quantum fluctuations of the traceless and divergent-free part of the metric. 

The mechanism of the quantum fluctuations amplification is easier to explain thanks to tensor perturbations. The perturbed Einstein equations give the evolution of the gravitational waves in a spatially-flat Universe as: 
\begin{equation}
(ah_{k})'' + \left( k^2 - \frac{a''}{a} \right)(ah_{k}) = 0,
\end{equation}
with $h_k$ the tensor modes in Fourier space and ${}^{'}$ standing for the derivative with respect to conformal time. The resulting tensor modes $ah_k$ therefore undergo different regimes regarding the sign of $k^2 - \frac{a''}{a}$. If this quantity is positive, the regime is oscillatory. On the contrary, if $k^2 - \frac{a''}{a} < 0$, an exponential growing mode is solution of the equation: the tensor modes are amplified. The exponential behaviour of the scale factor $a(t)$ during the inflation ensures both $\frac{a''}{a}$ to be positive valued and to grow rapidly. As inflation goes, $k^2 - \frac{a''}{a}$ becomes negative and the tensor perturbations are therefore generated. 
A similar analysis explains the generation of the density perturbations.

The primordial power spectra of the scalar perturbations $\pk_\sa(k)$ and of the gravitational waves $\pk_\te(k)$ are parametrised following a power law:
\begin{eqnarray}
\pk_\sa(k) & = & \mathcal{A}_{S} \times (\frac{k}{k_0})^{n_\sa - 1}, \\ \nonumber
\pk_\te(k) & = & \mathcal{A}_{T} \times (\frac{k}{k_0})^{n_\te}, 
\label{eq:scalpert}
\end{eqnarray} 
with $\mathcal{A}_{S(T)}$ the amplitude of the scalar (tensor) modes, $n_{S(T)}$ the spectral index of the scalar (tensor) modes and $k_0$ the so-called \textit{pivot scale}, an arbitrary scale at which the perturbation amplitude is evaluated.  

In the scope of a slow-rolling inflation, the primordial power spectrum parameters are related to the slow roll parameters $\epsilon_V$ and $\eta_V$. The scalar modes amplitude and scalar index are given by:
\begin{eqnarray}
\mathcal{A}_S & \thickapprox & \frac{V}{24\pi^2M_{Pl}^4\epsilon_V}. \\
n_\sa - 1 & \thickapprox & 2\eta_V - 6\epsilon_V, 
\end{eqnarray}
The tensor modes parameters write:
\begin{eqnarray}
\mathcal{A}_T & \thickapprox & \frac{2V}{3\pi^2M_{Pl}^4}. \\
n_\te & \thickapprox & -2 \epsilon_V,
\end{eqnarray}

The tensor primordial power spectrum is usefully expressed by introducing the \textit{tensor-to-scalar ratio} $r$. It quantifies the ratio between the tensor and scalar modes amplitude at the pivot scale: 
\begin{equation}
r = \frac{\mathcal{A}_T(k_0)}{\mathcal{A}_S(k_0)} \propto \epsilon_V.
\label{eq:r}
\end{equation} 
This phenomenological parameter is convenient to check the \textit{consistency relation}:
\begin{equation} 
r \thickapprox - 8 n_\te. 
\end{equation}
Its value also characterised the shape of the potential $V(\phi)$. 
Consequently, the scalar and tensor primordial power spectra are directly related to the potential of the scalar field and its derivatives. A measurement of the primordial power spectra would thus determine the characteristics of the inflationary period such as its energy. Fortunately, both the generated density fluctuations and the gravitational waves leave their imprints during the photons released at the recombination that form today cosmic microwave background (CMB). For this reason, the CMB is an open window on the first instants of the Universe.


\section{The Cosmic Microwave Background}

\subsection{Temperature anisotropies}

The CMB is the remaining light from the first instants of the Universe, its first detection was consequently a milestone in favour of the hot Big Bang model. Before the CMB emission, the density and temperature were such that the Universe was opaque due to the short mean free path of the light. They were indeed constantly interacting with the free electrons of the primordial plasma via Thomson scattering\footnote{Scattering of photons by a free charged particle.}. However, as the Universe was expanding, the temperature was decreasing and when it reached $T \sim 3000 K$, the nuclei and the electrons could form the first lightest atoms as previously mentioned in the main steps of the Universe thermal story. The light could then propagate freely, without collapsing with free electrons, and formed the so-called \textit{cosmic microwave background}.

\begin{figure}[h!]
\begin{center}
	\includegraphics[scale=0.35]{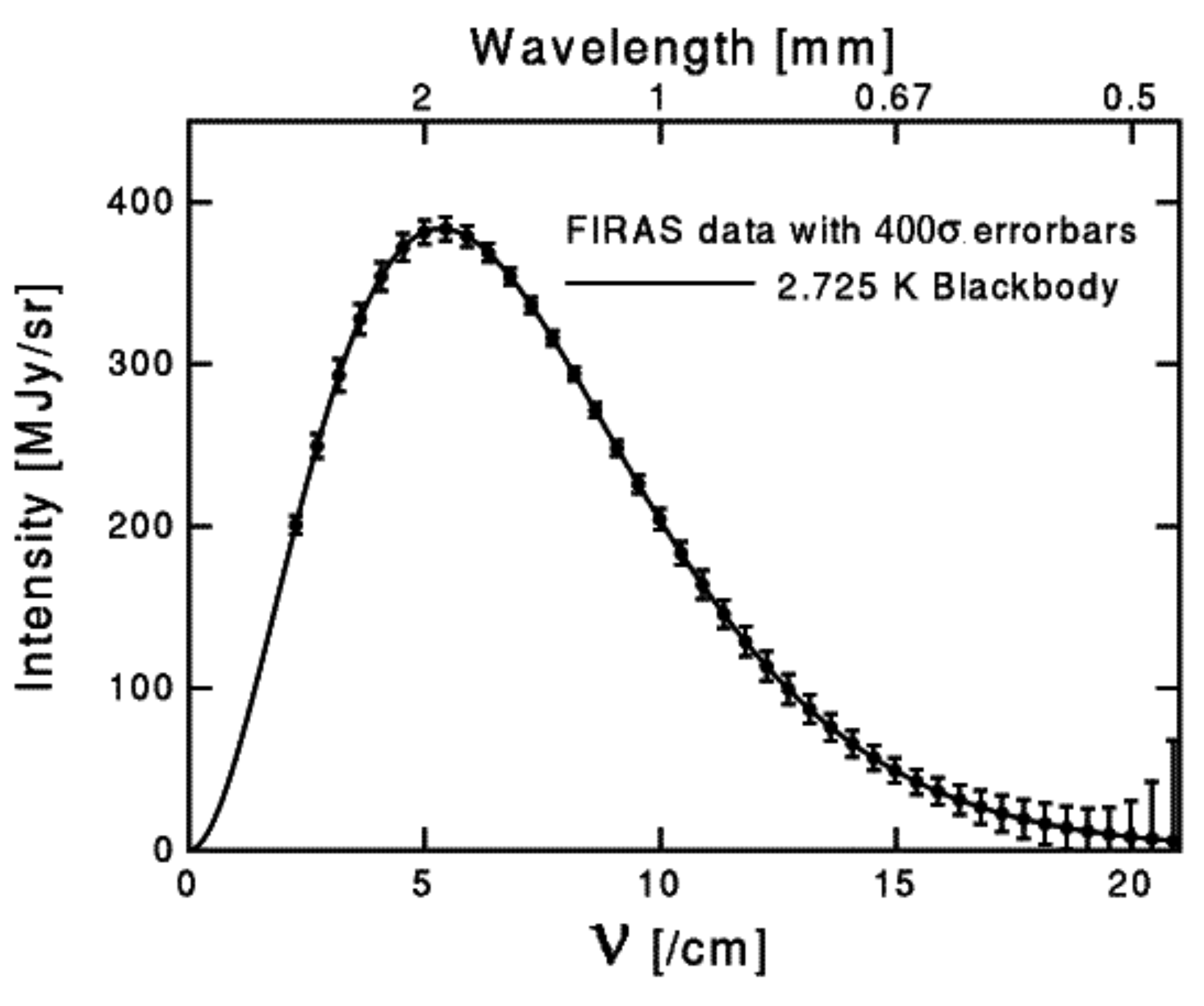}
	\caption{The black body spectrum of the CMB measured by the \textit{FIRAS} instrument of \textit{COBE} satellite (\cite{mather_1994}).}
	\label{fig:bbcmb}
\end{center}
\end{figure}

As the photons were tightly coupled to the matter before being release, they have a black body spectrum distribution with a today temperature of $T_{CMB} = 2.7260 \pm 0.0013 K$ (\cite{fixsen_2009}) as shown in Fig.~\ref{fig:bbcmb}. The frequency peak is at $\nu = 160 GHz$ that is to say photons belong to the microwave domain. Moreover, this light appears to be isotropic in the sky once we have removed the so-called dipole due to the motion of the Sun in the CMB photon bath. 

\begin{figure}[h!]
\begin{center}
	\includegraphics[scale=0.35]{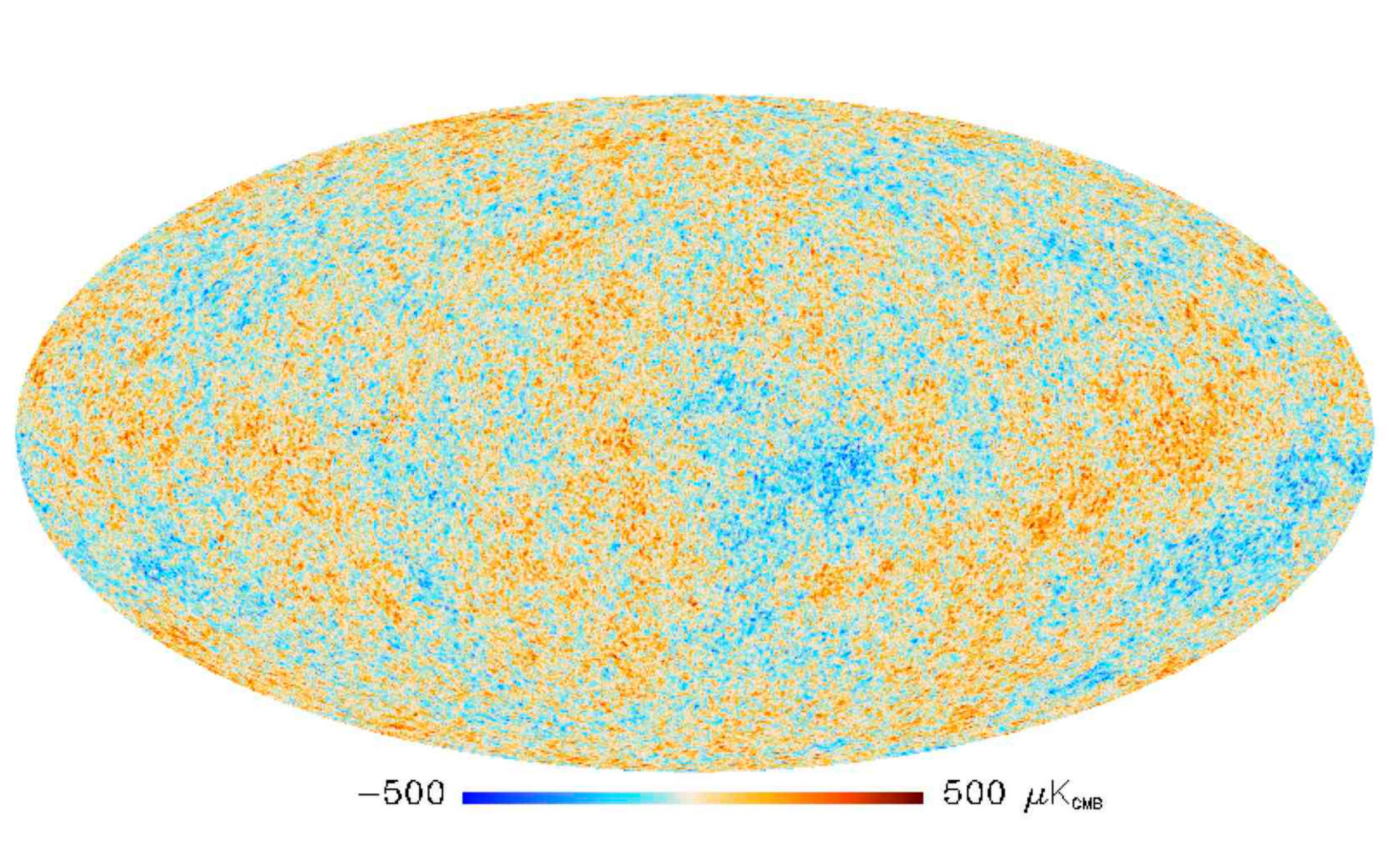}
	\caption{Mollweide projection of the CMB temperature fluctuations detected by team Planck.}
	\label{fig:cmbplanck}
\end{center}
\end{figure}

At first, the CMB appears to be roughly homogeneous over the whole sky. The COBE satellite however revealed the presence of tiny fluctuations of the CMB temperature. These fluctuations are known to have an amplitude of: 
\begin{equation} 
\frac{\Delta T}{T_{CMB}} = 10^{-5}.
\end{equation}
Their spatial correlations give a prodigious amount of information on the Universe history either on its primordial or in its late time state. The figure~\ref{fig:cmbplanck} displays a Mollweide projection\footnote{Also known as the Babinet projection, the proportions in areas are conserved as much as possible.} of the most precise CMB temperature fluctuations map to date which has detected by the Planck satellite in \cite{Planck_tmap_2013}.

\underline{Primary anisotropies}

The \textit{primary anisotropies} emerge from the properties of the primordial baryonic-photon fluid before the emission of the CMB. They were sourced by the primordial fluctuations described in the previous section. The amplitude and form of the temperature fluctuations are driven by a competition of various processes such as the Doppler effect or the Saches-Wolfe effect. An utter description of these anisotropies statistics is reviewed in \cite{hu_1997}.

\underline{Secondary anisotropies}

During the journey of the CMB photons from the last scattering surface to our instruments, they have faced different cosmological phenomenons such as the reionisation or the gravitational potential of large scale structures. The latter alter the photons' energy through, for instance, the Sunyaev-Zel'dovich effect or the gravitational lensing. These alterations must be corrected to fully characterise the primordial effects hidden in the CMB. Also, they can be very useful to probe the late time Universe. A complete review of those \textit{secondary anisotropies} can be found in \cite{aghanim_2008}.  
 
\subsection{Statistics of the CMB temperature anisotropies}

Instead of handling the large temperature fluctuations map, the CMB temperature anisotropies are usually decomposed on the spherical harmonics: 
\begin{equation} 
T(\theta,\phi) = \sum_{\ell = 1}^{\infty} a^T_{\ell m}(\vec{n}) Y_{\ell m}(\theta,\phi).
\label{eq:tempfluc}
\end{equation} 

The temperature can indeed be expanded in modes on the sphere with the multipoles $\ell$ - analog to the Fourier wavenumber $|\vec{k}|$ - the inverse of the scale between two points on the sky and $m$ the orientation on the sphere. Small $\ell$ values therefore correspond to large angular scales on the celestial sphere. The coefficients of this decomposition contain all the information on the amplitude of the fluctuations for a given scale. 

These temperature fluctuations contain a remarkable amount of information on the state of the scalar and tensor perturbations at the time of the last scattering surface. To extract as much information as possible, we deal with their statistics. From the second moment of their statistics, the \textit{angular power spectrum} is built as: 
\begin{equation} 
\left<a_{\ell m}a_{\ell' m'}^*\right> = C_{\ell}\delta_{\ell\ell'}\delta_{mm'},
\label{eq:cell}
\end{equation} 
with $\left<.\right>$ standing for an average over all the possible universe realisations for a given repartition of the $a_{\ell m}$. The $C_{\ell}$ angular power spectrum is independent of $m$ due to statistical isotropy of the temperature fluctuations. A peculiar feature arises from this definition: as we only have access to one observable universe, we only do have a finite amount of information to sample the $a_{\ell m}$ distribution. This inherent uncertainty on the observed angular power spectrum is called the \textit{cosmic variance} and is such as: $\frac{2C_{\ell}^{TT2}}{2\ell+1}$. For a given $\ell$, there are indeed $2\ell + 1$ independent modes, so that the cosmic variance is larger at low $\ell$. The chapter~\ref{Chapter4} is dedicated to the properties of the angular power spectrum estimation. 

\begin{figure}[h!]
\begin{center}
	\includegraphics[scale=0.45]{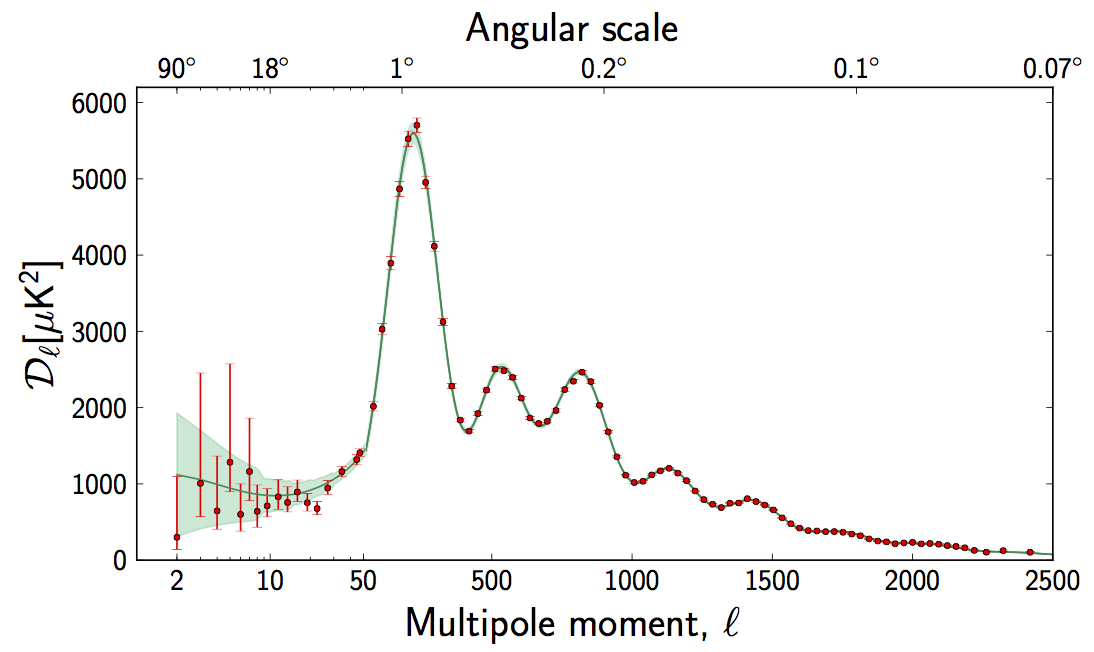}
	\caption{The angular power spectrum of the CMB observed by the Planck satellite from \cite{Planck_cl_2013}. The red points stand for the data while the best fit model is displayed as a green line. The green shaded area indicates the cosmic variance magnitude. }
	\label{fig:cell}
\end{center}
\end{figure}

The angular power spectrum $C_{\ell}$ reconstructed from the temperature CMB map of Fig.~\ref{fig:cmbplanck} imaged by the Planck satellite is displayed as red points in Fig.~\ref{fig:cell}. The green line represents the best-fit angular power spectrum of the standard model of cosmology: the model and the data are incredibly consistent. The green area symbolises the \textit{sampling variance} that is the previously explained cosmic variance applied to an incomplete sky coverage. The Planck satellite has been designed to get the best sensitivity by the use of fifty-two cooled down bolometers. The obtained CMB temperature map is therefore said to be the ultimate one as its precision is only limited by the sampling variance. 

\underline{Probing primordial perturbations}

As explained in Sec.~\ref{ssec:primpert}, the temperature fluctuations of the CMB are directly linked to the primordial scalar and tensor perturbations in the Universe. In pratice, the CMB temperature angular power spectrum $C_{\ell}$ is directly related to the initial power spectrum $\pk(k)$ modulated by a transfer function $\Delta_{\ell,,\sa/\te}$ including the information on the evolution of the matter power spectrum up to today and on radiative transfer:
\begin{eqnarray} 
C_{\ell} & = & \displaystyle\int dk \left[\Delta_{\ell,\sa}^2(k,\eta_0)\pk_\sa(k) + \Delta_{\ell,\te}^2(k,\eta_0)\pk_\te(k) \right],
\label{eq:pktocell}
\end{eqnarray} 
with $\eta$ the line of sight and $\Delta_{\ell,S(T)}$ the transfer function of the scalar (tensor) perturbations. 

The transfer functions and the primordial power spectra lean on the considered cosmology and thus depend on a set of cosmological parameters. Therefore, fitting the data with a predicted angular power spectrum allows to set constraints on the cosmological parameters. 

\subsection{Constraints on cosmological parameters} 

The standard model of cosmology is described by a set of cosmological parameters. The main ones are: \\
- the density of Universe contents: $\Omega_b$, $\Omega_{\Lambda}$, $\Omega_{DM}$ are respectively the density of baryonic matter, dark energy and dark matter; \\
- the optical depth due to the reionisation of the Universe: $\tau$; \\
- the amplitude of the primordial scalar fluctuations $A_S$ and its spectral index $n_\sa$; \\
- the tensor-to-scalar ratio $r$ and the spectral index of the primordial tensor fluctuations power spectrum $n_\te$.	

A large set of other cosmological parameters such as the Hubble constant $H_0$ can be deduced from them. Their current tightest constraints mainly come from the CMB detection from the Planck satellite in \cite{Planck_param_2013} whose observations have permitted the elaboration of a precise temperature power spectrum from which shape can be derived the cosmological parameters.

\begin{figure}[h!]
\begin{center}
	\includegraphics[scale=0.2]{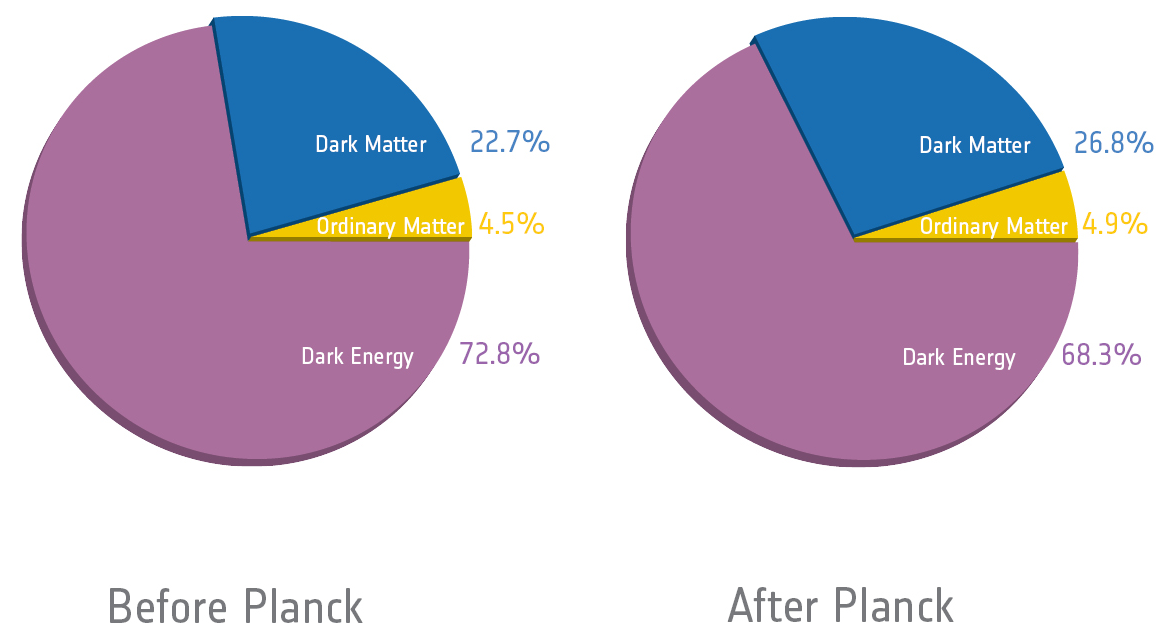}
	\caption{The universe contents as derived by the preceding experiments on the left and by the Planck satellite on the right. Picture taken from \cite{camembert} (The scientific results can be found in~\cite{Planck_param_2013}).}
	\label{fig:camembert}
\end{center}
\end{figure}

In 2013, the Planck collaboration has released a large set of results concerning the primordial and the late-time Universe. The results relevant for the present manuscript concern the constraints on inflation. However, it is worth mentioning here that the constraints on the universe contents are slightly changed compared to the results provided by WMAP, the former satellite dedicated to CMB detection, as shown in Fig.~\ref{fig:camembert}.

\begin{figure}[h!]
\begin{center}
	\includegraphics[scale=0.4]{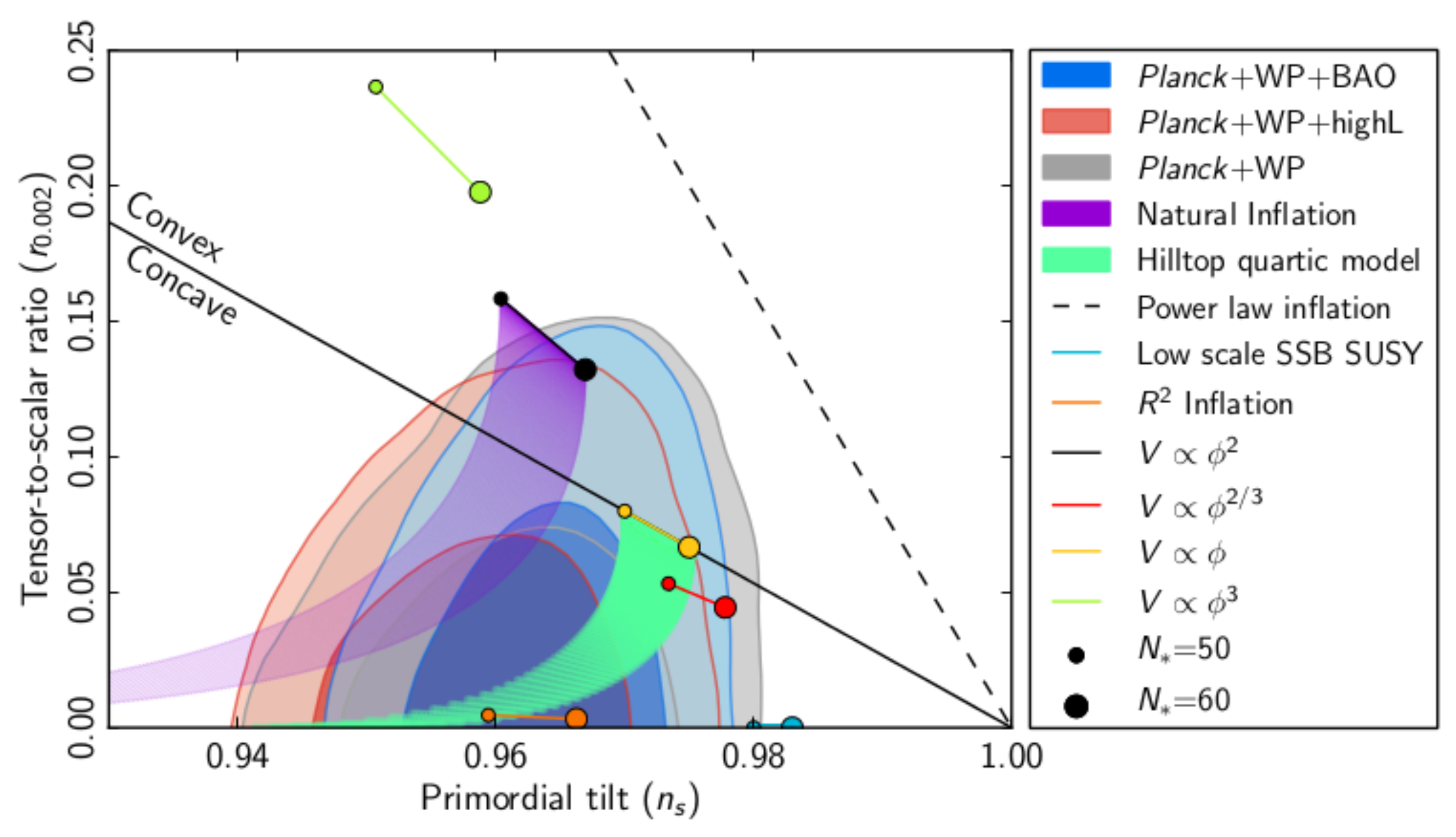}
	\caption{Behaviour of $r$ in function of $n_\sa$ for different inflation models (colored lines) along with the constraints coming from the Planck satellite and other observations (in shaded areas). Picture taken from~\cite{Planck_infl_2013}.}
	\label{fig:rns}
\end{center}
\end{figure}

In \cite{Planck_infl_2013}, precise constraints on the tensor-to-scalar ratio $r$ and the scalar spectral index $n_\sa$ are calculated. With $n_\sa = 0.960 \pm 0.0073$, the scale invariance of the primordial power spectrum of the scalar perturbations is excluded at more than $5 \sigma$. This strong statement favours the inflation mechanism as a generation of the density perturbations and the gravitational waves. Moreover, $r$ is upper bounded by 0.11 at $3 \sigma$ for a pivot-scale $k_0 = 0.002 Mpc^{-1}$. The setting constraints enable to discriminate between the abundant inflationary models. The figure~\ref{fig:rns} from \cite{Planck_infl_2013} shows the main inflationary models on a $(r,n_\sa)$ graph along with their constraints. The ruled out models are clearly represented on the same graph. As an example, an inflationary model with a potential in $\phi^3$ is clearly disfavoured.

In addition, the Planck satellite has provided a tremendous amount of cosmological results. Among others, no deviation from gaussianity of the primordial temperature fluctuations has been shown in~\cite{Planck_nongauss_2013}.  

\section*{Conclusion} 
The Planck satellite has provided an utter knowledge on the CMB temperature anisotropies and subsequently on the cosmological parameters. The $\Lambda$CDM remains nowadays the favoured model to describe the evolution of the Universe. Nonetheless the CMB temperature power spectrum fails to constrain the whole physics of inflation, such as its energy scale. However as explained in Chapter~\ref{Chapter1}, the CMB polarisation might hold some valuable information on the primordial universe. The next chapter consequently details the characteristics of the CMB polarisation.  

\chapter{CMB Polarisation} 

\label{Chapter3} 

\lhead{Chapter 3. \textit{CMB Polarisation}} 
\noindent \hrulefill \\
\textit{The CMB polarisation comes in two flavours: the $E$ and $B$ modes. The $E$ modes have been detected twelve years ago by the Dasi team while the $B$ modes are still imperceptible due to their faintness. A variety of experiments were nonetheless dedicated to their detection as they are a unique probe of the very first instants of the Universe. This effort have been rewarded by a great stride during this year 2014. For the first time, a direct detection of the lensed $B$-modes has been claimed by the {\sc polarbear} experiment, and only one week later a direct detection of the primordial $B$ modes was announced by the {\sc bicep2} experiment. The latter detection have been widely relayed on the web and in the newspapers (the importance of the detection has even been explained in a comic strip\footnote{\url{http://phdcomics.com/comics.php?f=1691}}). However, since then, the primordial $B$ modes detection remains controversial and the BICEP2 team has tempered their conclusions on this detection.} 
\noindent \hrulefill \\


The Planck satellite has recently given the ultimate map of the CMB temperature anisotropies in \cite{Planck_tmap_2013}. The extracted information are the basis of the current standard model of cosmology describing the evolution of the Universe and its contents. However, an important piece of information on the primordial Universe is not reachable using the CMB temperature power spectrum alone. Indeed, at large angular scales, the imprints of the gravitational waves are hidden owing to the cosmic variance. As highlighted in a generic case in chapter~\ref{Chapter1}, the CMB polarisation may be a very useful tool giving access to different physical processes that cannot be obtained when observing only the CMB intensity. The CMB is indeed polarised due to the state of the primordial plasma and the perturbed background. The present chapter starts with a qualitative description of the origin of these polarised anisotropies. Because it is linear, the CMB polarisation can be described only by the two Stokes parameters $(Q,U)$, but a description in $E$ and $B$ modes is preferred since they are physically more relevant. I will therefore focus on the link between $(Q,U)$ and $(E,B)$, first within a harmonic approach, and secondly in the real space, and eventually on the CMB polarised power spectra. The last section of this chapter is dedicated to the current and forthcoming experiments aiming at detecting the CMB polarisation and subsequently setting constraints on cosmological parameters such as the tensor-to-scalar ratio $r$.   

\section{Origins of the CMB Polarisation}

\subsection{Thomson scattering}

Before the release of the CMB photons, they were tightly coupled to the matter via the Thomson scattering. As the cross section of this interaction scales as the inverse of the squared mass of the scattering particle, the photons were mostly scattered by the free electrons of the primordial plasma. They have kept trace of this scattering witnessed by the temperature fluctuations today observed in the CMB as seen in the previous Chapter~\ref{Chapter2}. We explained in Chapter~\ref{Chapter1} that the scattering is responsible for the polarisation of the light. Thus the Thomson scattering also determines the vector description of the scattered light. The Thomson scattering differential cross section between a monochromatic electromagnetic plane wave on an electron is written as:
\begin{equation}
\frac{d\sigma}{d\Omega} = \frac{3\sigma_T}{8\pi} |\vec{\epsilon} . \vec{\epsilon}'|^2
\label{eq:thomson}
\end{equation}
with $\sigma_T$ the total Thomson cross-section, $d\Omega$ the elementary solid angle and $\vec{\epsilon}$ and $\vec{\epsilon}'$ the polarisation state of the incoming and scattered photon respectively. 

This equation comes from the re-radiation along the direction $\Omega$ of the electron accelerated by the incoming light. Thomson scattering of photons is accurately analysed in terms of Stokes parameters in \cite{Collett_1992}.

\begin{figure}[!h]
\begin{center}
	\includegraphics[scale=0.6]{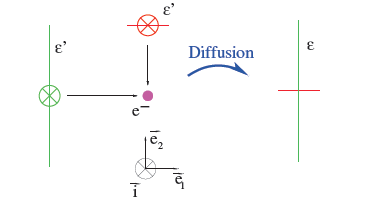}\includegraphics[scale=0.6]{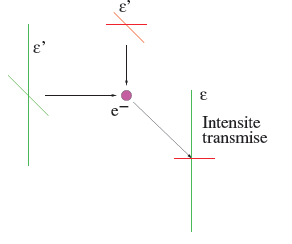}
	\caption{Transmitted intensity after scattering of photons presenting a quadrupolar anisotropy on a free electron. On the left panel, two incident perpendicular light beams with different intensity and the resulting intensity are depicted as projected on the plane orthogonal to the line of sight. The right panel displays the same scheme in pseudo-perspective. Picture taken from \cite{Ponthieu_2003}.}
	\label{fig:mwmag}
\end{center}
\end{figure}

This way, the Thomson scattering selects a polarisation vector even if the incoming light is not polarised. Moreover, the polarisation degree of the scattered light is driven by the observation angle of the scattered light. 
In practice, the electric field of a non polarised light can be decomposed over two orthogonal axes as shown in Eq.~(\ref{eq:light}) of Chapter~\ref{Chapter1}. If such a light meets an electron, the latter will be accelerated along the direction of the electric field and will therefore radiate light -- the scattered light. The scattered light intensity, polarisation direction and degree depend on the observation angle. In reality, the CMB photons can be polarised due to Thomson scattering on the free electrons. However, in the primordial Universe made of the plasma, the light is coming from every direction before reaching the scattering electron. Thus, Thomson scattering of an isotropic light beam on an electron would not select a specific polarisation direction. In order to observe a net polarisation of the CMB light, it is necessary for the incoming light to be anisotropic for the scattering electron. To explain this effect, we will first consider that the electron scatters light coming from four orthogonal direction denoted by the cardinal points: on Fig~\ref{fig:mwmag}, North (South) stands for the light beam along $\vec{e_2}$ ($-\vec{e_2}$) and West (East) for the light beam along $\vec{e_1}$ ($-\vec{e_1}$). Also, for convenience sake, the line of sight is along the $i$ axis orthogonal to the figure plane.

First of all, according to Eq.~(\ref{eq:thomson}), the light coming from the North or South will give a WE contribution to the polarisation of the scattered light observed along our line of sight. Indeed, the scattered light does not have any component on the $i$ axis, which is the direction of propagation. In the same way, the light from the West and East will only give NS contribution. This consideration acts as a rule of thumb which will make the conclusions easier to draw. 

The simplest pattern of incoming light on the electron is an isotropic pattern: the light coming from the four directions have the same intensity. The contributions from each light beams will give the same contribution on the NS and WE axis of the scattered light. Or equivalently, each light beams will accelerate the electron by the same amount thus the outgoing light electric field has no favoured direction. As a consequence, the scattered light is not polarised. 

The light can now present a dipole pattern: the intensity of the North (respectively South) light is greater (resp. lower) than the West and East light beams which have the same intensity. Using the rule of thumb, the light coming from West and East will induce a NS contribution. The North light beam will provide a greater contribution on the WE components of the scattered light. However, it will be compensated for by the lower contribution from the light coming from the South direction. In other words, the induced acceleration of the electron is the same in all directions: no polarisation is produced. 

A more elaborated pattern is the quadrupole: the North and South light have a greater intensity than the West and East ones. Automatically, we deduce that the selected polarisation has not the same magnitude in the NS direction than the WE one. Since the electron is more accelerated in the WE direction, the scattered light is then polarised. 

As a result, an incoming light showing a quadrupolar anisotropy of its intensity induces a net polarisation of the scattered light.
The same kind of analysis shows that there is no other pattern which can produce polarised scattered light. To better understand the process of polarisation production, we assumed the electron is surrounded by four orthogonal light beams as an illustration. The results obtained in this example remain true for a continuous pattern. 

Also, we have considered that the line of sight is perpendicular to the plane of the quadrupolar anisotropy. In the general case, we have to integrate over all the different lines of sight which leads to a modulation of the signal.
The quadrupolar origin of the polarisation can be explicitly derived as in \cite{Kosowsky_1996} (and interestingly interpreted in \cite{Ponthieu_2003}) by computing the Stokes parameters for Thomson scattering using a harmonic decomposition of the intensity. A quadrupole corresponds to the $\ell = 2$ components in the spherical harmonics expansion. Thus the contribution from $Y_{\ell = 2,m}$ (with $m \in [-2;2]$) of the intensity decomposition alone therefore causes the CMB polarisation. It is noticeable that the phase of the electromagnetic wave does not play a role in the equation~(\ref{eq:thomson}). And consequently, the Thomson scattering does not produce any circular polarisation.

As a conclusion, the CMB can be linearly polarised if the light intensity presents a quadrupolar anisotropy around the scattering electron. The following section is dedicated to the explanation of the existence of such anisotropies in the primordial plasma. 

\subsection{Quadrupolar anisotropies}

In the electron reference frame, only the quadrupolar component $Y_{\ell = 2,m}$ of the intensity decomposition on the spherical harmonics contributes to the CMB polarisation. There exists several configurations that coincide with a quadrupolar pattern corresponding to the different values of the azimuthal number $m$ ($m \in [-2;2]$). Consequently, the perturbations of the background sourcing quadrupolar anisotropies come in three flavours: scalar ($m = 0$), vector ($m = \pm 1$) and tensor ($m = \pm 2$). Each of them cause distinct polarisation pattern.

\underline{$m = 0$: scalar perturbations} 

The scalar perturbations are fluctuations of energy density which are translated into potential fluctuations. At scales where the gravitation exceeds the pressure, over(under)-densities do attract (respectively repel) the surrounding matter. As the gravity has a radial symmetry, this system has an azimuthal symmetry as shown in the left panel of Fig.~\ref{fig:density} displaying a density perturbation in the primordial plasma. As a consequence, this radial case corresponds to the $Y_{20}$ component of the decomposition on spherical harmonics.

In the case of an over-density, an electron falling into the gravitational potential is accelerating towards the centre. The forward plasma is thus falling faster than the electron while the backward plasma is falling slower than the electron. Therefore, the electron sees the forward and backward plasma receding from him. Moreover, the surrounding plasma in iso-latitude annulus will appear flowing towards the electron. The photons being tightly coupled to the plasma, the intensity distribution of the incoming light gets the same pattern
The electron thus sees light showing a quadrupolar anisotropy as sketched in the left panel of Fig.~\ref{fig:density}. 

\begin{figure}[!h]
\begin{center}
	\includegraphics[scale=0.6]{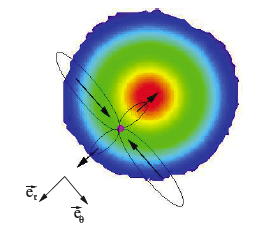}\includegraphics[scale=0.6]{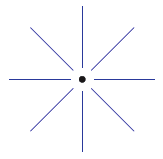}
	\caption{Over-density in the primordial plasma. Around a given electron, the light intensity is quadrupolar. On the right panel, the polarisation pattern induced by the density perturbation. Image taken from \cite{Ponthieu_2003}.}
	\label{fig:density}
\end{center}
\end{figure}

Moreover, in this case, the polarisation pattern will necessarily be radial. Indeed, by applying to the present case the rule of thumb seen in previous section and assuming that the line of sight is perpendicular to the quadrupolar plane, the scattered light is polarised in the direction orthogonal to the incoming light. As the electron sees a more intense light in the direction tangent to its trajectory, the outgoing light is polarised along the radius of the perturbation as shown in the right panel of Fig.~\ref{fig:density}. In the same way, the polarisation pattern is tangential for an under-density. The symmetry of the perturbation is then memorised at the level of the polarisation pattern.

\underline{$m = \pm 1$: vector perturbations} 

The vector perturbations are characteristics of vortical perturbation -- \textit{i.e.} presenting a null-divergence but a non zero curl component of the velocity -- in the cosmological fluids, they correspond to the $Y_{2, \pm1}$ component. Such perturbations are negligible as they do not outlast the inflationary phase due to their amplitude being proportional to the inverse of the scale factor, $a(t)$. They thus will not be taken into account in the present manuscript.

\underline{$m = \pm 2$: tensor perturbations} 

\begin{figure}[!h]
\begin{center}
	\includegraphics[scale=0.3]{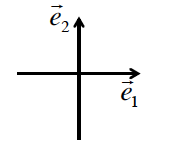}\includegraphics[scale=0.5]{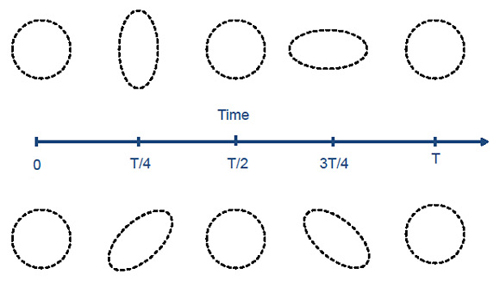}\includegraphics[scale=0.35]{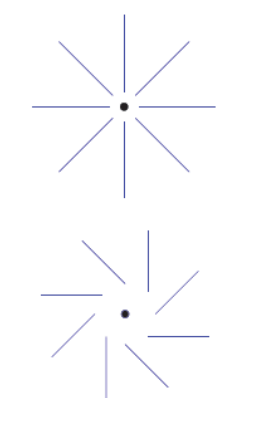}
	\caption{A gravitational wave passing through a circle of motionless test particles. On the upper (lower) panel, the gravitational wave have a `+' (`$\times$') polarisation. The induced polarisation patterns corresponding to each gravitational wave polarisation are shown on the right planel.}
	\label{fig:gw}
\end{center}
\end{figure}

The tensor perturbations of the perturbed metric stands for gravitational waves. When a gravitational wave go through a circle of motionless test particles, the circle is deformed. The photons, coupled to the plasma, are therefore redshifted in one direction while they are blueshifted in the orthogonal direction. An electron being localised at the centre of the test particles therefore sees quadrupolar anisotropy of the light intensity. The figure~\ref{fig:gw} shows such a process for the plus- ($+$) and cross- ($\times$) polarisations of the gravitational waves in the upper and lower panel respectively. The induced polarisation pattern has a radial and a curl component. 

To summarise, the Thomson scattering induces a linearly polarised light if a quadrupolar anisotropic light scatters on electrons in the primordial plasma. Because quadrupolar anisotropies are generated by scalar and tensor perturbations, the CMB is linearly polarised. Moreover, the imprint of the CMB polarisation is made at the \textit{last scattering} of the CMB photons on the primordial electrons. Indeed, before the recombination, no quadrupole anisotropy could remain for there were too many Thomson scatterings. Also, after the recombination, there is no more free electrons left to permit the scattering of photons. The thickness of the last scattering surface leads to a low polarisation degree of the CMB: $p = 10\%$. 
 
\begin{figure}[!h]
\begin{center}
	\includegraphics[scale=0.15]{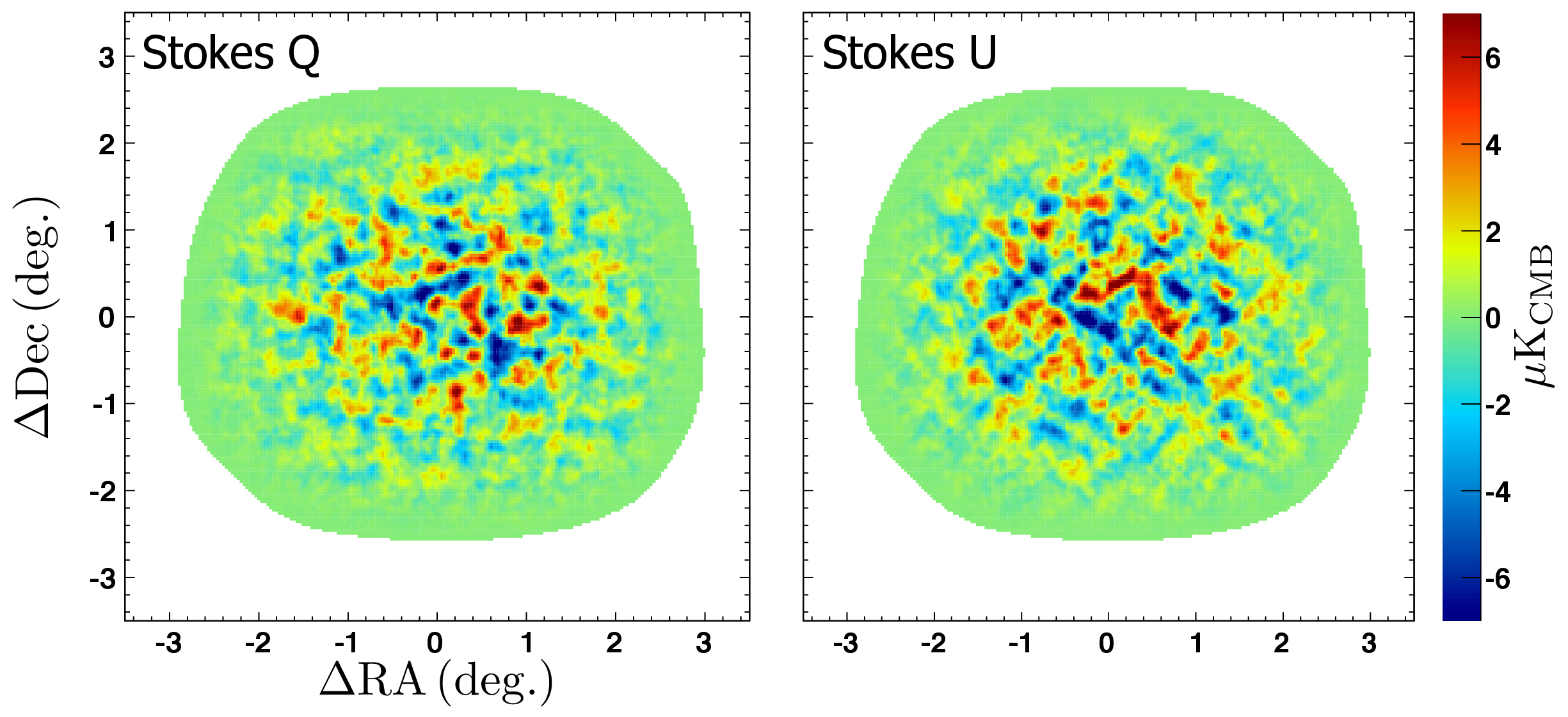}
	\caption{$Q$ and $U$ maps from \cite{Polarbear_2014}.}
	\label{fig:qu}
\end{center}
\end{figure}

Being linearly polarised, the CMB polarisation can be described only by the two Stokes parameters $(Q,U)$ introduced in the Chapter~\ref{Chapter1}. The Figure~\ref{fig:qu} shows maps of the observed CMB $Q$ and $U$ Stokes parameter derived from the POLARBEAR CMB observations in \cite{Polarbear_2014}. 

\section{Statistics of CMB Polarisation}

The CMB polarisation is a stupendous property of the primordial Universe because it traces the density perturbations and the primordial gravitational waves amplitude. In the same manner as temperature, the polarisation angular power spectra would be a useful tool directly extract the cosmological information. Nonetheless, the $(Q,U)$ Stokes parameters define a spin-$(\pm 2)$ field and therefore depend on a change of coordinate system: constructing their power spectra is consequently intricate but doable as in \cite{Melchiorri_1997}. Constructing temperature-like quantity from $(Q,U)$ parameters could avoid this complexity. I will expose how to build such temperature-like quantities, which are the so-called $E$ and $B$ modes, firstly in the harmonic domain and secondly in the real domain. The CMB polarised power spectra are then easily built from these scalar expressions and show peculiar feature that I will develop later.

\subsection{Harmonic approach: $E$ and $B$ modes}

In this section, I will detail the main steps to derive the harmonics of the scalar fields deduced from the $(Q,U)$ Stokes parameters following \cite{Zaldarriaga_1997}. 

From the $(Q,U)$ Stokes parameters, the polarisation fields $P_{\pm 2}$ along the line of sight $\vec{n}$ that encompass all the information about polarisation are defined: 
\begin{equation}
P_{\pm 2}(\vec{n}) = Q(\vec{n}) \pm i U(\vec{n}).
\end{equation}
The equation~(\ref{eq:stokesrot}) in Chapter~\ref{Chapter1} shows that, under a rotation of an angle $\Psi$, such a field $P_{\pm 2}$ is transformed into $P'_{\pm 2}$ following: 
\begin{equation}
P_{\pm 2}'(\vec{n}) = e^{\pm2i\Psi}P_{\pm 2}(\vec{n}).
\end{equation}
This is why the polarisation field is by definition a spin-$(\pm2)$ field. It thus can be decomposed over the basis of the spin spherical harmonics: 
\begin{equation}
P_{\pm 2}(\vec{n}) = \sum_{\ell m} {}_{\pm 2}a_{\ell m}~{}_{\pm 2}Y_{\ell m}(\vec{n}),
\end{equation}
or equivalently:
\begin{equation}
{}_{\pm 2}a_{\ell m} = \int d\Omega~{}_{\pm 2}Y_{\ell m}(\vec{n})^*P_{\pm 2}(\vec{n}).
\label{eq:multipole_p2}
\end{equation}

The spin-raising $\partial$ and spin-lowering $\bar{\partial}$ operators, built from the derivatives on the sphere, respectively increase or decrease the spin of a unity as shown in~\cite{Goldberg_1967} and~\cite{Zaldarriaga_1997}. The spin-$(\pm 2)$ spherical harmonics, ${}_{\pm 2}Y_{\ell m}$, are linked to the standard spherical harmonics following:
\begin{eqnarray}
{}_{2}Y_{\ell m} & = & \frac{1}{\sqrt{\alpha_{\ell}}}\partial\partial Y_{\ell m}, \nonumber \\ 
{}_{-2}Y_{\ell m} & = & \frac{1}{\sqrt{\alpha_{\ell}}}\bar{\partial}\bar{\partial} Y_{\ell m},
\end{eqnarray}
with $\alpha_{\ell} = \sqrt{\frac{(\ell+2)!}{(\ell-2)!}} $

A spin-0 quantity can thus be deduced from the spin-$(\pm 2)$ polarisation field multipoles in Eq.~\ref{eq:multipole_p2} using twice the spin-raising $\partial$ and spin-lowering $\bar{\partial}$ operators and by integrating by part:
\begin{eqnarray}
{}_{2}a_{\ell m} & = \frac{1}{\alpha_{\ell}} \int d\Omega Y_{\ell m}^*(\vec{n}) ~\bar{\partial}\bar{\partial} P_{2}(\vec{n}), \\
{}_{-2}a_{\ell m} & = \frac{1}{\alpha_{\ell}} \int d\Omega Y_{\ell m}^*(\vec{n}) ~\partial\partial P_{-2}(\vec{n}). 
\label{eq:spinharm}
\end{eqnarray}

From the spin multipole coefficients ${}_{\pm 2}a_{\ell m}$, two new relevant multipoles are introduced:
\begin{eqnarray}
a^E_{\ell m} & = & -\frac{1}{2}[~{}_{2}a_{\ell m} + {}_{-2}a_{\ell m}~], \nonumber \\ 
a^B_{\ell m} & = & \frac{i}{2}[~{}_{2}a_{\ell m} - {}_{-2}a_{\ell m}].
\end{eqnarray}

The $E$ and $B$ multipoles are characterised by their behaviour under parity change. The Eq.~(\ref{eq:1.14}) of Chapter~\ref{Chapter1} recalls the expression of the $(Q,U)$ Stokes parameters. If the coordinate system $(\vec{e_x}$,$\vec{e_y})$ undergoes a parity transformation in $(\vec{e_x}',\vec{e_y}') = (\vec{e_x}$,$-\vec{e_y}$), then the Stokes parameters are straightforwardly expressed as $Q' = Q$ and $U' = -U$. Consequently, the $E$ multipoles $a^E_{\ell m}$ are not changed under this transformation while the $B$ modes multipoles $a^B_{\ell m}$ become $-a^B_{\ell m}$. The behaviour of $E$ and $B$ multipoles under parity change is the reason for their denomination recalling the electric and magnetic field properties. 

The built $E$ and $B$ multipoles, $a_{\ell m}^{E}$ and $a_{\ell m}^B$ respectively, are the coefficients of the scalar $E$ and $B$ modes fields: 
\begin{eqnarray}
E(\vec{n}) & = & \sum_{\ell m} a^E_{\ell m} Y_{\ell m}(\vec{n}), \nonumber \\
B(\vec{n}) & = & \sum_{\ell m} a^B_{\ell m} Y_{\ell m}(\vec{n}).
\end{eqnarray}

Alternatively, the $E$ and $B$ modes have a unequivocal correspondence with respectively divergent- and curl-like quantity as shown in \cite{Kamionkowski_1997}. As a result, the decomposition in the harmonic domain of the spin-$(\pm 2)$ polarisation field allows the construction of the scalar quantities $E$ and $B$.

\subsection{Real space approach: $\chi^E$ and $\chi^B$ fields}

Otherwise, it is possible to adopt a real space approach to construct scalar quantities from the $(Q,U)$ Stokes parameters as in \cite{Zaldarriaga_1997}. To this purpose, the spin-raising and spin-lowering operators are applied twice to the polarisation field, thus defining two new scalar quantities denoted $\chi^{E/B}$ related to the $E/B$ modes:  
\begin{eqnarray}
\chi^E(\vec{n}) & = & -\frac{1}{2} [\bar{\partial}\bar{\partial} P_{2}(\vec{n}) + \partial \partial P_{-2}(\vec{n})],  \\ \nonumber
\chi^B(\vec{n}) & = & \frac{i}{2} [\bar{\partial}\bar{\partial} P_{2}(\vec{n})	- \partial \partial P_{-2}(\vec{n})].
\end{eqnarray}

The $\chi^{E/B}$ fields are thus scalar maps which contain all about the polarisation information. Moreover, they give a local description of the polarisation field. 

Using the previous equations along with Eqs.~(\ref{eq:spinharm}), it can be easily shown that the $\chi^{E/B}$ field is directly related to the $E/B$ modes in an unambiguous way by: 
\begin{eqnarray}
\chi^E & = & \sum_{\ell m} \alpha_{\ell} a^E_{\ell m} Y_{\ell m}, \\
\chi^B & = & \sum_{\ell m} \alpha_{\ell} a^B_{\ell m} Y_{\ell m}.
\end{eqnarray}

Therefore, the $\chi^{E/B}$ fields are equivalent to the $E$ and $B$ modes as derived in the harmonic domain, though $\chi^{E/B}$ field power spectra differ from the $E$ and $B$ ones by a factor of $\sim \ell^4$.

As a consequence, although the observables of the polarisation field are the $(Q,U)$ Stokes parameters, the decomposition in $E$ and $B$ modes in the harmonic space or in $\chi^E$ and $\chi^B$ fields in the real space is more convenient. Indeed, these quantities are scalars that enable to build the polarisation power spectra. In addition, the decomposition in $E$ and $B$ fields has also an underlying benefit: they are directly related to the physics of the primordial universe as described in the following section.

\subsection{$E$ and $B$ modes physical interpretations}

In this section, we will consider the $E$ and $B$ modes as constructed in the harmonic space albeit the same conclusions can be drawn for the $\chi^{E/B}$ fields. The main issue for interpreting the $E$ and $B$ modes is that they are not locally related to the $(Q,U)$ Stokes parameters. We cannot deduce the value for $E$ or $B$ modes on a given pixel from the observed $(Q,U)$ parameters on the same pixel. Nonetheless, the $E$ and $B$ modes have characteristic polarisation patterns. As shown above, the $E$ modes are an even quantity therefore the corresponding polarisation pattern should be also parity invariant. On the contrary, the $B$ modes are an odd quantity as their sign change under a parity transformation. 

From previous considerations, we have acknowledged that the scalar perturbations always produce a symmetric polarisation pattern, so they only account for the even $E$ modes. The figure~\ref{fig:gw} shows that the gravitational waves are partly invariant \textit{and} partly variant under a parity transformation. In the reference frame $(O,\vec{e}_1,\vec{e}_2)$ in Fig.~\ref{fig:gw}, the $+$ polarisation gravitational waves displayed in the upper panel are indeed even while the $\times$ polarisation gravitation waves on the lower panel are odd. The even gravitational waves can thus induce both temperature ($T$ modes) and $E$ modes. The odd gravitational waves can however only generate $B$ modes pattern in the CMB, unlike scalar perturbations. In other words, the $B$ modes are only a signature of the tensor perturbations.

However, from the last scattering surface, the CMB photons have crossed gravitational potentials implying that they are deflected, which results in $E$ modes deformation. The distorted $E$ modes behave like $B$ modes and are thus called the \textit{lensed $B$ modes}. Fortunately, this only affects the small angular scales of the $B$ modes pattern while the primordial signal in $B$ modes is expected to be predominant on the largest scales. 

Thus, the large scales $B$ modes are a powerful probe of the primordial universe as they are a signature of the primordial gravitational waves.
The latter also affect the temperature and $E$ modes anisotropies power spectra. Nonetheless, the contribution from the scalar perturbations overwhelms the tensor perturbations presumed to be low. Thus, the $B$ modes power spectra is a key quantity to target the primordial universe. 

\subsection{Polarisation power spectra}

The CMB temperature and polarisation power spectra are built thanks to the scalar description of the polarisation field. Similarly to the temperature, the polarised power spectra are defined by: 
\begin{eqnarray}
<a_{\ell m}^Ta_{\ell' m'}^{T *}> = C^{TT}_{\ell}\delta_{\ell \ell'}\delta_{m m'},  \nonumber \\
<a_{\ell m}^Ea_{\ell' m'}^{E *}> = C^{EE}_{\ell}\delta_{\ell \ell'}\delta_{m m'}, \nonumber \\
<a_{\ell m}^Ba_{\ell' m'}^{B *}> = C^{BB}_{\ell}\delta_{\ell \ell'}\delta_{m m'},  \nonumber \\
<a_{\ell m}^Ta_{\ell' m'}^{E *}> = C^{TE}_{\ell}\delta_{\ell \ell'}\delta_{m m'}, \nonumber\\ 
<a_{\ell m}^Ta_{\ell' m'}^{B *}> = C^{TB}_{\ell}\delta_{\ell \ell'}\delta_{m m'},  \nonumber \\
<a_{\ell m}^Ea_{\ell' m'}^{B *}> = C^{EB}_{\ell}\delta_{\ell \ell'}\delta_{m m'}.
\end{eqnarray}
with $a^{T/E/B}_{\ell m}$ the coefficients of the decomposition of the $T$,$E$ and $B$ modes on the spherical harmonics $Y_{\ell m}$ and ${}_{\pm 2}Y_{\ell m}$. 
We point out that the $\chi^{E/B}$ can also be used to build the power spectra but, as mentioned upwards, the $E$ and $B$ modes power spectra are more convenient. Furthermore, as the $E$ modes are even and the $B$ modes are odd, the $TB$ and $EB$ cross-correlations are expected to vanish in the standard model of cosmology as the universe is parity invariant. It is worth reminding that although the $B$ modes are odd, their power spectrum is even because it involves squared quantities.

\begin{figure}[!h]
\begin{center}
	\includegraphics[scale=0.5]{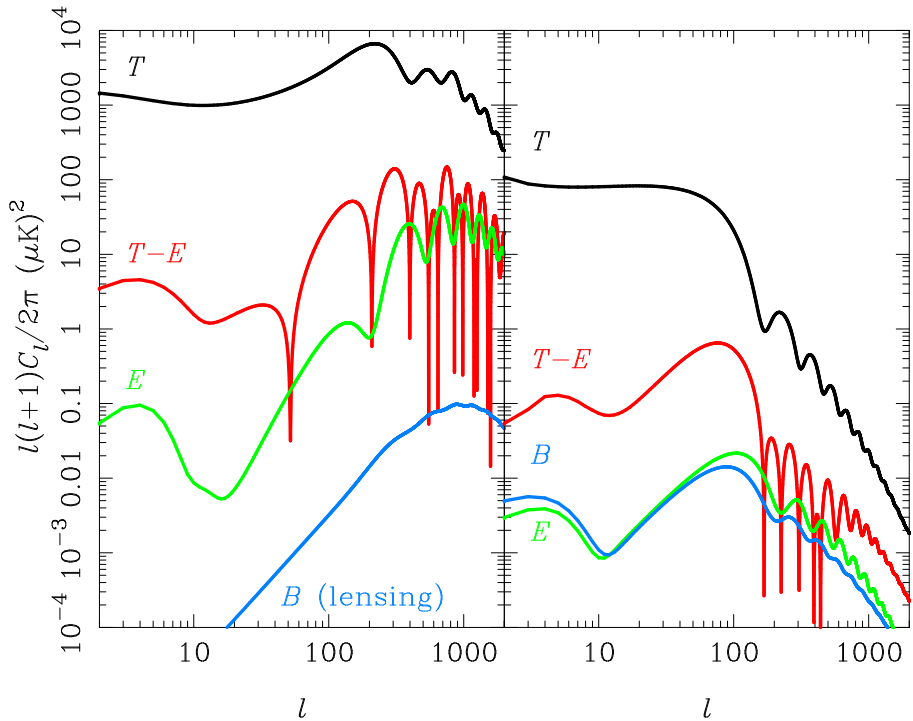}
	\caption{Scalar (tensor) contributions to the power spectra $T$, $E$ and $B$ modes and to the $TE$ correlations on the left (right) panel (from \cite{Challinor_2013}).}
	\label{fig:cl}
\end{center}
\end{figure}

The figure~\ref{fig:cl} shows the scalar (left panel) and tensor (right panel) part of the temperature as well as the polarised power spectra, from \cite{Challinor_2013}. The temperature is, as expected, at least an order of magnitude higher than the polarisation. Furthermore, the temperature and the $E$ modes are anti-correlated because of the $E$ modes amount for the velocity gradient in the primordial plasma, whereas the $T$-modes are only sensitive to the velocity itself. Moreover, the bump at $\ell \lesssim 10$ in the polarised power spectra is the signature of a second scattering process during the reionisation, the formation of the first stars. Besides, the $B$ modes power spectra is decomposed on its expected primordial tensor part, peaking at low $\ell$, and its lensing scalar part, dominating at $\ell \sim 1000$. In the end, the detected power spectra are the sum of the two, scalar and tensor, contributions. Nevertheless, the temperature, $E$ modes power spectra and $TE$ correlations are dominated by the scalar contribution which is at least one order of magnitude higher than the tensor contribution, as clearly shown on Fig.~\ref{fig:cl}. It establishes the $B$ modes as a unique signature of the tensor perturbations at low $\ell$. 

Furthermore, the angular power spectra $C_{\ell}$ directly depend on the primordial power spectra $P(k)$. The peculiar case of the temperature power spectrum of the Eq.~(\ref{eq:pktocell}) formula is generalized following: 
\begin{equation} 
C_{\ell}^{XY} = \displaystyle\int dk \left[ {\Delta_{\ell,\sa}^{X}}(k,\eta_0){\Delta_{\ell,\sa}^{Y}}(k,\eta_0)\pk_\sa(k) + 
				\Delta_{\ell,\te}^{X}(k,\eta_0)\Delta_{\ell,\te}^{Y}(k,\eta_0)\pk_\te(k)			\right],
\label{eq:pktocellpolar}
\end{equation} 
where $X$ and $Y$ stand for $T$, $E$ or $B$ modes. The tensor perturbations being the only responsible for the $B$ modes existence, in this case we have the equation:
\begin{equation} 
C_{\ell}^{BB} = \displaystyle\int dk \left[{\Delta_{\ell,\te}^{B}}^2(k,\eta)\pk_\te(k)\right].
\label{eq:pktocellBB}
\end{equation} 
The tensor primordial power spectrum $\pk_\te(k)$ therefore drives the $B$ modes angular power spectrum shape which thus can be potentially used to constrain the tensor parameters $r$ or $n_{\te}$.

\underline{parameter constrains}

As just mentioned, the angular power spectra are sensitive to the cosmological parameters. The $E$ modes are caused by the same perturbations than the ones causing temperature fluctuations, they therefore offer a redundancy on their determination. However, they also enable to break degeneracy between cosmological parameters. In particular, the $E$ modes give an important constraint on the reionisation optical depth $\tau$ thus breaking the degeneracy between for instance $\tau$ and the scalar spectral index $n_\sa$.

The origin of the $B$ modes being the tensor perturbations, their detection at large angular scales would provide the best constraints on the tensor-to-scalar ratio $r$, which gives the scale of the energy scale of inflation. The temperature and $E$ modes power spectra are indeed not sufficient to set tight constraints on $r$ as the signal is overwhelmed by the higher scalar contribution as shown in Fig.~\ref{fig:cl}. Their detection can therefore only help to set upper bounds on $r$ since the tensor contribution is within the cosmic variance. Moreover, the $B$ modes power spectrum is also affected by the reionisation, it would then help to constrain $\tau$. At smaller scales, the detected $B$ modes power spectrum enables to check our knowledge on the gravitational potential field in the universe thus providing information about the great structures formation.  

The detection of the CMB $B$ modes is thus essential in modern observational cosmology for the study of the first instants of the universe. Numerous kinds of experiments aiming at detecting either its primordial or lensing part are ongoing or being developed. The low expected signal is however an instrumental challenge to overcome which requests a meticulous design.


\section{CMB Polarisation Detection}

\begin{figure}[!h]
\begin{center}
	\includegraphics[scale=0.5]{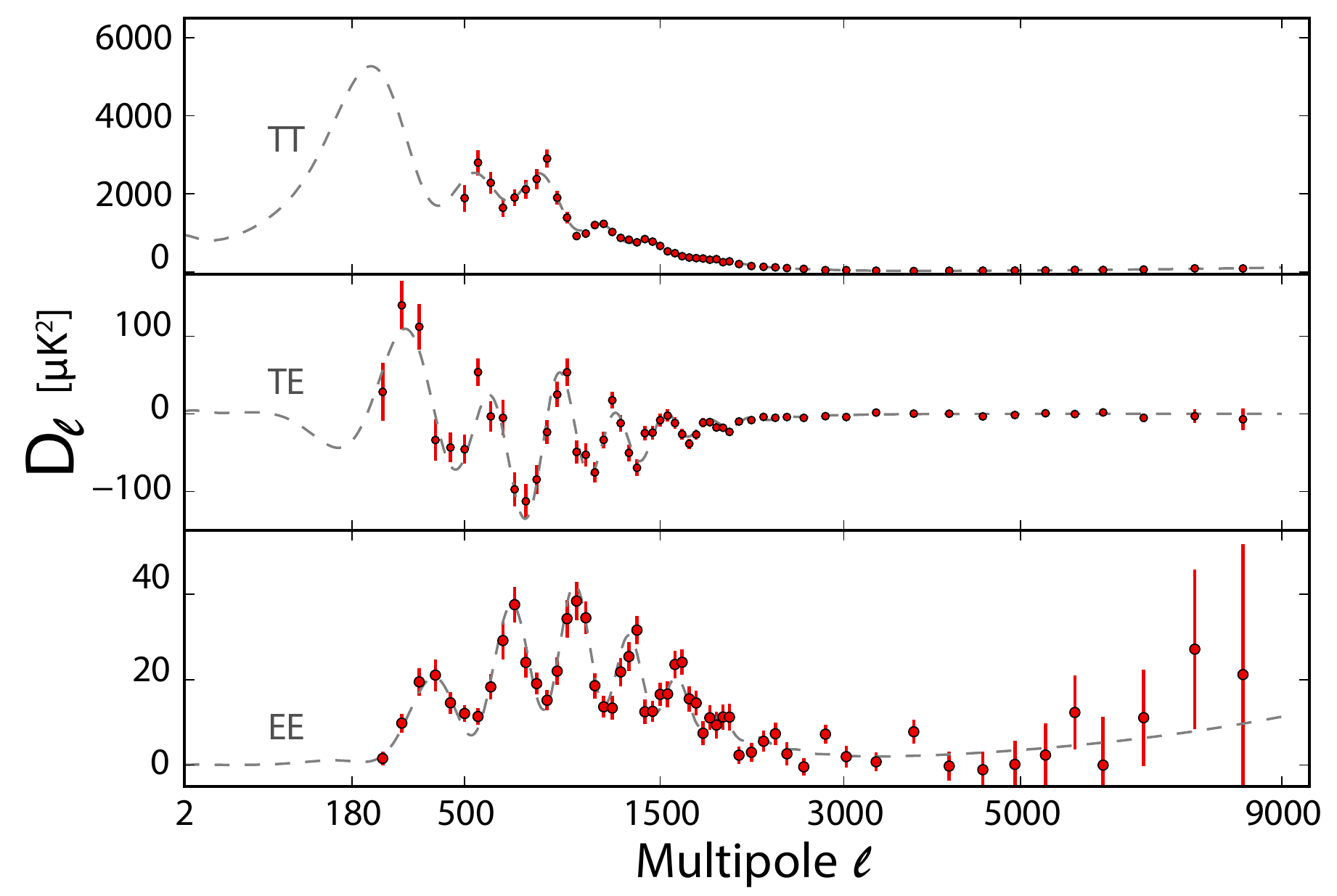}
	\caption{$TT$, $EE$ and $TE$ power spectra from the ACTPol experiment. Picture taken from \cite{ACTPol_2014}.}
	\label{fig:actpol}
\end{center}
\end{figure}

The era of CMB polarisation observation has begun with the first detection of the $E$ modes by \cite{DASI_2002}, \cite{DASI_2002b}. Ever since, important improvements have been made on the instrumental aspect as well as on the data analysis domain to answer the challenge that represents the CMB polarisation detection. 
A profusion of experiments has been designed to obtain the $E$ modes power spectrum. The {\sc ACTPol} collaboration has recently provided an accurate reconstruction of the temperature, $E$ modes and their correlations at small scales ($\ell \in [200;9000]$) as shown in Fig.~\ref{fig:actpol} from \cite{ACTPol_2014}.

\begin{figure}[!h]
\begin{center}
	\includegraphics[scale=0.04]{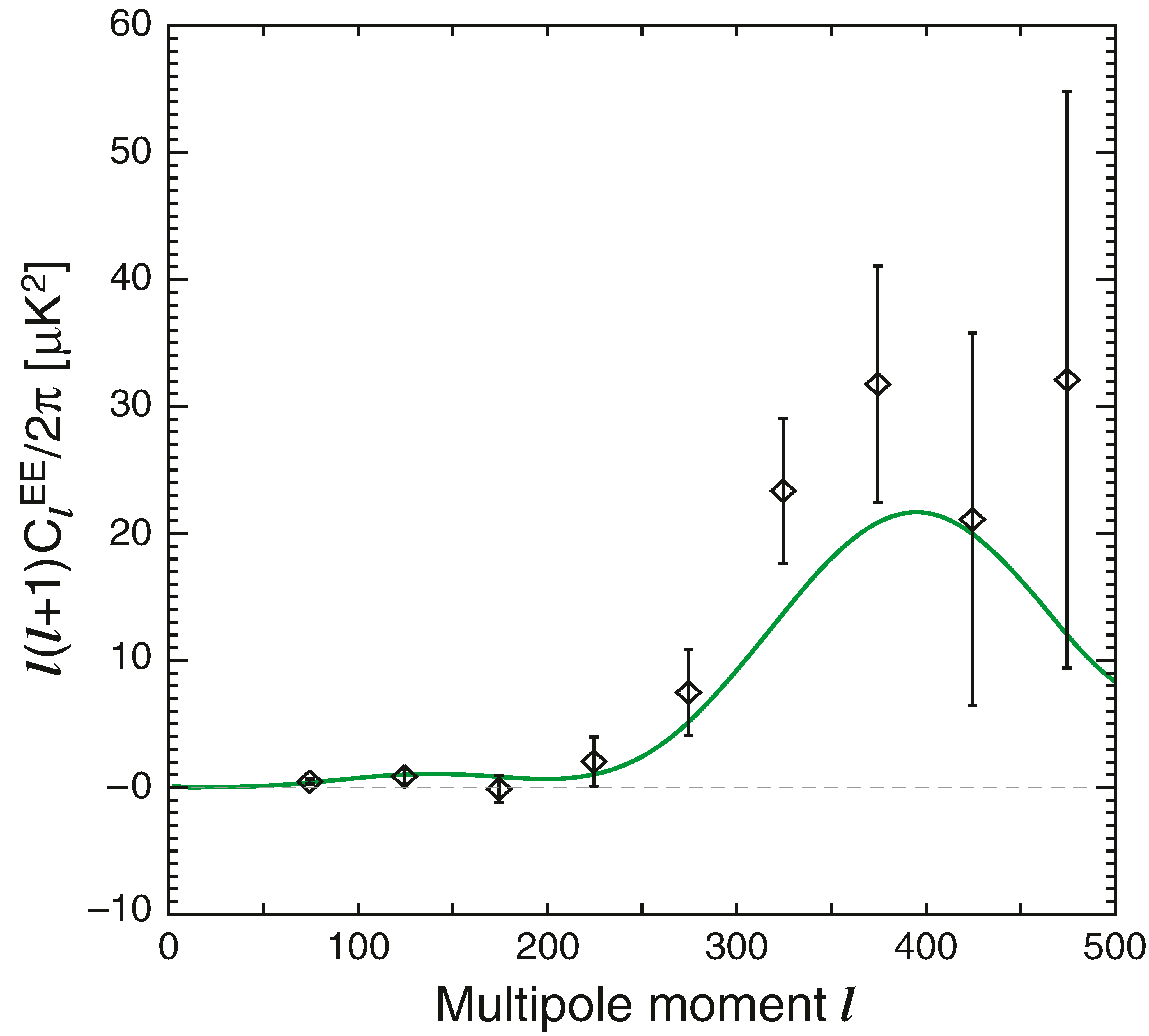}
	\caption{$E$ modes power spectrum from the 7-yr WMAP experiment (from \cite{larson_2011}.)}
	\label{fig:wmapee}
\end{center}
\end{figure}

The {\sc WMAP} satellite had access to the largest scales allowing for a reconstruction of the $E$ modes power spectrum for $\ell \in [26;500]$ as shown in Fig.~\ref{fig:wmapee}. The data from the Planck satellite are however expected to provide better constraints especially at the lower multipoles. 

\begin{figure}[!h]
\begin{center}
	\includegraphics[scale=0.3]{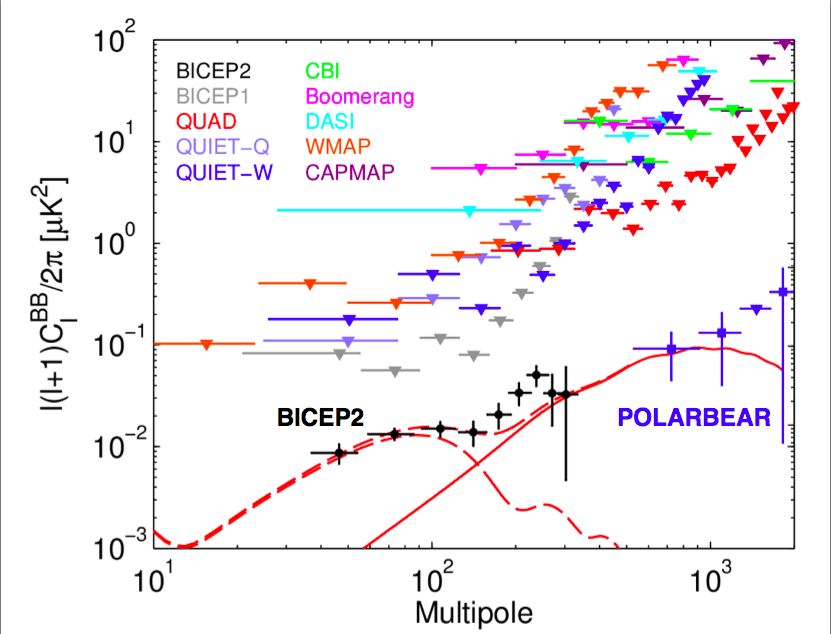}
	\caption{Earlier constraints and current measurements of the $B$ modes power spectrum.}
	\label{fig:Bconstrain}
\end{center}
\end{figure}

The $B$ modes power spectrum is a key issue for the completion of the standard model of cosmology. The motivation for its detection therefore leads the way to designing a large set of ground based spatial or balloon experiments. The figure~\ref{fig:Bconstrain} displays the past and current constraints on the $B$ modes power spectrum. The $B$ modes were still imperceptible to our instruments up to the beginning of 2014. Indeed, the POLARBEAR experiment has directly detected the lensed $B$-modes at small angular scales ($\ell \in [500;2100]$) for the first time in~\cite{Polarbear_2014}. The reconstructed lensed $B$ modes power spectrum is displayed on the figure~\ref{fig:Bconstrain} as blue points.

In March 2014, the BICEP2 team claimed the very first detection of the primordial $B$ modes in the multipole range of $\ell \in [30;150]$ in \cite{BICEP2_2014}. The corresponding data points are displayed in black on the Fig.~\ref{fig:Bconstrain}. This observational achievement sets the current best constraints on the tensor-to-scalar $r$ to $r = 0.2^{+0.07}_{-0.05}$. However, their results are controversial mainly due to the way they remove the foregrounds, that are thought to be underestimated (see \cite{Mortonson_2014} or \cite{Flauger_2014} for instance). These results, though impressive, should be regarded with precautions: confirmation or invalidation from the Planck satellite is expected in the next few months since it will provide a measurement of the foreground contamination in the BICEP2 field of view. Contrary to BICEP2 experiment that only observes at 150 GHz, the Planck satellite has nine frequency channels which enables to efficiently estimate the galactic foregrounds, and especially of the galactic dust emission thanks to the HFI instrument.

Observational and theoretical cosmology are consequently living an exciting era as for the first time, the primordial universe is \textit{directly} observable. This year 2014 is indeed a pivot in the $B$ modes detection and the obtained results are a good augur for the exploitation of the $B$ modes power spectrum. Furthermore, the current experiments dedicated to the $B$ modes detection among which POLARBEAR 2 (see \cite{POLARBEAR2_2012}) or the Keck array (see \cite{Keck_2012}) will be upgraded in the coming years. Also the QUBIC experiment (\cite{QUBIC_2011}) based on bolometric interferometry is one of the promising forthcoming $B$ modes experiments to be built. Moreover, $B$ modes observation over the full celestial sphere would give access to the crucial low multipoles, including the reionisation bump, of the $B$ modes power spectrum. Several spatial experiments such as LiteBird (which design is described in \cite{LiteBird_2013}) or a Core-like satellite (see \cite{COrE_2011}) are under studies to be proposed in the coming years to the spatial agencies.

\section*{Conclusion}

The $B$ modes detection is part of current observational cosmology challenges as they are a unique open window on the primordial universe physics. To answer these, a large set of experiments is currently acquiring data or being upgraded and some new experiments are being designed. However, the $B$ modes signal is low compare to the $T$ or $E$ modes. The uncertainties on its detection have therefore to be fully understood in order to set the best constraints on the primordial universe parameters. The noise level of the CMB experiments is lowering thanks to the instrumental improvements. The data analysis have therefore to be very accurate to take benefit of the high quality data taken by the instruments. Nonetheless, any kind of experiment provides noisy (but still at a low level), pixelised CMB maps covering only a small part of the celestial sphere. Crucial issues such as the $E$-to-$B$ leakage arise from these inherent experimental features and can ruin all the work of map making or foregrounds cleaning.

The following part~\ref{part2} is therefore dedicated to my work on accurate estimation of the $B$ modes power spectrum taking into account the complications in analysis of the CMB maps. In the part~\ref{part3}, I will expose the constraints on the primordial physics such as inflation, parity violation or magnetic field that one can put from the properly estimated $B$ modes power spectrum in the case of current and forthcoming CMB experiments.

\part{CMB Polarised Power Spectra Estimation}
\label{part2}

\chapter{Power Spectrum Estimation} 
\label{Chapter4} 

\lhead{Chapter 4. \textit{Power Spectrum Estimation}} 
\noindent \hrulefill \\
\textit{Cosmology is a science like no other: we study a unique realisation of a process, our only and unique Universe in which we live. The cosmic variance originates from this peculiar feature. Also, I like to see cosmology as a science of variances as the main information we have from the CMB is contained in the angular power spectrum, which is nothing less than the variance of the ${a}_{\ell m}$ distribution.}
\noindent \hrulefill \\


\section{A Brief Overview of Data Analysis}
 
In order to efficiently extract their cosmological information, the acquired CMB data undergo a long process which is described in \cite{Tristram_2007}. The current experiments dedicated to the CMB polarisation observation (such as the HFI instrument of Planck satellite for instance) are based on bolometers sensitive to polarisation. These detectors provide raw data which are then transformed in time-ordered data (TOD) removing parasitic signals from electronic or external sources such as the glitches. The TOD go through an important pipeline including a step which consists in subtracting the systematics such as thermal fluctuations or $1/f$ noise present in the TOD. Afterwards, the detectors features such as its gain, its pointing or its response to a point source are reconstructed. The next key step is the noise power spectrum estimation which has to be properly done to ensure the cleanliness of the data. The data analysis pipeline goes on with the map-making: the clean data are projected on a pixelised map of the sky. The pixelisation used in the present work is the common one: the HEALPIX pixelisation from \cite{Gorski_2005} in which the pixels have the same area and are distributed in iso-latitude rings. The obtained CMB maps thus contain the CMB signal along with the one from astrophysical foregrounds such as dust emission, a dominant foreground beyond $100 GHz$, or point sources emissions superposed to the CMB signal. Taking advantage of their spectral and spatial distributions, different methods are constructed to separate them from the CMB signal. However, the galactic emission -- at least its most intense part -- and the point sources are usually simply masked reducing the observed fraction of the sky. Once clean CMB maps are obtained, the very last step of the analysis is the power spectrum estimation which has to be properly carried out in order to profit from all the data analysis pipeline. 

The main goal of the present chapter is precisely to carefully estimate the CMB polarised power spectra on masked CMB maps which have gone through all the previous pipeline (including foregrounds subtraction). The standard way of reconstructing the CMB polarisation in the harmonic domain indeed leads to a problematic effect in the case of an incomplete sky coverage: the $E$-to-$B$ leakage. This effect can be disastrous for the CMB polarisation detection even if the previous pipeline provides the cleanest CMB $(I,Q,U)$ maps. Diverse methods which circumvent the $E$-to-$B$ leakage have been proposed in the literature in the recent years. During my PhD I choose to focus only on the ones based on the pseudospectrum estimation for they are expected to be fast in analysing the large amount of produced CMB maps. I therefore firstly give a brief overview on `cosmostatistics' paying particular attention to the pseudospectrum estimation in Sec.~\ref{sec:cosmostat}. Secondly, the section~\ref{sec:stand} is dedicated to the principle of the standard method to reconstruct the CMB polarisation, highlighting the $E$-to-$B$ leakage. Thirdly, the section~\ref{sec:opt} give hints on the optimal estimator which can finally be well approximated by pseudospectrum approaches. Three of them correcting for the leakage at all the order moments of the statistics are exposed in Sec.~\ref{sec:free}.   

All along the analysis, the binary mask (equal to 1 on observed pixels and 0 if not observed) will be denoted $M(\vec{n})$ while the window function, taking its values between 0 and 1 will be referred to as $W(\vec{n})$. 

  
\section{Harmonic Approach of the CMB Statistics}\label{sec:cosmostat}

The current and forthcoming experiments dedicated to the CMB detection are producing substantial amount of data: the Planck satellite has for instance provided TeraBytes of raw data to be analysed. In order to extract the cosmological information from the CMB maps, the usual procedure is to perform the data analysis in the harmonic domain instead of the pixel space. A map of $N_{pix}$ pixels ($N_{pix} \sim 10^5 (10^7)$ for small (large) scale experiment) is indeed equivalent to a set of about $\sqrt{N_{pix}}$ numbers in the harmonic domain which contains all the information assuming isotropy and gaussianity of the CMB anisotropies ($\ell_{max}$ $C_{\ell}$ with $\ell_{max} \sim \sqrt{N_{pix}}$).

The founding principle for the extraction of information from the CMB $(I,Q,U)$ maps in the harmonic domain is thus the construction of the power spectrum $C_{\ell}$. It takes advantage of the statistical properties of the CMB as its statistical isotropy. Nevertheless, the CMB being a non reproducible experiment, peculiar statistical features arise and affect the reconstruction of the power spectra. In order to present the main ideas of these features, I will first take the example of a perfect detection of the CMB anisotropies followed by a realistic detection.

\subsection{An ideal detection}

In this section, a perfect and ideal detection of the CMB temperature anisotropies -- \textit{i.e.} over all the celestial sphere and with no instrumental noise -- is assumed. 

Usually, a power spectrum is built from the Fourier transform of the signal. However, the CMB anisotropies potentially contain information at all scales on the celestial sphere, its accurate expansion is then made on the orthonormal basis on the sphere, equivalent to the Fourier transform: the spherical harmonics $Y_{\ell m}$. The CMB temperature and polarised anisotropies give:
\begin{eqnarray}
T(\vec{n}) & = & \sum_{\ell m} a_{\ell m}^T Y_{\ell m},  \nonumber \\
P_{\pm 2}(\vec{n}) & = & -\frac{1}{2} \sum_{\ell m} (a^E_{\ell m} \pm ia^B_{\ell m}) {}_{\pm 2}Y_{\ell m}(\vec{n}),
\label{eq:decomp}
\end{eqnarray}
with $\vec{n}$ the vector defining the line of sight direction and $P_{\pm 2}(\vec{n}) = (Q\pm i U)(\vec{n}) $.

All the statistics of the CMB anisotropies is consequently contain in the $a_{\ell m}^T$, $a^E_{\ell m}$ and $a^B_{\ell m}$ distribution. Due to the physics of the CMB anisotropies, their distribution is thought to be a Gaussian with zero mean. The cosmological information consequently lies in the variance of this process. The true CMB power spectrum $C_{\ell}^{XY}$ is then related to the multipole variance as: 
\begin{equation}
\left<{a}_{\ell m}^X{a}_{\ell' m'}^{Y*}\right> = C_{\ell}^{XY}\delta_{\ell \ell'}\delta_{m m'},
\label{eq:cell}
\end{equation}
with $X,Y$ either the temperature $T$ or the $E$ and $B$ modes. If $X = Y$, $C_{\ell}^{XX}$ is called the auto-spectrum and cross-spectrum if not. 

As explained in the Chapter~\ref{Chapter2}, the brackets $\left<.\right>$ stand for an ensemble average over all the possible realisations of the universe, or in other words over all the sets of multipoles $a_{\ell m}$. Carrying out such an ensemble average is equivalent to averaging over the CMB anisotropies observations seen by all possible observers in our universe. This operation obviously can not be achieved as we observe only one realisation of our Universe. Nevertheless, the ergodic principle enables to partly circumvent this observational limitation. Indeed, the continuous average over the possible realisations of our universe can be translated into a discrete arithmetic average on independent directions on the sky. The obtained quantity is the so-called \textit{estimator} of the CMB $XY$ power spectrum:  
\begin{equation}
\hat{C_{\ell}}^{XY} = \frac{1}{2\ell+1}\sum_{m=-\ell}^{\ell} {a}_{\ell m}^X{a}_{\ell m}^{Y*}.
\label{eq:est}
\end{equation} 

The estimator is thus the observable we have access to, the issue is now to know how it is related to the underlying \textit{true} CMB power spectrum $C_{\ell}^{XY}$. We would especially expect the estimator to coincide with the true power spectrum in a particular way. To answer this question, the statistics of the CMB power spectrum estimator have to be studied. 

First of all, using the above equations (\ref{eq:cell}) and (\ref{eq:est}), the first statistical moment of the CMB estimator $\hat{C}_{\ell}^{XY}$ is:
\begin{eqnarray}
\left<\hat{C_{\ell}}^{XY}\right> & = & \frac{1}{2\ell+1} \sum_{m=-\ell}^{\ell} \left<{a}_{\ell m}^X{a}_{\ell m}^{Y*}\right> \nonumber \\
				  	  & = & \frac{1}{2\ell+1} \sum_{m=-\ell}^{\ell} C_{\ell}^{XY} \delta_{\ell \ell'}\delta_{m m'}  \nonumber \\
					  & = & \frac{1}{2\ell+1}\times (2\ell+1) C_{\ell}^{XY} = C_{\ell}^{XY}. 
\label{eq:meanest}
\end{eqnarray}
The mean of the CMB power spectrum estimator is therefore the true CMB power spectrum itself: the estimator is said to be \textit{unbiased}. Consequently, the estimator of the CMB anisotropies we have access to does correctly reconstruct the true CMB power spectrum on average. However, an accurate reconstruction of the CMB anisotropies power spectrum can be ruined by the uncertainties on the CMB power spectrum estimator. 

Using Eq. (\ref{eq:est}), the full covariance of the CMB $XY$ power spectrum estimator $\hat{C}_{\ell}^{XY}$ is:
\begin{eqnarray}
\Sigma_{\ell_1 \ell_2}^{XY,X'Y'} & = & \left<\hat{C}_{\ell_1}^{XY}\hat{C}_{\ell_2}^{X'Y'}\right> - \left<\hat{C}_{\ell_1}^{XY}\right>\left<\hat{C}_{\ell_2}^{X'Y'}\right> \nonumber \\
								& = & \frac{1}{2\ell_1+1}  \frac{1}{2\ell_2+1} \sum\limits_{m_1=-\ell_1}^{\ell_1} 
								\sum\limits_{m_2=-\ell_2}^{\ell_2} \left<{a}_{\ell_1 m_1}^X{a}_{\ell_1 m_1}^{Y*} {a}_{\ell_2 m_2}^{X'}{a}_{\ell_2 m_2}^{Y'*}\right> - \left<\hat{C}_{\ell_1}^{XY}\right>\left<\hat{C}_{\ell_2}^{X'Y'}\right>. 
\label{eq:varest}
\end{eqnarray}

As the estimator is unbiased, the last term of the right-hand side of this equation (\ref{eq:varest}) is directly $C_{\ell_1}^{XY}C_{\ell_2}^{XY}$. The four-point correlation function in the first term of the right hand side of Eq. (\ref{eq:varest}) can be expressed thanks to Wick theorem (as the CMB anisotropies are a Gaussian process) as:
\begin{eqnarray}
\left<{a}_{\ell_1 m_1}^X{a}_{\ell_1 m_1}^{Y*} {a}_{\ell_2 m_2}^{X'}{a}_{\ell_2 m_2}^{Y'*}\right> & = & \left<{a}_{\ell_1 m_1}^X {a}_{\ell_1 m_1}^{Y*}\right>\left< {a}_{\ell_2 m_2}^{X'}{a}_{\ell_2 m_2}^{Y'*}\right>   \nonumber \\
& + & \left<{a}_{\ell_1 m_1}^X{a}_{\ell_2 m_2}^{X'}\right>\left<{a}_{\ell_1 m_1}^{Y*}{a}_{\ell_2 m_2}^{Y'*}\right> \nonumber \\
& + & \left<{a}_{\ell_1 m_1}^X{a}_{\ell_2 m_2}^{Y'*}\right>\left<{a}_{\ell_1 m_1}^{Y*}{a}_{\ell_2 m_2}^{X'}\right>.
\end{eqnarray}

By definition of the unbiased power spectrum estimator, the first term of this summation gives: 
\begin{eqnarray}
\frac{1}{2\ell_1+1}  \frac{1}{2\ell_2+1} \sum\limits_{m_1 = -\ell_1}^{\ell_1} \sum\limits_{m_2 = -\ell_2}^{\ell_2} \left<{a}_{\ell_1 m_1}^{X} {a}_{\ell_1 m_1}^{Y*}\right>\left<{a}_{\ell_2 m_2}^{X'}{a}_{\ell_2 m_2}^{Y'*}\right>   \nonumber \\
= \frac{1}{2\ell_1+1}  \frac{1}{2\ell_2+1}(2\ell_1+1)(2\ell_2+1) C_{\ell_1}^{XY} C_{\ell_2}^{X'Y'} = C_{\ell_1}^{XY} C_{\ell_2}^{X'Y'}.
\end{eqnarray}
This term therefore cancels with the last term of the right-hand side of equation~(\ref{eq:varest}). Besides, making use of the property of the multipole stating that $a_{\ell (-m)} = (-1)^m a_{\ell m}^*$ and noticing that the summation is symmetric, the second term of the four-point correlation function expression can be written as:
\begin{eqnarray}
& & \frac{1}{2\ell_1+1}  \frac{1}{2\ell_2+1} \sum\limits_{m_1 = -\ell_1}^{\ell_1} \sum\limits_{m_2 = -\ell_2}^{\ell_2} \left<{a}_{\ell_1 m_1}^X{a}_{\ell_2 m_2}^{X'}\right>\left<{a}_{\ell_1 m_1}^{Y*}{a}_{\ell_2 m_2}^{Y'*}\right> \nonumber \\
& = & \frac{1}{2\ell_1+1}  \frac{1}{2\ell_2+1} \sum\limits_{m_1 = -\ell_1}^{\ell_1} \sum\limits_{m_2 = -\ell_2}^{\ell_2} C_{\ell_1}^{XX'}\delta_{\ell_1 \ell_2} \delta_{m_1 m_2} C_{\ell_2}^{YY'} \delta_{\ell_1 \ell_2} \delta_{m_1 m_2} \nonumber \\
& = & C_{\ell_1}^{XX'}C_{\ell_2}^{YY'} \delta_{\ell_1 \ell_2}.
\label{eq:term1}
\end{eqnarray}

Additionally, proceeding as for the first term, the third term of the four-point correlator can be simplified into: 
\begin{eqnarray}
&  & \frac{1}{2\ell_1+1}  \frac{1}{2\ell_2+1}  \sum\limits_{m_1 = -\ell_1}^{\ell_1} \sum\limits_{m_2 = -\ell_2}^{\ell_2} \left<{a}_{\ell_1 m_1}^X{a}_{\ell_2 m_2}^{Y'*}\right>\left<{a}_{\ell_1 m_1}^{Y*}{a}_{\ell_2 m_2}^{X'}\right> \nonumber \\
& = & \frac{1}{2\ell_1+1}  \frac{1}{2\ell_2+1}  \sum\limits_{m_1 = -\ell_1}^{\ell_1} \sum\limits_{m_2 = -\ell_2}^{\ell_2} C_{\ell_1}^{XY'} \delta_{\ell_1 \ell_2}\delta_{m_1 m_2} C_{\ell_1}^{YX'} \delta_{\ell_1 \ell_2}\delta_{m_1 m_2} \nonumber \\
& = & C_{\ell_1}^{XY'}C_{\ell_2}^{YX'}\delta_{\ell_1\ell_2}.
\label{eq:term2}
\end{eqnarray}

Thus, adding Eqs.~(\ref{eq:term1}) and~(\ref{eq:term2}), the generic expression for the covariance of the CMB $XY$ power spectrum estimator results in:
\begin{eqnarray}
\Sigma_{\ell_1 \ell_2}^{XY,X'Y'} = \frac{C_{\ell_1}^{XX'}C_{\ell_2}^{YY'} + C_{\ell_1}^{XY'} C_{\ell_2}^{YX'}}{2\ell_1+1}  \delta_{\ell_1 \ell_2}, 
\label{eq:cosmicvar}
\end{eqnarray}
the covariance is therefore diagonal. 

As an example, the covariance of the $B$ modes power spectrum estimator has the following expression: 
\begin{equation}
\Sigma_{\ell_1 \ell_2}^{BB,BB} = \frac{2 \delta_{\ell_1 \ell_2}}{2\ell_1+1}  {C_{\ell_1}^{BB}}^2.
\end{equation}

The full covariance of the CMB $XY$ power spectrum estimator is therefore depending on the true CMB power spectrum itself. This unusual behaviour is owing to the lack of independent directions to average on at large scales: the lower is $\ell$, the higher is the covariance of the estimator. In other word, the $a_{\ell m}$ distribution for a given $\ell$ is not sampled enough for low $\ell$ to effectively reconstruct its true distribution. The resulting covariance is called the \textit{cosmic variance}.

As a consequence, the estimator $\hat{C}_{\ell}^{XY}$ of the CMB anisotropies is unbiased but its reconstruction accuracy is limited by the inherent cosmic variance. Besides, the assumption of a perfect detection is ideal: the construction of the CMB power spectrum estimator has to be adjusted in the presence of experimental effects.

\subsection{A noisy sky}

In this section, the detection is supposed to be made over all the celestial sphere but with an instrumental noise. The CMB anisotropies observations are thus altered by irreducible experimental effects. Two main effects will be considered: the instrumental noise and the experimental main beam function, whose angular power spectrum can be derived. As shown in \cite{Tegmark_1997}, the multipole $b_{\ell}$ of a Gaussian beam function is indeed well approximated by:
\begin{equation}
b_{\ell} = e^{\theta_b^2\ell(\ell+1)/2},
\end{equation}
with $\theta_b$ the standard deviation of the Gaussian beam.

Furthermore, as derived in \cite{Kamionkowski_1997} and \cite{Zaldarriaga_1997}, a uniform instrumental noise can be considered as an additional random field with a power spectrum such as:
\begin{equation}
N_{\ell} = \frac{4\pi \sigma^2}{N_{pix}}, 
\end{equation}
with $\sigma$ the root mean square of the noise in each of the $N_{pix}$ pixels.

In this frame, the estimator of the CMB $XY$ power spectrum is changed regarding the previous ideal case and is given by:
\begin{equation}
\hat{C}_{\ell}^{XY} = \frac{1}{2\ell+1} \sum_m ({b_{\ell}a_{\ell m}^X} + {n_{\ell m}^{X}}) (b_{\ell}a_{\ell m}^Y + n_{\ell m}^{Y})^*,
\label{eq:noisy_est}
\end{equation}
with $n_{\ell m}$ the harmonic coefficients of the noise. 
Under the realistic assumption of uncorrelated CMB signal and noise, the correlator $\left<a_{\ell m}^X n_{\ell m}\right>$ cancels. As the signal estimator is unbiased (see previous section), the mean of the power spectrum estimator is therefore simplified into:
\begin{eqnarray}
\left<\hat{C}_{\ell}^{XY}\right> =  b_{\ell}^2C_{\ell}^{XY} + N_{\ell}^{XY},
\end{eqnarray}
with $N_{\ell}$ defined as: $N_{\ell} = \frac{1}{2\ell + 1} \sum\limits_m \left< n_{\ell m} n_{\ell m}^*\right>$. 

If the noise and beam power spectra are known, the estimator thus can be debiased. 

As previously computed, the variance of the estimator $\hat{C}_{\ell}^{XY}$ is expressed as: 
\begin{eqnarray}
\Sigma_{\ell\ell'}^{XY,X'Y'} = \frac{\delta_{\ell\ell'} }{2\ell+1} [(C_{\ell}^{XX'}+\frac{N_{\ell}^{XX'}}{b_{\ell}^2})(C_{\ell}^{Y'Y'}+\frac{N_{\ell}^{YY'}}{b_{\ell}^2}) \nonumber \\
+ (C_{\ell}^{X'Y}+\frac{N_{\ell}^{X'Y}}{b_{\ell}^2})(C_{\ell}^{XY'}+\frac{N_{\ell}^{XY'}}{b_{\ell}^2})].
\end{eqnarray}

The estimator of the total CMB power spectrum is therefore unbiased and its variance is a combination of the cosmic variance, noise and beam effects. Nonetheless, an unavoidable issue still have to be taken into account: the partial sky coverage of any CMB experiment.

\subsection{A masked sky}

The observations of the CMB anisotropies on the whole celestial sphere are in practice infeasible. In the case of suborbital experiments (ground-based or balloon borne), \cite{Jaffe_2000} indeed showed that only few percent of the sky are required for an optimal $B$ modes detection, for a fixed time of observation and sensitivity. 

Besides, a carefully chosen mask -- choose as a compromise between foregrounds contamination removal and loss of statistics -- is generally applied to the CMB maps provided by a satellite mission. In this case, the effective sky coverage can fall from the expected $100\%$ to $\sim 70 \%$ of the sky. We intuitively expect this effect to be dramatic for the CMB power spectrum reconstruction especially for the large angular scales. The estimator introduced in equation \ref{eq:noisy_est} can be debiased from this effect. An approximation of the resulting covariance matrix is obtained by simply counting the accessible modes and neglecting the correlations between modes. Thus the covariance matrix writes:
\begin{eqnarray}
\Sigma_{\ell\ell'}^{XYX'Y'} = \frac{\delta_{\ell\ell'} }{(2\ell+1)f_{\mathrm{sky}}} [(C_{\ell}^{XX'}+\frac{N_{\ell}^{XX'}}{b_{\ell}^2})(C_{\ell}^{YY'}+\frac{N_{\ell}^{YY'}}{b_{\ell}^2}) \nonumber \\
+ (C_{\ell}^{X'Y}+\frac{N_{\ell}^{X'Y}}{b_{\ell}^2})(C_{\ell}^{XY'}+\frac{N_{\ell}^{XY'}}{b_{\ell}^2})].
\label{eq:sampl}
\end{eqnarray}
This variance will be refer to the \textit{mode-counting variance} in the following. The variance obtained without instrumental noise is called the \textit{sampling variance} and is thus written as Eq.~(\ref{eq:cosmicvar}) replacing $2\ell+1$ by $(2\ell+1)f_{\mathrm{sky}}$. As expected, the equation~(\ref{eq:sampl}) shows that the low $\ell$ range of the true CMB power spectrum is less precisely reconstructed in the case of a small scale survey.

This reasoning however does not explicitly take into account all the issues arose by a partial sky coverage. An appropriate estimator and the non idealised variance has to be derived to ensure a proper reconstruction of the true CMB power spectrum on a masked sky. 


\section{The \textit{standard} pseudospectrum approach}
\label{sec:stand}

In order to correctly reconstruct the true CMB power spectrum on a portion of the celestial sphere, an approach built on previous discussions, the so-called \textit{pseudospectrum} approach, is usually employed. In this section, I will first expose the principle of this standard approach in the case of the CMB temperature anisotropies. The standard approach consists in decomposing the \textit{masked} CMB anisotropies on the spherical harmonics and then correcting for these masking issues at the level of the power spectrum mean (see MASTER method in \cite{Hivon_2002}).  
The more difficult case of the CMB polarisation anisotropies will be treated afterwards highlighting the issue of the $E$-to-$B$ leakage.

\subsection{CMB temperature}

The coefficient of the masked temperature map decomposition on the basis of the spherical harmonics are called the \textit{pseudomultipoles}. They are expressed as: 
\begin{equation}
\tilde{a}_{\ell m}^T = \int_{\Omega} T(\vec{n}) M(\vec{n}) Y_{\ell m}^*(\vec{n})d\vec{n},
\end{equation}
with $\vec{n}$ the vector defining the line of sight direction and $\Omega$ the observed fraction of the sky (thus $\Omega < 4\pi$). $M$ is the binary mask which is 1 in the observed or kept-in-analysis pixels while 0 if unobserved. 

Using the decomposition of the true temperature anisotropies on the $Y_{\ell m}$ in equation (\ref{eq:decomp}), the temperature pseudomultipoles are related to the true temperature multipoles as: 
\begin{eqnarray}
\tilde{a}_{\ell m}^T & = & \int_{\Omega} \sum\limits_{\ell' m'} M(\vec{n}) a_{\ell' m'}^T Y_{\ell' m'} Y_{\ell m}^*d\vec{n}  
\nonumber \\
& = & \sum\limits_{\ell' m'}  K_{\ell m, \ell' m'} a_{\ell' m'}^T,
\label{eq:pseudostandalm}
\end{eqnarray}
with $K_{\ell m, \ell' m'} = \int_{\Omega} M(\vec{n}) Y_{\ell' m'} Y_{\ell m}^* d\vec{n}$, the convolution kernel only depending on the applied mask. It drives the induced coupling between the modes $(\ell,m)$ and $(\ell',m')$. The origin of such a convolution kernel can be understood as the $Y_{\ell m}$ not being orthogonal on a portion of the sphere. 

By correlating the pseudomultipoles, the temperature \textit{pseudospectrum} is built as: 
\begin{equation}
\tilde{C}_{\ell}^{TT} = \frac{1}{2\ell+1} \sum_{m} \tilde{a}_{\ell m}^T \tilde{a}_{\ell m}^{T*}.
\end{equation}

The pseudospectrum can be expressed as a function of the true multipoles thanks to the equation~(\ref{eq:pseudostandalm}) as: 
\begin{equation}
\tilde{C}_{\ell}^{TT} = \frac{1}{2\ell+1} \sum_{m} \sum_{\ell'' m''}  \sum_{\ell' m'} K_{\ell m,\ell' m'} K_{\ell m,\ell'' m''}^* a_{\ell' m'}^T  a_{\ell'' m''}^{T*}.
\end{equation}

In order to recover the underlying true CMB temperature multipoles, we therefore need to invert the convolution kernels for each $m$ and $m'$ for each multipole $\ell$ thus asking for heavy computations. However, carrying out an ensemble average on the pseudospectrum gives a more straightforward and computationally tractable relation:
\begin{equation}
<\tilde{C}_{\ell}^{TT}> = \sum_{\ell'} K_{\ell \ell'} C_{\ell'}^{TT},
\end{equation}
where the convolution kernel $K_{\ell \ell'}$ stands for: $K_{\ell \ell'} = \frac{1}{2\ell+1} \sum\limits_{m m'} |K_{\ell m \ell' m'}|^2$. Nonetheless, as we have underlined it in the section \ref{sec:cosmostat}, an ensemble average \textit{cannot} be performed on all the possible realisations of the Universe. As a consequence, we define an estimator $\hat{C}_{\ell}^{TT}$ of the true angular power spectrum $C_{\ell}^{TT}$, called the \emph{pseudo-$C_{\ell}$ estimator} as:
\begin{equation}
\tilde{C}_{\ell}^{TT} = \sum_{\ell'} K_{\ell \ell'} \hat{C}_{\ell'}^{TT}.
\end{equation}
The interest of the defined estimator is that it is unbiased, the previous expression indeed straightforwardly gives: 
\begin{equation}
<\hat{C}_{\ell}^{TT}> = C_{\ell}^{TT}.
\end{equation}

An estimator of the temperature pseudospectrum accounting for the issue of the partial sky coverage has consequently been built. In the realistic case of an experimental CMB detection, the estimator $\hat{C}_{\ell}^{TT}$ is defined by inverting the following system: 
\begin{equation}
\tilde{C}_{\ell}^{TT} = \sum_{\ell'} K_{\ell \ell'} b_{\ell'}^2 \hat{C}_{\ell'}^{TT} + N_{\ell},
\end{equation}
with $b_{\ell}^2$ and $N_{\ell}$ the beam and noise power spectrum respectively. 

The pseudospectrum $\tilde{C}_{\ell}^{TT}$ is debiased firstly from the noise by subtracting its power spectrum, secondly from the beam and finally from the mode mixing by inversion of the $K_{\ell \ell'}$ matrix. Besides, if the power spectrum estimator is built by correlating two sets of $a_{\ell m}^{T,(A)}$ and $a_{\ell m}^{T,(B)}$ coming from two different uncorrelated detectors or experiments $(A)$ and $(B)$, the noise correlation $\left<n_{\ell m}^{(A)}n_{\ell m}^{(B)}\right>$ cancels. Such a power spectrum estimator is called a \textit{cross-spectrum} (see \cite{Tristram_2005}).    
The same reasoning is usually applied to the case of polarisation, a more subtle issue as the polarisation field is a linear combination of the two Stokes parameters $Q$ and $U$. 

\subsection{CMB polarisation}

In the same way as for the temperature, the masked CMB polarisation field $M(\vec{n}) P_{\pm 2}$ can be decomposed on the spin-${\pm2}$ spherical harmonics leading to the definition of $E$ and $B$ modes pseudomultipoles:
\begin{eqnarray}
\tilde{a}_{\ell m}^E & = & -\frac{1}{2} \int_{\Omega} M(\vec{n}) [ P_{+2}(\vec{n}) {}_{+2}Y_{\ell m}^*(\vec{n}) + P_{-2}(\vec{n}) {}_{-2}Y_{\ell m}^*(\vec{n}) ]d(\vec{n}),  \nonumber \\
\tilde{a}_{\ell m}^E & = &  ~~\frac{i}{2} \int_{\Omega} M(\vec{n}) [ P_{+2}(\vec{n}) {}_{+2}Y_{\ell m}^*(\vec{n}) + P_{-2}(\vec{n}) {}_{-2}Y_{\ell m}^*(\vec{n}) ]d(\vec{n}).
\end{eqnarray}

We recall the expression for the expansion of the polarisation field $P_{\pm 2} = Q \pm i U$ on the spin-${\pm2}$ spherical harmonics: 
\begin{equation}
P_{\pm 2}(\vec{n}) = -\frac{1}{2} \sum_{\ell m} (a^E_{\ell m} \pm ia^B_{\ell m}) {}_{\pm 2}Y_{\ell m}.
\end{equation}

By making use of this expansion and appropriate use of the spin-${\pm2}$ spherical harmonics properties, the $E$ and $B$ modes pseudomultipoles are related to the true CMB polarisation multipoles $a_{\ell m}^{E/B}$ as: 
\begin{eqnarray}
\tilde{a}_{\ell m}^E & = & \sum\limits_{\ell' m'} [K_{\ell m, \ell' m'}^{\mathrm{std},+} a_{\ell' m'}^E ~~+ i K_{\ell m, \ell' m'}^{\mathrm{std},-} a_{\ell' m'}^B], \label{eq:pseudostdalm} \nonumber \\
\tilde{a}_{\ell m}^B & = & \sum\limits_{\ell' m'} [-i K_{\ell m, \ell' m'}^{\mathrm{std},-} a_{\ell' m'}^E + K_{\ell m, \ell' m'}^{\mathrm{std},+} a_{\ell' m'}^B],
\label{eq:standebpseudomultipole}
\end{eqnarray}
with $K_{\ell m,\ell' m'}^{\mathrm{std},\pm}$ the convolution kernels (also called the mixing matrices) defined as: 
\begin{eqnarray}
K_{\ell m, \ell' m'}^{\mathrm{std},\pm} = \int_{4\pi} M(\vec{n}) \left [ {}_{2}Y_{\ell' m'}^* {}_{-2}Y_{\ell m} \pm {}_{-2}Y_{\ell' m'}^* {}_{2}Y_{\ell m} \right ] d\vec{n}.
\end{eqnarray}

The expression of the $E$ and $B$ pseudomultipoles in Eq.~(\ref{eq:standebpseudomultipole}) shows that both the $E$ and $B$ modes contribute to the $E$ and $B$ pseudomultipoles. The spin-weighted spherical harmonics does not form an orthogonal basis on a portion of the sky. Besides, the amount of mixing between the $E$ and $B$ modes is driven by the convolution kernel $K_{\ell m,\ell' m'}^{\mathrm{std},\pm}$ (a more detailed interpretation of this effect is exposed in the section~\ref{sec:interpret}).    

This effect would potentially lead to a laborious recovery of the true $B$ ($E$) modes power spectrum if the $E$ ($B$) modes weighted by the mixing kernel are higher than the $B$ ($E$) modes. In practice, the $E$ modes amplitude is at least 10 times higher than the expected $B$ modes. Therefore a small fraction of leakage is sufficient to spoilt the $B$ modes signal. This problematic contribution of the $E$ modes to the $B$ modes signal is known as the \textit{$E$-to-$B$ leakage}. This masking effect can be corrected for on average. As for the temperature anisotropies, the correlation of the pseudomultipoles indeed provides the following $E$ and $B$ modes pseudospectra:
\begin{eqnarray}
\tilde{C}_{\ell}^{EE} & = & \frac{1}{2\ell+1} \sum\limits_{m} \tilde{a}_{\ell m}^E \tilde{a}_{\ell m}^{E*},  \nonumber \\
\tilde{C}_{\ell}^{BB} & = & \frac{1}{2\ell+1} \sum\limits_{m} \tilde{a}_{\ell m}^B \tilde{a}_{\ell m}^{B*}.
\end{eqnarray}

The relation between the defined $E$ and $B$ modes pseudospectra and their true power spectra appears by making use of the equation~(\ref{eq:pseudostdalm}) and performing an ensemble average:
\begin{eqnarray}
\left ( \begin{array}{c} \left<\tilde{C}_{\ell}^{EE}\right>  \\ \left<\tilde{C}_{\ell}^{BB}\right>  \end{array} \right ) = \sum\limits_{\ell'}
\left ( \begin{array}{cc} K_{\ell \ell'}^{\mathrm{std},+} & i K_{\ell \ell'}^{\mathrm{std},-} \\ -i K_{\ell \ell'}^{\mathrm{std},-} &  K_{\ell \ell'}^{\mathrm{std},+} \end{array} \right ) 
\left ( \begin{array}{c} C_{\ell'}^{EE} \\ C_{\ell'}^{BB}  \end{array} \right ),
\label{eq:pseudo2true}
\end{eqnarray}
with the convolution kernels $K_{\ell \ell'}^{\mathrm{std},\pm}$ defined as: 
\begin{equation}
K_{\ell \ell'}^{std,\pm} = \frac{1}{2\ell+1} \sum_{mm'} |K_{\ell m,\ell' m'}^{\mathrm{std},\pm}|^2.
\end{equation}

In the case of a realistic CMB detection, following the same procedure as for the temperature power spectrum estimator, the estimators $\hat{C}_{\ell}^{E/B}$ of the CMB polarisation power spectra are defined as: 
\begin{equation}
\left ( \begin{array}{c} \tilde{C}_{\ell}^{EE} \\ \tilde{C}_{\ell}^{BB} \end{array} \right )
= \sum_{\ell'} b_{\ell'}^2 \left ( \begin{array}{cc} K_{\ell \ell'}^{\mathrm{std},+} & i K_{\ell \ell'}^{\mathrm{std},-} \\ -iK_{\ell \ell'}^{\mathrm{std},-} & K_{\ell \ell'}^{\mathrm{std},+}  \end{array} \right ) \left ( \begin{array}{c} \hat{C}_{\ell'}^{EE} \\ \hat{C}_{\ell'}^{BB} \end{array} \right ) + \left ( \begin{array}{c} N_{\ell}^{EE} \\ N_{\ell}^{BB} \end{array} \right ).
\end{equation}

By inverting the previous system and employing the relation~\ref{eq:pseudo2true}, we confirm that the estimator is constructed to be unbiased: $<\hat{C}_{\ell}^{EE/BB}> = C_{\ell}^{EE/BB}$. The estimator is therefore free from any polarisation modes mixing at the level of the mean, the first order moment of the statistics. 

However, the variance of the estimator still includes a contribution from leaked polarisation mode. For its explicit expression, we refer the reader to \cite{Tristram_2005}. In practice, three distinct source of variance contribute to the variance on the estimator of the $B$ modes $\mathrm{Var}(\hat{C}_{\ell}^{BB})$ and can be understood as this non rigorous expression: 
\begin{equation}
\mathrm{Var}(\hat{C}_{\ell}^{BB}) = \mathrm{Var}(\tilde{C}_{\ell}^{BB}) + \mathrm{Var}(\tilde{C}_{\ell}^{EE \rightarrow BB}) + \mathrm{Var}(\mathrm{noise}), \nonumber
\end{equation}
with $\mathrm{Var}(\tilde{C}_{\ell}^{BB})$ the $B$ modes sampling variance, $\mathrm{Var}(\tilde{C}_{\ell}^{EE \rightarrow BB})$ the contribution to the variance of the leaked $E$ modes and $\mathrm{Var}(\mathrm{noise})$ the noise variance.

Following previous reasoning on the pseudomultipoles, the high $E$ modes amplitude leads to high variance induced by the $E$-to-$B$ leakage on $B$ modes power spectrum. This could be dramatic for their reconstruction, specially if $\mathrm{Var}(C_{\ell}^{BB}) > C_{\ell}^{BB}$. Of course, the $E$ modes estimation is also affected by the $B$ modes signal but the effect is much less significant due to the low expected level of $B$ modes amplitude. 

To conclude, a peculiar feature appears carrying out the standard pseudospectrum approach on a masked sky to reconstruct the polarisation field. The polarisation $E$ and $B$ modes are indeed tightly entangle, their mixing being driven by a convolution kernel depending on the applied mask. An unbiased estimator of the CMB polarisation is easily constructed. The main issue is thus to cancel the contribution of the leaking modes in the \textit{variance}, the second moment of the statistics, of the estimator. The standard approach to estimate the CMB polarisation is therefore not efficient, specially to reconstruct the low amplitude signal of $B$ modes: we need to construct new estimators to get rid of the polarisation modes mixing issue.

\subsection{A word on the convolution kernels}

The $K_{\ell m, \ell' m'}$ convolution kernels amount the coupling between the modes $(\ell, m)$ and $(\ell', m')$. In the quantum formalism, the convolution kernel $K_{\ell_1 m_1, \ell_2 m_2}$ can be seen as the probability of having an angular momentum $(\ell_3, m_3)$ given by the coupling of the momentum $(\ell_1 m_1)$ of the window function and the momentum $(\ell_2 m_2)$.  

The mixing kernels only depend on the mask $M(\vec{n})$ applied to the CMB map. The $K_{\ell m, \ell' m'}$ are therefore very long to compute and in any case not invertible due to their oscillations. The matrix $K_{\ell \ell'}$ has thus been introduced: it is invertible although its computation scales as $O(N_{pix}^2)$. However, the $K_{\ell \ell'}$ convolution kernels appear when taking the mean of the power spectrum. Therefore, it guarantees the correction of the leakage only at the first order moment of the statistics. The variance is spoilt by the $E$-to-$B$ leakage because the pseudomultipoles themselves are not pure in $E$ or $B$ modes. This demonstrates the need for correcting for the leakage at the level of the pseudomultipoles.

\subsection{Interpreting the $E$ to $B$ leakage}
\label{sec:interpret}

The $E$-to-$B$ leakage is a phenomenon arising from masking the CMB maps. In this case, the decomposition in pure $E$ and $B$ modes is not direct any more. In addition to the $E$ and $B$ subspaces, there exists a third subspace of the so-called \textit{ambiguous} modes which can not be determined as $E$ or $B$ modes as explained in \cite{Bunn_2003} and \cite{Smith_2007}. They indeed satisfy the conditions of both modes. In one dimension, it can be understood as modes at all scales with infinite extension, for which we would have only access to on a finite interval. Only a sub-part of the modes will cancel at the boundary of this interval. The modes can thus be fully characterised as a $E$ or $B$ modes. However, the modes non vanishing at the edge of the interval are equivocal as there is not enough information to fully characterise them. The modes cancelling at the edges of the interval are the analogous of pure $E$ and pure $B$ modes while the ones non vanishing are the so-called \textit{ambiguous} modes which fulfil both the conditions to be $E$ or $B$ modes. Following this reasoning, on the complete celestial sphere, the polarisation field has a unequivocal decomposition in $E$ and $B$ modes. If only a part of the sky is accessible, the polarisation field is thus decomposed on pure $E$ modes, pure $B$ modes and ambiguous modes containing both $E$ and $B$ modes. These modes thus cause the $E$-to-$B$ leakage. In the standard method, pure $B$ modes and the ambiguous modes contribute to the reconstructed, and therefore polluted, $B$ modes signal. Discarding the ambiguous modes would ensure the leakage removal but at the same time a part of information on the $E$ and $B$ modes would be lost.     


\section{Minimal Variance Quadratic Estimator}\label{sec:mvqe}

The best estimator of the angular CMB power spectra has to satisfy two conditions: it has to be unbiased \textit{and} give the smallest variance. Besides, a good estimator should also be fast to compute and loss as less information as possible. \cite{tegmark_estim_1997} and \cite{tegmark_2001} derive the best estimator for the temperature and the polarisation CMB power spectra, the so-called \textit{minimal variance quadratic estimator}. It is built by requiring to minimise the variance. Here we propose its reconstruction in the pseudospectrum-like frame. 

A given polarised data set $\left( Q(i) ~U(i) \right)^T$ (in the $i^{th}$ pixel) has a covariance matrix $\mathrm{C}_{ij}$ such as: 
\begin{eqnarray}
\mathrm{C}_{ij} & = \left< \left( \begin{array}{c} Q(i) \\ U(i) \end{array} \right) \left( Q(j)~U(j)\right) \right>  \\
& = \left( \begin{array}{cc} \left<Q(i)Q(j)\right> & \left<Q(i)U(j)\right> \\  \left< U(i)Q(j)\right> & \left<U(i)U(j)\right> \end{array} \right),
\end{eqnarray}
containing signal and noise. 

In order to construct an optimal estimator, the data has first to be multiplied by the inverse of the covariance matrix. In that way, the data are optimally filtered:
\begin{equation} 
\left( \begin{array}{c} Q \\ U \end{array} \right)^{(opt)} = \mathrm{C}^{-1} \left( \begin{array}{c} Q \\ U \end{array} \right).
\end{equation}

An optimal polarisation field $P_{\pm 2}^{opt}$ can be constructed from this optimal vector of Stokes parameters. Optimal $E$ and $B$ modes multipoles can thus be derived by projecting $P_{\pm 2}^{opt}$ on the spin-2 spherical harmonics ${}_{\pm 2}Y_{\ell m}$: 
\begin{eqnarray}
a_{\ell m}^E & = & -\frac{1}{2}  \int_{4\pi}  \left [  P_2^{(opt)} {}_{2}Y_{\ell m}^* + P_{-2}^{(opt)} {}_{-2}Y_{\ell m}^*  \right ]d\vec{n}, \nonumber \\
a_{\ell m}^B & = & \frac{i}{2} ~~ \int_{4\pi} \left [  P_2^{(opt)} {}_{2}Y_{\ell m}^* - P_{-2}^{(opt)} {}_{-2}Y_{\ell m}^*  \right ]d\vec{n},
\end{eqnarray}

From these multipoles, optimal pseudospectra can be defined as: 
\begin{equation}
\tilde{C}_{\ell}^{XY,(opt)} = \frac{1}{2\ell+1}\sum_m a^X_{\ell m} a^{Y*}_{\ell m}, 
\end{equation}
with $X$, $Y$ standing for $E$ or $B$ modes.

Correcting for the remaining leakages which amplitude is driven by a mixing kernel $[F]_{\ell \ell'}^{XY}$, the optimal estimator (formerly noise debiased) is obtained by inverting the following system: 
\begin{eqnarray}
\left( \begin{array}{c} \tilde{C}_{\ell}^{EE, (opt)} - N_{\ell}^{EE} \\ \tilde{C}_{\ell}^{BB, (opt)} - N_{\ell}^{BB} \end{array} \right) 
= \frac{1}{2} \sum_{\ell' m'} \left( \begin{array}{cc} [\mathrm{F}]_{\ell \ell'}^{EE} & [\mathrm{F}]_{\ell \ell'}^{EB} \\ \left[\mathrm{F}\right]_{\ell \ell'}^{BE} & \left[\mathrm{F}\right]_{\ell \ell'}^{BB}  
\end{array} \right)  \left( \begin{array}{c} \hat{C}_{\ell}^{EE, (opt)} \\ \hat{C}_{\ell}^{BB, (opt)} \end{array} \right ),
\label{eq:mvqelinsys}
\end{eqnarray}
with the Fisher matrices $[\mathrm{F}]_{\ell \ell'}^{XY}$ defined as (see Chapter~\ref{Chapter6} for a description of the Fisher formalism): 
\begin{equation}
[\mathrm{F}]_{\ell \ell'}^{XY} = \frac{1}{2} \mathrm{Tr}\left[ \frac{\partial \mathrm{C}}{\partial C_{\ell}^{XX}}\mathrm{C}^{-1} \frac{\partial \mathrm{C}}{\partial C_{\ell}^{YY}}\mathrm{C}^{-1}  \right].
\end{equation}

The constructed estimator $\hat{C}_{\ell}^{XY,(opt)}$ is unbiased and has a variance of: $\mathrm{Cov}(\hat{C}^{XY,(opt)}_{\ell},\hat{C}^{XY,(opt)}_{\ell'}) = [\mathrm{F}^{-1}]_{\ell \ell'}^{XY}$. By definition of the Fisher matrix, the obtained estimator therefore has the lowest uncertainties and is thus optimal. 

It was shown in \cite{Bond_1998} that this optimal estimator coincide with the estimator obtained by maximum likelihood at the likelihood maximum. The minimal variance quadratic estimator is therefore a powerful quantity as it is optimal and approach a maximum likelihood method. However the main issue is that it requires the knowledge and the inversion of the data covariance matrix $\mathrm{C}$. The pseudospectrum methods can be preferred as they do require less stringent prior on the signal. The minimal variance quadratic estimator is a guideline for the construction of these pseudospectrum approaches. The latter can indeed be seen as filtering data with a diagonal matrix which has to be close to the inverse of the covariance matrix to ensure an efficient estimation. The following section is dedicated to the description of three pseudospectrum approaches aiming at correcting for the $E$-to-$B$ leakage.


\section{Pseudospectrum Approaches to Correct for the $E$-to-$B$ Leakage}
\label{sec:free}

The issue of the $E$-to-$B$ leakage lies in the definition of the polarisation estimators. The standard estimator tackles the problem at the level of the mean but not at the level of the higher order moments of the statistics, in particular the variance. A remedy is therefore to construct an estimator free from any leakage at the level of the pseudomultipoles. In this perspective, three pseudospectrum approaches have been proposed in the literature: their goal is to construct pseudomultipoles pure in the $E$ and $B$ modes. The $\chi^{E/B}$ fields containing only $E/B$ modes respectively offer a wonderful quantity to build a leakage-free estimator. However, by definition, their direct reconstruction requires to derive the noisy CMB polarisation maps. This operation is indeed laborious owing to the presence of the noise. The core of the considered methods thus lies in the reconstruction of the $\chi^{E/B}$ fields avoiding the direct derivation of the CMB polarisation maps.

For each of the proposed methods, the key issue is to ensure a $B$ modes pseudomultipole free from any leakage by cancelling the convolution $K_{\ell m,\ell' m'}^{\mathrm{method},-}$ in the following system: 
\begin{equation}
\left ( \begin{array}{c} \tilde{a}_{\ell m}^{E} \\ \tilde{a}_{\ell m}^B \end{array} \right )
= \sum_{\ell' m'} \left ( \begin{array}{cc} H_{\ell m,\ell' m'}^{\mathrm{method},+} & H_{\ell m,\ell' m'}^{\mathrm{method},-} \\ K_{\ell m,\ell' m'}^{\mathrm{method},-} & K_{\ell m,\ell' m'}^{\mathrm{method},+}  \end{array} \right ) \left ( \begin{array}{c} a_{\ell' m'}^E \\ a_{\ell' m'}^B \end{array} \right ),
\label{eq:keylinsys}
\end{equation}
where $K_{\ell m,\ell' m'}^{\mathrm{method},\pm}$ are the convolution kernels mixing the $E$ and $B$ modes in the $B$ modes pseudomultipoles. They have different expressions regarding the pseudospectrum methods and can potentially differ from the $H_{\ell m,\ell' m'}^{\mathrm{method},\pm}$ which quantifies the contribution to the $E$ modes. The masks applied to the CMB maps for the $E$ modes reconstruction can indeed differ from the ones used for the $B$ modes reconstruction leading to different mixing kernels.

Including the effect of the beam $b_{\ell}$ and the noise $N_{\ell}$, the system to be inverted at the level of the power spectrum is: 
\begin{equation}
\left ( \begin{array}{c} \tilde{C}_{\ell}^{E} - N_{\ell}^E \\ \tilde{C}_{\ell}^B - N_{\ell}^B \end{array} \right )
= \sum_{\ell'} B_{\ell'}^2 \left ( \begin{array}{cc} H_{\ell,\ell'}^{\mathrm{method},+} & H_{\ell,\ell'}^{\mathrm{method},-} \\ K_{\ell,\ell'}^{\mathrm{method},-} & K_{\ell,\ell'}^{\mathrm{method},+}  \end{array} \right ) \left ( \begin{array}{c} \hat{C}_{\ell'}^E \\ \hat{C}_{\ell'}^B \end{array} \right )
\label{eq:keylinsysCl_main}
\end{equation}
with $K_{\ell \ell'}^{\mathrm{method},\pm}$ such as: 
\begin{equation}
K_{\ell \ell'}^{\mathrm{method},\pm} = \frac{1}{2\ell + 1} \sum_{m = -\ell}^{\ell} \sum_{m' = -\ell'}^{\ell'} |K_{\ell m,\ell' m'}^{\mathrm{method},\pm}|^2
\end{equation}

The present section is consequently dedicated to the theoretical principle of each of pseudospectrum method correcting for the leakage in the perspective of their numerical implementations, driving some choices. The first technique operates in the harmonic domain while the two last are pixel based.  

\subsection{The pure method}

A convenient decomposition of the polarisation field in the harmonic domain has been proposed in \cite{Smith_2006} and then in \cite{Smith_2007}, which approach will be refer to as the \textit{pure} method in this manuscript. Instead of expanding the polarisation field on the basis of the spin-$(\pm2)$ spherical harmonics, the point of this method is to decompose the polarisation field on the spherical harmonics weighted by a window function $W$ ensuring to be on the \textit{pure} $B$ modes basis. The so-called \textit{pure pseudomultipoles} are then defined by: 
\begin{eqnarray}
\tilde{a}_{\ell m}^{E} & = & -\frac{1}{2\alpha_\ell} \int_{\Omega} [P_{2}(\vec{n}) (\partial\partial W(\vec{n})Y_{\ell m}(\vec{n}))^*  \nonumber \\
& & + P_{-2}(\vec{n})(\bar{\partial}\bar{\partial}W(\vec{n})Y_{\ell m}(\vec{n}))^*]d\vec{n},  \nonumber \\
\tilde{a}_{\ell m}^{B} & = & \frac{i}{2\alpha_\ell} \int_{\Omega} [P_{2}(\vec{n}) (\partial\partial W(\vec{n})Y_{\ell m}(\vec{n}))^*  \nonumber \\
& & - P_{-2}(\vec{n})(\bar{\partial}\bar{\partial}W(\vec{n})Y_{\ell m}(\vec{n}))^*]d\vec{n},  
\label{eq:pseudopurealm}
\end{eqnarray}
with $\Omega$ the observed sky fraction, thus $\Omega < 4\pi$. We also recall the expression of the $\alpha_\ell$ which has been introduced in Chapter~\ref{Chapter3}: $\alpha_{\ell} = \sqrt{\frac{\ell+2}{\ell-2}}$.

A sufficient condition on $W$ to warrant the decomposition on the pure basis is to satisfy the Dirichlet and Neumann boundary conditions \textit{i.e.} at the boundaries:
\begin{equation}
\left\{ \begin{array}{l} W = 0, \\
\partial W = 0. \end{array} \right .
\end{equation} 
 
Assuming these conditions, the system~(\ref{eq:pseudopurealm}) can indeed be integrated by part twice. The pure $B$ modes pseudomultipoles are thus expressed as the following (not mathematically rigorous) expression:  
\begin{eqnarray}
\tilde{a}_{\ell m}^B & \propto & [\partial(WY)^*P_{2}]_{\mathcal{C}} - \int_{\Omega}\partial P_2(\partial WY_{\ell m})^* 
- [\bar{\partial}(WY)^*P_{-2}]_{\mathcal{C}} + \int_{\Omega}\bar{\partial} P_{-2}(\bar{\partial} WY_{\ell m})^*  \nonumber \\
&\propto &[\partial(WY)^*P_{2}]_{\mathcal{C}} + [(WY)^* \partial P_{2}]_{\mathcal{C}} - \int_{\Omega}\partial \partial P_2(WY_{\ell m})^*   \nonumber \\
& + & [\bar{\partial}(WY)^*P_{-2}]_{\mathcal{C}} - [(WY)^*\bar{\partial}P_{-2}]_{\mathcal{C}} + \int_{\Omega}\bar{\partial} \bar{\partial} P_{-2}(WY_{\ell m})^*, 
\end{eqnarray}
where $\mathcal{C}$ indicates that the expression has to be evaluated on the boundaries of the mask. In the end, the pure $E$ and $B$ pseudomultipoles are consequently expressed as: 
\begin{eqnarray}
\tilde{a}_{\ell m}^B & = & ~ \frac{i}{2 \alpha_{\ell}} \int_{\Omega} WY_{\ell m}^* [\bar{\partial}\bar{\partial} P_2(\vec{n}) - \partial\partial P_{-2}(\vec{n})] 
\nonumber  \\
& = & \frac{1}{\alpha_{\ell}} \tilde{\chi}_{\ell m}^B.
\label{eq:pseudochiBpure}
\end{eqnarray} 

In a similar way, for the $E$ modes: 
\begin{eqnarray}
\tilde{a}_{\ell m}^E & = & \frac{-1}{2\alpha_{\ell}} \int_{\Omega} WY_{\ell m}^* [\bar{\partial}\bar{\partial} P_2(\vec{n}) + \partial\partial P_{-2}(\vec{n})] \nonumber \\
 & = & \frac{1}{\alpha_{\ell}} \tilde{\chi}_{\ell m}^E.
\label{eq:pseudochiEpure}
 \end{eqnarray}

As a result, the pure $B$ ($E$) pseudomultipoles are the pseudomultipoles of the $\chi^B$ ($\chi^E$ resp.) field including a normalisation factor: they therefore contain only $B$ ($E$ resp.) modes. Window functions satisfying the Dirichlet and Neumann conditions enables a decomposition of the CMB polarisation field on the basis of pure $B$ and $E$ modes and thus the construction of leakage-free pseudomultipoles. The pure estimator and \textit{all} its statistical moments are therefore theoretically free from any polarisation modes leakage. The convolution kernel $K_{\ell m,\ell' m'}^{\mathrm{pure},-}$ is therefore exactly equal to zero. The expression of the mixing kernels can be found in Appendix~\ref{AppendixA}


\subsection{The \zb~approach}

The principle of this method lies in the direct reconstruction of the $\chi^{E/B}$ fields. Its procedure has been proposed in \cite{Zhao_2010} and can be seen as the pure method performed in the real space. The method will be refer to as the \zb~method, from the initial of the authors of the original article. 

For an incomplete sky coverage, the pseudo-$\chi^{E/B}$ fields are defined as:
\begin{eqnarray}
\tilde{\chi}^{E}(\vec{n}) & = & -\frac{1}{2} [\bar{\partial}\bar{\partial} (W(\vec{n})P_{2}(\vec{n})) + \partial\partial(P_{-2}(\vec{n}) W(\vec{n}) ) ], \nonumber \\
\tilde{\chi}^{B}(\vec{n}) & = & ~~ \frac{i}{2} ~ [\bar{\partial}\bar{\partial} (W(\vec{n})P_{2}(\vec{n})) - \partial\partial(P_{-2}(\vec{n}) W(\vec{n}) ) ],
\end{eqnarray}
with $W$ a window function cancelling outside the observed region of the sky. 

Following the reasoning of \cite{Zhao_2010}, the pseudo-$\chi^{E/B}$ fields are related to the underlying $\chi^{E/B}$ fields. The masked $\chi^{B}$ field expression is recalled: 
\begin{eqnarray} 
W(\vec{n}) \chi^{B}(\vec{n}) = \frac{i}{2} \left[ W\bar{\partial}\bar{\partial}(P_{2}(\vec{n})) - W\partial\partial(P_{- 2}(\vec{n})) \right].
\end{eqnarray}

As this expression includes derivation of the noisy polarisation field, the masked $\chi^B$ field can be rewritten as the contribution of the pseudo-$\chi^B$ field and counter terms:
\begin{eqnarray}
	W(\vec{n})\chi^B(\vec{n})&=&\frac{i}{2}\left[\bar\partial\bar\partial\left(WP_{2}\right)-\partial\partial\left(WP_{-2}\right)\right] 
	\label{eq:chiB}
	 \nonumber \\
	&-&i\left[\frac{\bar\partial W}{W}\bar\partial\left(WP_{2}\right)-\frac{\partial W}{W}\partial\left(WP_{-2}\right)\right] \nonumber \\
	&-&\frac{i}{2}\left[\left(\bar\partial\bar\partial W\right)P_{2}-\left(\partial\partial W\right)P_{-2}\right]  \nonumber \\
	&+&i\left[\frac{\left(\bar\partial W\right)^2}{W}P_{2}-\frac{\left(\partial W\right)^2}{W}P_{-2}\right].
\end{eqnarray}

The derivatives of the weighted polarisation field $\partial^n(WP_{\pm 2})$ or $\bar{\partial}^n(WP_{\pm 2})$ are computable, therefore their linear combination intervening in equation~(\ref{eq:chiB}) can \textit{a fortiori} be derived. Moreover, to ensure the continuity of the window function $W$ which is differentiated twice, its value and first derivative have to cancel on the mask boundaries.

From the masked polarisation field $WP_{\pm2}$, it is thus possible to reconstruct the $\chi^B$ field but only in the observed pixels. Two ways to reconstruct the true underlying $\chi^B$ field on the whole celestial sphere are manageable. The first one originally proposed in \cite{Zhao_2010} consists in dividing the right hand side of the equation~(\ref{eq:chiB}) by $W$ and then compute the pseudomultipoles of $\chi^B$ field. However, the masked $\chi^B$ field would be divergent at the edges of the mask as the window function $W$ is cancelling on the boundaries. We therefore have to be careful not to divide by the window function. In \cite{Zhao_2010}, the procedure applied to avoid this problem is to define a new binary mask $M'$ where the edge of the window function $W$ have been removed. The $\chi^{E/B}$ pseudomultipoles $\tilde{a}_{\ell m}^{\chi^{E/B}}$ are then defined by projecting the reconstructed $M' \times \chi^{B}$ map on the spherical harmonics:
\begin{equation}
\tilde{a}_{\ell m}^{\chi^{E/B}} = \int_{\Omega} M'(\vec{n}) \chi^{E/B}(\vec{n}) Y_{\ell m}^*(\vec{n}) d\vec{n}.
\label{eq:pseudochimultipole_zb1}
\end{equation}

Nevertheless, as the binary mask $M'$ is smaller than the originally observed part of the sky, information have been lost. It can be dramatic specially when the signal to be detected is expected to be as low as the primordial $B$ modes. 

In order to circumvent this issue, we have improved on the proposed method and propose to reconstruct an apodised version of the $\chi^B$ field and to derive its pseudomultipoles. Following this procedure, we would aim at recovering the masked field $W\chi^B$ directly. However, as the right-hand side of the equation~(\ref{eq:chiB}) shows a division by the window function $W$ which cancels at the edges, $W\chi^B$ is undefined. We consequently choose to reconstruct the masked $W^2\chi^B$ field to avoid any singularities issue:    
\begin{eqnarray}
W^2(\vec{n})\chi^B(\vec{n})&=&\frac{i}{2}W\left[\bar\partial\bar\partial\left(WP_{2}\right)-\partial\partial\left(WP_{-2}\right)\right] 
	\label{eq:w2chiB}
	\nonumber \\
	&-&i\left[\bar\partial W\bar\partial\left(WP_{2}\right)-\partial W\partial\left(WP_{-2}\right)\right] \\ \nonumber 
	&-&\frac{i}{2}W\left[\left(\bar\partial\bar\partial W\right)P_{2}-\left(\partial\partial W\right)P_{-2}\right] \\ \nonumber 
	&+&i\left[\left(\bar\partial W\right)^2P_{2}-\left(\partial W\right)^2P_{-2}\right]. \nonumber
	\label{eq:eqzb}
\end{eqnarray}
In this approach, the Eq.~(\ref{eq:pseudochimultipole_zb1}) is modified and the $\chi^{E/B}$ field pseudomultipoles $\tilde{a}_{\ell m}^{\chi^{E/B}}$ are consequently determined by: 
\begin{equation}
\tilde{a}_{\ell m}^{\chi^{E/B}} = \int_{\Omega} W(\vec{n})^2 \chi^{E/B}(\vec{n}) Y_{\ell m}^*(\vec{n}) d\vec{n}.
\end{equation}

The here defined pseudomultipoles of the $\chi^{B(E)}$ fields contain only $B$ ($E$ respectively) modes: the constructed polarisation fields are free from any leakage. The $E$ and $B$ modes pseudospectrum are then simply constructed from these pseudomultipoles. Therefore, by construction, the pure and \zb~methods are equivalent \textit{in theory} as they both consists in constructing the $W\chi^{E/B}$ fields, in the harmonic domain for the former one and in the pixel domain for the latter. In the same way as the pure method, the convolution kernel $K_{\ell m, \ell' m'}^{zb,-}$ is consequently exactly vanishing.


\subsection{The kn approach}

A second real space pseudospectrum approach was proposed in \cite{Kim_2010} and re-address in \cite{Kim_2011}. It takes advantage of the fact that the $E$-to-$B$ leakage is very localised in the real space. Its main principle is to estimate the $\chi^{E/B}$ fields on the masked sky after the rejection of the pixels plagued by the $E$-to-$B$ leakage. This technique will be refer to as the \kn-method in the present manuscript. 

The $\chi^{E/B}$ fields being defined by differentiating twice the polarisation map, their decomposition on the spherical harmonics gives:
\begin{eqnarray}
\chi^E & = & \sum_{\ell m} \alpha_{\ell} \times a_{\ell m}^E Y_{\ell m}, \\
\chi^B & = & \sum_{\ell m} \alpha_{\ell} \times a_{\ell m}^B Y_{\ell m}.
\end{eqnarray}

Besides, we recall the relation between the $E$ and $B$ modes multipoles and the polarisation field $P_{\pm 2}$:
\begin{eqnarray}
a_{\ell m}^E & = & \int_{4\pi} -\frac{1}{2} \left [  P_{2} \times {}_{-2}Y_{\ell m}^* + P_{-2} \times {}_{2}Y_{\ell m}^*   \right ]d\vec{n}, \nonumber \\
a_{\ell m}^B & = & \int_{4\pi} ~~\frac{i}{2} \left [  P_{2} \times {}_{-2}Y_{\ell m}^* - P_{-2} \times {}_{2}Y_{\ell m}^*   \right ]d\vec{n}.
\end{eqnarray}

Inserting this equation in the expressions of the $\chi^{E/B}$ maps gives: 
\begin{eqnarray}
\chi^E(\vec{n}) & = & -\frac{1}{2} \left( \int_{4\pi} \left[ F_{+}(\vec{n},\vec{n}') P_{-2}(\vec{n}') + F_{-}(\vec{n},\vec{n}') P_{2}(\vec{n}') \right] \right) d\vec{n}',  \\
\chi^B(\vec{n}) & = & ~ \frac{i}{2} ~ \left( \int_{4\pi} \left[ F_{+}(\vec{n},\vec{n}') P_{-2}(\vec{n}')  -  F_{-}(\vec{n},\vec{n}') P_{2}(\vec{n}') \right] \right) d\vec{n}',
\label{eq:deffullskychi}
\end{eqnarray}
with the pixel-filter function $F_{\pm}(\vec{n},\vec{n}')$ defined as: 
\begin{eqnarray}
F_{\pm}(\vec{n},\vec{n}') = \sum_{\ell m} \alpha_{\ell}\times{}_{\pm 2} Y_{\ell m}^{}(\vec{n}')Y_{\ell m}^*(\vec{n}).
\end{eqnarray}

\begin{figure}[!h]
\begin{center}
	\includegraphics[scale=0.35]{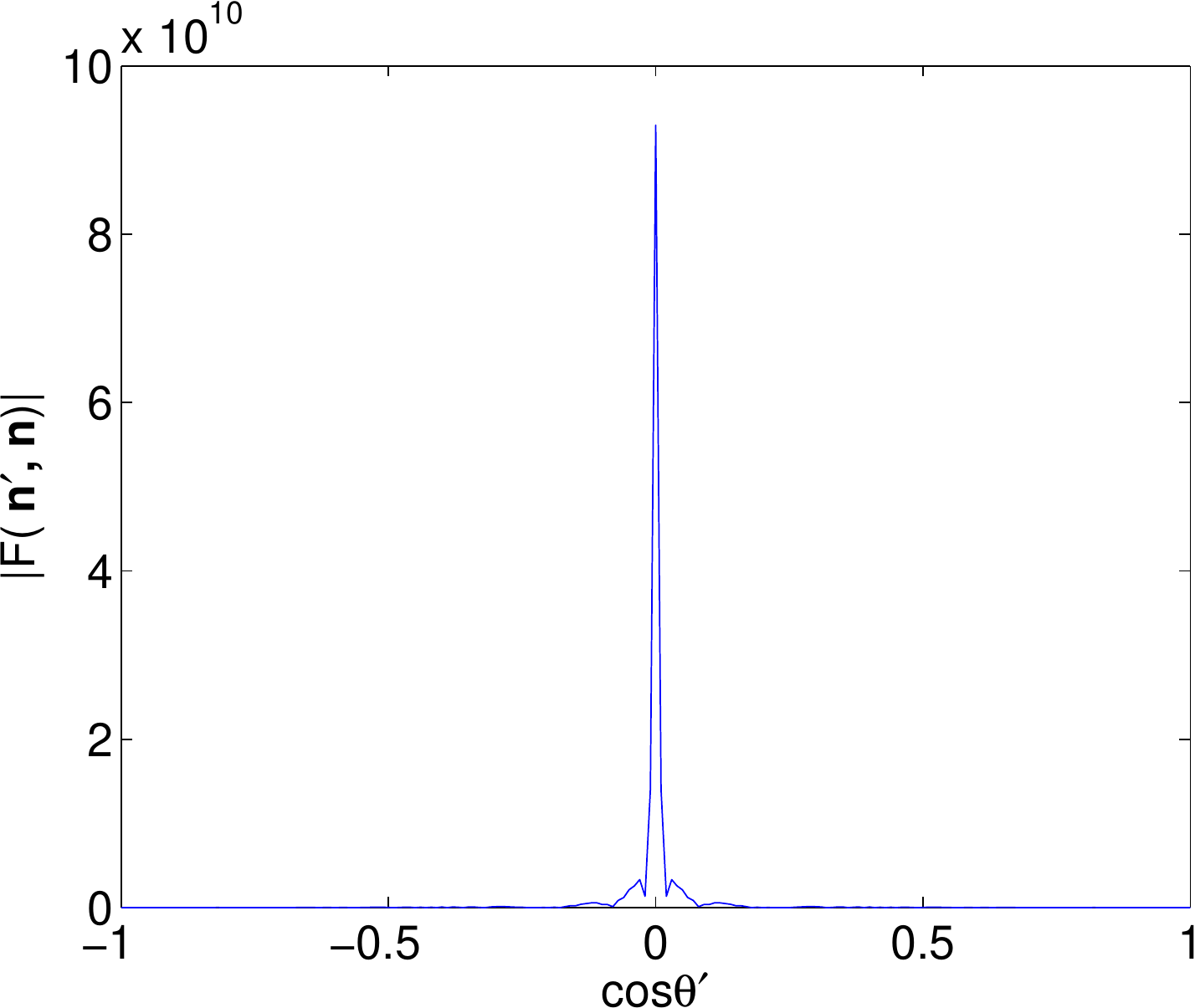}
	\caption{The modulus of the convolution kernel $F_{\pm}(\vec{n})$ for a given $\vec{n}$ such as $\theta = \frac{\pi}{2}$ and $\phi = 0$, from \cite{Kim_2010}. This filter is highly peak for $\cos(\theta) = 0$.} 
	\label{fig:Fmod}
\end{center}
\end{figure}

This convolution function appears to be highly peak in the direction $\vec{n}$ as shown in the figure \ref{fig:Fmod} taken from \cite{Kim_2010}: it selects the pixels near from the direction of $\vec{n}$. It however has a non zero extension in the case of pixelised maps as we will see in the next chapter. 

The above procedure only applies in the case of a CMB detection on all the celestial sphere. In the case of a masked sky, the $\chi^{E/B}$ reconstruction can be done thanks to their pseudomultipoles defined as: 
\begin{eqnarray} 
\tilde{\chi}^E(\vec{n}) & = & -\frac{1}{2} \left( \int_{\Omega} M(\vec{n}') \left[ F_{+}(\vec{n},\vec{n}') P_{-2}(\vec{n}') + F_{-}(\vec{n},\vec{n}') P_{2}(\vec{n}') \right] \right), \label{eq:pseudochi} \nonumber \\
\tilde{\chi}^B(\vec{n}) & = & ~~ \frac{i}{2}  \left( \int_{\Omega} M(\vec{n}')\left[ F_{+}(\vec{n},\vec{n}') P_{-2}(\vec{n}')  -  F_{-}(\vec{n},\vec{n}') P_{2}(\vec{n}') \right]. \right) 
\label{eq:pseudochikn}
\end{eqnarray}

The $B$ modes pseudopower spectrum resulting from the auto-correlation of pseudo-$\chi^B$ multipoles does also contain $E$ modes: the induced $E$-to-$B$ leakage can therefore be quantified. The next step is crucial and consists in flagging the pixels where this leakage is high compared to a carefully chosen value.

First of all, the equation~(\ref{eq:pseudochi}) can be re-expressed in terms of the usual $E$ and $B$ modes true multipoles. In particular the pseudo-$\chi^B$ field writes: 
\begin{eqnarray}
\tilde{\chi}^B(\vec{n}) & = & -2i \sum_{\ell m} \left [ K_{\ell m}^-(\vec{n}) a_{\ell m}^E + i K_{\ell m}^+(\vec{n}) a_{\ell m}^B \right ] \nonumber \\
& = & \tilde{\chi}^{E \rightarrow B}(\vec{n}) + \tilde{\chi}^{B \rightarrow B}  (\vec{n})
\end{eqnarray} 
with $K_{\ell m}^{\pm}(\vec{n}) = \int_{4\pi} M(\vec{n}') [ F_+(\vec{n},\vec{n}') {}_{-2}Y_{\ell m}(\vec{n}') \pm F_-(\vec{n},\vec{n}') {}_{2}Y_{\ell m}(\vec{n}') ]d\vec{n}'$. The term $\tilde{\chi}^{E (B) \rightarrow B}$ stands for the contribution of the $E$ ($B$ respectively) modes in the $B$ modes. The convolution kernels $K_{\ell m}^{\pm}$ amount the level of the contribution of the $E$ modes from potentially the whole celestial sphere to the obtained pseudo-$\chi^B$ field in a given direction. If $M$ is constant, the pseudo-$\chi^{E/B}$ fields boil down to the expression of $\chi^{E/B}$ fields for a full sky coverage in Eq.~(\ref{eq:deffullskychi}), the convolution kernel being highly peaked in this case. Therefore, the leakage is expected to be higher on an extended layer around the edge of the mask where $M$ is varying abruptly from $0$ to $1$ because of the oscillating behaviour of the convolution kernels. 

Secondly, the pixels plagued by this leakage expected to be on the edge of the mask can be flagged by quantifying the ratio $R(\vec{n})$ between the leaking $E$ modes and the $B$ modes given by:
\begin{equation} 
R(\vec{n}) = \frac{<|\chi^{E\rightarrow B}(\vec{n})|^2>}{<|\chi^{B \rightarrow B}(\vec{n})|^2>}.
\end{equation}
The comparison of this ratio with a carefully chosen threshold would lay down the pixels rejection or admission. If the value of $R(\vec{n})$ in a given pixel is too high regarding the threshold, the pixel will be considered as `ambiguous' and then actually be rejected. The $\chi^B$ field will consequently be reconstructed on the remaining pixels. In practice, \cite{Kim_2010} proposed to perform Monte Carlo simulations for a given cosmological model in order to amount $R(\vec{n})$. The kept pixels are considered to be `pure' in $B$ modes and consequently verifies the following condition:
\begin{equation}
R(\vec{n}) < \frac{r_c}{r},
\end{equation} 
where $r$ is the value of tensor-to-scalar ratio of the input cosmological model and $r_c$ is the level of $B$-modes due to the leaking $E$-modes. $r_c$ plays the role of the threshold. If it is chosen as too low, a large amount of pixels would be considered as ambiguous and subsequently the final mask will cover a small part of the sky, leading to a significant loss of cosmological information. However, if $r_c$ is too high, the kept pixels might not be as pure as required. By minimizing the variance of the $B$ mode power spectrum, \cite{Kim_2010} derived the equation to be verified by $r_c$ which gives the best compromise between the loss of information and the pixels cleanliness.

The $\chi^B$ field is then reconstructed from the pseudo-$\chi^B$ estimated on the cut binary mask containing only `pure' pixels.

A second strategy to reject the leakage has been proposed in \cite{Kim_2011}. Instead of choosing a binary mask, an apodised window function $W(\vec{n})$ is applied to the maps. In this case, the convolution kernel becomes:
\begin{equation}
K_{\ell m}^{\pm}(\vec{n}) = \int_{4\pi} W(\vec{n}') [ F_+(\vec{n},\vec{n}') {}_{-2}Y_{\ell m}(\vec{n}') \pm F_-(\vec{n},\vec{n}') {}_{2}Y_{\ell m}(\vec{n}') ]d\vec{n}'.
\end{equation}
The window function $W$ is a function which is constant in the centre of the observed region of the sky and smoothly decreases to zero at the edge of the mask. The edge effects are therefore expected to be smaller than in the case of a binary mask $M$, so that the leakage is expected to be minimised. 
This second approach will be used in the following as the criterion on the apodisation length of $W$ is directly found, the leakage being concentrated on the edges where $W$ varies.

In both approaches, the power spectrum estimator can be defined by auto-correlating the obtained pseudo-$\chi^{E/B}$ field. The $\chi^B$ on the built mask $M_{W^{\chi^B}}$ containing only `pure' pixels, poorly affected by the leakage, is therefore constructed. Similarly to the previously described methods, the convolution kernel $K_{\ell m,\ell' m'}^{kn,-}$ exactly cancels.

\section{Visualisation of the leakage}\label{sec:leakmap}

The three presented pseudospectrum approaches offer an expression for $\chi^{E/B}(\vec{n})$ on a masked sky. They thus enable to map the $E$-to-$B$ leakage. Such maps are convenient to comprehend the localisation of the leakage and its sensitivity regarding the choice of the method. The leakage is obtained by setting the input $B$ modes to zero and subsequently constructing the $\chi^B$ map. The resulting map therefore directly traces the $E$-to-$B$ leakage. The issue rests on the construction of the $\chi^B$ field.

The pure method $B$ pseudomultipoles are proportional to the ones of $W\times\chi^B$ as shown in Eqs.~(\ref{eq:pseudochiBpure}). Consequently, masked map of $\chi^B$ is constructed by projecting the pure polarised pseudomultipoles weighted by the $\alpha_{\ell}$ factor on the spherical harmonics $Y_{\ell m}$. Moreover, the masked $\chi^B$ field can be reconstructed straightforwardly in the frame of the \zb~technique by applying Eq.~(\ref{eq:eqzb}). In the \kn~approach, the $B$ modes maps are made by the use of the pseudo-$\chi^{B}$ fields in Eq.~(\ref{eq:pseudochikn}). The involved window function $W$ can be tuned to remove the aliased pixels. In the peculiar case of $W = M$ with $M$ a binary mask, the \kn~approach boils down to the standard method. 

Thanks to the implementation of their computation in each approach, the masked $\chi^B$ fields are numerically constructed for an input power spectrum with $E$ modes power spectrum from WMAP-7yr and \textit{no} input $B$-modes. In practice, the RMS over the simulated $\tilde{\chi}^B$ maps give a good insight on the resulting leakage map (the mean being equal to zero by construction). The maps shown afterwards are thus the RMS of the simulated maps:  
\begin{equation}
\sigma_{\tilde{\chi}^{B}} = \sqrt{\left<  \tilde{\chi}^{B^2} \right>_{MC}}
\end{equation}
with $ \left< .\right>_{MC}$ the average on the Monte Carlo simulations.

\begin{figure}[!h]
\begin{center}
	\includegraphics[scale=0.35]{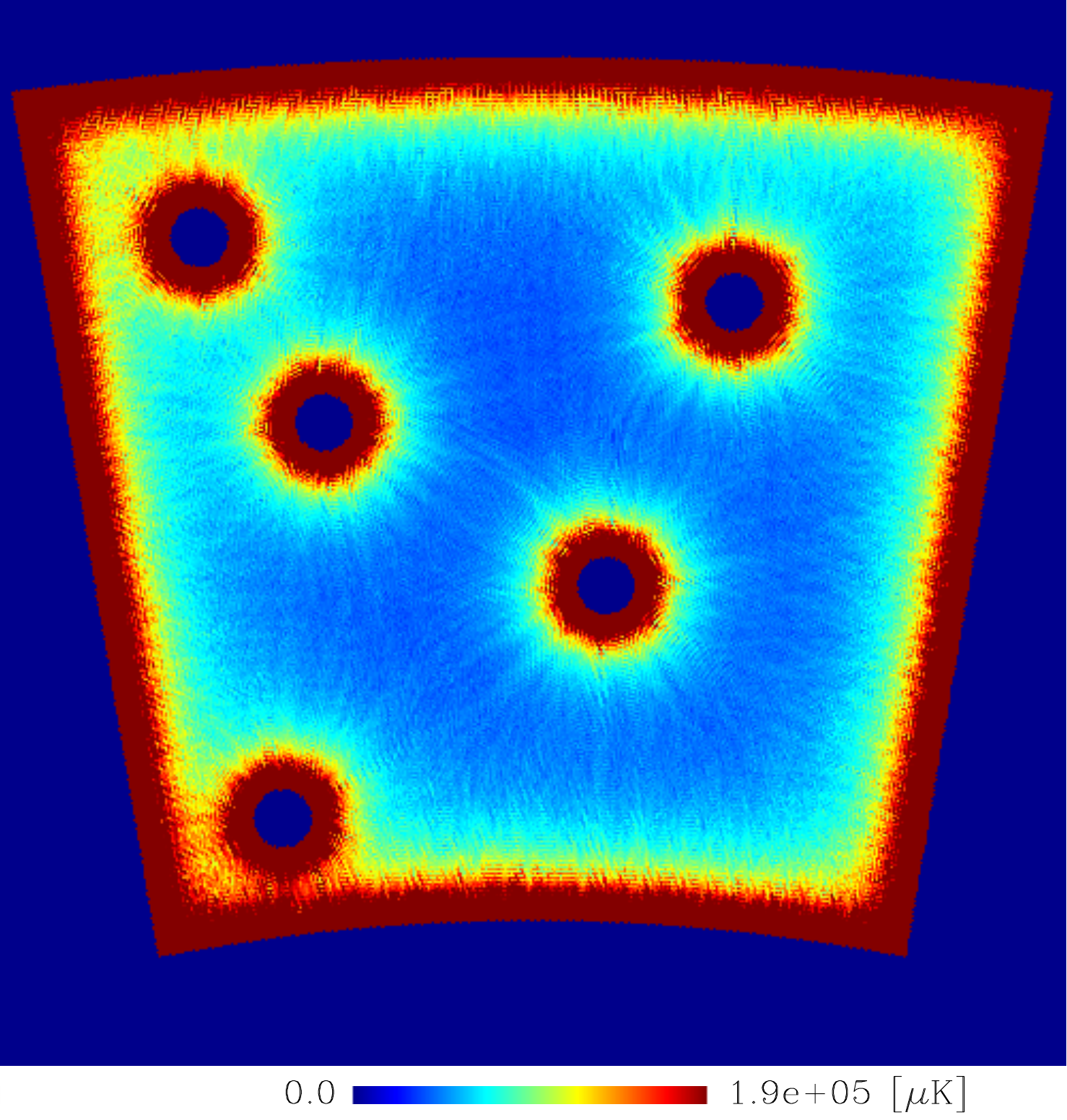}
	\caption{Leakage map obtained using the standard pseudospectrum reconstruction. The redder areas are the most aliased pixels.} 
	\label{fig:chimapstand}
\end{center}
\end{figure}

By way of example, leakage maps are constructed in the frame of the different methods for a small scale experiment covering $1\%$ of the sky. Holes are present to mask polarised point sources. As a benchmark, the leakage map obtained in the standard method is displayed in Fig.~\ref{fig:chimapstand}. The leakage is concentrated at the edges of the mask as expected. Nonetheless, the inner part of the observed sky patch is also spoilt by the $E$-to-$B$ leakage. 

\begin{figure}[!h]
\begin{center}
	\includegraphics[scale=0.15]{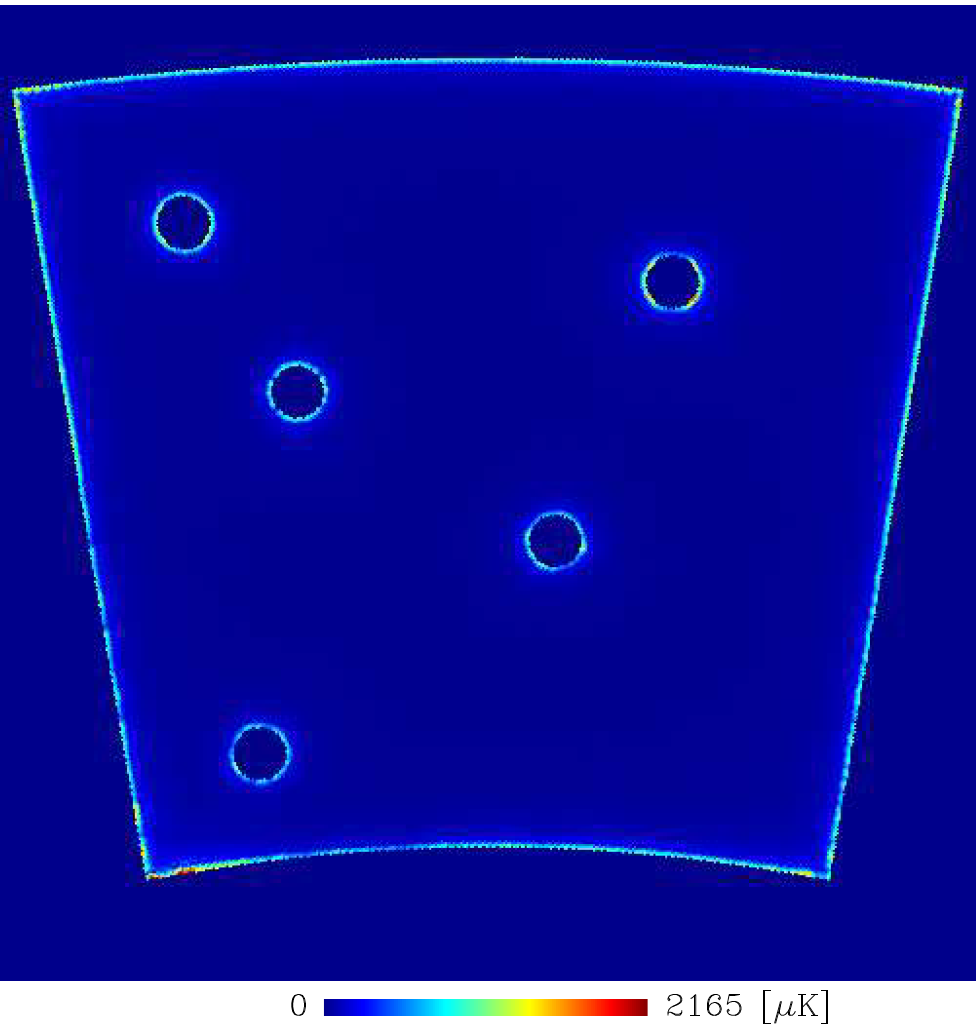}~~~~\includegraphics[scale=0.14]{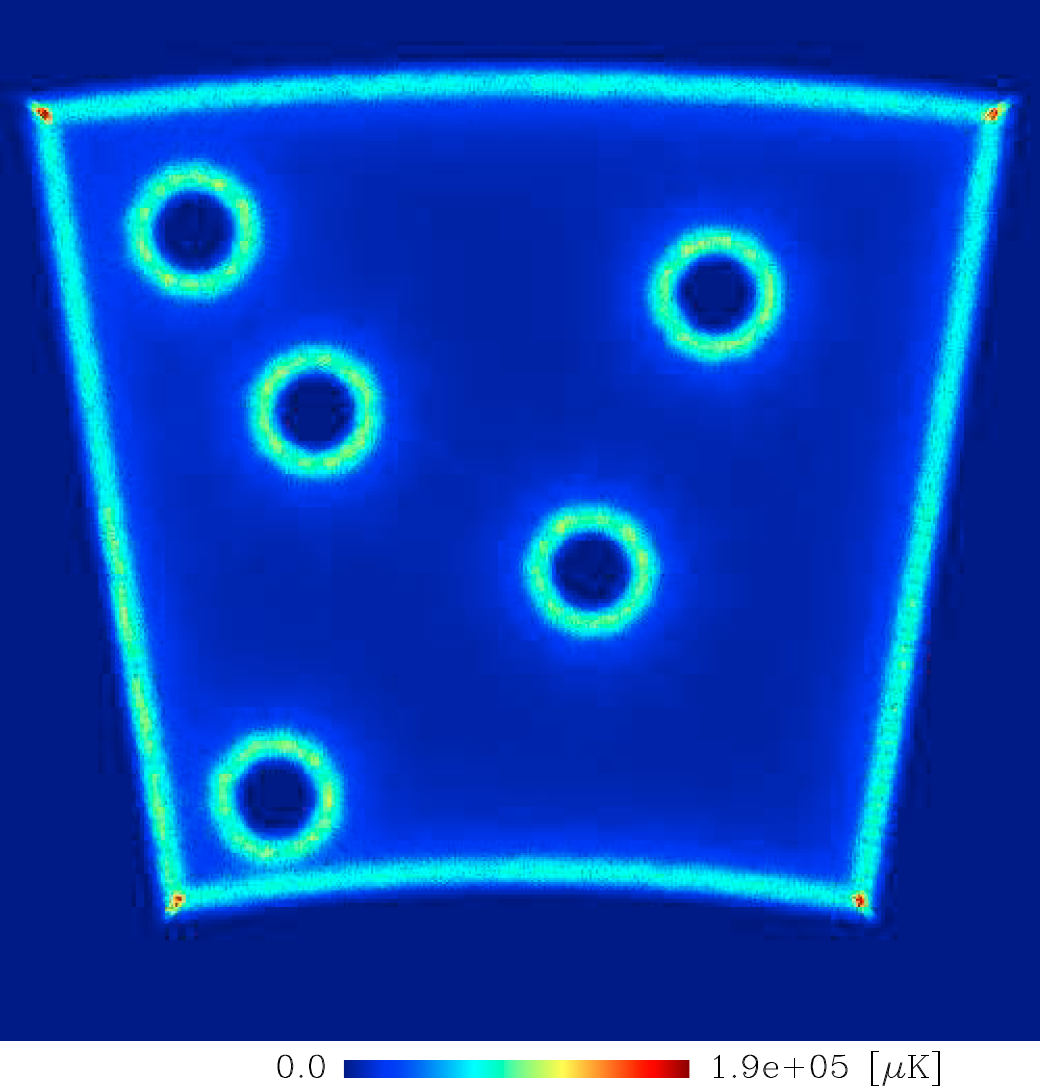}
	\caption{Leakage map in the frame of the pure (\zb) pseudospectrum reconstruction in the left (right) panel.} 
	\label{fig:chimapszb}
\end{center}
\end{figure}

In the frame of the pure and \zb~approaches, the leakage maps are shown respectively on the left and right panel in Fig.~\ref{fig:chimapszb}. The results are striking: the leakage is at the same time reduced and more localised on the edges of the maps compared to the standard method. The leakage is however more important and extended around the mask contours in the \zb~approach than in the pure method. Moreover, the leakage does not fall down exactly to zero as expected because of pixelisation effects. 

\begin{figure}[!h]
\begin{center}
	\includegraphics[scale=0.35]{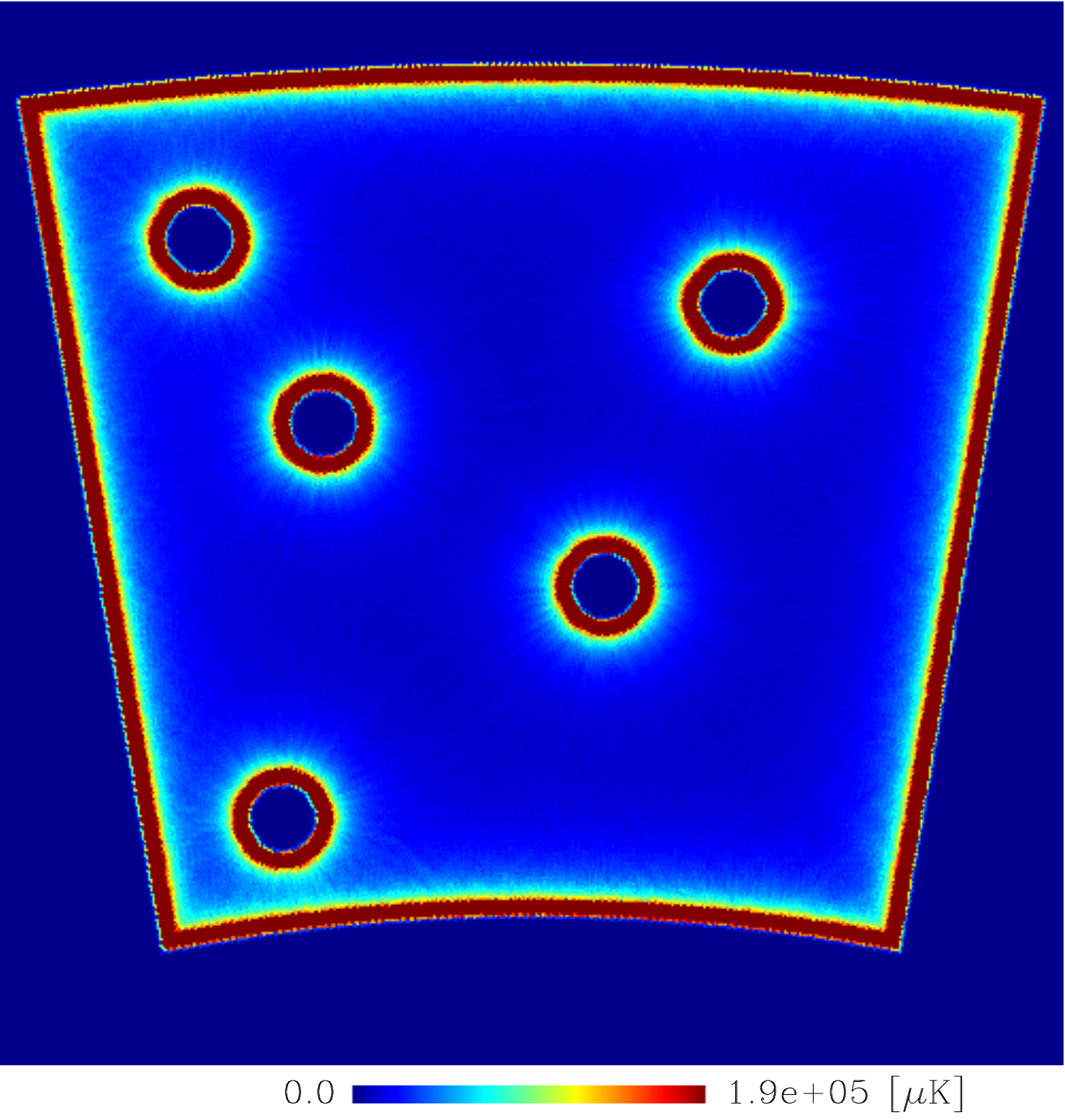}~~~~\includegraphics[scale=0.35]{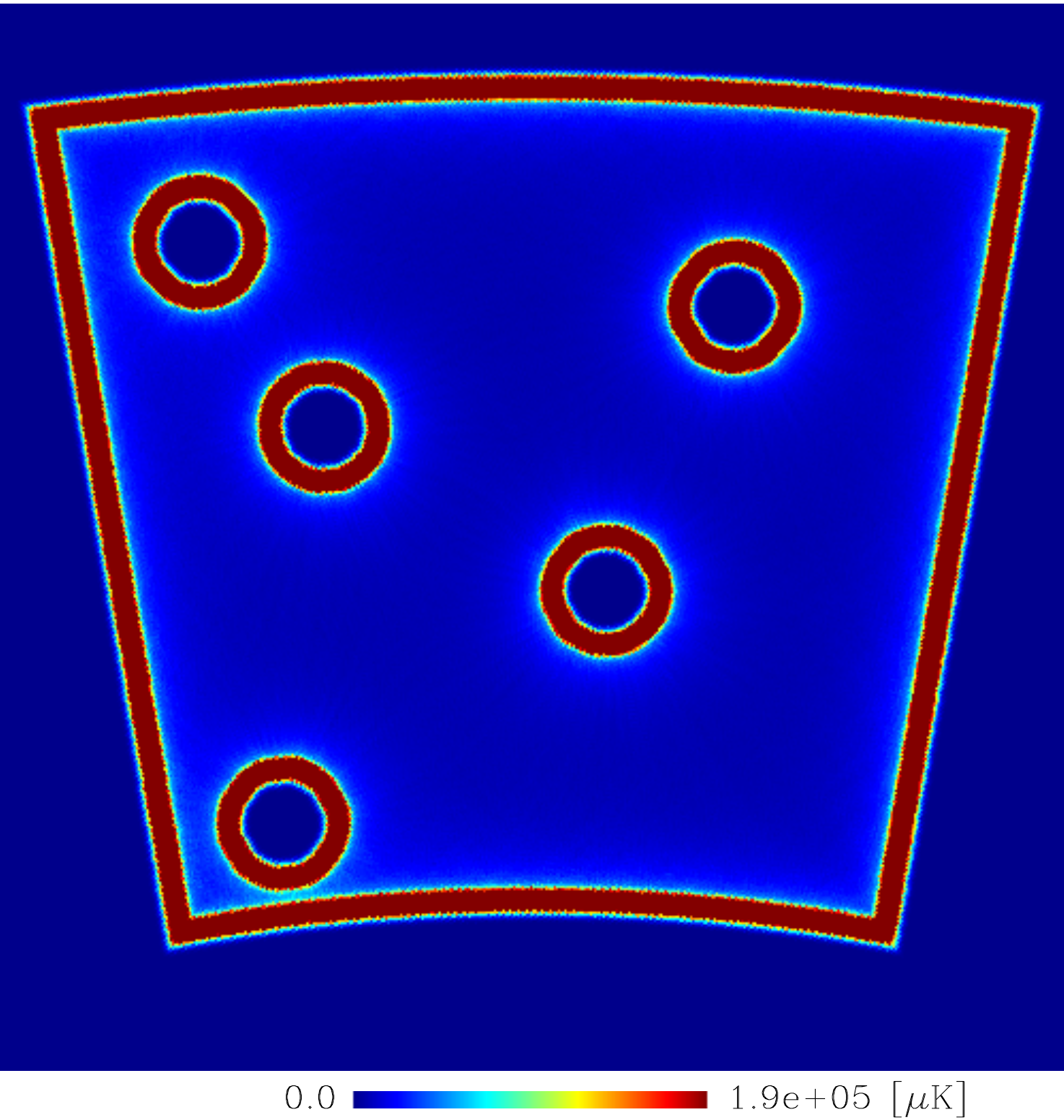}
	\caption{Leakage map reconstructed via \kn pseudspectrum reconstruction. The map in the left (right) panel is obtained using an apodised window function with an apodisation length of $\theta_{apo} = 0.5^o$ ($\theta_{apo} = 1^o$).} 
	\label{fig:chimapkn}
\end{center}
\end{figure}

The window function involved in the \kn~approach has to be designed to reduce the leakage. The left panel of Fig.~\ref{fig:chimapkn} shows the leakage map obtained with an apodised window function of apodisation length $\theta_{apo} = 30$ arcmin. The resulting map for a larger apodisation length of $\theta_{apo} = 1^o$ is displayed in the right panel. The leakage appears to have a different behaviour regarding the chosen apodisation length. For $\theta_{apo} = 0.5^o$ the leakage is more extended although less powerful than for $\theta_{apo} = 1^o$. The leakage therefore appears to be located where $W$ is varying giving insight of which pixels have to be removed. Also, a residual leakage is present in the inner part of the patch but lowered for larger values of $\theta_{apo}$. Moreover, the leakage is more present in these maps than the ones obtained in the pure and \zb~methods.

To summarise, $E$-to-$B$ leakage is intrusive and can potentially spoil an important amount of pixels over all the observed sky, although highly concentrated on the edges of the mask. By the use of the leakage free methods to reconstruct the leakage map, the leakage is diminished and more localised. However, its power and localisation depend on the chosen methods and on the window function shape. And last but not least, the obtained leakage maps are obviously not zero contrary to one would expect from the theory. The issue of reducing the $E$-to-$B$ leakage is thus more subtle than foreseen if applied to realistic CMB pixelised maps. The present analysis can not conclude on the respective efficiency of the different approaches as only the $B$ modes power spectrum reconstruction is relevant to compare performance on $B$ modes estimation.

\section*{Conclusion}

The pseudospectrum estimation offers a fast and reliable method to construct the CMB power spectra on a masked sky. The standard method is nonetheless not optimal as the resulting variance on the estimated spectra are high due to the $E$-to-$B$ leakage only corrected for in the mean. It can significantly damage the measurement of the polarisation power spectra, specially in the case of the faint $B$ modes. Thus, the use of leakage free methods is mandatory to get the most precise detection. Three pseudospectrum methods proposed in the literature has been exposed: they all exactly correct for the leakage in theory. They however seem to give different performances and to depend on the window function shape. Moreover, the $E$-to-$B$ leakage is not the only source of variance in the power spectra reconstruction as at the end, the minimal \textit{global} variance (including the noise) on the estimated power spectra is required. A numerical exploration of the efficiency of the different methods is thus necessary in order to study the obtained total variance on the reconstructed power spectra. 
\setcounter{mtc}{11}
\chapter{Numerical Results on $B$ Modes Estimation} 
\label{Chapter5} 

\lhead{Chapter 5. \textit{Numerical Results}} 
\noindent \hrulefill \\
\textit{`Another reason for the recent focus on analysis is one I hope to convey in this chapter: analysis is exciting.' \newline
S. Dodelson in \cite{dodelson_2003}.
} 
\noindent \hrulefill \\
\minitoc 
The founding principles of three different pseudospectrum approaches (the so-called pure, \zb~ and \kn~methods) built to accurately reconstruct the CMB $B$ modes have been exposed in the previous chapter. \textit{In theory}, these methods are constructed to completely correct for the $E$-to-$B$ leakage. As they involve distinct calculations, their numerical implementation differs thus leading to difference in the reconstructed power spectrum \textit{in practice}. A residue of the leakage mainly owing to the pixelisation of the CMB maps indeed remains and could potentially invalidate all the beforehand theoretical work. The variances on the estimated angular power spectra can consequently be higher than expected from the calculations. The issue addressed in the present chapter is therefore to test the efficiency of each method on realistic cases in the scope of $B$ modes angular power spectrum estimation, specially its variance. 

In this purpose, the \zb~ and \kn~techniques have been implemented and intensely tested, along with the pure method which implementation is presented in \cite{Grain_2009}.  Their performances are examined thanks to numerical simulations applied to two fiducial experimental set-ups, one characteristic of current ground based or balloon-borne experiment (referred to as \textit{small scale survey}) and one typical for potential satellite mission (referred to as \textit{large scale survey}). For the three considered techniques, the best reconstruction is ensured by both a good choice of the window function $W$ and the construction of $B$ modes multipoles free from any leakage. In this chapter, the issue of the window function choice is therefore  briefly tackle. In the second section, the main step of the methods implementations and the used supercomputer system are exposed. The chapter ends on 
the performances of the three considered first at the level of the pseudospectrum for a first glimpse on the leakage and last but not least, on the $B$ angular power spectrum reconstruction. 


\section{Binning Power Spectra}

The considered pseudospectrum approaches allow for a reconstruction of the angular power spectrum multipole-by-multipole. In practice, the power spectra are estimated in multipole band (or band power) of width $\Delta_b$, the power spectra is thus said to be \textit{binned}. This width $\Delta_b$ is set by the experimental multipole resolution which scales as the inverse of the largest angular scales accessible from a given observed region of the sky (in the simplified case of an observed spherical cap, this would scales as the inverse of the square root of the sky fraction). The binning process has two opportune side effects. Firstly, the inversion of the mixing kernel is simplified. 

Secondly, the correlation between the multipoles $\ell$ are reduced. Indeed, the mixing kernel $K_{\ell \ell'}^{\mathrm{method},\pm}$ induces a contribution from the multipole $\ell'$ to the estimated power spectrum at the multipole $\ell$. Averaging over $\ell$ thus reduces these correlations: the correlations between two adjacent bins are lower than the correlation between two adjacent $\ell$.

The pseudospectrum $\tilde{C}_{\ell}$ and the estimated angular power spectrum $\hat{C}_{\ell}$ are averaged over $\ell$ in band power as: 
\begin{eqnarray}
\tilde{C}_{\ell} & = & \sum_{\ell \in b} P_{b\ell}\tilde{C}_{\ell}, \\ \nonumber
\hat{C}_{\ell} & = & \sum_{\ell \in b} P_{b \ell}\hat{C}_{\ell},
\end{eqnarray}
with the binning operator $P_{b \ell}$. Its inverse $Q_{b \ell}$, that can be understood as the interpolation in a band power, is defined as: 
\begin{equation}
P_{b \ell} = \left\{ \begin{array}{cc} \frac{S_{\ell}}{\ell_{max}^b - \ell_{min}^b}, & \ell \in [\ell_{min}^b;\ell_{max}^b], \\
0, & \ell \notin [\ell_{min}^b;\ell_{max}^b], \end{array} \right . 
\end{equation}
\begin{equation}
Q_{b \ell} = \left\{ \begin{array}{cc} \frac{1}{S_{\ell}}, & \ell \in [\ell_{min}^b;\ell_{max}^b], \\
0, & \ell \notin [\ell_{min}^b;\ell_{max}^b]. \end{array} \right . 
\end{equation}

In a first approximation, the CMB angular power spectra roughly behave as $\frac{1}{\ell^2}$, the operator $S_{\ell}$ is consequently expressed as: 
\begin{equation}
S_{\ell} = \frac{\ell(\ell + 1)}{2\pi}.
\end{equation}
This choice makes the power spectra to be averaged over flat band power thus allowing for an even contribution of the $\ell$ within the band power.

As a consequence, the linear system~(\ref{eq:keylinsysCl_main}) to be inverted boils down to: 
\begin{equation}
\left ( \begin{array}{c} \tilde{C}_{b}^{E} - N_{b}^E \\ \tilde{C}_{b}^B - N_{b}^B \end{array} \right )
= \sum_{b'} \left ( \begin{array}{cc} H_{bb'}^{\mathrm{method},+} & H_{bb'}^{\mathrm{method},-} \\ K_{bb'}^{\mathrm{method},-} & K_{bb'}^{\mathrm{method},+}  \end{array} \right ) \left ( \begin{array}{c} \hat{C}_{b'}^E \\ \hat{C}_{b'}^B \end{array} \right ) ,
\label{eq:keylinsysCl}
\end{equation}
with $K_{bb'}^{\mathrm{method},\pm}$ such as: 
\begin{equation}
K_{bb'}^{\mathrm{method},\pm} = \sum_{\ell,\ell'} P_{b\ell} K_{\ell \ell'}^{\mathrm{method},\pm}B_{\ell'}^2Q_{b' \ell'}.
\end{equation} 

In the present analysis, the chosen bandwidth $\Delta_{\ell}$ is $\Delta_{\ell}\sim40$ ensuring a large enough band to reduce the correlation between multipoles. 

\section{Apodised Window Functions}

In order to retrieve leakage free polarised power spectra, the pseudomultipoles are built on a map weighted by a window function $W$. This window function is \textit{a priori} arbitrary, nevertheless it may have to be properly tuned for an efficient reconstruction of the power spectrum. As shown in the previous chapter, the window function $W$ in the case of all the methods has to verify the Dirichlet and Neumann conditions: $W = 0$ and $\partial W = 0$ at the edges of the mask. It indeed guarantees the projection on pure basis and warrants $W$ continuity. An apodised window function varying smoothly from 0 on the unobserved pixels to 1 on the observed region of the sky verifies this condition. The apodisation shape can be chosen in two ways that are exposed in this section. A window function with a simple \textit{analytic} expression which fulfils the Dirichlet and Neumann conditions can indeed be fashioned. Another strategy consists in \textit{optimising} the shape of the window function in the perspective of minimising the variance on the reconstructed $B$ modes power spectrum. 

\subsection{Analytic apodisation}
\label{sec:AnaApo}

A possible procedure to build a window function is to set its boundaries layer equal to an analytic function which verifies the required Dirichlet and Neumann conditions. The window function is consequently fast to compute and do not need any preconception on the CMB signal. A cosine function is usually chosen as in \cite{Smith_2006}. Besides \cite{Grain_2009} have proposed an analytic expression of $W$ such that its second derivative is also vanishing at the boundaries of the observed patch. Such a condition is expected to give better performance as second derivative of $W$ enters in the computation of the $B$ modes pseudomultipoles. The window function is in this case given by:
\begin{equation}
W =  \left\{ \begin{array}{cc}  -\frac{1}{2\pi} \sin(2\pi\frac{\delta_i}{\delta_c}) - \frac{\delta_i}{\delta_c}, & \text{if~} \delta_i < \delta_c, 
\\ 1, & \text{if} ~ \delta_i > \delta_c,
\end{array} \right . 
\label{eq:w_anaapo}
\end{equation}
and with $W = 0$ outside of the observed patch of the sky. The index $i$ stands for the $i^{th}$ pixel of the map, $\delta_i$ is the distance of the $i^{th}$ pixel to the closest boundary and $\delta_c$ represents the width of the boundary layer in which the window function is smoothly decreasing from 0 to 1, also called the apodisation length. This analytically apodised window function has been made using the implementation by M. Betoule\footnote{\url{http://www.apc.univ-paris7.fr/betoule/doku.php?id=fr:software}}. 

As mentioned in the previous chapter, the selection of the apodisation length is a competition between the loss of information on $B$ modes and the $E$-to-$B$ leakage decrease. In this analysis, Monte Carlo (MC) simulations have been carried out in order to explore the performance of the $B$ modes power spectrum reconstruction for different apodisation lengths as detailed in the next section.

This method has consequently the benefit to be quick and to satisfy the required conditions on $W$ for the methods to be reliable. In addition, another strategy have been proposed for the peculiar case of the pure method. 

\subsection{Variance-optimised apodisation}

A family of window functions adapted to the pure $B$ modes estimation have been introduced in \cite{Smith_2007}. The apodisation is optimised regarding the obtained variance on the estimated power spectrum. This apodisation can be either \textit{pixel-based} or performed in the \textit{harmonic} domain as exposed in \cite{Grain_2009}. A qualitative insight on both window function computations are exposed in the following and the reader is referred to the aforementioned articles which give substantial details and tests.

The central idea for the computation of the variance-optimised window functions is to minimise the \textit{global} variance on the reconstructed angular power spectrum. The three pseudospectrum approaches under scrutiny have been shown to theoretically correct for the leakage, thus cancelling the contribution of the leaked $E$ modes to the variance on the $B$ modes power spectrum. The polarisation field is indeed projected on the pure $E$ and $B$ modes basis, removing the ambiguous modes. Although damaging for $B$ modes reconstruction, the ambiguous modes nonetheless hold a part of the $B$ modes signal. They therefore represent a loss of information which could potentially increase the variance on $B$ modes reconstruction. As a consequence, cancelling the leakage does not warrant the minimisation of the global variance. The issue tackled down in the window function computation is thus more subtle and comes down to a competition between the minimisation of the $E$-to-$B$ leakage and the loss of information on the $B$ modes.

As expounded in Chapter~\ref{Chapter4}, the best and lossless quadratic estimator $\hat{C}_{\ell}^{XY}{}^{(opt)}$ for the polarisation angular power spectrum is such as: 
\begin{equation}
\left ( \begin{array}{c} \tilde{C}_{\ell}^{EE, (opt)} - N_{\ell}^{EE} \\ \tilde{C}_{\ell}^{BB, (opt)} - N_{\ell}^{BB} \end{array} \right ) 
= \frac{1}{2} \sum_{\ell'} \left ( \begin{array}{cc} \mathrm{F}_{\ell \ell'}^{EE EE} & \mathrm{F}_{\ell \ell'}^{EE BB} \\ 
\mathrm{F}_{\ell \ell'}^{BB EE} & \mathrm{F}_{\ell \ell'}^{BB BB}  \end{array} \right ) \left ( \begin{array}{c} \hat{C}_{\ell'}^{EE}{}^{,(opt)} \\ \hat{C}_{\ell'}^{BB}{}^{,(opt)} \end{array} \right ) ,
\end{equation}
with $F_{\ell \ell'}^{XY}$ the Fisher matrix of the $C_{\ell}^{XY}$ power spectrum and where the data have been filtered by the inverse of the covariance matrix. 
The performance of a given estimator \textit{i.e.} its ability to give the smallest uncertainties on the $B$ modes reconstruction, therefore boils down to know its discrepancy from the best estimator $\hat{C}_{\ell}^{XY}{}^{,(opt)}$. The pure estimator gives the opportunity to tune the window function $W$ to make the estimator as close as possible to the optimal one. It amounts to the inversion of the following system:
\begin{equation}
\sum_{i = 1}^{N_{obs}} C_{ij}P_{ij}^{b}W_i = 1 \text{~for all j},
\label{eq:linsyst}
\end{equation} 
where $i$,$j$ denote the pixel index, $N_{obs}$ stands for the number of observed pixels, $C_{ij}$ is the covariance matrix of the data, $P_{ij}^{(b)}$ the projection matrix averaged on a band power and $W_i$ the optimised window function.

The pixel-based approach consists in solving for the linear system Eq.~(\ref{eq:linsyst}) in the pixel domain dealing with the non-diagonal matrix $P_{ij}^{b}$. The implementation made by \cite{Grain_2009} is based on an iterative preconditioned conjugate gradient method. Such a procedure enables to solve this inversion problem in a minimum of iterations thanks to a preconditioner already close to the inverse of $C_{ij}P_{ij}^{b}$. The pixel-based variance-optimised window functions will be denoted PCG window function. 

The crucial point of this approach is that it allows for flexibility in the computation of the window function. The spin-s window functions $W_s$ are related to spin-0 window function $W$ and its derivative following:
\begin{eqnarray}
W_0 & = & W, \\ \nonumber
W_1 & = & \partial W, \\ \nonumber
W_2 & = & \partial \partial W.
\label{eq:Ws}
\end{eqnarray} 
As developed in the following section, the construction of the pure pseudomultipoles does not explicitly require these conditions to be exactly verified. The pixel-base approach precisely authorises $W_0$, $W_1$ and $W_2$ to be independent. The PCG window function can consequently be applied to the pure $B$ modes estimation. Relaxing these constraints allows for the minimisation of the global variance by means of remaining ambiguous modes, thus avoiding an excessive loss of information on the $B$ modes. Although this PCG computation requires numerical time of about $n_{iter}n_{pix}^2$ with $n_{ites} \propto 100$ (few thousands) for a noise of $5.75 ~(1.) \mu K$-arcmin, it can be performed in a reasonable amount of time with current supercomputer. This approach has the advantage to be systematised and very robust. 

When the noise is uncorrelated and homogeneous over the observed part of the sky, the inversion of Eq.~(\ref{eq:linsyst}) can be done in the harmonic domain. The linear system is reduced to a division in the harmonic domain, the numerical time therefore scales as $N_{pix}^{3/2}$. In this case, the conditions on the window function lead to constraints on the mask and its contour harmonic representation. From those quantities, the spin-1 an spin-2 window functions are constructed. This approach has the merit to be fast as it basically only involves spherical harmonics transforms (SHT). However, numerical issues such as the SHT of the contour of the mask lead to small discrepancy from the Dirichlet and Neumann conditions. This issue is handling by adding an extra-apodisation on the window function which makes it slightly suboptimal with respect to the previous pixel based approach.

\begin{figure}[!h]
\begin{center}
	\includegraphics[scale=0.17]{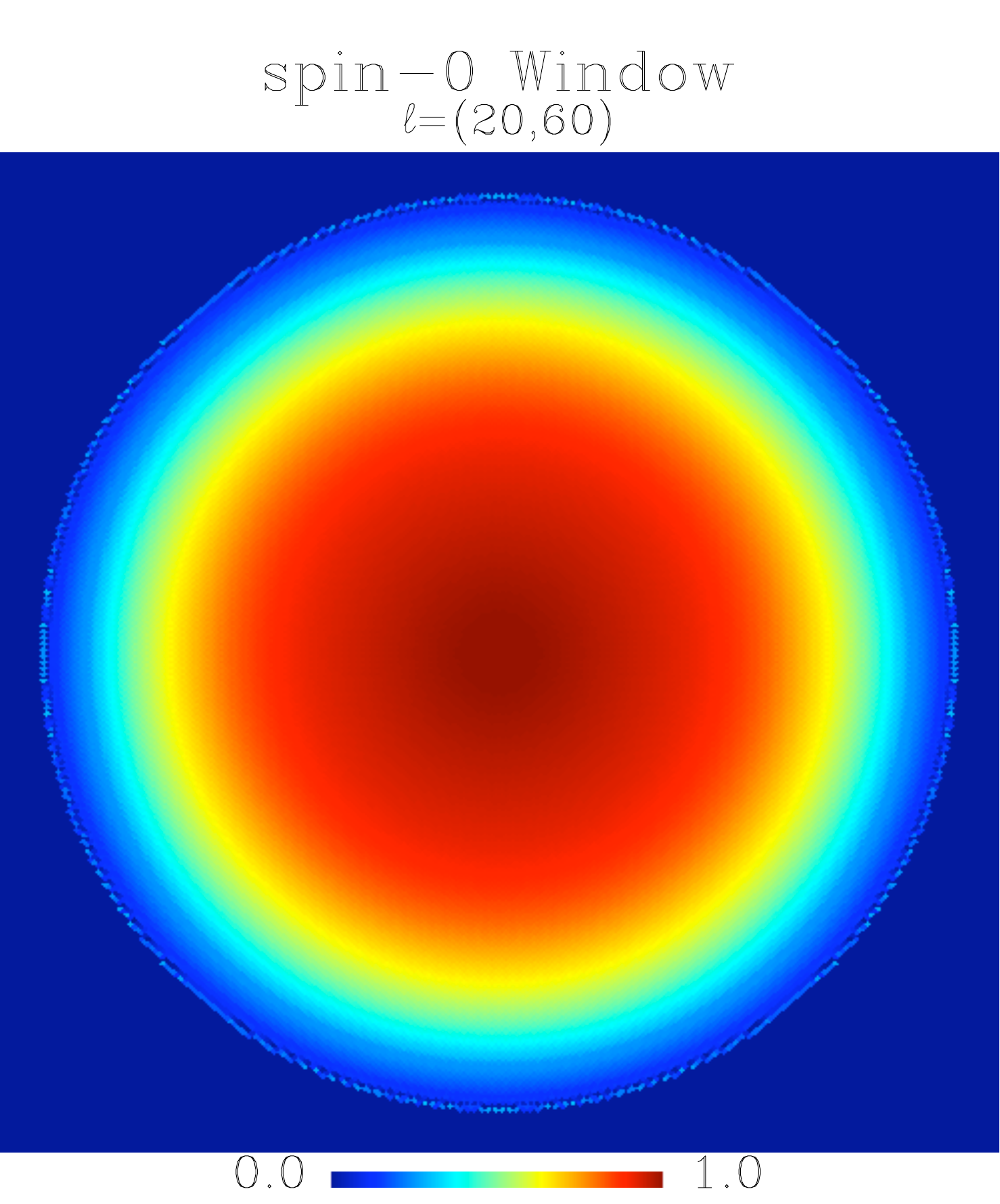} \includegraphics[scale=0.17]{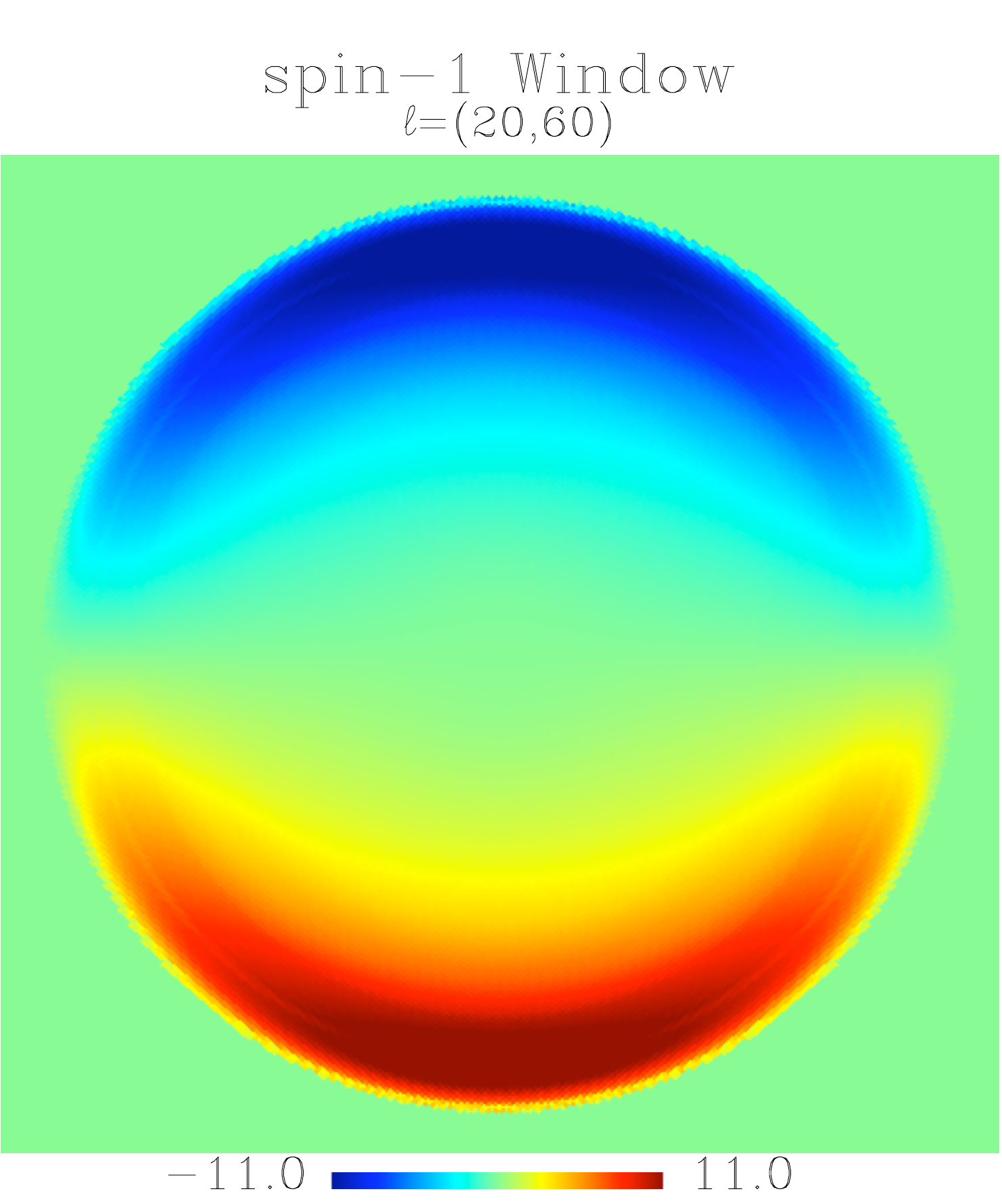} 
	\includegraphics[scale=0.17]{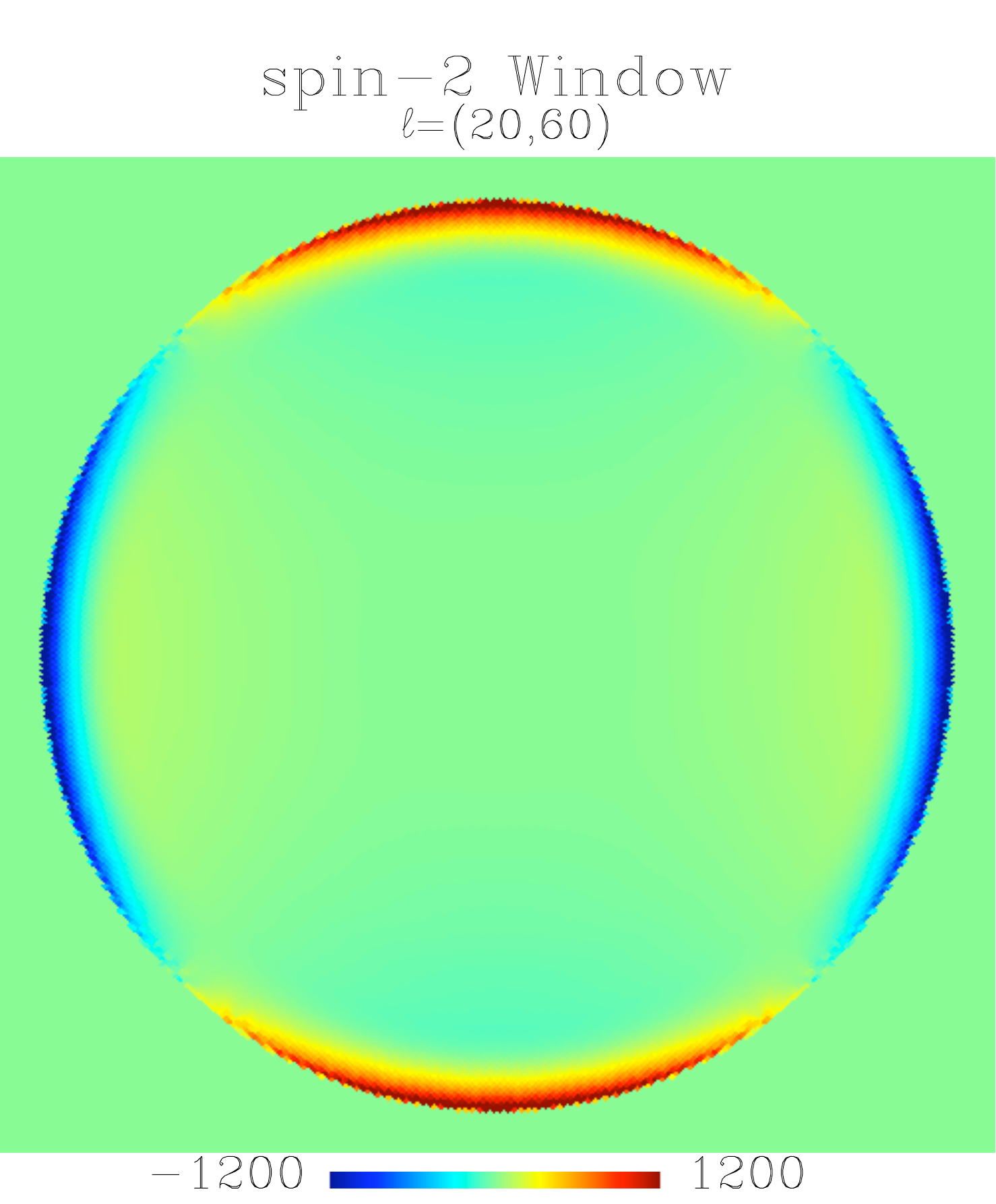} \\
	\includegraphics[scale=0.17]{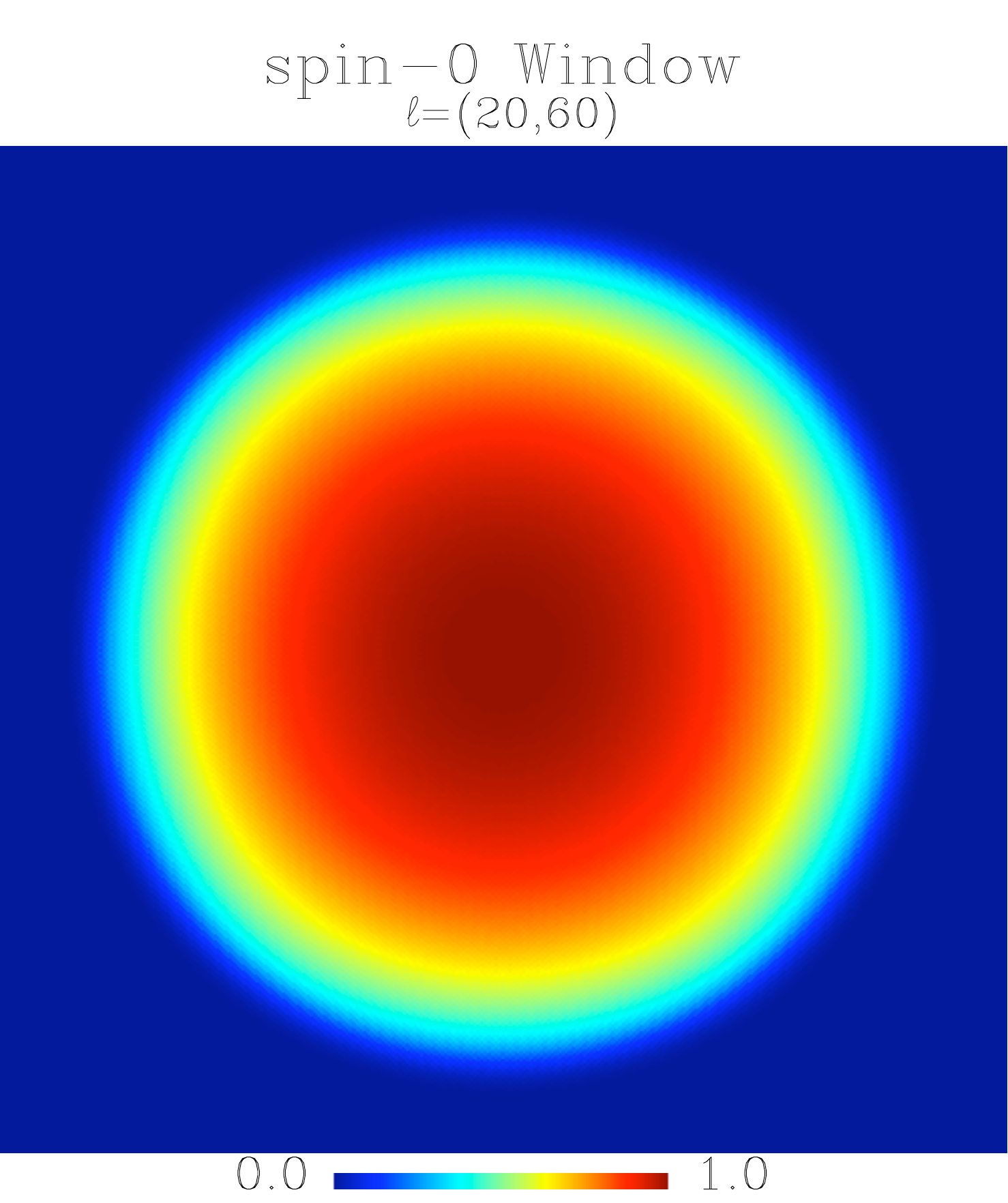} \includegraphics[scale=0.17]{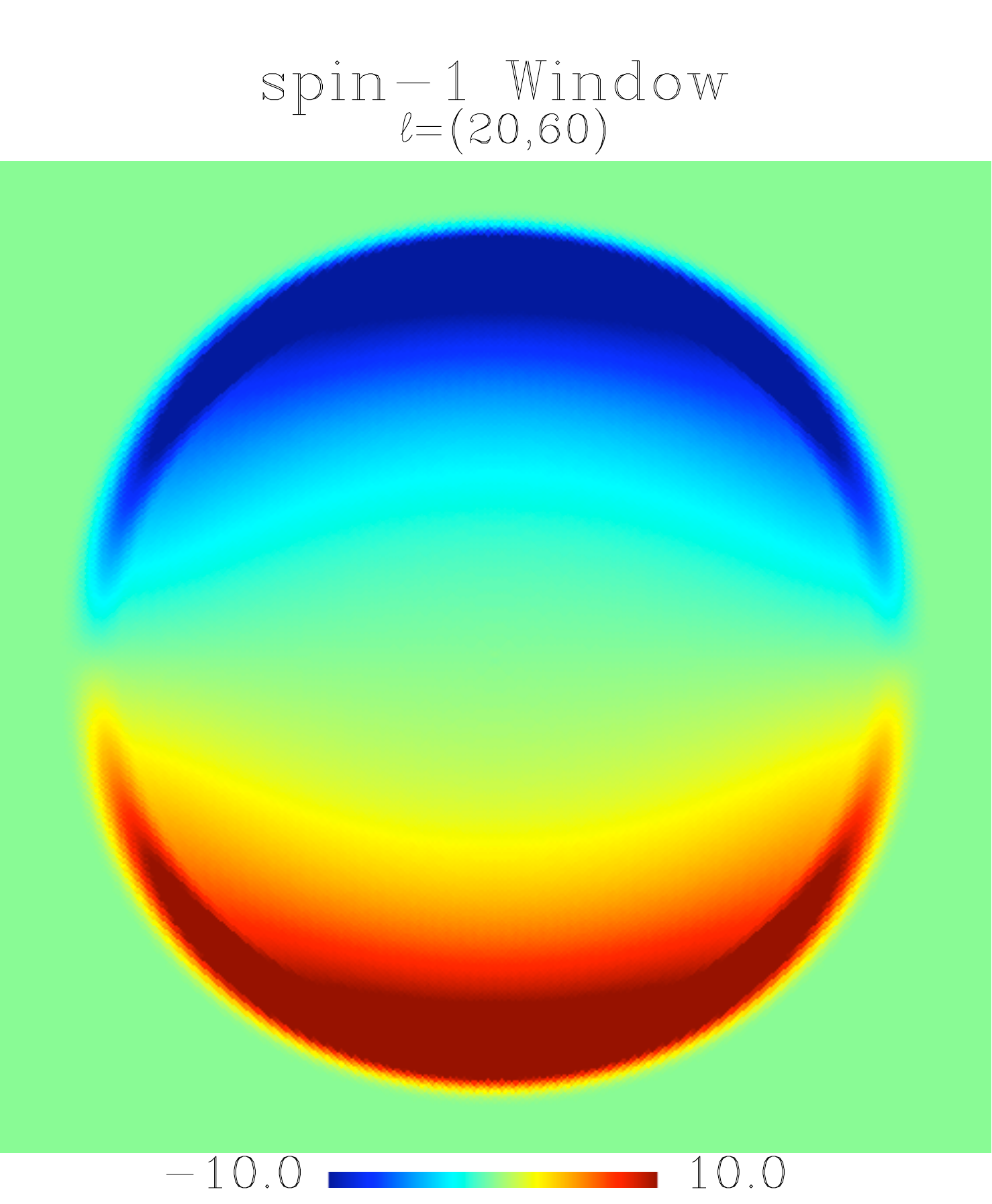} 
	\includegraphics[scale=0.17]{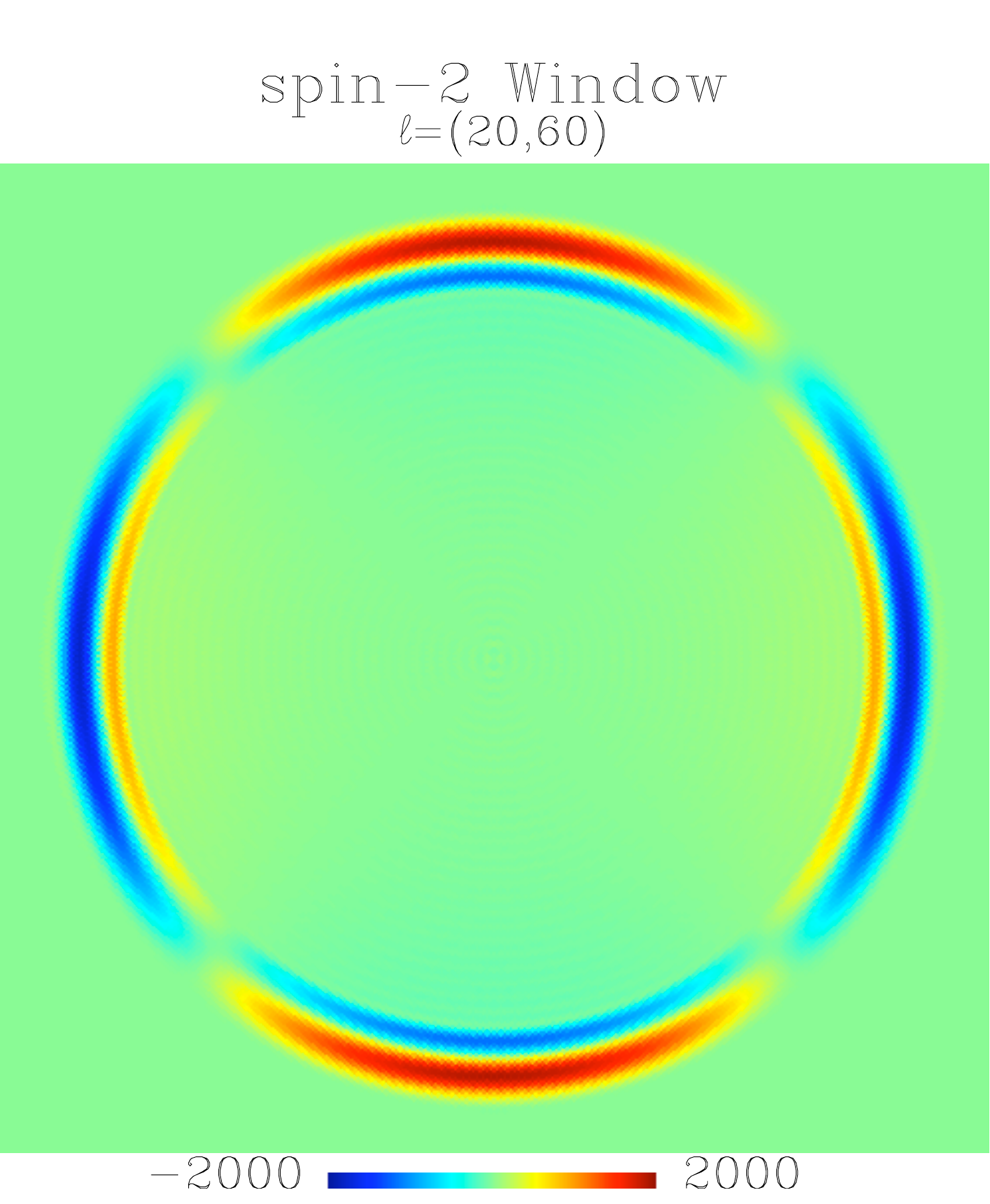} \\
	\includegraphics[scale=0.17]{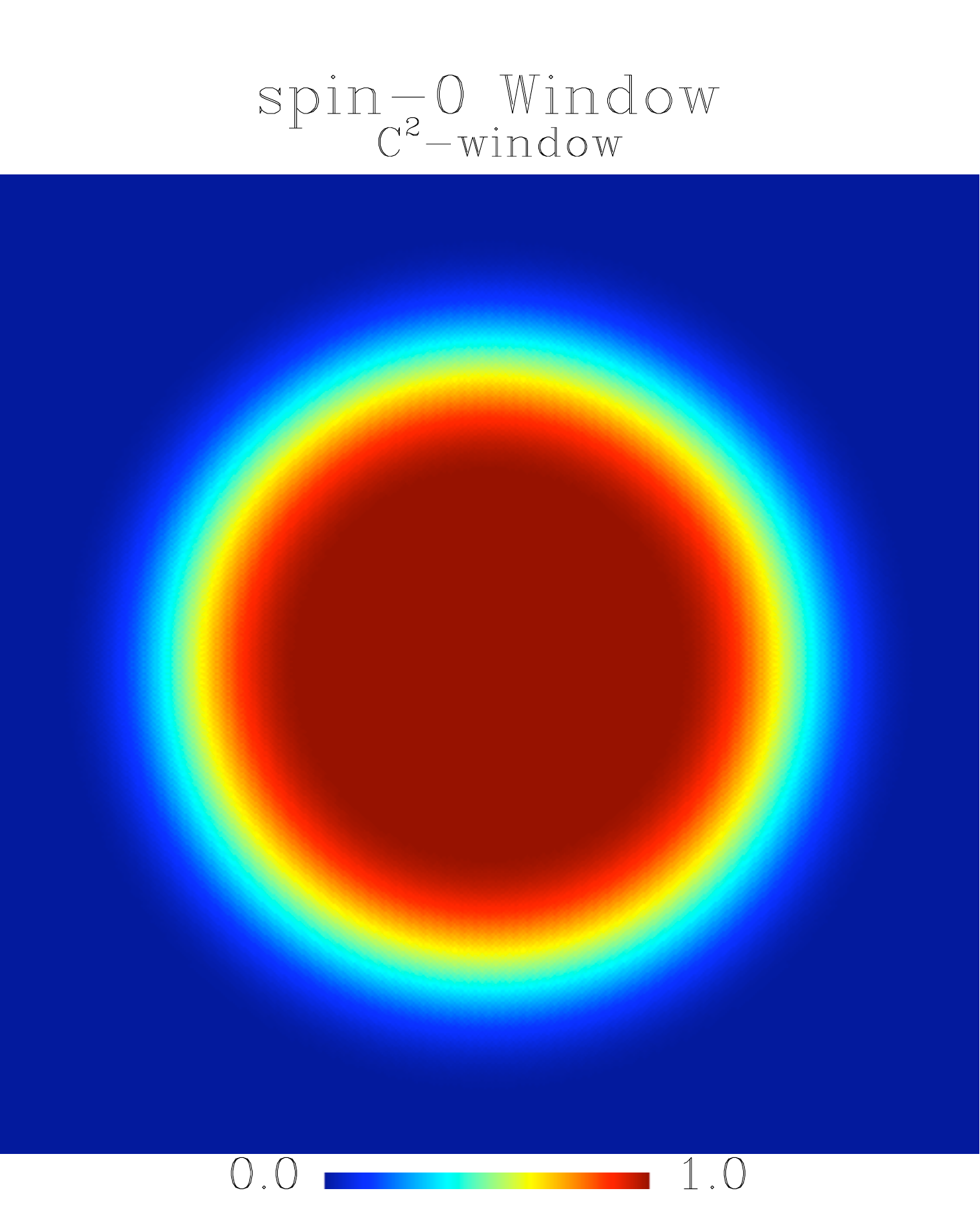} \includegraphics[scale=0.17]{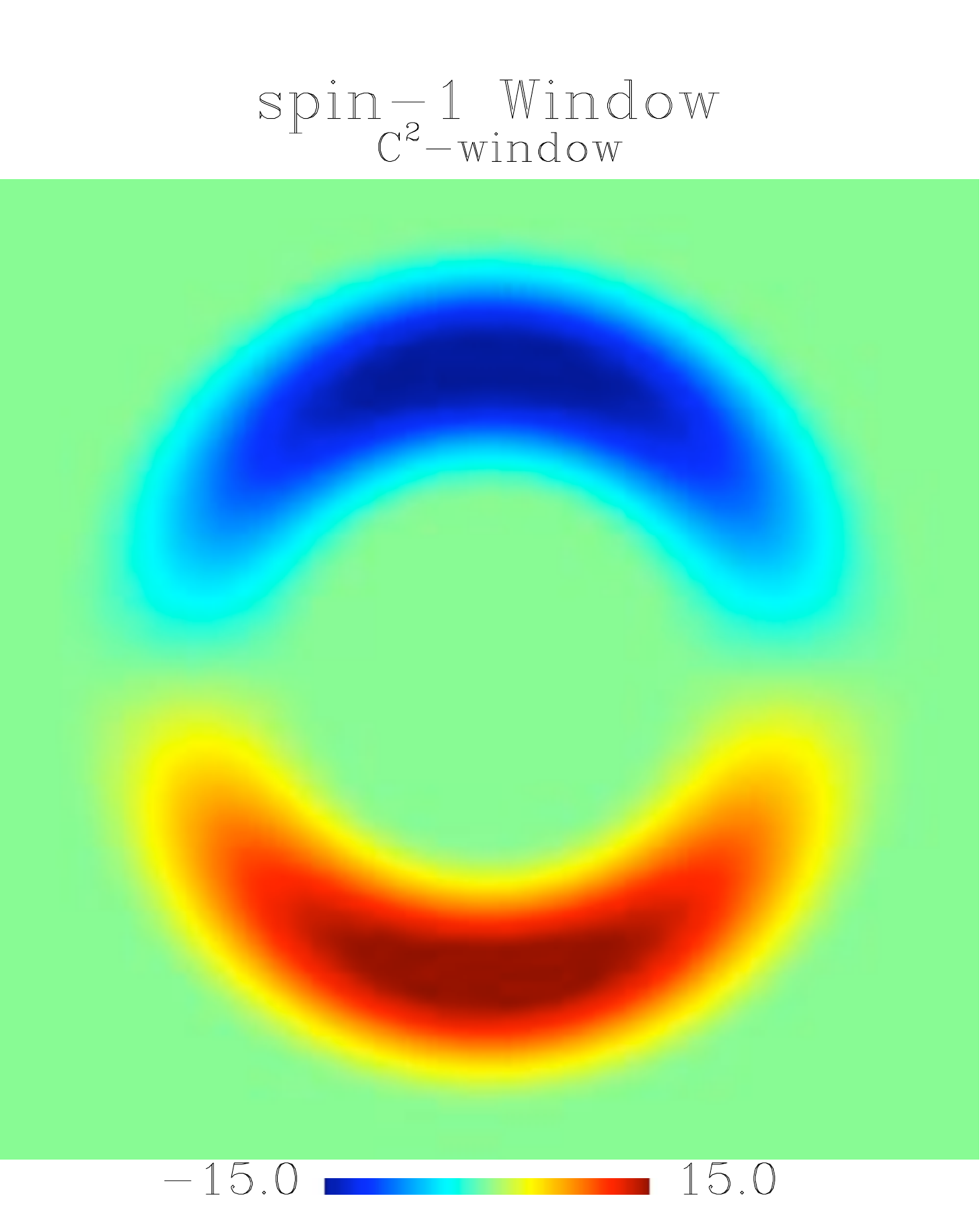} 
	\includegraphics[scale=0.17]{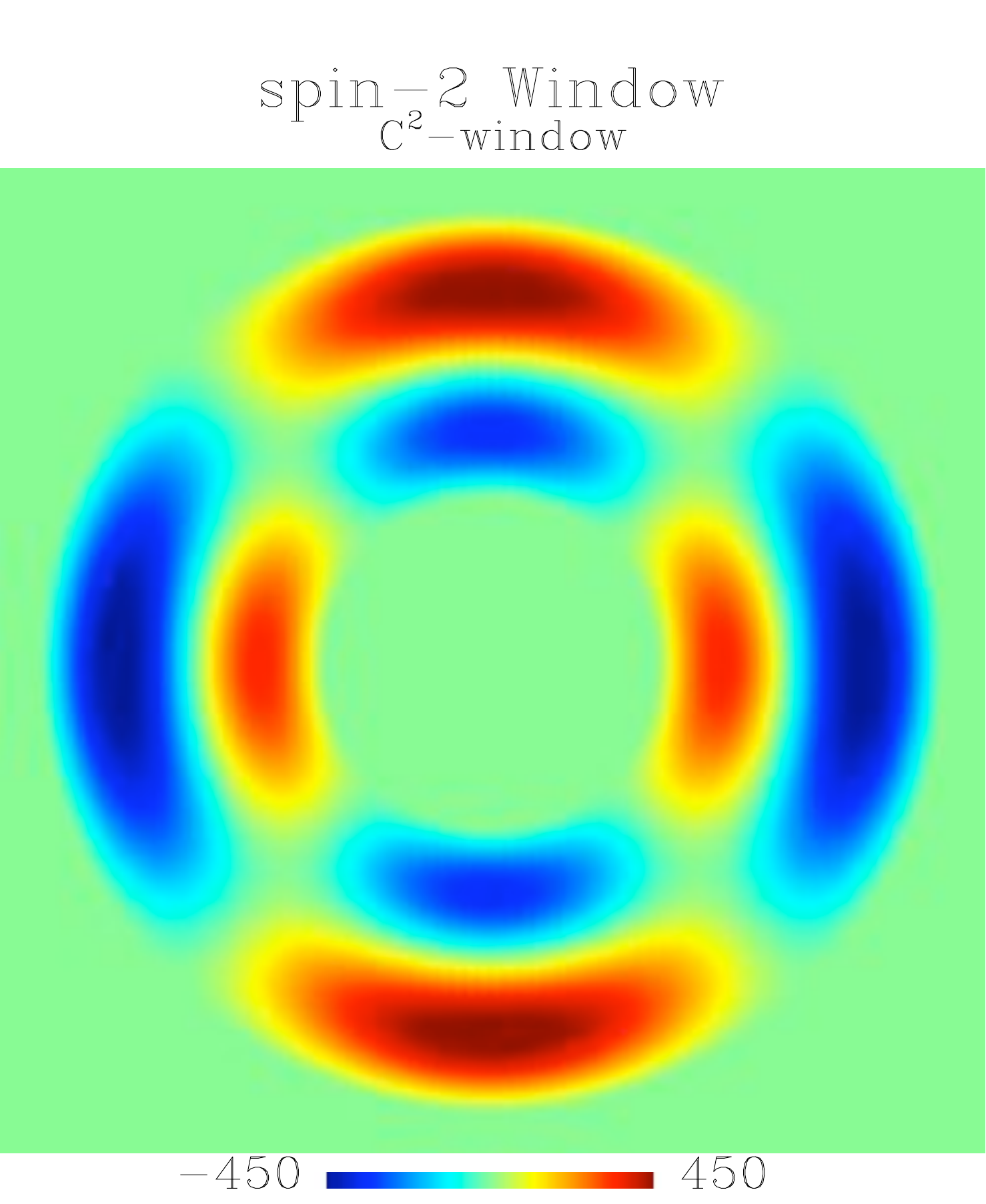} \\
	\caption{From left to right: the spin-0, spin-1 and spin-2 window functions of a spherical cap of radius $11^o$. The input signal is $E$ modes from WMAP-5yr and the induced lensing $B$ modes, assuming a primordial $B$ modes signal with $r = 0.05$ and a noise level of $5.75 \mu K$-arcmin. Only the real part of the spin-1 ans spin-2 window functions are depicted. The top and middle rows show the spin-weighted optimized window function respectively in the pixel and the harmonic space, for the bin $\ell \in [20,60]$. The bottom row illustrates the window function computed with an analytic apodisation. The figures are taken from \cite{Grain_2009}.} 
	\label{fig:windowfct}
\end{center}
\end{figure}

Moreover, as shown in Eq.~(\ref{eq:linsyst}), the covariance matrix $C_{ij}$ intervenes in the window function optimisation. It therefore requires a prior on the signal to be reconstructed. The impact of the prior on the window function optimisation has therefore been extensively explored in \cite{Grain_2009}. The $E$-modes and the induced lensed $B$-modes being well known, the main issue is the prior on the primordial $B$ modes. The drawn conclusion is that the derived optimised window function shape is fortunately little affected by the \textit{a priori} assumption on the primordial $B$ modes signal.

To summarise, examples of the window functions for a spherical cap of radius $11^o$ computed using the three aforementioned approaches are depicted in Fig.\ref{fig:windowfct}. The figures are taken from \cite{Grain_2009} where they assume a $E$ mode signal from WMAP-5yr and the induced lensing $B$ modes, along with a primordial signal with a tensor-to-scalar ratio $r = 0.05$. They show different features albeit they satisfy the similar conditions.

The pixel-based variance-optimised window functions enable to find a compromise between the cancellation of $E$-to-$B$ leakage and the induced loss of information on $B$ modes information. The independence of the spin-weighted window functions is indeed permitted leading to a better exploration of the $W$ apodisation which provides the minimal global variance on the reconstructed $B$ modes. Therefore, the PCG window functions are expected to give the best performance in the perspective of $B$ modes reconstruction, thanks to the flexibility of its implementation. However, the computation in the harmonic domain has the advantage of being fast and might give the same results as the PCG window functions for simple contour of the observed sky patch. 

The question that has now to be addressed is: which strategy should we choose to properly estimate the $B$ modes for a given experiment? The theoretical investigation developed in the previous chapter tells us that they are conceptually equivalent. Their numerical implementations might however show different behaviour as foreseen through the study of the leakage maps (in Chapter~\ref{Chapter4}) and the different usable families of window functions.  


\section{Numerical Implementations}

The pure method has been implemented and intensely studied in \cite{Grain_2009} specially in the case of a small scale survey and was extended to $TB$ and $EB$ correlations in \cite{Grain_2012}. The main ideas of its implementation and operating system are described in the present section. Our own implementation of the \zb~ and \kn~methods rests upon the same design as the pure method although their practical computations are different, essentially in the computation of the pseudomultipoles and the convolution kernels. The principles of the produced codes will be exposed afterwards. For any selected strategy, a large numbers of CPUs and parallelisation implementation (chosen to be made in MPI) are nonetheless required. We will therefore first dwell on the description of the used supercomputer system.

\subsection{The Hopper system at NERSC and numerical tools}

The National Energy Research Scientific Computing Center (NERSC) is a division of the Lawrence Berkeley National Laboratory and is based in Oakland. Its goal is to provide one of the world best computer facilities for scientific research. The computational power and storage system are colossal, the NERSC distinctive feature being its attentive maintenance. The facilities attract diverse field of research from climatology to biology or high energy physics. The NERSC has 6 computational systems including one of the most powerful, the Hopper system used for this analysis. 

Hopper is a CRAY XE6 system made of a total of 153,216 cores with the peak performance of 1287 TFlops/sec. The Hopper system was therefore ranked as the 28th of TOP500 list of November 2013 which classifies the world most powerful computer system.

\begin{figure}[!h]
\begin{center}
	\includegraphics[scale=0.33]{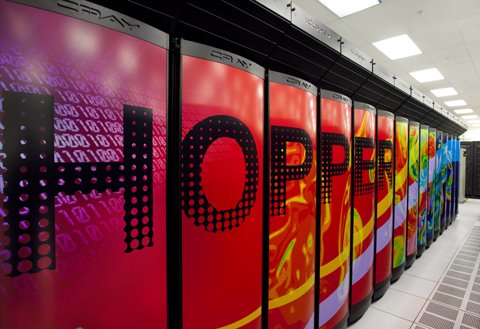}
	\caption{The hopper system (from \cite{nersc}).} 
	\label{fig:hopper}
\end{center}
\end{figure}

The architecture of the Hopper system is a key concern for the speed of the calculation time. The system is composed of 6,384 nodes, each of them organised in 2 twelve-cores processors. The transfer between cores in a given node being faster than between two nodes, the required number of processors for numerical computation will have to be carefully chosen in order to minimised the computation time. 

Both the computation of the window function and the simulations for the estimation of the CMB power spectra have been implemented on the Hopper system for it provides very robust and efficient facilities. The main numerical issues reside in the spherical harmonic transforms computations, solving a linear system and inverting large matrices. The use of the MPI library is therefore require to take advantage of the supercomputer system.  
Nonetheless, the available tools for SHT were not adapted to parallelised computation: the {\sc s$^2$hat} spherical harmonics transform library has been therefore implemented by \cite{Stompor_2011}, a description of which can be found in \cite{s2hat}. This library is implemented to be optimal in the distribution in the harmonic and in the pixel spaces and has the benefit to be automatised. It was also adapted in the p{\sc s$^2$hat} library for the pure polarised multipoles, its specifications can be found on the \cite{ps2hat}.

\subsection{The pure method}

As mentioned above, the pure method implementation is at the heart of \cite{Grain_2009}. The implemented code is denoted by {\sc x$^2$pure} and is made of three main steps. 

\underline{Step 1: pseudomultipoles}

In the scope of the implementation of the pure method, \cite{Smith_2006,Smith_2007,Grain_2009} propose to compute the pure pseudomultipoles formulated as Eq.~(\ref{eq:pseudopurealm}) in Chapter~\ref{Chapter4} as they avoid a direct derivation of the noisy maps $P_{\pm 2}$. We recall the formulation for the $B$ modes: 
\begin{eqnarray}
\tilde{a}_{\ell m}^{B} & = & ~ \frac{i}{2} \sqrt{\frac{1}{\alpha_{\ell,2}}} \int_{\Omega} [P_{2}(\vec{n}) (\partial\partial W(\vec{n})Y_{\ell m}(\vec{n}))^* - P_{-2}(\vec{n})(\bar{\partial}\bar{\partial}W(\vec{n})Y_{\ell m}(\vec{n}))^*],
\label{eq:purebmode}
\end{eqnarray}
remembering that $\alpha_{\ell,s} = \frac{\sqrt{\ell+s}}{\sqrt{\ell-s}}$.

For convenience, the following spin-weighted window functions are introduced:
\begin{eqnarray}
W_0 = W,~W_1 = \partial W,W_2 = \partial \partial W.
\end{eqnarray}
The spin-1 and spin-2 window functions are potentially complex but as $W$ is real, we have: $W_{s}^* = W_{-s}$.
Three spin-weighted apodised maps are also introduced:
\begin{eqnarray}
\mathcal{P}_{0} = W_0P_{\pm2},~ \mathcal{P}_{\pm 1} = W_{\pm 1} P_{\pm2},~ \mathcal{P}_{\pm 2} = W_{\mp2} P_{\pm 2}
\end{eqnarray}

Consequently, by performing the spin-lowering or -raising operators and using the properties of the window function, the Eq.~(\ref{eq:purebmode}) is written as:
\begin{equation}
\tilde{a}_{\ell m}^B = \frac{1}{\alpha_{\ell,2}} (\mathcal{B}_{0,\ell m} + 2\alpha_{\ell,1}\mathcal{B}_{1,\ell m} + \alpha_{\ell,2}\mathcal{B}_{1,\ell m})
\label{eq:keyeq}
\end{equation}
with $\mathcal{B}_{s,\ell m}$ standing for the $B$ modes pseudomultipoles of the spin-weighted polarised maps:
\begin{equation}
\mathcal{B}_{s,\ell m} = \frac{i}{2} \int \left [ \mathcal{P}_{+s}(\vec{n}) {}_{s}Y_{\ell m}^*(\vec{n}) - (-1)^s \mathcal{P}_{-s}(\vec{n}) {}_{-s}Y_{\ell m}^*(\vec{n}) \right ] d\vec{n}.
\end{equation}

The previous Eq.~(\ref{eq:keyeq}) is the implemented equation to build the pseudomultipoles\footnote{The $W\chi^{E/B}$ maps (such as the ones in Sec.~\ref{sec:leakmap} in Chapter~\ref{Chapter4}) are constructed by projecting these pseudomultipole on the spherical harmonics.} and then straightforwardly the corresponding pseudospectrum: $\tilde{C}_{\ell}^B = \frac{1}{2\ell + 1}\sum\limits_m \tilde{a}_{\ell m}^B\tilde{a}_{\ell m}^{B*}$.

\underline{Step 2: mixing kernels}

The mixing kernels $K_{\ell \ell'}^{\mathrm{pure},+/-}$ embody the relation between the pseudospectrum and the true power spectrum estimator and only depend on the window function. Therefore, from the multipoles of the window function and the computation of the Wigner symbol, the convolution kernel are directly computed and their expressions are shown in Appendix~\ref{AppendixA}. Their explicit computations is possible thanks to the independence of the spin-weighted window functions $W_s$. The mixing kernel $K_{\ell \ell'}^{\mathrm{pure},-}$ (vanishing in theory) can therefore be evaluated and thus amount the residual leakage due to the pixelisation.

\underline{Step 3: power spectrum}

The last step of the implementation of the pure method consists in the inversion of the linear system of Eq~\ref{eq:keylinsys} in Chapter~\ref{Chapter4} relating the true power spectrum to its estimator.  
\newpage
\underline{Window function apodisation}

The three kinds of apodised window functions, analytic and the two variance-optimised apodisations, can be use in the scope of the pure CMB power spectra estimation.

\subsection{The \zb~method}

\underline{Step 1: pseudomultipoles}

The masked $W^2\chi^B$ field is reconstructed on the observed part of the sky as explicitly formulated in the Eq.~(\ref{eq:eqzb}) of Chapter~\ref{Chapter4}. The explicit derivative operations are performed in the harmonic domain for more manageability. The pseudomultipoles of the computed field are then straightforwardly carried out by projecting the masked $\chi^{E/B}$ on the spherical harmonics: $\tilde{a}_{\ell m}^{E/B} = \int W^2 \chi^{E/B} Y_{\ell m}$.

\underline{Step 2: mixing kernels}

The $W^2\chi^B$ field is a scalar field, the mixing kernel $K_{\ell \ell'}^{\mathrm{zb},\pm}$ is therefore computed in the same way as for temperature, which expression is shown in Appendix~\ref{AppendixA}. The mixing kernel $K_{\ell \ell'}^{\mathrm{zb},-}$ is set to zero and thus ignores the remaining leakage coming from the pixelisation.

\underline{Step 3: power spectrum}

Ultimately, the polarisation power spectrum is retrieve by inverting the pseudospectrum linear system Eq.~(\ref{eq:keylinsys}) in Chapter~\ref{Chapter4}.

\underline{Widow function apodisation}

The computation of the pseudomultipole is such as the equalities in Eq.~(\ref{eq:Ws}) cannot be relaxed, thus preventing for the use of the PCG window function. The harmonic variance-optimised along with the analytically apodised window functions are suitable.

\subsection{The \kn~method}

\underline{Step 1: pseudomultipoles}

The $\tilde{\chi}^B$ field is first computed on the sky weighted by a window function $W$ with an apodisation length $\theta_{apo}$. The estimation is performed afterwards on a mask $M^{W\chi^B}$ cutting a layer of width $\theta_{cut}$ which contains only pixels poorly affected by leakage.   

\underline{Step 2: mixing kernels}

The mixing kernels are computed in the same way as for the \zb~method also setting $K_{\ell \ell'}^{\mathrm{kn},-}$ equal to zero.

\underline{Step 3: power spectrum}

As previously, the $B$ modes power spectrum is retrieved by inverting Eq.~\ref{eq:keylinsys} in Chapter~\ref{Chapter4}.
\newpage
\underline{Window function apodisation}

This approach only allows for the use of analytic apodisation. In \cite{Kim_2011}, a window function apodisation with a Gaussian profile was used. We however choose apodisation such as Eq.~\ref{eq:w_anaapo} in our analysis. Besides, our implementation is made to test various apodisation lengths at the same time. In that way, the apodisation length $\theta_{apo}$ is tuned by performing Monte Carlo simulations in the perspective of the performance on the bias and variance on the reconstructed $B$ modes power spectrum. In the analysis, the results for three values of $\theta_{apo}$ are shown. Also, we have seen that cut such as $\theta_{cut} = \theta_{apo}$ ensures a `pure' enough binary mask.

\subsection{Inputs}

In our analysis, the implementations of the three methods have been used in a simulation mode. Schematically, from an input power spectrum $C_{\ell}^{theo}$, the $a_{\ell m}$ are drawn from a Gaussian distribution of mean zero and variance $C_{\ell}^{theo}$. The procedures described above are applied to reconstruct the input $C_{\ell}^{theo}$ for each method. As we do not have an analytic expression for the variances, Monte Carlo (MC) simulations are performed in order to estimated the error bars on the power spectrum reconstruction. The speed of pseudospectrum approaches enable to carry out such simulations. In this analysis, $N_{sim} = 500$ simulations were required in order to minimise the scattering variance. 

In practice, the inputs of the codes are divided to in two parts. The first part is dedicated to the computation of the convolution kernel $K_{\ell,\ell'}^{\mathrm{method},\pm}$. The required inputs are the: \newline
- spin-0, spin-1 and spin-2 window functions $W_0$, $W_1$ and $W_2$; \newline
- maximal multipoles $\ell_{max}$ up to which the convolution kernel are computed.

The second part consists in the $N_{sim}$ CMB temperature and polarisation angular power spectra reconstruction. The inputs of the codes for all the methods are: \newline
- the input theoretical CMB temperature and polarisation power spectra $C_{\ell}^{theo}$ (or $I,Q,U$ map); \newline
- above computed mixing kernels for a given window function $W$; \newline
- noise level of the temperature and polarisation maps;\newline
- The spin-0, spin-1 and spin-2 window functions $W_0$, $W_1$ and $W_2$; \newline
- Beam function $b_{\ell}$; \newline
- The maximal multipoles $\ell_{max}^{sim}$ up to which the simulations are performed and $\ell_{max}^{est}$ up to which the spectra are estimated.


\section{Numerical Results: Pseudospectra and Angular Power Spectra}

The pure, \zb~and \kn~methods are theoretically constructed to give the same $B$ modes power spectrum reconstruction free from any leakage. They may nonetheless give different results due to their distinct numerical implementation. The practical behaviour of each of these pseudospectrum approaches and their relative efficiency is the driving question of the present section. At the end, we will be able to conclude on which power spectrum reconstruction is most appropriate for CMB data analysis. 

All along the analysis, the input signal for the simulations is a $E$ modes signal according to the parameters obtained by WMAP-7 years in \cite{larson_2011}. The $B$ modes signal includes the lensing part and a fiducial primordial contribution with r = 0.05.

\subsection{Fiducial experimental set-ups}

For an utter study of the efficiency of the different strategies applied to realistic CMB maps, we have designed two fiducial experimental set-ups mainly differing on their observed sky fraction: a small and a large scale survey. Both set-ups have indeed distinct issue specially regarding to the mask shape. Albeit they are idealised, their outlines are based on current or forthcoming CMB experiments dedicated to $B$-modes.

\begin{figure}[!h]
\begin{center}
	\includegraphics[scale=0.2]{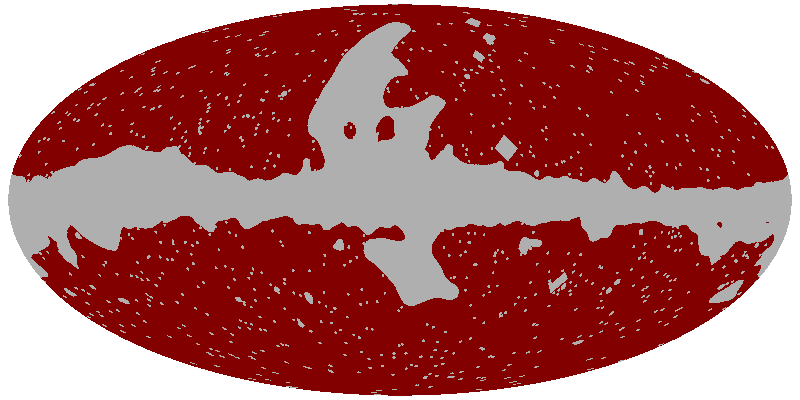}\includegraphics[scale=0.2]{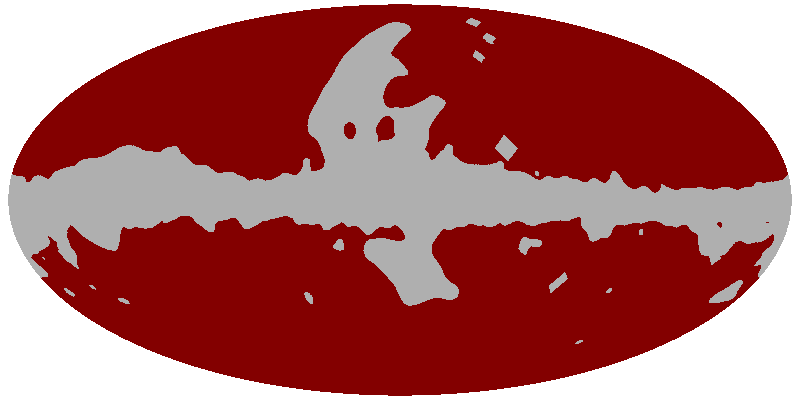}
	\caption{\textit{Left panel}: The binary mask of the fiducial satellite-like experiment designed from the polarised galactic mask of WMAP-7yr release with the corresponding polarised point sources catalogue removal. The (grey) red area is the (non) observed region of the sky. \textit{Right panel}: same with no point sources removal.} 
	\label{fig:maskwmap}
\end{center}
\end{figure}

The large scale experiment imitates is characteristic of a satellite-like experiment dedicated to the $B$ modes detection with the specifications of the potential forthcoming satellite such as EPIC-2m described in \cite{EPIC_2008}. It is typical of Stage IV (expected around the year 2020) experiments as described in \cite{Abazajian_2014}. The beam is supposed to be a Gaussian with a FWHM of 8 arcmin while the noise level is 2.2 $\mu K$-arcmin. In order to simulate a realistic large scale coverage, the galactic mask from WMAP-7yr release adding its point source catalogue mask is used. The obtained sky fraction is then about $71\%$ of the celestial sphere and the shape of the mask is displayed in Fig.\ref{fig:maskwmap}. As the large angular scales are the main interest in this analysis, the chosen HEALPIX pixelisation $N_{side}$ is of 512 corresponding to a pixel resolution of 7 arcmin. A sub-case of the satellite-like experiment was also defined with the same instrumental characteristics but now without masking the point sources. The mask therefore boils down to the galactic mask alone as shown in the right panel of Fig.~\ref{fig:maskwmap}. Such a sky coverage can be achieved by performing an in-painting of the holes masking the point sources. This set-up will be useful for the following analysis in order to study the effect of the holes in the mask.

\begin{figure}[!h]
\begin{center}
	\includegraphics[scale=0.25]{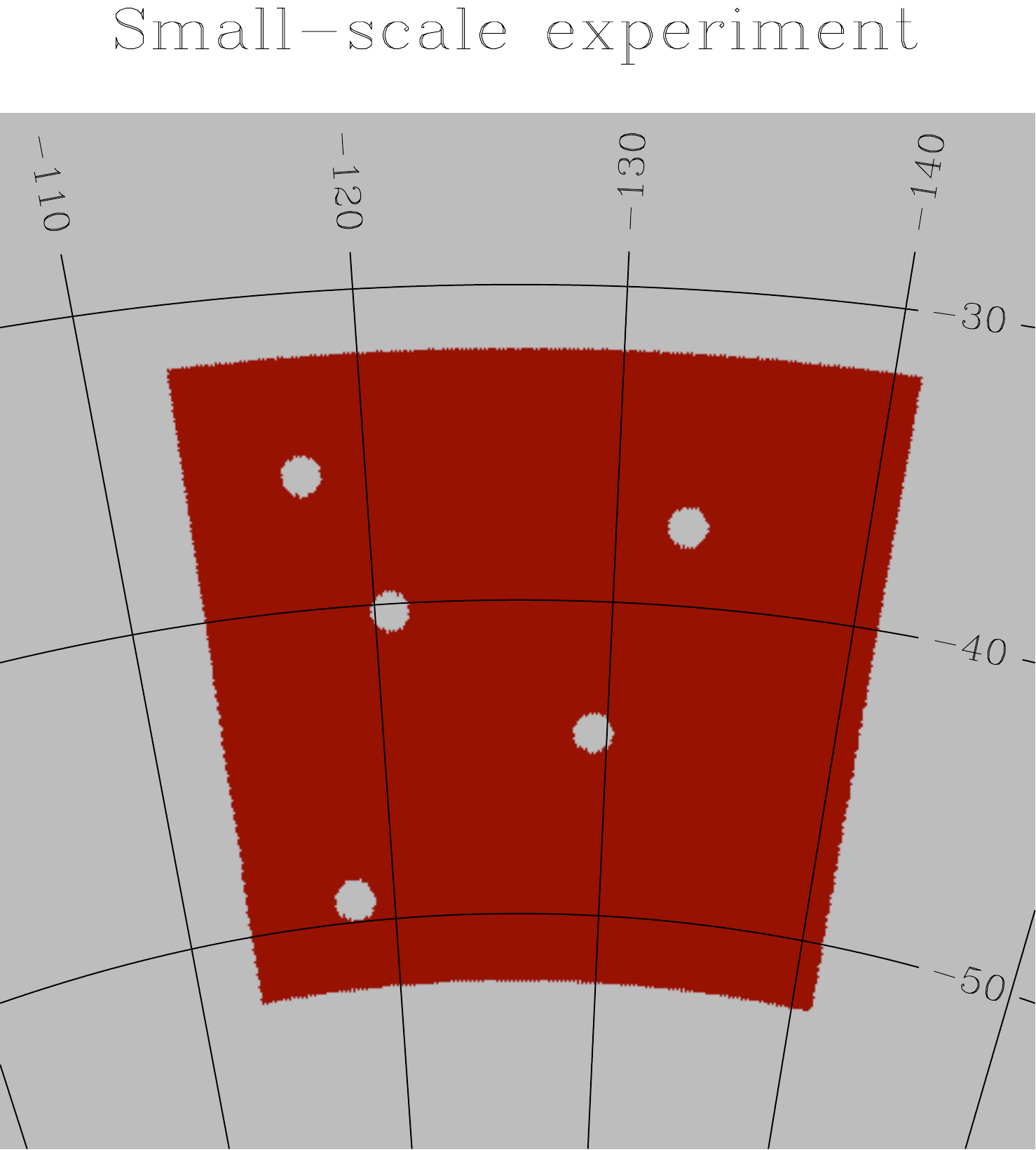}
	\caption{Binary mask of the fiducial small scale experiment covering $1\%$ of the sky. The holes in the mask corresponds to masked point sources. The (grey) red area is the (non) observed region of the sky. } 
	\label{fig:maskebex}
\end{center}
\end{figure}

The designed fiducial small scale experiment is inspired by the balloon-borne experiment {\sc EBEX} described in \cite{ebex} which has flew in 2013. The observed sky coverage is $f_{sky} = 1\%$ with a Gaussian beam of $8$ arcmin and an homogeneous noise level of 5.75$\mu K$-arcmin. The designed mask, shown in Fig.~\ref{fig:maskebex}, is a square of area $400$ square degrees including 5 holes standing for possible cut out of polarised foregrounds. In the {\sc healpix} convention, the chosen $N_{side}$ is of $1024$ leading to a pixel size of roughly 3.5 arcmin.

The fiducial experiments being designed, we are now able to test the different approaches of $B$ modes reconstruction for the two distinct scanning strategies. A first hint of the method efficiency is the resulting pseudospectra which quantify the remaining $E$-to-$B$ leakage.


\subsection{At the pseudospectrum level: looking at the $E$-to-$B$ leakage}

For both fiducial experimental set-ups, the conclusions drawn on the leakage minimisation by the use of the three described approaches are expected to be different, the mask shape being deeply dissimilar. The remaining leakage will therefore be under scrutiny in the case of a satellite-like experiment first where the leakage is presumed to be low, followed by the case of the small-scale survey.

The $B$-modes pseudospectrum $\tilde{C}_{\ell}^{BB}$ is straightforwaldy built from the $B$ pseudomultipoles $\tilde{a}_{\ell m}^B$ as computed above in the case of the standard, pure, \zb~and \kn~approaches: 
\begin{equation}
\tilde{C}_{\ell}^{BB} = \frac{1}{2\ell +1}\sum_m \tilde{a}_{\ell m}^B\tilde{a}_{\ell m}^{B^*}.
\end{equation}

The pseudospectra are easily calculated by performing a mean on the different simulated pseudospectra. Moreover, we point out that at this stage the binning is not required as no matrix inversion is involved. 

This analysis consists in measuring the residual leakage which boils down to amount the ratio between the two following relevant quantities. The first quantity to evaluate is the $B$ modes pseudospectrum only coming from the leaked $E$ modes. In practice, $\tilde{C}_{\ell}^{E\rightarrow B} = \sum\limits_{\ell'}K_{\ell \ell'}^{\mathrm{method},-}C_{\ell}^{EE}$ is calculated with solely $E$-modes in input (no $B$ modes). The second quantity amounts the $B$ modes contribution to the $B$ modes pseudospectrum. It is quantified thanks to the $B$ modes pseudospectrum $\tilde{C}_{\ell}^{B\rightarrow B} = \sum\limits_{\ell'}K_{\ell \ell'}^{\mathrm{method},-}C_{\ell}^{BB}$ produced with only $B$ modes in input. Comparing these quantities or their ratio gives an estimate of the left-over leakage.


\subsubsection*{Standard method: large scale and small scale surveys}

\begin{figure}[!h]
\begin{center}
	\includegraphics[scale=0.4]{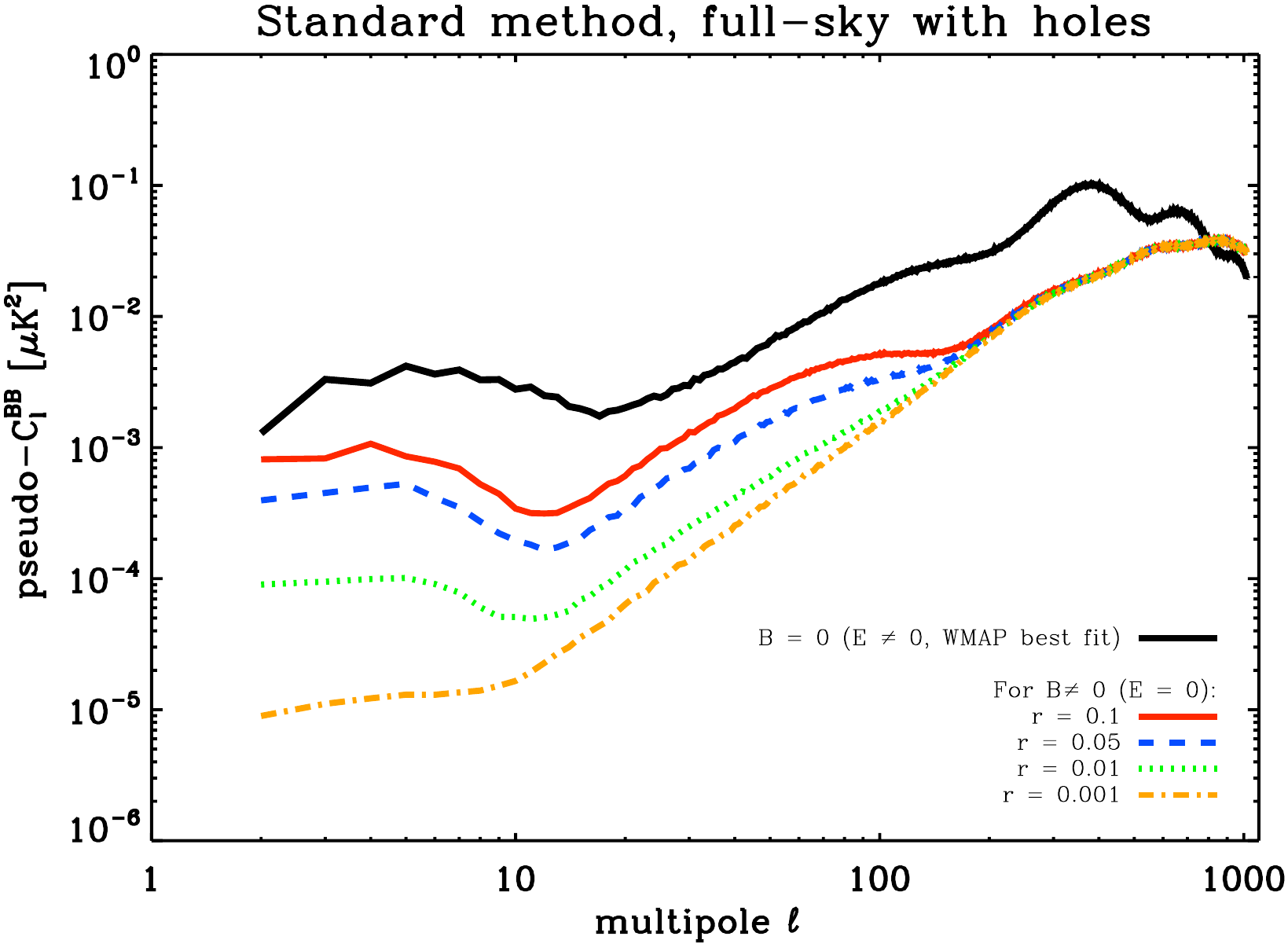}\includegraphics[scale=0.44]{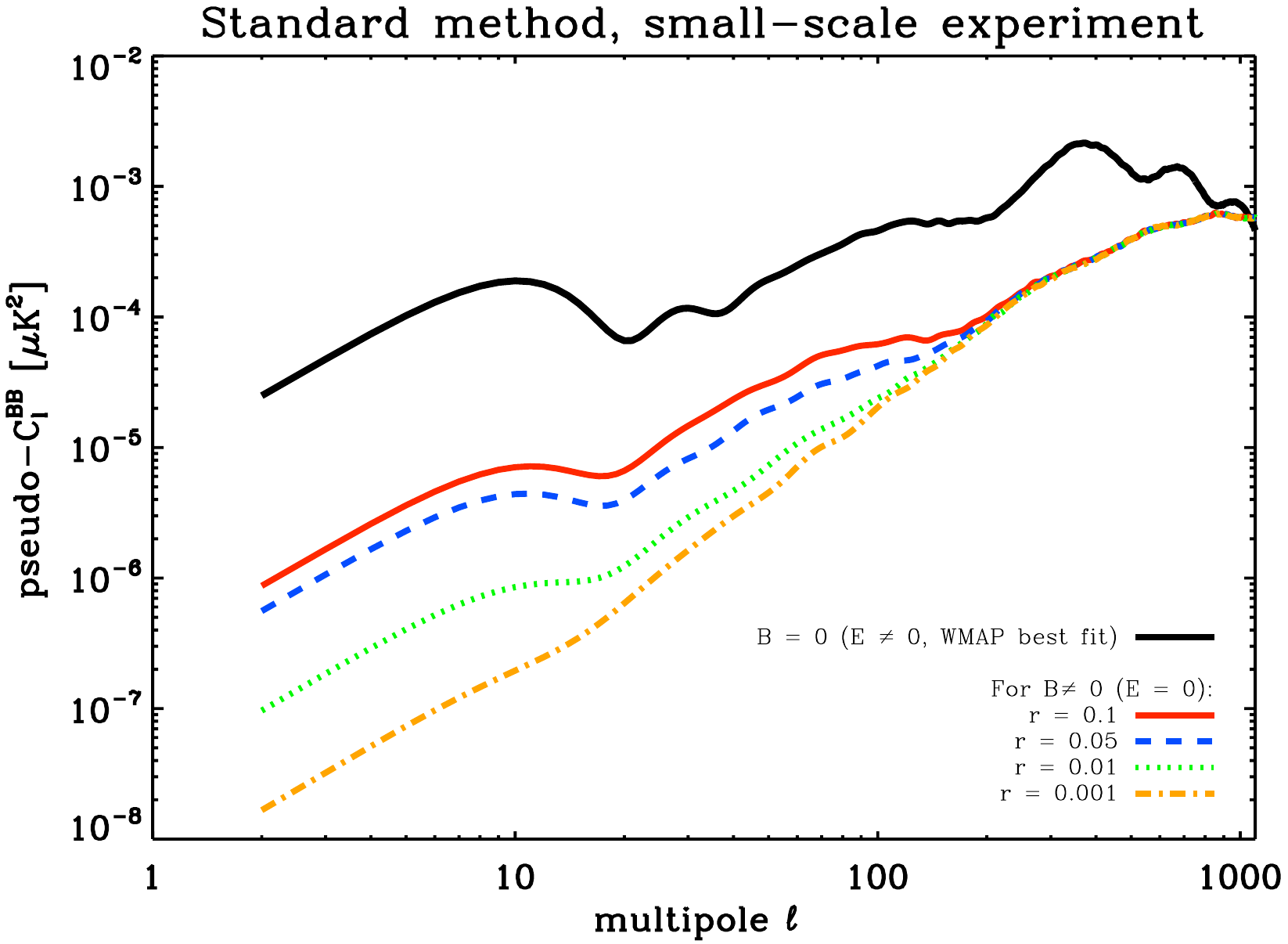}
	\caption{\textit{Left panel:} The pseudospectrum $\tilde{C}_{\ell}^{E \rightarrow B}$ obtained with no input $B$ modes and $E$ modes power spectrum from WMAP-7yr best fit in solid black line obtained using the standard method for a large scale experiment. The coloured curves stand for the pseudospectrum $\tilde{C}_{\ell}^{B \rightarrow B}$ with no input $E$ modes and a theoretical $B$ modes power spectrum for different values of $r$ ($r = 0.001,0.01,0.05,0.1$) along with the lensing contribution.\textit{Right panel:} Same for the small scale experiment.}
	\label{fig:pseudo_compar_std}
\end{center}
\end{figure}

The first outlook to appreciate the control of the leakage by the different approaches is to reconstruct the $B$ modes pseudospectrum in the standard method. By virtue of its non vanishing mixing kernels, the standard method is expected to lead to high amount of leakage. The resulting pseudospectra are shown in Fig.~\ref{fig:pseudo_compar_std} in the case of a large scale survey (left panel) and of a small scale survey (right panel). The obtained contribution\footnote{Notice that the $B$ modes pseudospectrum $\tilde{C}_{\ell}^{B \rightarrow B}$ does not have the shape of the input $B$ modes angular power spectrum as one would expect. Indeed the pseudospectrum is related to the true angular power spectrum by a convolution which affects the spectrum shape.} from the leakage to the $B$ modes, $\tilde{C}_{\ell}^{E\rightarrow B}$, is displayed in solid black line. The contribution from the $B$ modes only, $\tilde{C}_{\ell}^{B \rightarrow B}$, is shown in coloured lines for different values of the tensor-to-scalar ratio $r$ ranging from 0.001 to 0.1. 

First of all, it is noticeable that the level of the factitious contribution $\tilde{C}_{\ell}^{E\rightarrow B}$ is higher, by roughly two orders of magnitude, than the one in the small scale survey. In a first approximation, the pseudospectrum level indeed scales with the observed sky fraction. This explains the discrepancy between the obtained $\tilde{C}_{\ell}^{E\rightarrow B}$ on the small and large scale surveys.
The amount of residual leakage is quantified by the ratio between the two contributions $\tilde{C}_{\ell}^{E\rightarrow B}$ and $\tilde{C}_{\ell}^{B\rightarrow B}$. As a first guess, this ratio is expected to be small in the case of large scale experiment as the accessible part of the sky is large. The obtained results displayed in Fig.~\ref{fig:pseudo_compar_std} are striking: the contribution from the $E$ modes to $\tilde{C}_{\ell}^{BB}$ is around one order magnitude higher than the one from the true $B$ modes, up to $\ell = 800$ for $r = 0.1$. This discrepancy is more marked in the case of a small scale experiment: $\tilde{C}_{\ell}^{E\rightarrow B}$ is higher than $\tilde{C}_{\ell}^{B \rightarrow B}$ up to $\ell = 1000$ with at least one order of magnitude. At low $\ell$ and for $r = 0.001$, the contribution is 3 orders of magnitude higher than the one from the $B$ modes due to the low signal of the primordial $B$ modes.

As a result, even for a detection of the CMB polarisation on $71\%$ of the sky, the leakage between polarisation modes is significant and has to be considered. The reason for such a high leakage seems to be due to the shape of the mask instead of its covered area. The WMAP-7yr polarised galactic mask shown in the right panel of Fig.~\ref{fig:maskwmap} and the holes masking the foregrounds are indeed associated to give a complicated and twisted contours of the mask.

\begin{figure}[!h]
\begin{center}
	\includegraphics[scale=0.4]{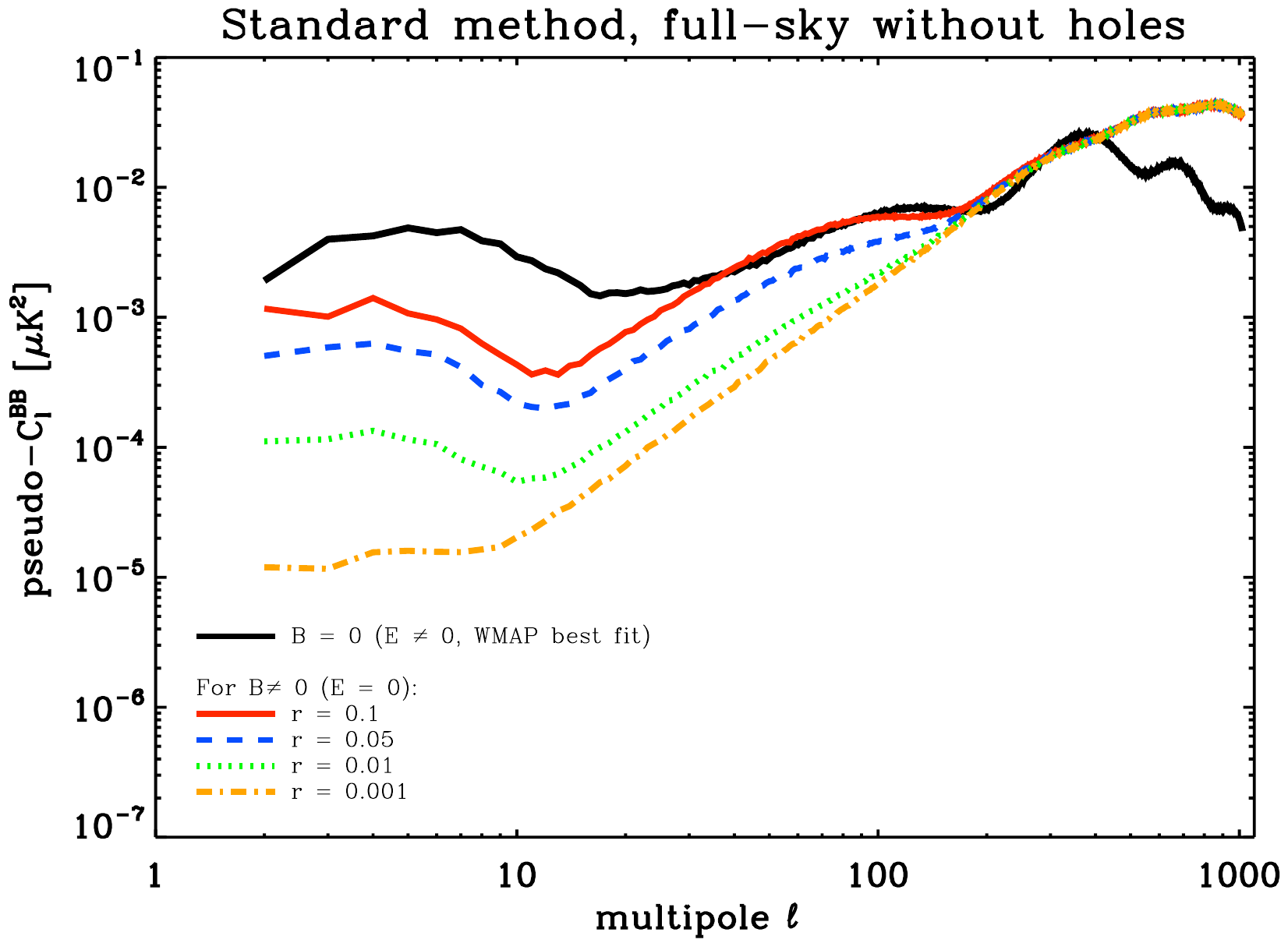}
	\caption{The pseudospectrum $\tilde{C}_{\ell}^{E \rightarrow B}$ obtained with no input $B$ modes and $E$ modes power spectrum from WMAP-7yr best fit in solid black line obtained using the standard method for a large sky coverage without holes. The colored curves stand for the pseudospectrum $\tilde{C}_{\ell}^{B \rightarrow B}$ with no input $E$ modes and a theoretical $B$ modes power spectrum for different values of $r$ ($r = 0.001,0.01,0.05,0.1$) along with the lensing contribution.} 
	\label{fig:pseudo_noholes_std}
\end{center}
\end{figure}

The shape of the mask consequently appears to be the key ingredient in the level of the $E$-to-$B$ leakage as indicated in \cite{Bunn_2003}. We therefore propose to quantify the impact of the holes on the overall leakage level. The corresponding mask is solely the galactic one as illustrated in the right panel of Fig.~\ref{fig:maskwmap} and covers $73\%$ of the celestial sphere. The resulting pseudospectra are displayed in Fig.~\ref{fig:pseudo_noholes_std} with the same conventions as previously. The $E$ modes contribution to the $B$ modes pseudospectrum is lower with respect to the case with holes. Nonetheless, the true $B$ modes contribution to $\tilde{C}_{\ell}^{BB}$ is higher and exceeds the leakage for $\ell$ higher than 200. In particular, for $r=0.1$, the $E$ modes contribution $\tilde{C}_{\ell}^{E \rightarrow B}$ is equivalent to the true signal $\tilde{C}_{\ell}^{B \rightarrow B}$ in the range of the recombination bump. The amount of $E$-to-$B$ leakage is therefore reduce when filling the holes of the mask. While masking the point sources only slightly reduces the observed sky fraction, it does have a substantial impact on the amount of leakage. The holes indeed add a lot of disjointed small edges to the mask thus increasing the source of $E$-to-$B$ leakage.

As a conclusion, this analysis does \textit{not} necessarily mean that the variance on the $B$ modes power spectrum reconstruction will be higher than its amplitude. However it demonstrates that the $B$ modes sampling variance is dominated by the contribution from the leaked $E$ modes. Moreover, the issue of the shape mask has been arose in the case of the full sky survey which appears to be more intricate than expected and thus have to be investigated. 


\subsubsection*{Leakage free methods: the satellite-like survey}

The pure, \zb~ and \kn~methods offer a way to correct for the leakage by cancelling the mixing matrix $K_{\ell \ell'}^{-}$.

\begin{figure}[!h]
\begin{center}
	\includegraphics[scale=0.4]{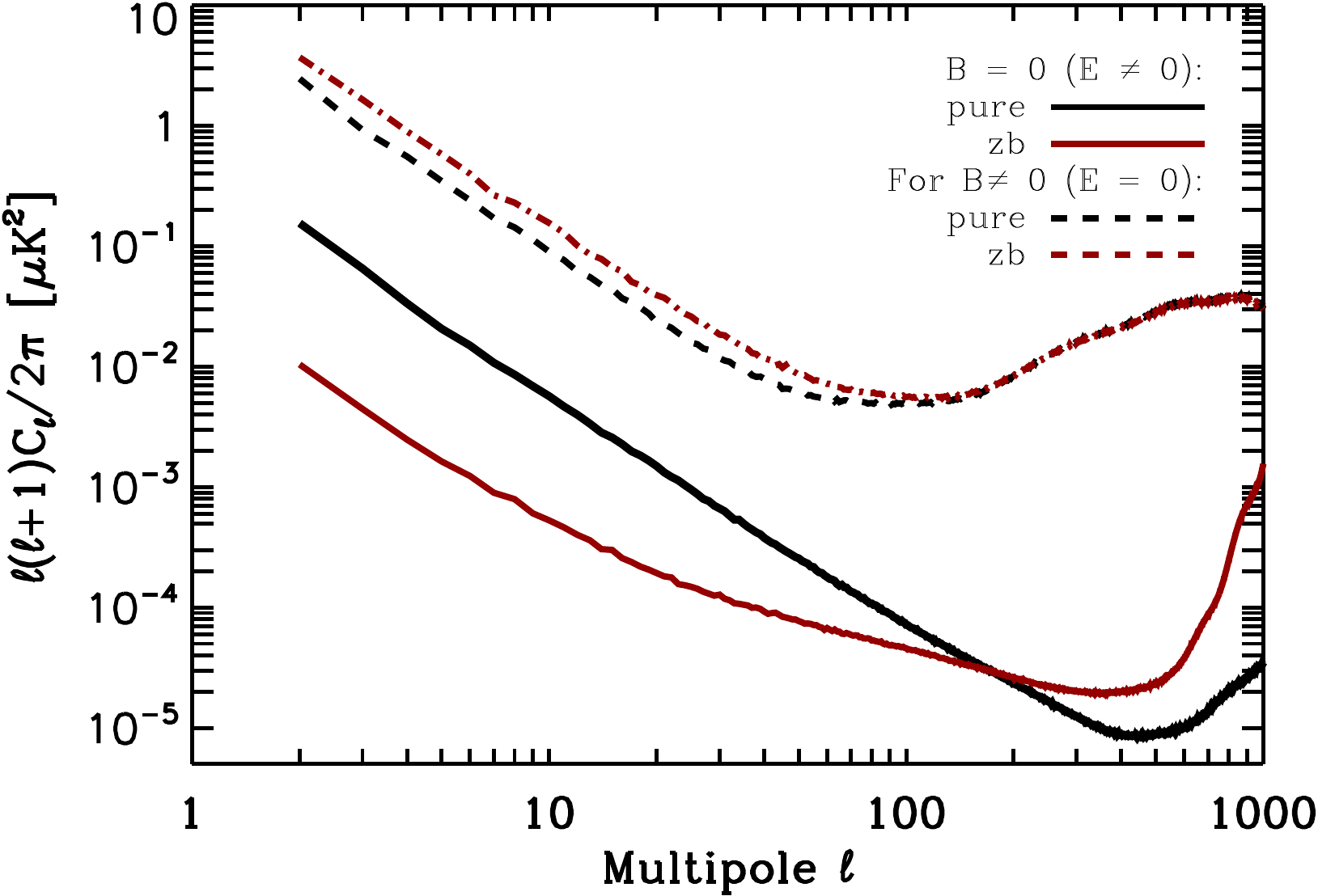}\includegraphics[scale=0.35]{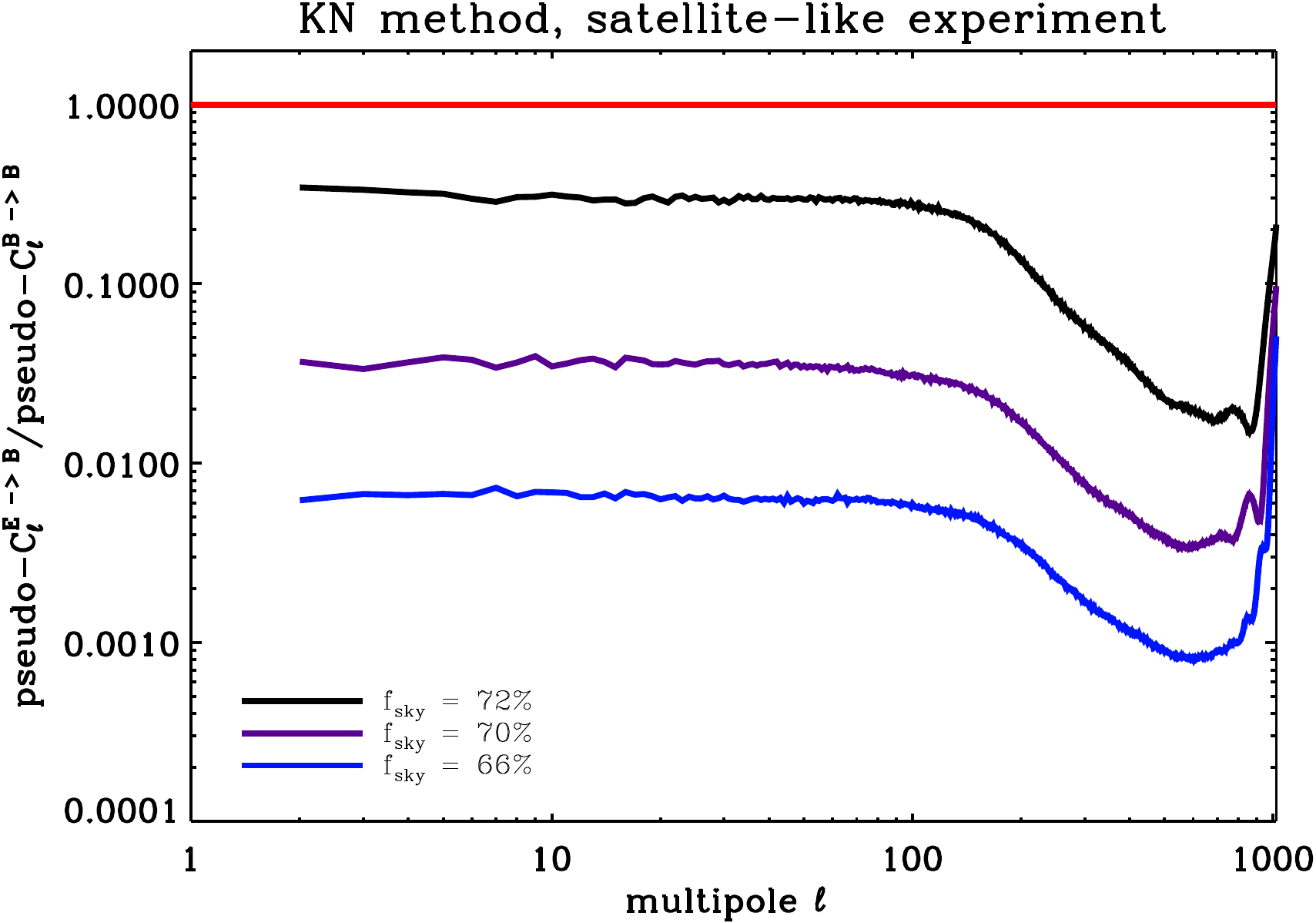}
	\caption{Pseudospectra built from leakage free methods in the case of a large scale experiment. \\
	\textit{Left panel:} The pseudospectrum $\tilde{C}_{\ell}^{E \rightarrow B}$ obtained with no input $B$ modes and $E$ modes power spectrum from WMAP-7yr best fit in solid (red) black line obtained using the pure (\zb) method. The pseudospectrum $\tilde{C}_{\ell}^{B \rightarrow B}$ obtained with no input $E$ modes and a theoretical $B$ modes power spectrum for $r = 0.05$ in dashed (red) black line obtained using the pure (\zb) method. \\ 
	\textit{Right panel:} The ratio $\tilde{C}_{\ell}^{E \rightarrow B}/\tilde{C}_{\ell}^{B \rightarrow B}$ between the $B$ modes pseudospectrum with no $B$ modes in input and no $E$ modes in input obtained in the \kn~method for different applied masks ($f_{sky} = 72\%$, $70\%$ and $66\%$ in black, purple and blue respectively).} 
	\label{fig:pseudo_wmap_all}
\end{center}
\end{figure}

Performing the same procedure as above in the case of the pure and \zb~methods for a large scale survey give the pseudospectra displayed in the left panel of Fig.~\ref{fig:pseudo_wmap_all}. The first conspicuous result is that the contribution from the $E$-to-$B$ leakage $\tilde{C}_{\ell}^{E \rightarrow B}$ (in solid lines) is not completely vanishing contrary to what was theoretically expected. The leakage coming from the mask being mainly correcting, the pseudospectra are now sensitive to more subtle effects. In particular, the pixelisation of the CMB maps is a source of $E$-to-$B$ leakage and is expected to affect the small scales. The pixel finite size is therefore thought to be the cause of the $\tilde{C}_{\ell}^{E \rightarrow B}$ high $\ell$ increase. 

At low $\ell$, the ratio between the true $B$ modes contribution and leakage is lower for the \zb~method with respect to the one obtained using the pure method, inversely to the high $\ell$. On the right panel of Fig.~\ref{fig:pseudo_wmap_all}, the ratios $\tilde{C}_{\ell}^{E \rightarrow B}/\tilde{C}_{\ell}^{B \rightarrow B}$ for the \kn~method are directly shown for various apodisation lengths of the mask. The highest apodisation length, $\theta_{apo} = 2^o$ ensures the best reduction of the leakage. However, for a window function covering $72\%$ of the sky, the ratio $\tilde{C}_{\ell}^{E \rightarrow B}/\tilde{C}_{\ell}^{B \rightarrow B}$ stays around 0.4 in the multipole range $\ell \in [2;100]$ while the same ratio evaluated thanks to the \zb~method stays below 0.01. Besides, the $B$ modes pseudospectra originating from the true $B$ modes obtained with the \kn~method have more power than the one obtained with the pure and \zb~method. A competition between the level of reconstructed power of the true $B$ modes and the leakage removal do arise. An analysis at the level of the angular power spectrum is thus required to state on the efficiency of each method. 

\subsubsection*{Influence of the holes in the mask}

\begin{figure}[!h]
\begin{center}
	\includegraphics[scale=0.4]{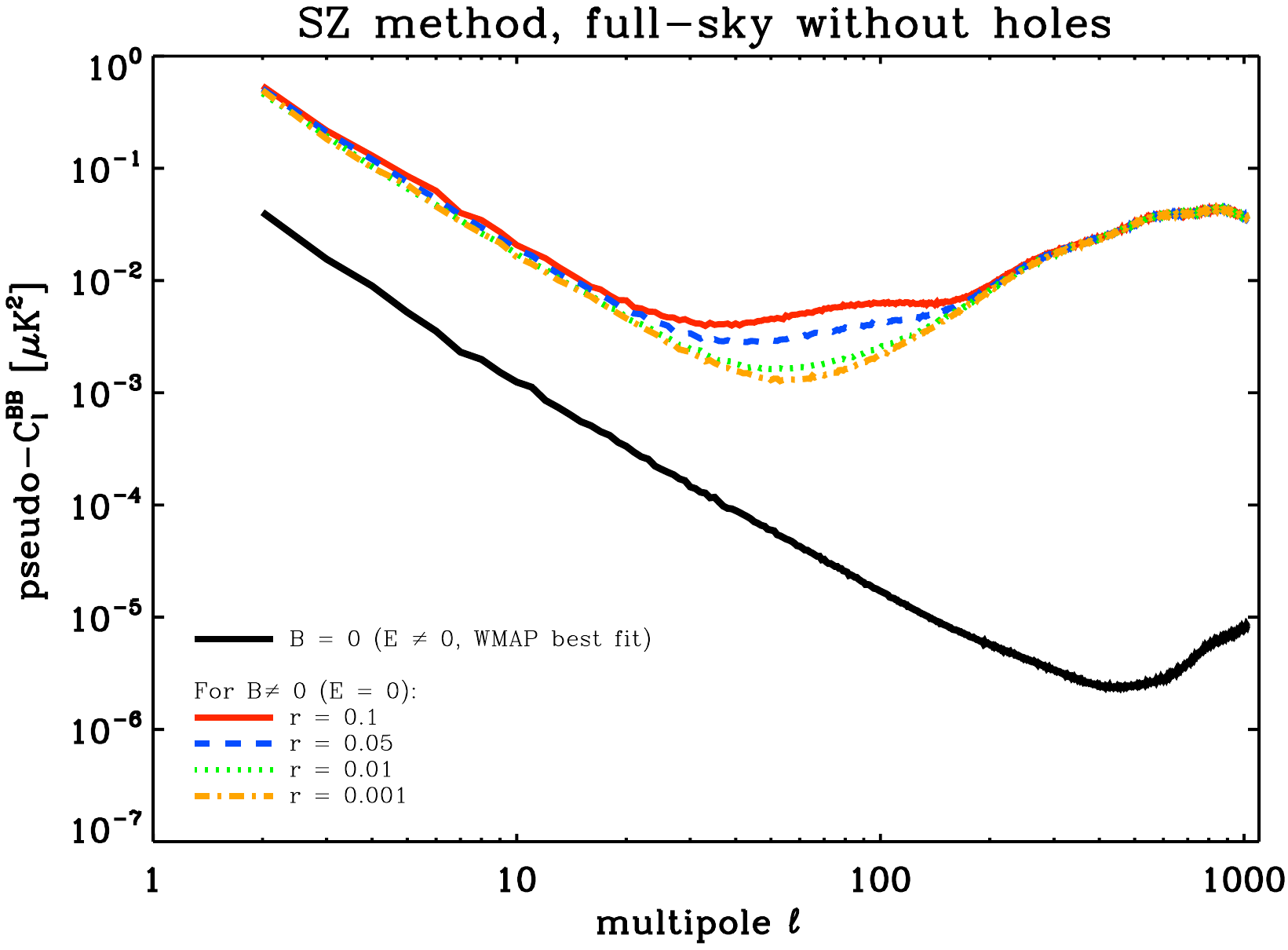} \\
	\caption{The pseudospectrum $\tilde{C}_{\ell}^{E \rightarrow B}$ obtained with no input $B$ modes and $E$ modes power spectrum from WMAP-7yr best fit in solid black line obtained using the pure method for a large sky coverage without holes. The colored curves stand for the pseudospectrum $\tilde{C}_{\ell}^{B \rightarrow B}$ with no input $E$ modes and a theoretical $B$ modes power spectrum for different values of $r$ ($r = 0.001,0.01,0.05,0.1$) along with the lensing contribution.} 
	\label{fig:pseudo_noholes_pure}
\end{center}
\end{figure}

The holes removal may have an impact on the efficiency of the leakage free $B$ modes pseudospectrum reconstructions. For clarity, only the results for the pure method are shown as the other methods show similar results. The figure~\ref{fig:pseudo_noholes_pure} shows the contribution from the leakage (true $B$ modes) to the $B$ modes pseudospectrum in black (coloured) lines. The leakage is lowered by one order of magnitude with respect to the case with holes in the mask. As a consequence, the presence of the holes also contribute to the amount of remaining leakage in the case of leakage free methods.   

\subsubsection*{Leakage free methods: the small-scale survey}

\begin{figure}[!h]
\begin{center}
	\includegraphics[scale=0.4]{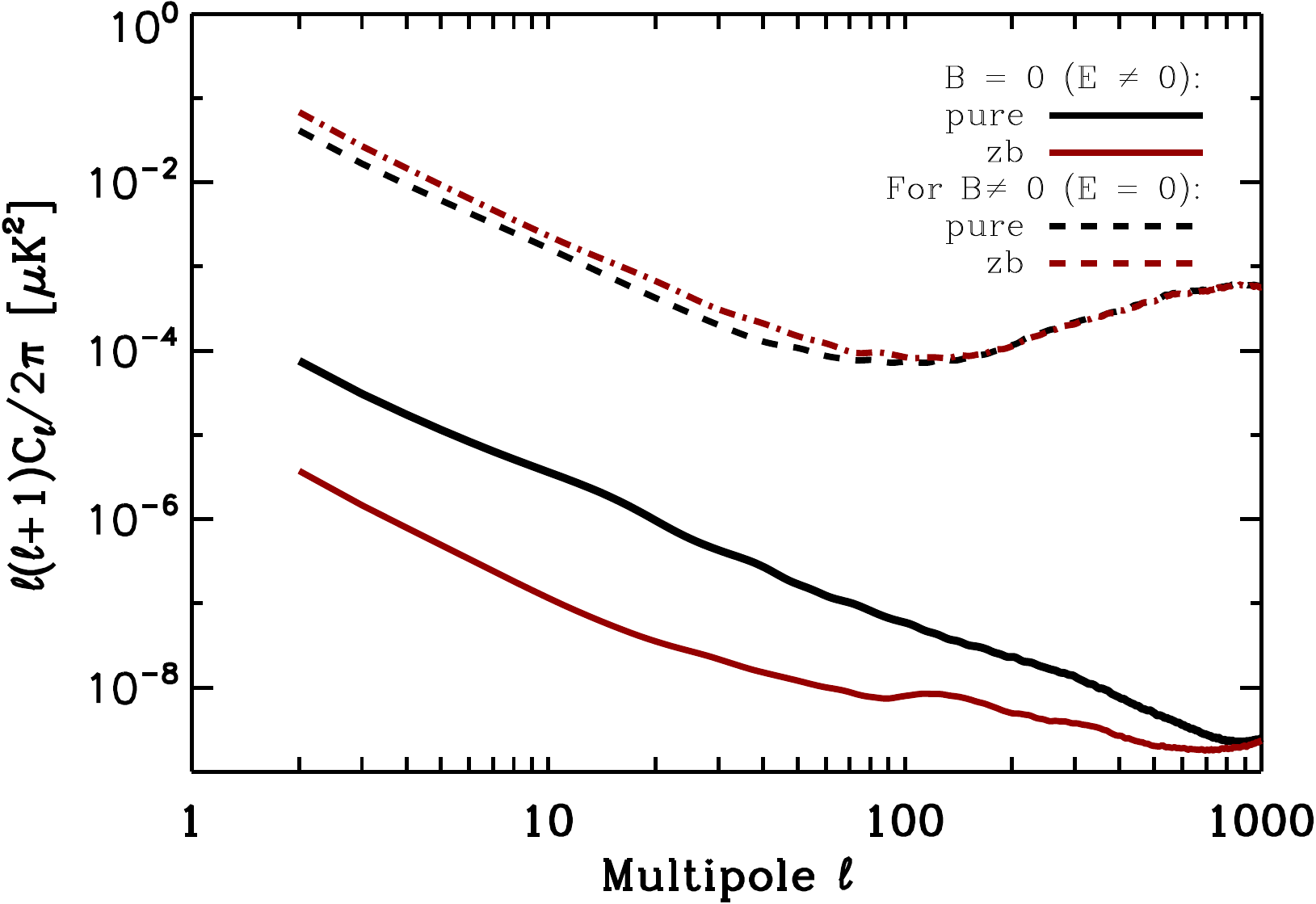}\includegraphics[scale=0.4]{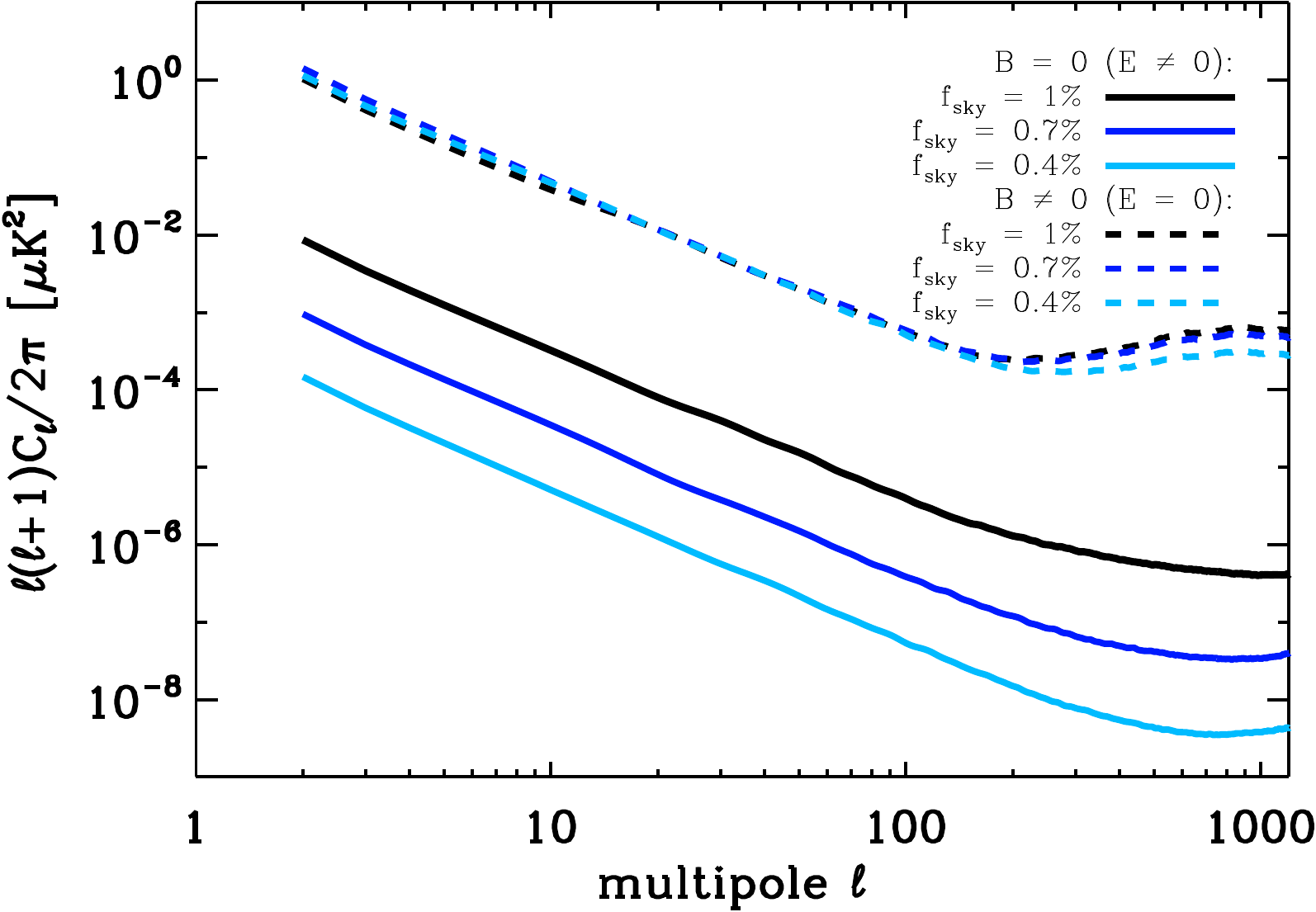}
	\caption{Pseudospectra built from leakage free methods in the case of a small scale experiment. \\
	\textit{Left panel:} The pseudospectrum $\tilde{C}_{\ell}^{E \rightarrow B}$ obtained with no input $B$ modes and $E$ modes power spectrum from WMAP-7yr best fit in solid (red) black line obtained using the pure (\zb) method. The pseudospectrum $\tilde{C}_{\ell}^{B \rightarrow B}$ obtained with no input $E$ modes and a theoretical $B$ modes power spectrum for $r = 0.05$ in dashed (red) black line obtained using the pure (\zb) method. \\ 
	\textit{Right panel:} The ratio $\tilde{C}_{\ell}^{E \rightarrow B}/\tilde{C}_{\ell}^{B \rightarrow B}$ between the $B$ modes pseudospectrum with no $B$ modes in input and no $E$ modes ($r = 0.05$) in input obtained in the \kn~method for different applied masks ($f_{sky} = 1\%$, $0.7\%$ and $0.4\%$ in black, blue and turquoise respectively).} 
	\label{fig:pseudo_ebex_all}
\end{center}
\end{figure}

Similarly, in the case of the small scale survey, the figure~\ref{fig:pseudo_ebex_all} shows the obtained results for the pure and \zb~methods in left panel and for the \kn~method in the right panel. In this case also, the true $B$ modes contribution $\tilde{C}_{\ell}^{B \rightarrow B}$ is much higher than the spurious $E$ modes contribution. For the pure method, the true pseudospectrum (in dashed lines) remains two orders of magnitude higher than the leakage at large angular scales, \textit{a priori} ensuring a confident reconstruction of the $B$ modes. The \zb~method shows again its ability to accurately reduce the leakage contribution to the $B$ modes pseudospectrum. The \kn~strategy appears to be more efficient in retrieving the $B$ modes as $\tilde{C}_{\ell}^{B \rightarrow B}$ is higher than in the pure and \zb~methods. Nonetheless, as in the previous cases, the leakage contribution also appears to be high compared to the one obtained in the two other approaches. A large apodisation length ($\theta_{apo} = 2^o$ corresponding to $f_{\mathrm{sky}} = 0.4\%$) is required for the leakage contribution to reach the level of the one obtained in the pure method. 

\underline{Intermediate conclusion:}

The performed analysis at the pseudospectrum level enables to catch a glimpse on the leakage effects for realistic experimental set-ups. First of all, the $E$-to-$B$ leakage is potentially a dramatic effect as, if not corrected, for a large scale and a small scale surveys, the $E$ modes contribution to the $B$ modes pseudospectrum overwhelms the one originating from the true signal respectively up to $\ell = 800$ and $\ell = 1000$. Moreover, we have shown that in the case of nearly full sky experiment the leakage is unexpectedly elevated: the mask complexity is a key ingredient of $B$ modes purity. For a clean $B$ modes reconstruction, the leakage free methods theoretically described in Chapter~\ref{Chapter4} are therefore required. They all result in a much lower level of the leakage, although unfortunately not completely vanishing. The residual leakage is thought to be due to the pixelisation and cannot be reduced in the current methods implementations. 

However, the amplitude of the remaining leakages differs regarding the used leakage free method. In both fiducial experiments, the pure method indeed leads to a higher residual compared to the one obtained with the \zb~method. The case of the \kn~strategy is more subtle as the final sky coverage is chosen \textit{a posteriori}. In this special case, the ratio between the $E$ modes contribution and the $B$ modes true signal is higher for higher value of the sky coverage due to the non removed ambiguous pixels. To conclude on the practical efficiency of each strategy to correct for the leakage, we must now study the reconstruction at the level of the $B$ modes angular power spectrum. 


\subsection{At the angular power spectrum level: large scale survey}

The two following sections are focused on the bias and variances on the $B$ power spectrum reconstruction by the three considered leakage free approaches. The bias is estimated by the deviation of the reconstructed power spectrum to the input power spectrum. The variance is derived by averaging over the simulated power spectra: 
\begin{equation}
\mathrm{Var}(\hat{C}_{\ell}^{BB}) = \frac{1}{N_{sim}-1} \left[\sum_{i = 1}^{N_{sim}} \hat{C}_{\ell}^{BB,i}\hat{C}_{\ell}^{BB,i} - \frac{1}{N_{sim}^2}\sum_{i = 1}^{N_{sim}} \hat{C}_{\ell}^{BB,i}\sum_{j = 1}^{N_{sim}} \hat{C}_{\ell}^{BB,j} \right]
\end{equation}
with $N_{sim}$ the number of simulation and $\hat{C}_{\ell}^{BB,i}$ the $i^{th}$ simulated reconstructed estimator of the $B$ modes power spectrum. The reference level is the mode-counting variance as introduced in Chapter~\ref{Chapter4}, the idealised variance on the $B$ modes power spectrum. All along the analysis, the power spectra are reconstructed with $N_{sim} = 500$ simulations for reliable results.

Minimal bias and variance are therefore the two requirements on which the following examination will focus on to settle on the leakage free method efficiency and robustness. The case of the fiducial balloon-borne and satellite-like experiments give rise to distinct issues. The results on a large scale experiment will thus be under scrutiny before the outcomes of the small-scale survey which was already deeply studied in the case of the pure method in \cite{Smith_2006}, \cite{Smith_2007}, \cite{Grain_2009}, \cite{Grain_2012}.

\begin{figure}[!h]
\begin{center}
	\includegraphics[scale=0.4]{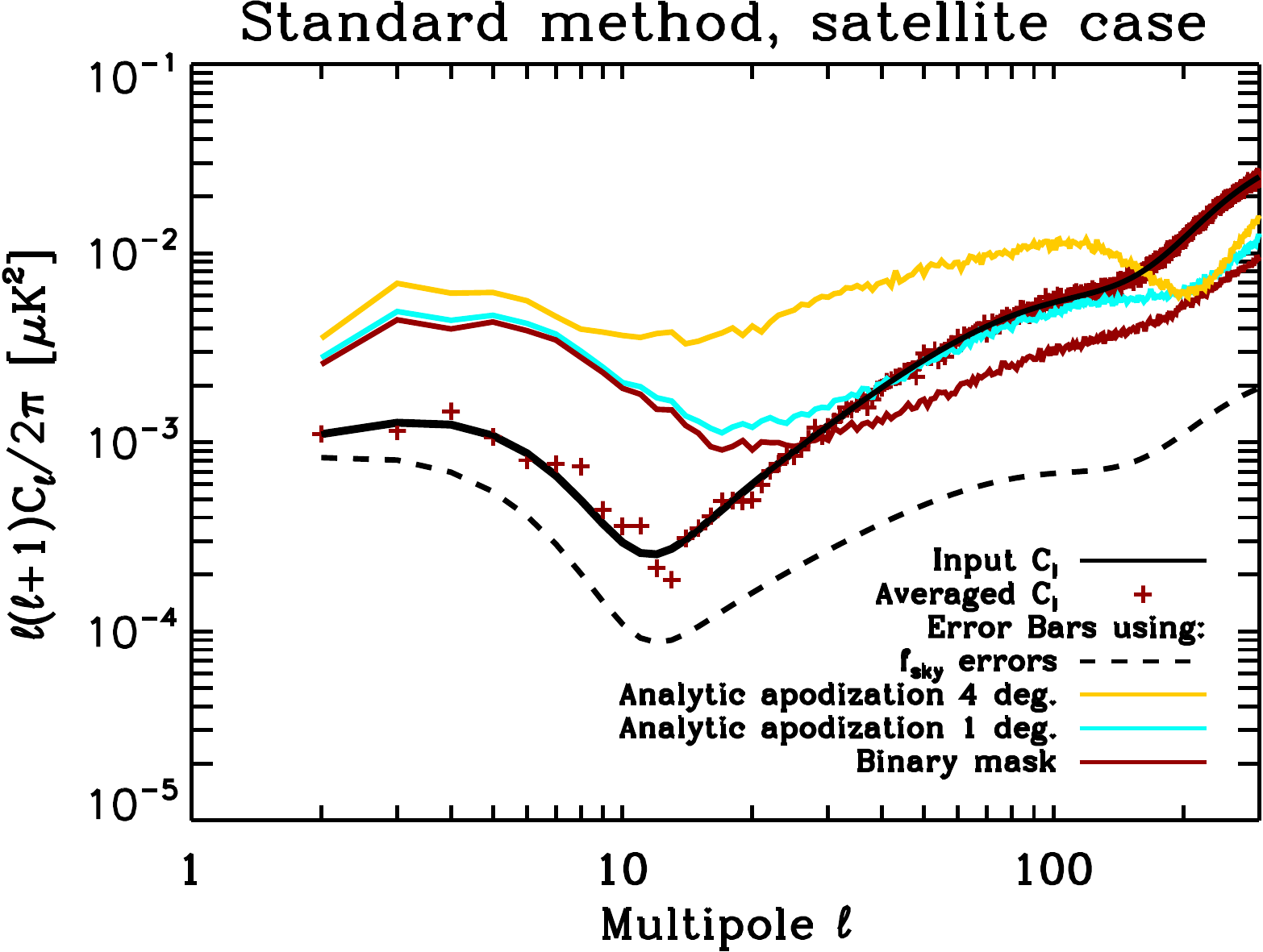} \\
	\caption{The red crosses stand for the $\ell$-by-$\ell$ reconstructed power spectrum using the standard method for a large scale experiment. The input $B$ modes power spectrum to be estimated is the solid black line. The dashed black line set a benchmark on the obtained uncertainties as it is the ideal mode counting ones. The error bars obtained using a binary mask, an analytic apodisation with an apodisation length $\theta_{apo}=1^o$ and $\theta_{apo}=4^o$ are displayed as red, blue and yellow curves respectively.}
	\label{fig:cl_wmap_std}
\end{center}
\end{figure}

The first investigation at the pseudospectrum level has given hint on the necessity of using a $B$ modes reconstruction correcting for the leakage even in the case of a large scale survey. To support this presumption, the $B$ modes angular power spectrum behaviour derived in the standard framework has to be explored. The standard method to estimate the $B$ modes is unbiased (as shown in Chapter~\ref{Chapter4}), we therefore expect the reconstructed simulated $B$ modes power spectrum to be unbiased. However, from the previous analysis at the level of the pseudospectrum, we predict a high variance induced by the leaked $E$ modes contribution. In Fig.~\ref{fig:cl_wmap_std}, the input angular power spectrum to be recover is shown in black solid line along with the mode counting error bars including noise and beam effects in black dashed line.

The red crosses depict the obtained $\ell$-by-$\ell$ $B$ power spectrum obtained via the standard method -- \textit{i.e.} without correcting for the $E$-to-$B$ leakage -- on a binary mask $M$. The crosses well follow the input power spectrum, showing nonetheless a slight scattering at low $\ell$. The latter is due to the finite amount of simulations adding to the fact that its variance appears to be higher than the power spectrum itself. As a result, the reconstructed power spectrum is unbiased: the theoretical standard calculations are corroborated by their numerical implementation. In particular, the convolution kernel $K_{\ell\ell'}^-$ is well computed. The coloured lines show the MC uncertainties for three apodisation lengths of the applied window functions. The red solid lines evince the lowest variance on $\hat{C}_{\ell}^{BB}$ power spectrum, obtained using a binary mask. For $\ell$ lower than 30, the variance exceeds the reconstructed $B$ modes preventing us from a genuine $B$ modes detection on large scales. The issue is dramatic especially since the low $\ell$ $B$ modes are the key observable to constrain the primordial universe. One may argue that the binary mask is not appropriate because of the sharp edges, I have therefore simulated standard $B$ modes reconstruction with apodised window functions with apodisation lengths of $1^o$ and $4^o$. The resulting variances are respectively shown in turquoise and yellow solid lines in Fig.~\ref{fig:cl_wmap_std}. They appear to give worth results than for the binary mask, the uncertainties level growing with the apodisation length. This is explained by the induced loss of information on the $B$ modes.

This preliminary work confirms the need for approaches correcting for the $E$-to-$B$ leakage, as presumed by analysing the pseudospectrum. The issue is particularly mandatory to detect the primordial $B$ modes as the error bars exceed the signal at least up to $\ell = 30$. For higher $\ell$, although the signal-to-noise ratio is greater than one, the obtained error bars remain higher than the mode counting ones. The power spectrum reconstruction can thus be improved. The three methods of interest are expecting to give better result and I propose to quantify it by performing the same procedure as above firstly focusing on the pure method performances followed by the \zb~ and \kn~ strategies.


\subsubsection*{Pure method}

\begin{figure}[!h]
\begin{center}
	\includegraphics[scale=0.4]{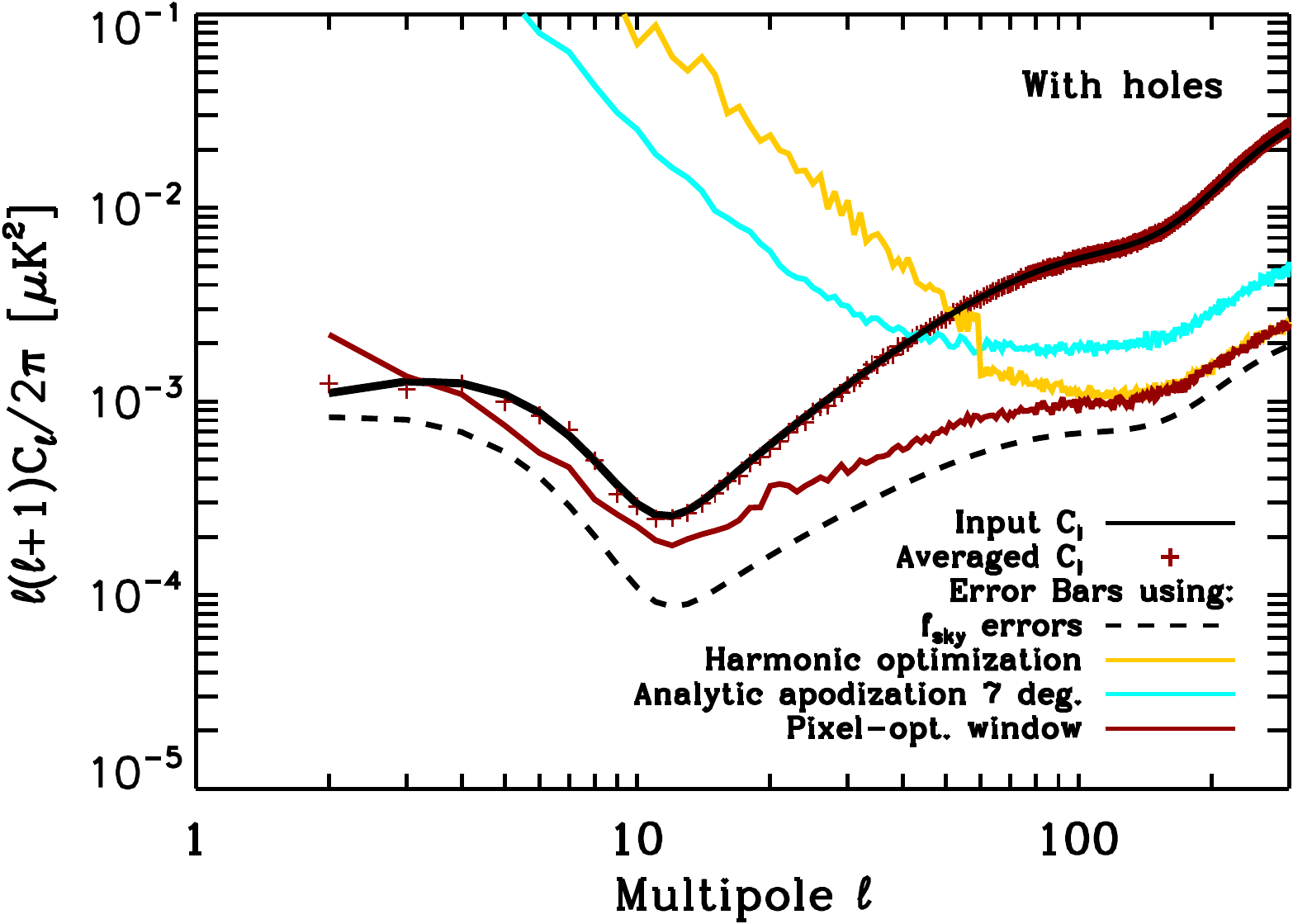} \\
	\caption{The red crosses stand for the $\ell$-by-$\ell$ reconstructed power spectrum using the standard method for a large scale experiment. The input $B$ modes power spectrum to be estimated is the solid black line. The dashed black line set a benchmark on the obtained uncertainties as it is the ideal mode counting ones. The error bars obtained using a binary mask, an analytic apodisation with an apodisation length $\theta_{apo}=7^o$ and an harmonic variance-optimised window function are displayed red, blue and yellow curves respectively.} 
	\label{fig:cl_wmap_pure_allW}
\end{center}
\end{figure}

The pure method enables a wide exploration of CMB $B$ modes estimation as its implementation allows for various kinds of window functions optimisation. In the scope of minimising variance on the recovered $B$ modes power spectrum, an $\ell$-by-$\ell$ reconstruction up to $\ell = 300$ is convenient as the leakage is the most intrusive in the range of low multipoles. 
At first, the window functions optimised in the harmonic domain seem to be the best solution as they are \textit{a priori} built to minimise the variance on the reconstructed $B$ modes and their computation is fast, taking advantage of the $s2hat$ library and homogeneous noise distribution. The Fig.~\ref{fig:cl_wmap_pure_allW} displays the reconstructed $B$ modes power spectrum from an input signal in solid black line with a mode counting variance in dashed black line. The uncertainties on the recovered $B$ modes power spectrum in red crosses obtained using the harmonic optimised window function is represented by the yellow line. The window functions were optimised bin-by-bin. The obtained performances strongly depend on the angular scales. The variance is indeed greater than the signal itself for $\ell < 50$ but quickly decreases towards the mode counting variance at high $\ell$. The variance obtained by an analytic apodisation of $7^o$ appears to be lower than the harmonic optimised window function up to $\ell = 60$. The endeavour for optimisation in the harmonic domain therefore seems to be useless for the low $\ell$ range.

However, the error bars obtained using PCG window functions are depicted as the red solid line in Fig.~\ref{fig:cl_wmap_pure_allW}. The conclusions are undeniable: the resulting uncertainties only eclipse the $B$ modes signal for $\ell < 3$ and rapidly decrease close to the mode counting variance. The obtained uncertainties are at least two orders of magnitude lower than the ones obtained using the other window function, for $\ell < 10$. It is thus a significant achievement especially in the perspective of setting constraints on the primordial physics. The price to pay is the numerical cost. In practice, a computation time of one hour and a half on 70 processors is indeed required to build the window function in the harmonic domain while more than 7 hours for one hundred processors are needed for pixel-based window function optimisation.

The efficiency of the PCG window functions is explained by its flexibility in the minimisation process. The release of the constrains on the spin-weighted window function actually allows for a better minimisation of the overall variance on the $B$ modes power spectrum. It indeed finds the compromise between correcting the $E$-to-$B$ leakage and the induced loss of information ensuring the lowest global uncertainties. The need for such an elaborated tool in the case of a large scale survey -- expected to be easier -- is a key issue raised during my PhD thesis. 

Furthermore, although the obtained variance is low, it stays as high as the signal up to $\ell \sim 20$. An appropriate binning would lower the variance on the recovered power spectrum of roughly a factor $\frac{1}{\Delta_{\ell}}$ with $\Delta_{\ell}$ the bin width. We therefore apply the binning process as exposed at the beginning of the present chapter. The first bin ranges between $\ell = 2$ up to $\ell = 20$ and $\Delta_{\ell} = 40$ for higher multipoles.
Such a binning is convenient for it reduces the variance but still keeps the shape of the $B$ modes spectrum (in particular, the recombination bump remains well sampled).

\begin{figure}[!h]
\begin{center}
	\includegraphics[scale=0.4]{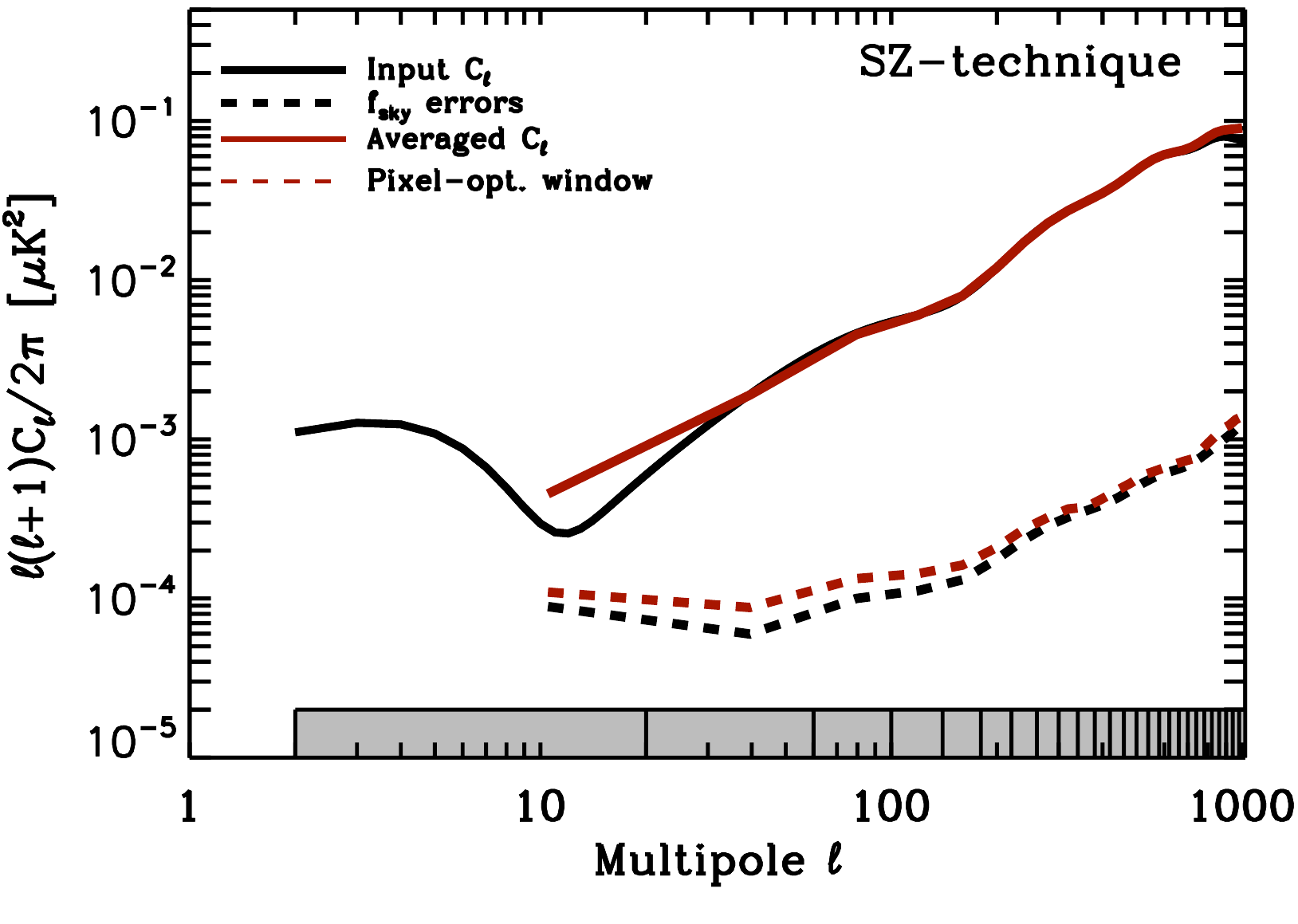} \\
	\caption{Power spectrum uncertainties on $B$-modes in dashed red line using the pure estimation for the case of a satellite-like experiment ($f_\mathrm{sky}\sim71\%$). The dashed black line displays the ideal mode counting uncertainties. The sold red curve stand for the reconstructed $B$ modes power spectrum of the input power spectrum in solid black curve. The grey shaded boxes represent the bins.} 
	\label{fig:cl_wmap_pure}
\end{center}
\end{figure}

Learning from the previous work, we have carried out the estimation of the CMB $B$ modes power spectrum in the pure method framework using PCG window functions optimised per bin. The resulting binned estimated $B$ modes power spectrum along with its uncertainties are displayed on Fig.~\ref{fig:cl_wmap_pure} by respectively a solid and a dashed red line. In order to appreciate the efficiency of the method, the binned mode-counting -- and therefore ideal -- uncertainties are shown in dashed black line as a benchmark. First of all, the reconstructed $B$ modes power spectrum is unbiased as theoretically presumed.

Second of all, the binned uncertainties on the reconstructed $B$ modes power spectrum tightly follow the mode-counting variance as indicated by the dashed lines in Fig.~\ref{fig:cl_wmap_pure}. It therefore ensure a high signal-to-noise ratio on $B$ modes detection -- the uncertainties are well below the $B$ modes powerspectrum. This demonstrates that the pure method is numerically efficient to minimise the $B$ modes variance and to reconstruct $B$ modes power spectrum over the whole multipole range.

As a result, thanks to the PCG window function optimisation, the pure method is proficient in estimating the genuine $B$ modes on the whole multipole range. The pure method is then promising for the CMB data analysis in the case of a satellite experiment dedicated to $B$ modes detection. 


\subsubsection*{\zb~ method}

The $B$ modes estimation using the pixel-based \zb~technique consists in reconstructing the masked $W^2\chi^B$ field. As explained at the beginning of the chapter, the \zb~technique allows for the use analytically apodised or harmonic-based variance-optimised window functions. The release of the conditions on the spin-weighted window functions allowed by the pixel-based implementation has been checked to be incompatible with the \zb~method. 

\begin{figure}[!h]
\begin{center}
	\includegraphics[scale=0.4]{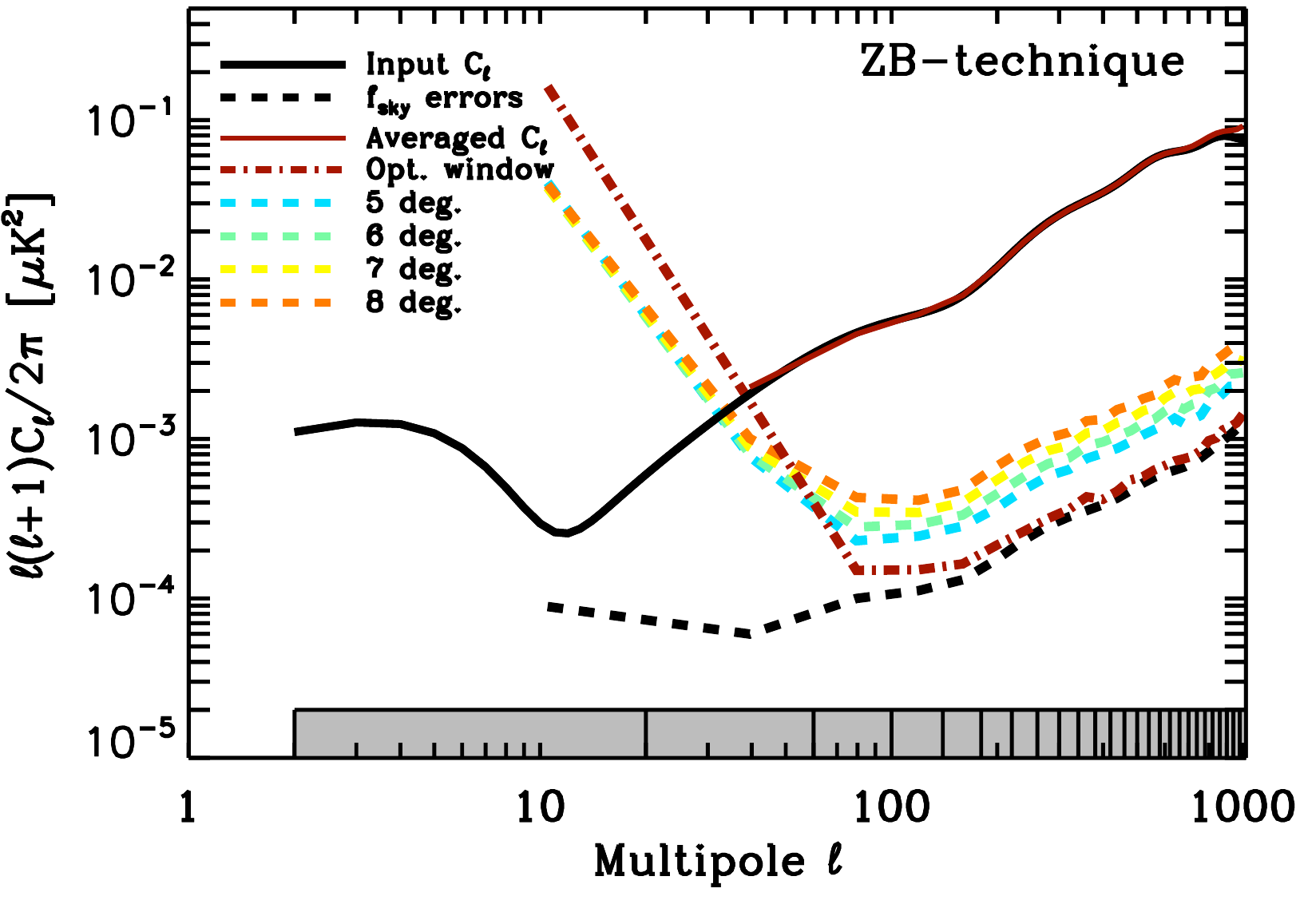} \\
	\caption{Power spectrum uncertainties on $B$-modes in dashed red line using the \zb~method thanks to an harmonic variance optimised window function for the case of a satellite-like experiment ($f_\mathrm{sky}\sim71\%$). The coloured dashed lines are the obtained error bars for different lengths of apodisation ranging from $\theta_{apo} = 5^o$ to $\theta_{apo} = 8^o$. The dashed black line displays the ideal mode counting uncertainties. The sold red curve stand for the reconstructed $B$ modes power spectrum of the input power spectrum in solid black curve. The grey shaded boxes represent the bins.} 
	\label{fig:cl_wmap_zb}
\end{center}
\end{figure}

The outcomes of the \zb~technique to estimate $B$ modes are illustrated in Fig.~\ref{fig:cl_wmap_zb} where the coloured dashed lines represent the variances on the estimated $B$ modes power spectrum displayed by the solid red curve. The first bin is manifestly plagued by a high variance, which swamps the signal, leading to a scattering of the estimated power spectrum. This explained the negative value of the estimated power spectrum in the first bin. However, it is not an impediment to the unbiased estimation as for the higher bins the power spectrum is consistent with the input power spectrum.   

Moreover, I have explored the \zb~method efficiency in the perspective of minimising the obtained variance. The two authorised families of window functions have different issues. On the one hand, the apodisation length of the analytic window functions have to be properly chosen via MC simulations. For clarity, only the results for four apodisation lengths $\theta_{apo}$ ranging from $5^o$ to $8^o$ are shown in Fig.~\ref{fig:cl_wmap_zb}. I have checked that higher or lower values of $\theta_{apo}$ give highest uncertainties. For the highest multipoles, the variance is simply increasing with the apodisation length because of the induced loss of cosmological information due to the lowering of effective sky fraction. Nonetheless, in the low $\ell$ range, the most appropriate apodisation length is the one resulting from the analytically apodised window function with $\theta_{apo} = 7^o$ which thus best lower the leakages. On the other hand, the harmonic-optimised window function give the uncertainties depicted by the dashed-dotted red line in the figure above. From the third bin, the obtained variance is below the one obtained via analytic apodisation and is following the mode-counting variance. In spite of this high-$\ell$ efficiency, the uncertainties on the reconstructed $B$ modes are higher than the error bars get from analytically apodised window function for $\ell < 60$ thus exceeding the $B$ modes amplitude. 

Being pixel-based, the \zb~method does not offer as much flexibility as the pure technique thus prohibiting the use of PCG window functions. Nonetheless, it gives competitive results for the highest multipoles $\ell \gtrsim 60$ as the obtained variances are very low. Using an harmonic optimisation or an analytically apodised window function, the $B$ modes power spectrum cannot be recovered in the first two bins \textit{i.e.} $\ell \in [2;60]$, where the primordial $B$ modes is peaking. 


\subsubsection*{\kn~method}

\begin{figure}[!h]
\begin{center}
	\includegraphics[scale=0.4]{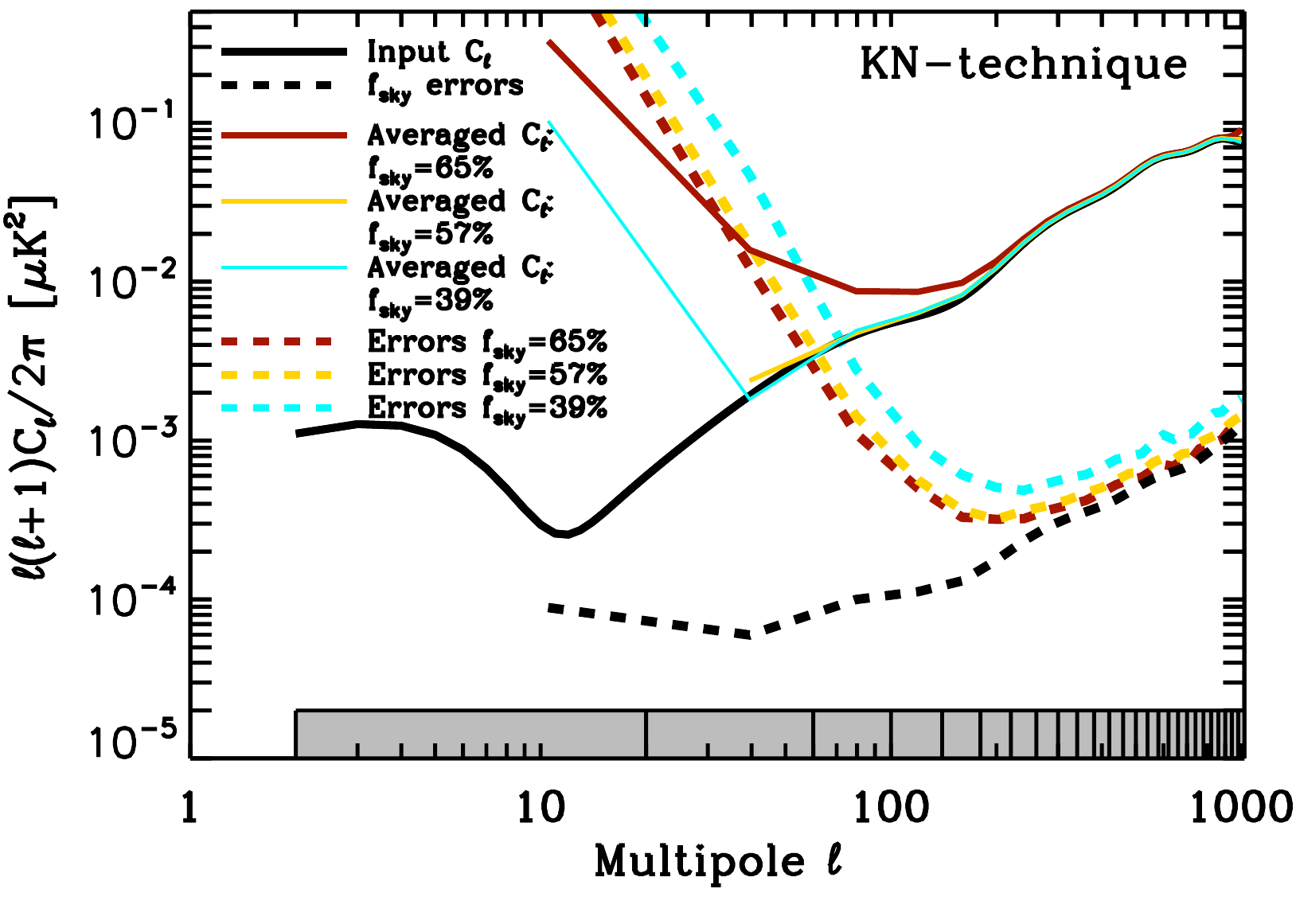} \\
	\caption{Power spectrum uncertainties on $B$-modes in dashed red, yellow and blue lines using the \kn~method thanks to different mask (with $f_{\mathrm{sky}} = 65\%, 57\%, 39\%$ respectively) in the case of a satellite-like experiment ($f_\mathrm{sky}\sim71\%$). The dashed black line displays the ideal mode counting uncertainties. The coloured solid lines are the obtained reconstructed power spectra for the different effective sky fraction. The input power spectrum is in solid black curve. The grey shaded boxes represent the bins.} 
	\label{fig:cl_wmap_kn}
\end{center}
\end{figure}

The \kn~method is expected to give worse performances than the two previous approaches as it is an approximation of the \zb~ technique. The Fig.~\ref{fig:cl_wmap_kn} displays the recovered $B$ modes power spectra in solid coloured lines along with their respective variances. 

Each power spectrum have been obtained by applying a $C^2$ apodised window function with width $\theta_{apo}$ and by removing external layers on a width $\theta_{cut} = \theta_{apo}$ where the window function is varying. The chosen apodisation lengths are $\theta_{apo} = 30',1^o, 2^o$, a compromise between a loss of information and an effective leakage reduction. The sky coverage where the power spectra are estimated is therefore respectively of $65\%$, $57\%$ and $39\%$.  

For $\theta_{cut} = 30'$, the reconstructed $B$ modes power spectrum is biased on the five first bins and reach a level two order of magnitude higher than the expected signal. This marked gap to the expected power spectrum is explained by the non vanishing convolution kernel $K_{\ell \ell'}^{kn,-}$, which is set to zero. The cut layer is indeed too narrow thus pixels plagued by leakage are still left in the analysis. For wider trimmed layer, the bias in the recovered power spectrum is less strong. The corresponding variances are displayed in coloured dashed lines while the mode counting variance for a sky coverage of $f_{sky} = 65\%$ is symbolised by the black dashed curve. The lowest variance is reach for the highest sky coverage \textit{i.e.} $\theta_{apo} = 30'$. Even in this optimistic case, the variance exceeds the signal itself in the first two bins, up to $\ell = 60$. From $\ell = 150$, the obtained variances behave as the mode counting variance.   

As a consequence, if the apodised window function is not properly chosen the \kn~technique leads to \textit{biased} estimated power spectrum. However, reducing the bias boils down to reducing the kept-in-analysis sky coverage meaning that the variance will rise. In any case, the \kn~method estimation does not enable to put constrain on the $B$ modes primordial part. 


\subsubsection*{Influence of the holes in the mask}

In order to quantify and explore the impact of the mask shape, the impact of the holes masking the polarised point sources on the $B$ modes reconstruction is investigated. This study comes in two correlated parts, the first regarding the choice of window function and the second focusing on the pseudospectrum methods. 

\begin{figure}[!h]
\begin{center}
	\includegraphics[scale=0.4]{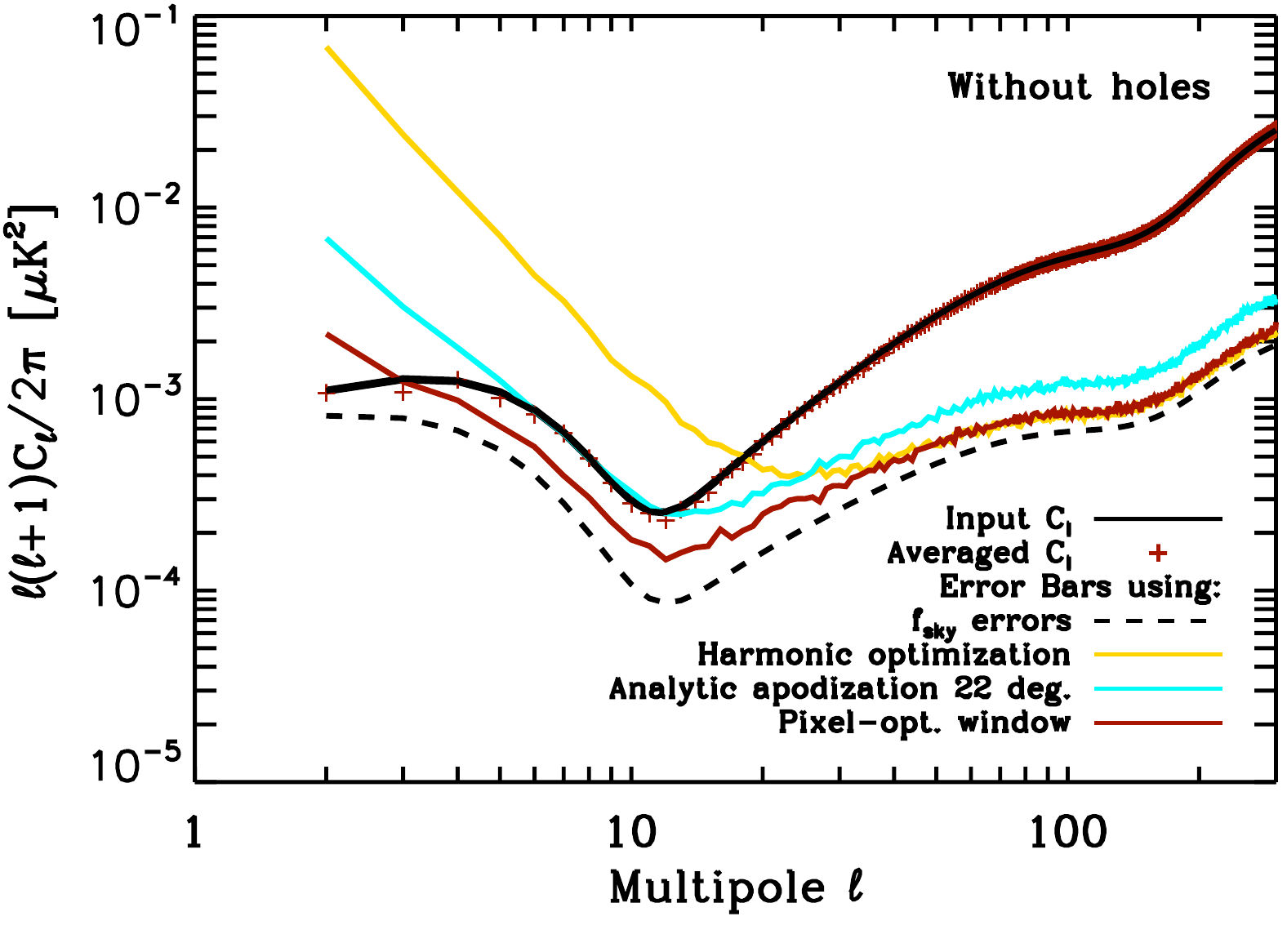} \\
	\caption{The red crosses stand for the $\ell$-by-$\ell$ reconstructed power spectrum using the standard method for a large scale experiment without holes. The input $B$ modes power spectrum to be estimated is the solid black line. The dashed black line set a benchmark on the obtained uncertainties as it is the ideal mode counting ones. The error bars obtained using a binary mask, an analytic apodisation with an apodisation length $\theta_{apo}=22^o$ and an harmonic variance-optimised window function are displayed red, blue and yellow curves respectively.} 
	\label{fig:cl_noholes_pure}
\end{center}
\end{figure}

We expect the various families of window function to behave differently regarding the mask shape. The pure method allows for the use of each kinds of window function. In the scope of exploring the chosen apodisation, we therefore reconstruct the $B$ modes power spectrum using the pure estimation on a sky where only the galactic emission is masked. The Fig.~\ref{fig:cl_noholes_pure} shows the $\ell$-by-$\ell$ reconstructed $B$ mode angular power spectrum in red crosses and its variance for various choices of window function. By comparison with Fig.~\ref{fig:cl_wmap_pure}, the use of PCG window functions, results in uncertainties level, depicted as the solid red curve, similar to the one obtained on $71\%$ of the celestial sphere. The PCG window functions thus well handle the complexity of the mask to recover the lowest variance. On the contrary, the uncertainties obtained for an analytically apodised window functions, in turquoise (with $\theta_{apo} = 22^o$, the lowest obtained variance) and for harmonic-based window functions in yellow are less accurate to reconstruct the $B$ modes power spectrum, especially at low $\ell$. It proves that the presence of holes in the mask affects the efficiency of both kinds of window functions in the scope of the pure estimation. The space between the holes can indeed be small consequently leading to a harder optimisation. As a result, the PCG window functions allow for a good treatment of the mask complexity while the other families of window function, being less flexible by construction, are sensitive to the mask shape. 

\begin{figure}[!h]
\begin{center}
	\includegraphics[scale=0.4]{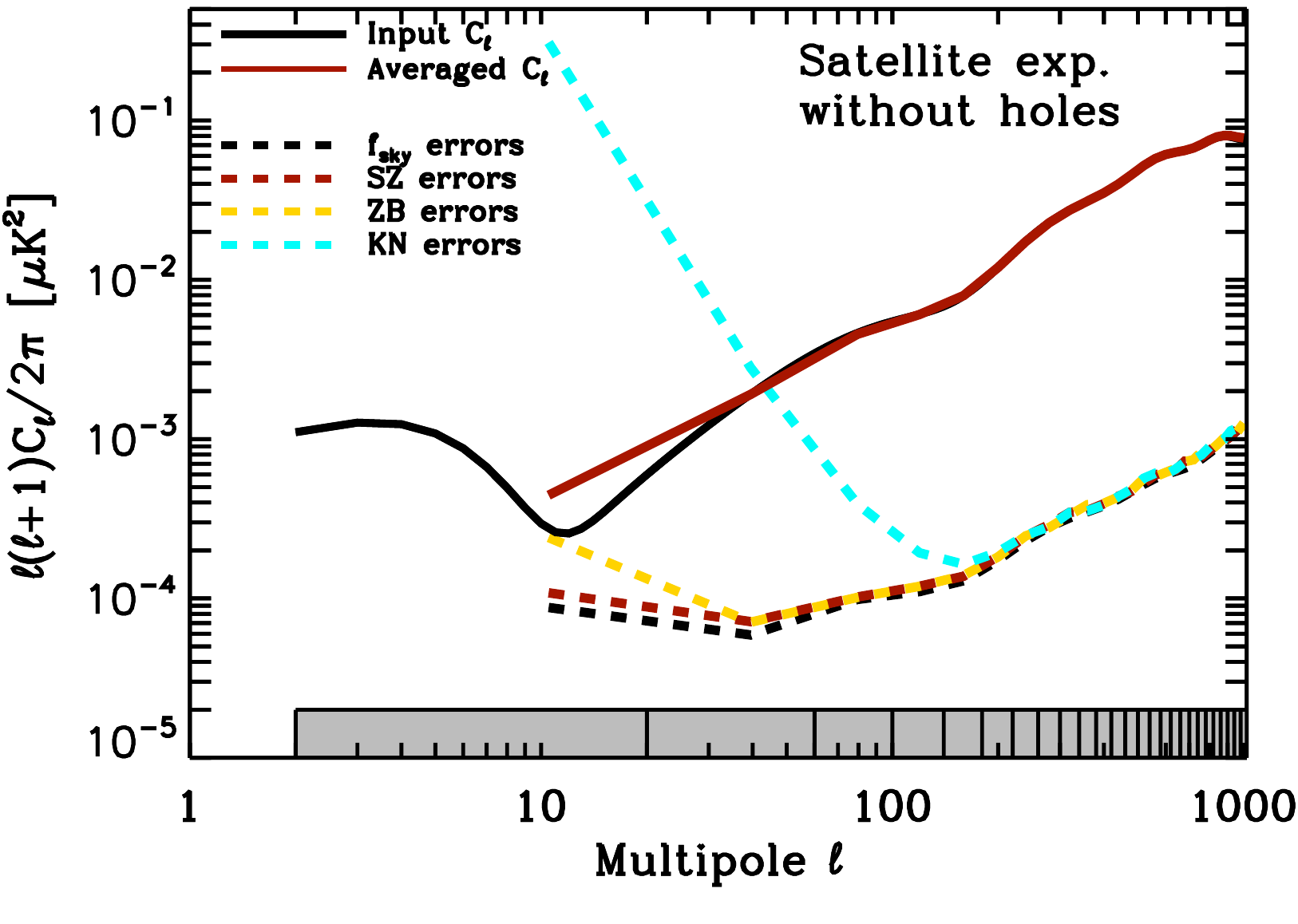} \\
	\caption{Summary of the power uncertainties obtained with a pure, \zb~and \kn~approach in dashed red, yellow ans blue curves respectively, in the case of a large scale survey without holes. The solid curve stands for the input power spectrum to be estimated. The grey shaded boxes represent the binning of the power spectra.} 
	\label{fig:cl_noholes_all}
\end{center}
\end{figure}

From this work on the impact of the choice of window function regarding the presence of holes in the mask, we expect the pure method to be the more robust approach to deal with the mask shape. The results of the $B$ modes reconstruction in the pure, \zb~and \kn~methods for a galactic mask only are shown in Fig.~\ref{fig:cl_noholes_all}. In particular, the induced uncertainties on $B$ modes power spectrum are depicted following the same convention as Fig.~\ref{fig:cl_wmap_pure}. A significant gain is obtained: the reconstruction starting from the second bin is possible now with the \kn~strategy and the whole power spectrum is recovered using the \zb~method. The pure estimation using PCG window function does not show any difference between the case with and without holes. As expected, the pure method used with the PCG window functions has the advantage not to be affected by the presence of the holes in the mask. 

As a result, the PCG window functions, optimised to give the lowest variance, are efficient to reconstruct the $B$ modes for all kinds of mask, in the pure estimation framework.  


\subsection{At the angular power spectrum level: small scale survey}

The $B$ modes reconstruction on a wide patch of the sky is expected to be less affected by polarisation modes mixing than on a small sky coverage. In particular, the sampling variance is such that the first bin ($\ell \in [2,20]$) cannot be recover. Besides, the mask used for our fiducial experiment is more simple than the ones of large scale coverage. Both statements imply that the issue of optimising the window function will not be as crucial as in the satellite-like experiment where the recovery of the primordial $B$ modes on the first multipoles was the ultimate goal. The analysis in the scope of a small-scale survey is therefore presumed to be more straightforward than in the case of a large scale coverage although the various $B$ modes estimations show distinct performances. The pure method has been designed specifically for small scale surveys in \cite{Smith_2006}, \cite{Smith_2007}, \cite{Grain_2009}, \cite{Grain_2012}. The \zb~method has only be tested in the case of simple spherical cap while the \kn~method has never be used to reconstruct angular power spectrum on small scale experiment.


\subsubsection*{Pure method}

\begin{figure}[!h]
\begin{center}
	\includegraphics[scale=0.4]{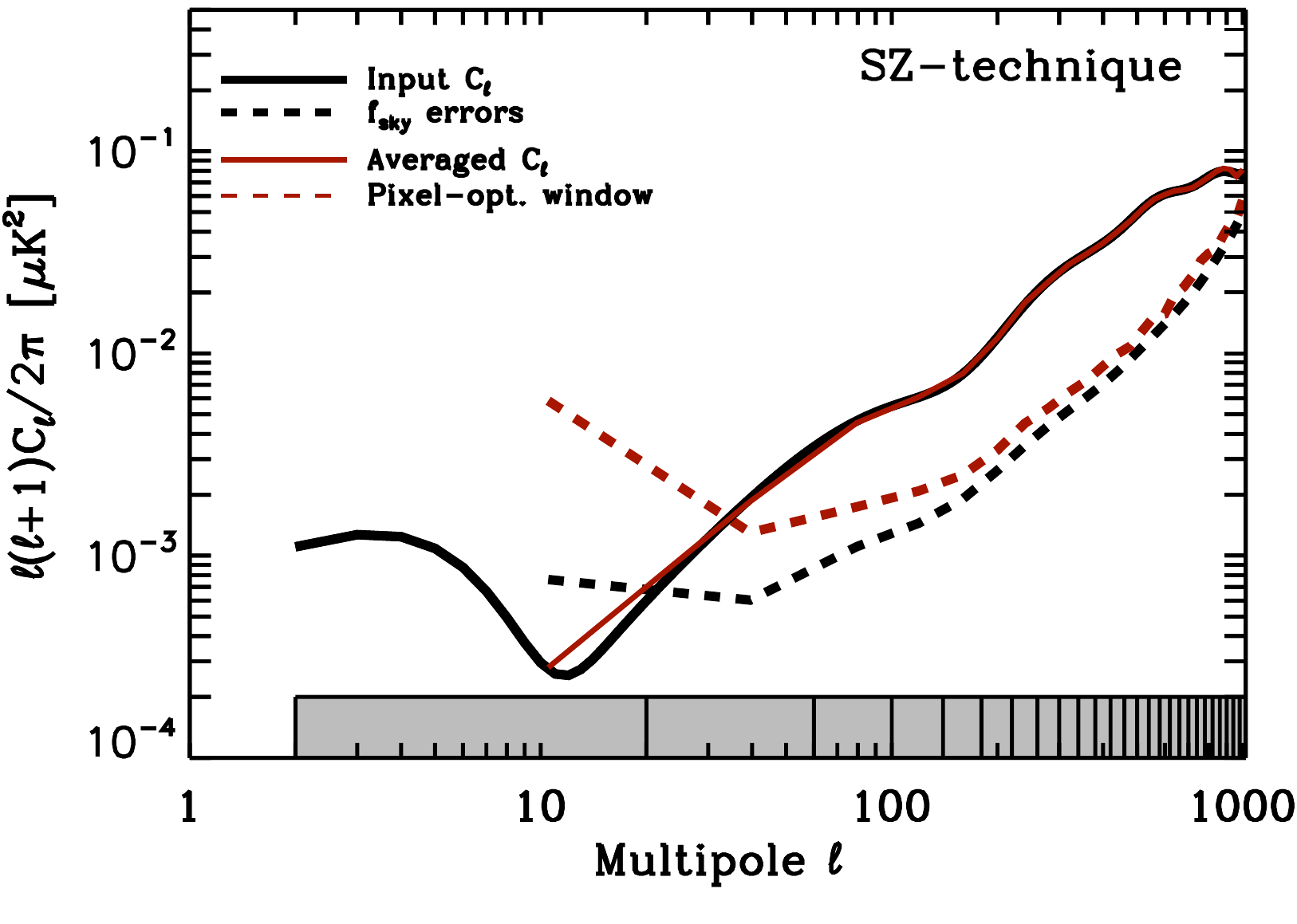} \\
	\caption{Power spectrum uncertainties on $B$-modes in dashed red line using the \zb~method for the case of a small scale experiment ($f_\mathrm{sky}\sim1\%$). The dashed black line displays the ideal mode counting uncertainties. The sold red curve stand for the reconstructed $B$ modes power spectrum of the input power spectrum in solid black curve. The grey shaded boxes represent the bins.} 
	\label{fig:cl_ebex_pure}
\end{center}
\end{figure}

The conclusions of the large scale survey analysis definitely state that the PCG window function is a powerful tool associated to the pure estimation for an efficient $B$ modes power spectrum reconstruction. Its efficiency in the case of a small scale experiment was also shown in \cite{Grain_2009}, \cite{Grain_2012}. The lowest variances are therefore expected to be achieved thanks to the PCG window functions, in the present case. The Fig.~\ref{fig:cl_ebex_pure} pictures the resulting variance in dashed red line on the estimated $B$ modes power spectrum. The first bin is obviously not recovered being one order of magnitude higher than the mode-counting variance in dashed black which is already exceeding the power spectrum. The variance on the $B$ modes power spectrum in the second bin ($\ell \in [20;60]$) is below the power spectrum consequently allowing for a detection in this bin. For the higher multipoles, the variance is closely behaving as the mode-counting variance indicating that the lowest variance level is achieved for this $\ell$ range. Furthermore, the numerical results confirm the unbiased construction of the estimator. 

Thus the pure estimation using PCG window function enables a clean recovery of the $B$ modes starting from $\ell = 20$, the reionisation bump being thus out of reach because of the small accessible sky fraction. 


\subsubsection*{\zb~ method}

\begin{figure}[!h]
\begin{center}
	\includegraphics[scale=0.4]{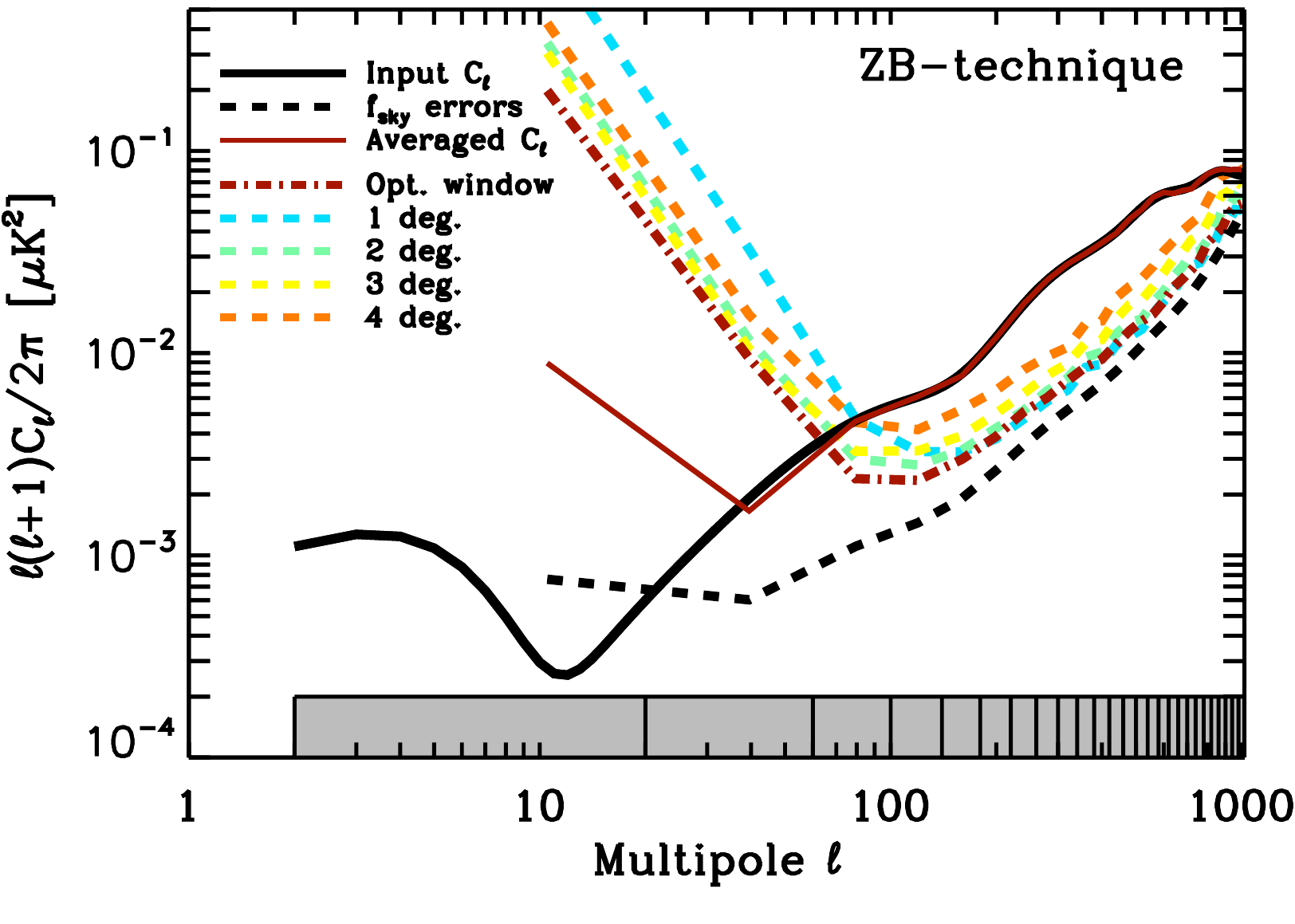} \\
	\caption{Power spectrum uncertainties on $B$-modes in dashed red line using the \zb~method thanks to an harmonic variance optimised window function for the case of a small scale experiment ($f_\mathrm{sky}\sim1\%$). The coloured dashed lines are the obtained error bars for different lengths of apodisation ranging from $\theta_{apo} = 1^o$ to $\theta_{apo} = 4^o$. The dashed black line displays the ideal mode counting uncertainties. The sold red curve stand for the reconstructed $B$ modes power spectrum of the input power spectrum in solid black curve. The grey shaded boxes represent the bins.} 
	\label{fig:cl_ebex_zb}
\end{center}
\end{figure}

The \zb~strategy offers the possibility of using two kinds of window function: the analytically apodised or the harmonic variance-optimised window functions. The resulting variance on $\hat{C}_{\ell}^{BB}$ for both estimation are displayed on Fig.~\ref{fig:cl_ebex_zb} respectively in coloured dashed and in red dotted-dashed. For both cases, the uncertainties are significantly higher than the input signal for $\ell \in [2;70]$, although the use of harmonic-based optimised window function gives the smallest error bars. In this multipole range, the analytic window function with $\theta_{apo} = 3^o$ leads to the lowest variance.  The variance level however scales with the used effective sky fraction for the highest mulitpoles. In this $\ell$ range, the variance obtained using $\theta_{apo} = 1^o$ follows the one obtained from an harmonic optimised window function, close to the mode-counting variance. Moreover, the estimation is unbiased albeit the reconstructed power spectrum shows a departure from the input power spectrum in the first bin due to the high variance. 

The lowest uncertainties on the $B$ modes power spectrum is therefore obtained by using the harmonic optimised window function in the \zb~estimation. This combination allows for a clean reconstruction on the multipoles starting from $\ell = 60$. 


\subsubsection*{\kn~ method}

In our implementation, the \kn~strategy is based on window function which apodisation length is tuned \textit{a posteriori}.

\begin{figure}[!h]
\begin{center}
	\includegraphics[scale=0.4]{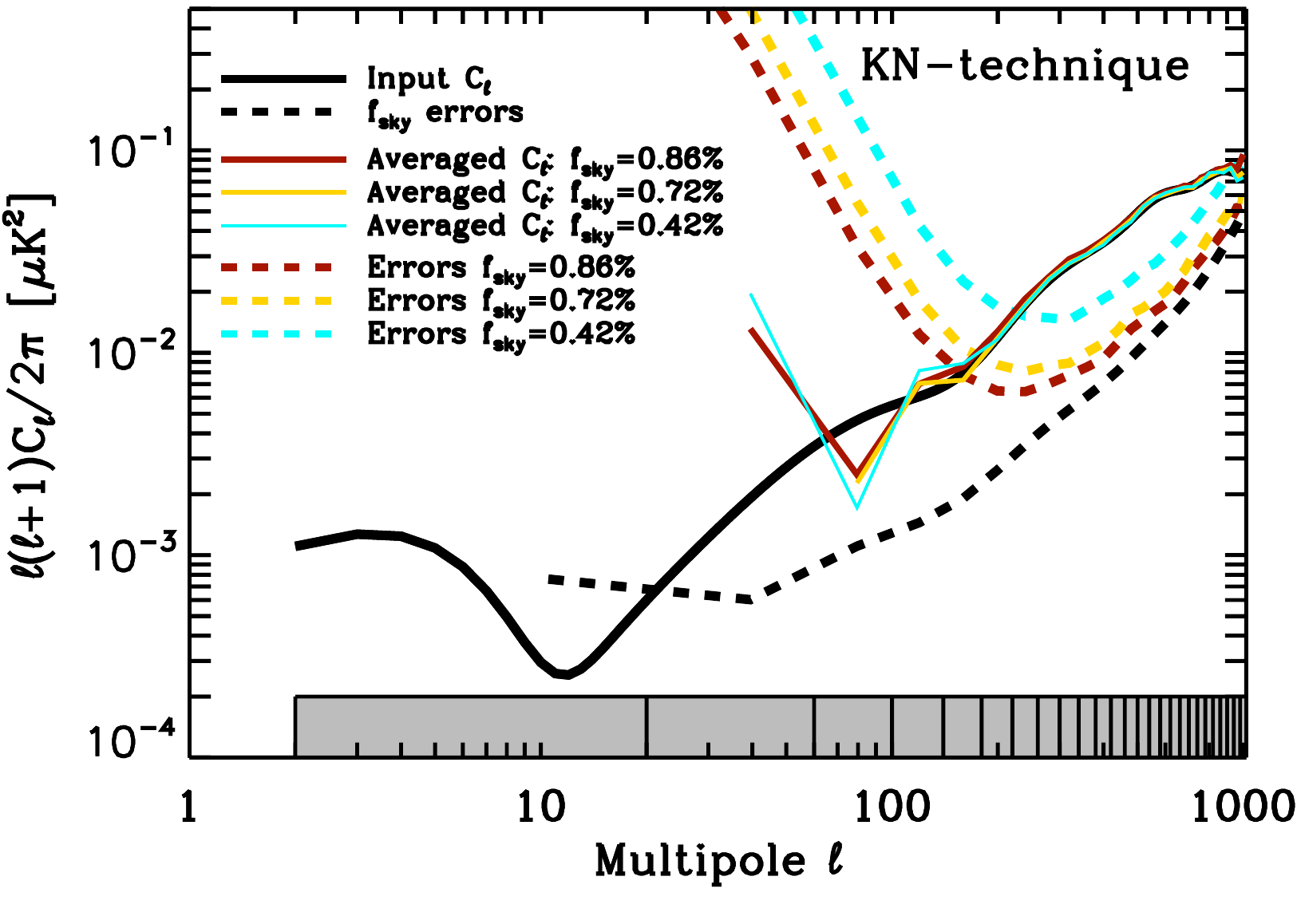} \\
	\caption{Power spectrum uncertainties on $B$-modes in dashed red, yellow and blue lines using the \kn~method thanks to different mask (with $f_{\mathrm{sky}} = 0.86\%, 0.72\%, 0.42\%$ respectively) in the case of a small scale experiment ($f_\mathrm{sky}\sim1\%$). The dashed black line displays the ideal mode counting uncertainties. The colored solid lines are the obtained reconstructed power spectra for the different effective sky fractions. The input power spectrum is in solid black curve. The grey shaded boxes represent the binning of the power spectra.} 
	\label{fig:cl_ebex_kn}
\end{center}
\end{figure}

In the present analysis, we also take window function with $\theta_{apo} = 30'$, $1^o$ and $2^o$. In Fig.~\ref{fig:cl_ebex_kn}, the error bars obtained for the various apodisation lengths are shown respectively in dashed red, yellow and turquoise. The lowest variance results from a window function with apodisation length of $\theta_{apo} = 30'$ and overreaches the estimated signal in the first five bins. For higher apodisation length, too much information is lost resulting in high variance for a wide range of $\ell$. For $\theta_{apo} = 2^o$ for instance, the $B$ modes detection is impossible for the seven first bins. Such high variance and loss of information leads to a bias of the estimated power spectrum, more significant for the highest apodisation length.

At best, the \kn~method recovers the $B$ modes power spectrum from $\ell = 180$, the primordial part including the reicombination bump ($\ell \sim 100$) is therefore out of reach, without ensuring an unbiased estimation for the lowest mulitpoles.

\subsection{The case of $TB$ and $EB$ correlations}

From the above work, the pure estimation appears to be the most efficient method for $B$ modes reconstruction. We expect the second polarisation mode -- the $E$ modes -- to be less sensitive to the mode mixing as the $B$ modes amplitude is presumed to be much weaker than the $E$ modes. In the scope of $E$ modes power spectrum, the standard method might be sufficient. However, the issue of the choice of a pseudospectrum method arises when estimating $TB$ and $EB$ cross correlations: should we use the standard or pure estimation ? \cite{Grain_2012} exposed their investigation to answer this question. In this article, the $EB$ and $TB$ power spectra are assumed to be zero as it is the case in standard model of cosmology. Nonetheless, a proper estimation of $TB$ and $EB$ is crucial \textit{e.g.} as a criteria to calibrate the detectors dedicated to CMB polarisation detection. 

\begin{figure}[!h]
\begin{center}
	\includegraphics[scale=0.4]{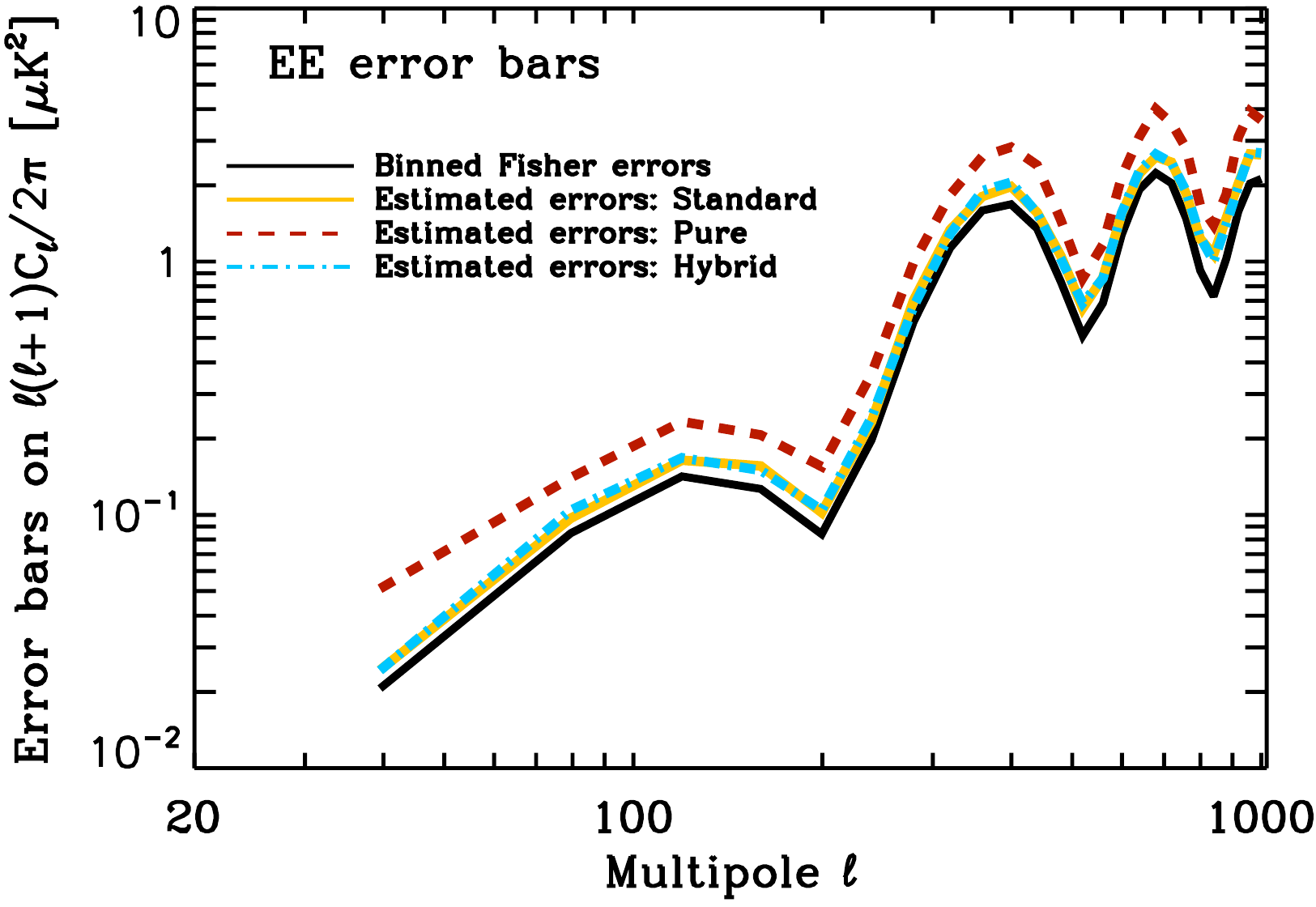}\includegraphics[scale=0.4]{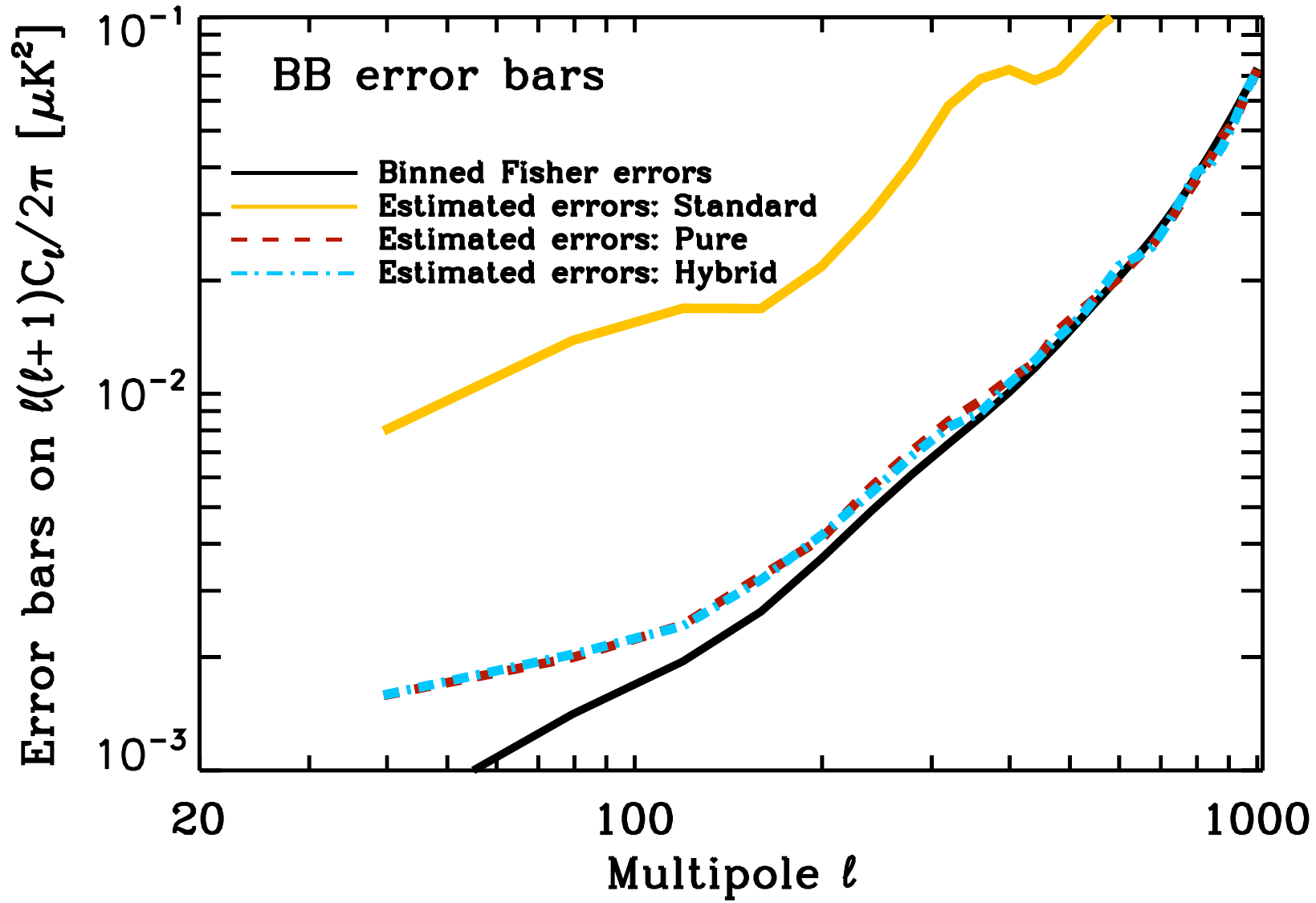} \\
	\includegraphics[scale=0.4]{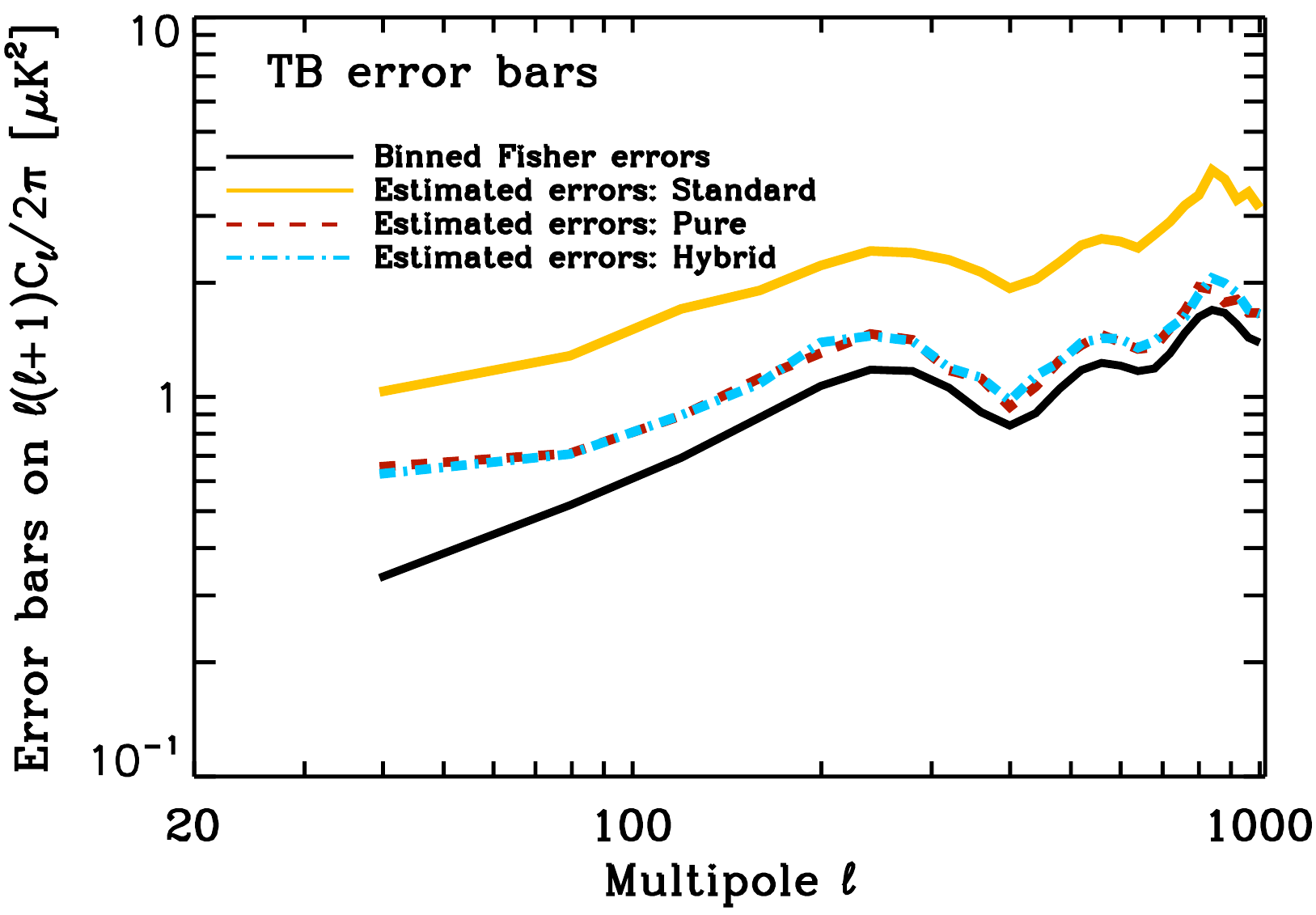}\includegraphics[scale=0.4]{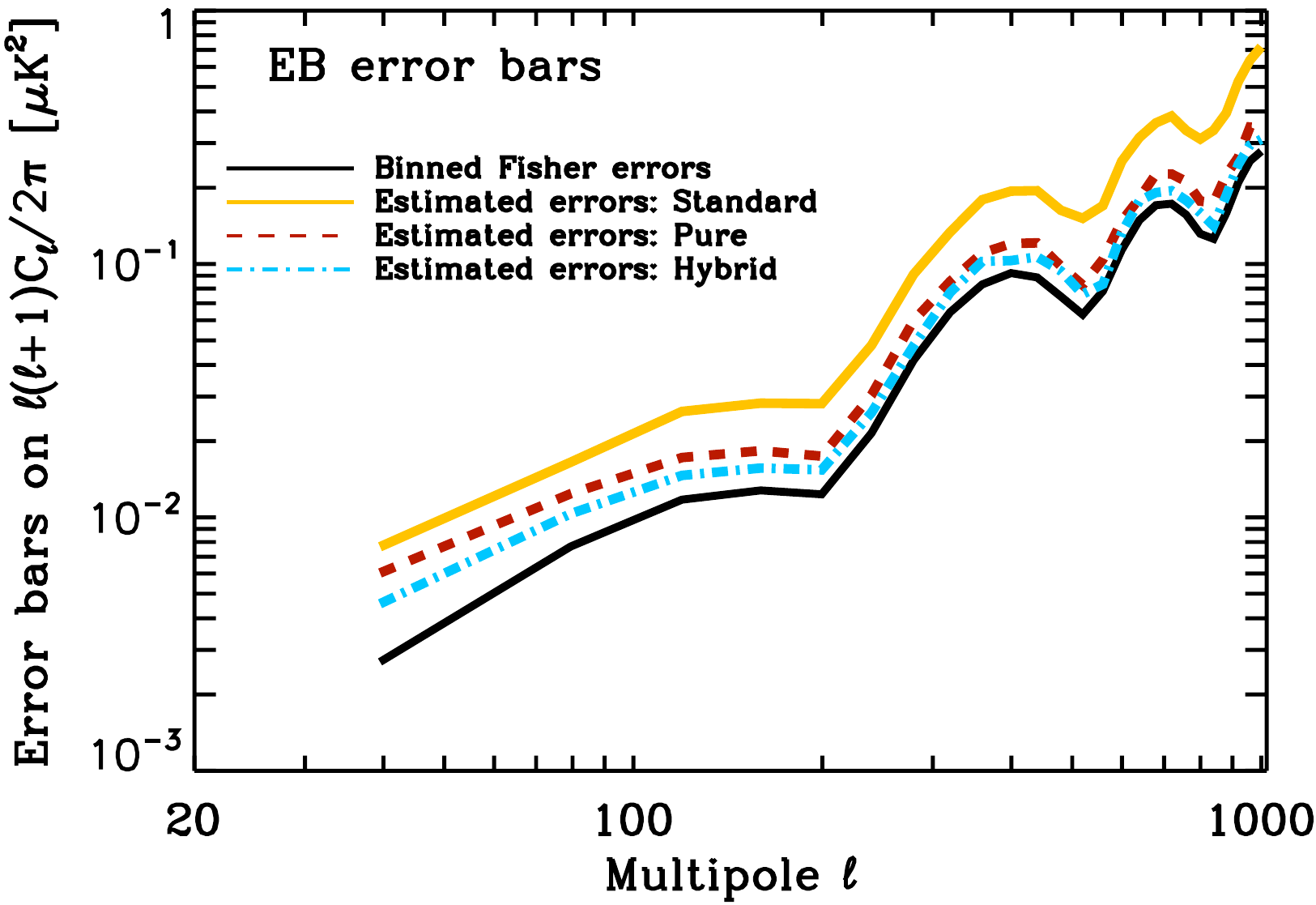} 
	\caption{Error bars on the reconstructed angular power spectra for each of the three formalisms (colored curves) alongside the naive (binned) mode counting estimate of such uncertainties for a small scale survey with inhomogeneous noise (from \cite{Grain_2012}).} 
	\label{fig:error_ebex_hybrid}
\end{center}
\end{figure}

It can be intuitively understood that the estimation of $E$ modes power spectrum, being poorly affected by the $B$-to-$E$ leakage, should be done using the standard method in order to keep the information on $E$ modes held in the ambiguous modes. On the contrary, the $B$ modes have to be estimated using the pure method as shown in this section. The $TB$ and $EB$ correlations would therefore be reconstructed using the standard (pure) method to reconstruct $E$ ($B$) modes power spectrum. This estimation of the $TB$ and $EB$ correlations is the so-called \textit{hybrid} method. The key drawn conclusion is that the hybrid estimation indeed gives the lowest uncertainties on the $EB$ and $TB$ cross-correlations. The Fig.~\ref{fig:error_ebex_hybrid} is taken from \cite{Grain_2012} and displayed the obtained variance on the $E$, $B$, $TB$ and $EB$ spectra using a pure, standard an hybrid estimation for a fiducial balloon-borne experiment. In each panel, the hybrid estimation shows its efficiency to give the smallest variance on the power spectra reconstruction.  This statement will be essential for the following part.

\section*{Conclusions}

The pure, \zb~and \kn~pseudospectrum methods were built to minimise the variance originating from the $E$-to-$B$ leakage on the reconstructed $B$ modes. They are representative of the family of leakage-free pseudospectrum estimators. The \kn~method is the most intuitive one as it consists in removing the pixels plagued by the leakage before the $B$ modes estimation. The \zb~method principle resides in the reconstruction of the masked $\chi^{B}$ field from which the polarised pseudospectra are deduced. The pure estimation sorts out the loss of orthogonality between the decomposition on $E$ and $B$ modes by introducing a new basis of pure $E$ and $B$ modes. 

The apodisation of the involved window functions $W$ to be applied to the CMB maps can be derived following three procedures: an analytic apodisation and variance-optimised window functions including a pixel- and an harmonic-based computation. They all differ in principle and can thus lead to different efficiency regarding $B$ modes estimation. In particular, the pixel-based variance-optimised window function is a powerful tool adapted to the pure method which allows for flexibility in its numerical computation. The numerical implementations of the three proposed methods and of the different kinds of window functions allowed for an utter exploration of the efficiency in $B$ modes power spectrum reconstruction.

\begin{figure}[!h]
\begin{center}
	\includegraphics[scale=0.4]{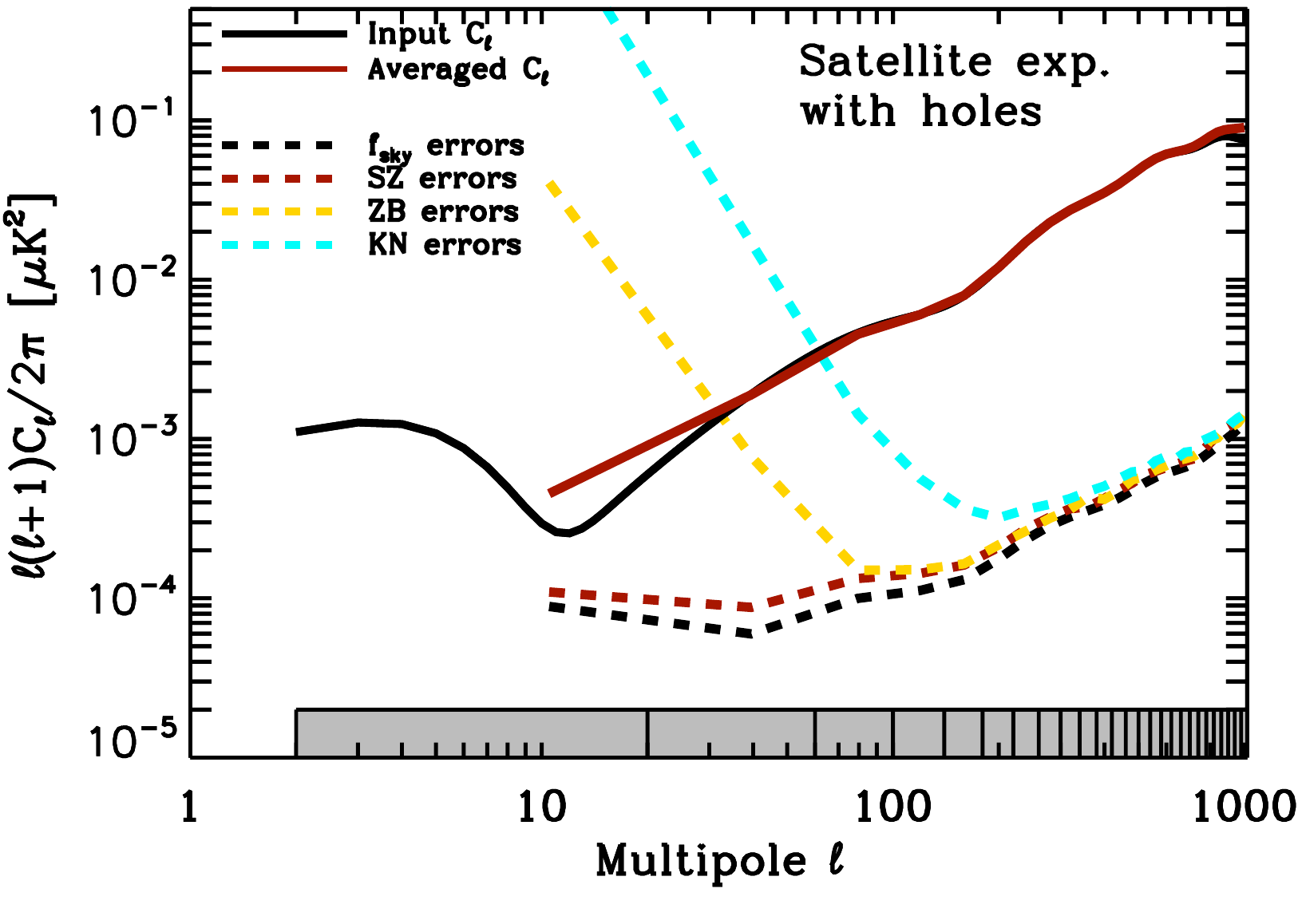}
	\caption{Power spectrum uncertainties on $B$-modes using cross-spectrum estimation for the case of satellite experiment with holes mimicking point sources-removal ($f_\mathrm{sky}\sim71\%$). The red dashed line represents the uncertainties obtained via pure method, the blue dashed line corresponds to \zb~method and at last the yellow dashed dotted to the \kn~method. The dashed-black curve stand for mode counting estimate of the error bars.} 
	\label{fig:cell_summary_wmap}
\end{center}
\end{figure}

In the scope of a CMB detection by a satellite-like mission (typical for Stage IV), the best results obtained for each method are diplayed in Fig.~\ref{fig:cell_summary_wmap}. A pure estimation of the $B$ modes gives extremely accurate results: both the reionisation and the recombination bumps are achievable. Moreover, the obtained error bars closely follow the ideal ones. The \zb~approach enables a detection of the recombination bump with a variance higher than the $B$ modes power at large angular scales. Furthermore, the \kn~method leads to a detection of the recombination bump only on a small multipole range.

Although the sky coverage is large, the $E$-to-$B$ leakage can be intrusive owing to the intricate shape of the mask and thus have to be corrected for. A key result of our analysis is that the choice of window function is crucial in this case. In particular, in the scope of $B$ modes pure estimation, the pixel-based variance-optimised window functions ensure an efficient recovery of the power spectrum. Its flexibility indeed allows for a minimisation of the global variance on the estimated $B$ modes. 

\begin{figure}[!h]
\begin{center}
	\includegraphics[scale=0.4]{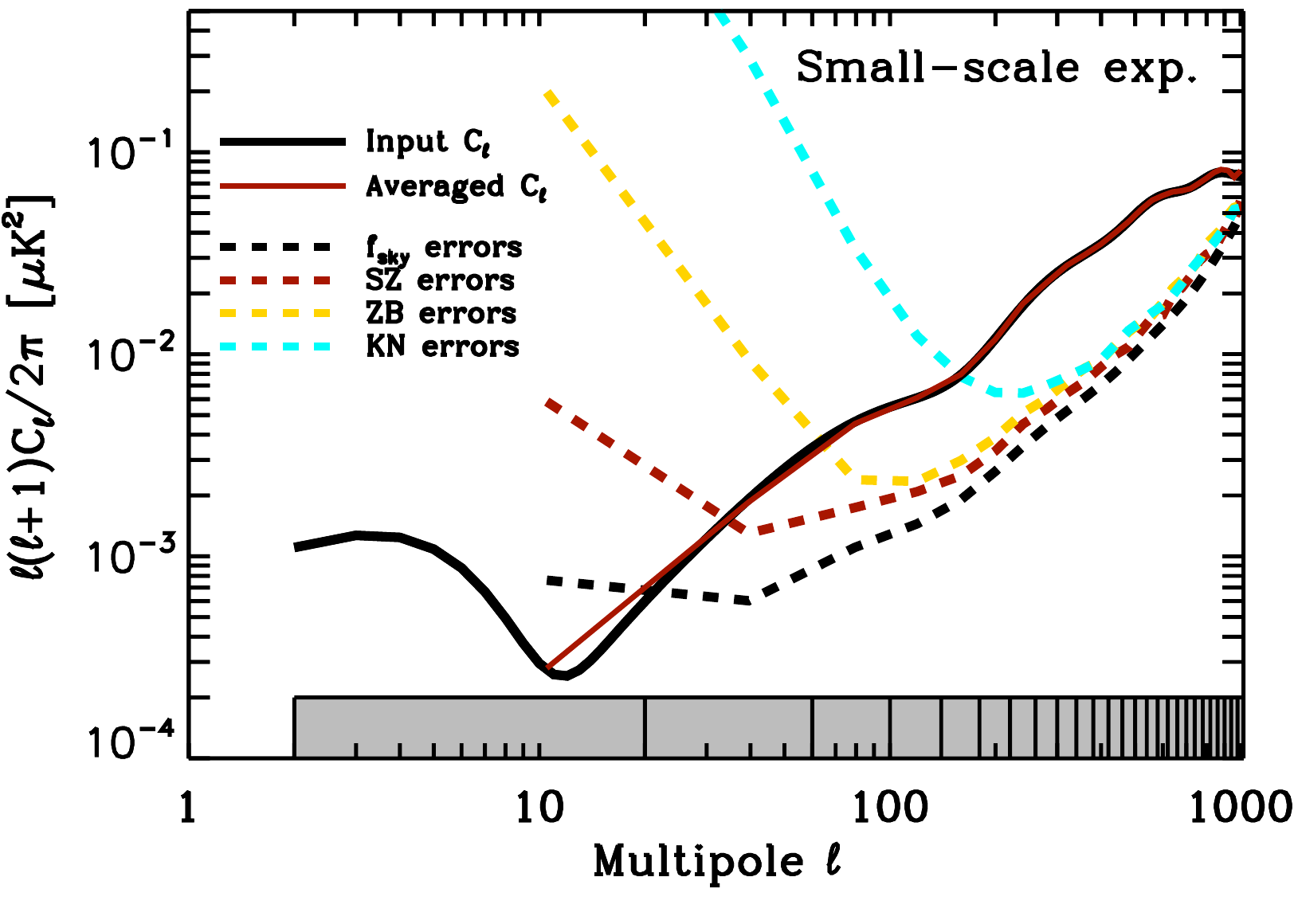}
	\caption{Power spectrum uncertainties for each of the three techniques for the case of a small-scale experiment with $f_\mathrm{sky}\simeq1\%$, a noise level of $5.75~\mu K$-arc minute and $\theta_{\mathrm{Beam}}=8~$arc minutes. Dashed-red, dashed-cyan and dashed-yellow curves are respectively for the pure, \zb~and \kn~techniques. The dashed-black curve stand for mode counting estimate of the error bars.} 
	\label{fig:cell_summary_ebex}
\end{center}
\end{figure}

The analysis were also performed in the case of a small scale experiment (typical of Stage II). A detection of the reionisation bump is infeasible for such a sky coverage as shown in the figure~\ref{fig:cell_summary_ebex} where the ideal error bars are displayed in the dashed black line. However, the pure method enables a detection of the recombination bump as the \zb~method although less efficient. The \kn~method does not give access to the primordial $B$ modes, only detecting the lensed $B$ modes.    

As a result, thanks to the pixel-based variance-optimised window function, the pure method is the most efficient estimation of the CMB $B$ modes. Also, the window function of a large scale survey have to be carefully optimised in order to efficiently reconstruct the primordial $B$ modes. The obtained results were published in \cite{Ferte_2013} which is found further to the present chapter. 

The implementation by \cite{Grain_2009} of the pure estimation of the $B$ modes power spectrum, the {\sc x2pure} code, is intensively used and is constantly developed either in the scope of performance forecasts or data analysis. The {\sc POLARBEAR} team in particular has used the {\sc x2pure} implementation for the estimation of the first direct detection of the lensing $B$ modes as exposed in \cite{Polarbear_2014}. Moreover, the implementation of the pure method is also expected to be applied to other CMB polarisation experiments such as the {\sc QUBIC} experiment. Furthermore, a flat-sky implementation of the pure method has been performed by \cite{Louis_2013} in the frame of the {\sc ACTPOL} experiment.

Finally, our analysis validates the use of the pure pseudospectrum method in order to efficiently reconstruct the $B$ modes angular power spectrum. It can thus be employed to perform \textit{realistic} forecasts on forthcoming CMB experiments. In the next chapters, the detectability of the primordial Universe physics -- such as the energy scale of inflation or a parity violation -- with the current and potential CMB polarisation experiments will be investigated.

\includepdf[pages=1,scale=1.,pagecommand={},offset=-60 -100]{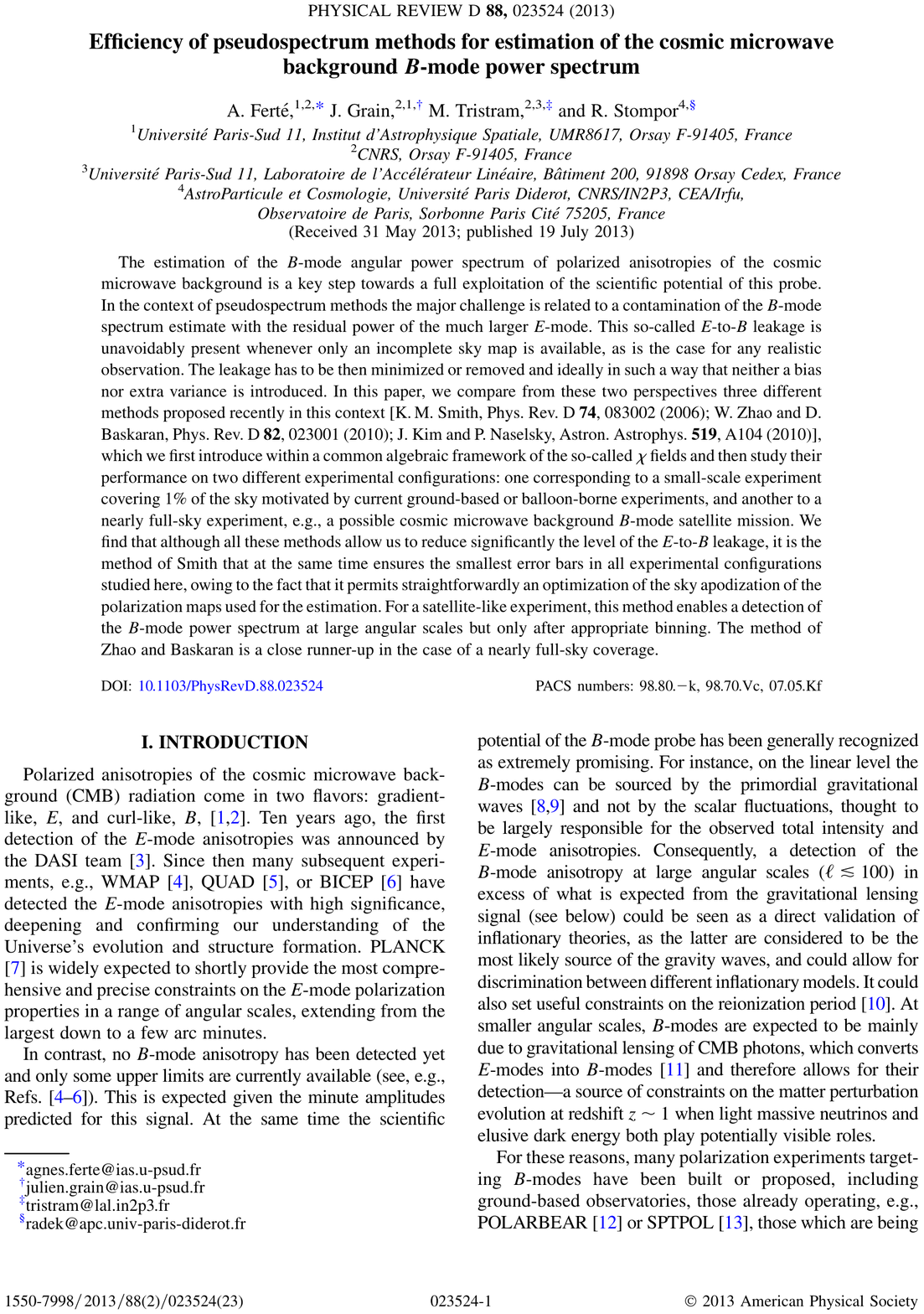}
\includepdf[pages=2,scale=1.,pagecommand={},offset=85 -100]{EBSep.pdf}
\includepdf[pages=3,scale=1.,pagecommand={},offset=-60 -100]{EBSep.pdf}
\includepdf[pages=4,scale=1.,pagecommand={},offset=85 -100]{EBSep.pdf}
\includepdf[pages=5,scale=1.,pagecommand={},offset=-60 -100]{EBSep.pdf}
\includepdf[pages=6,scale=1.,pagecommand={},offset=85 -100]{EBSep.pdf}
\includepdf[pages=7,scale=1.,pagecommand={},offset=-60 -100]{EBSep.pdf}
\includepdf[pages=8,scale=1.,pagecommand={},offset=85 -100]{EBSep.pdf}
\includepdf[pages=9,scale=1.,pagecommand={},offset=-60 -100]{EBSep.pdf}
\includepdf[pages=10,scale=1.,pagecommand={},offset=85 -100]{EBSep.pdf}
\includepdf[pages=11,scale=1.,pagecommand={},offset=-60 -100]{EBSep.pdf}
\includepdf[pages=12,scale=1.,pagecommand={},offset=85 -100]{EBSep.pdf}
\includepdf[pages=13,scale=1.,pagecommand={},offset=-60 -100]{EBSep.pdf}
\includepdf[pages=14,scale=1.,pagecommand={},offset=85 -100]{EBSep.pdf}
\includepdf[pages=15,scale=1.,pagecommand={},offset=-60 -100]{EBSep.pdf}
\includepdf[pages=16,scale=1.,pagecommand={},offset=85 -100]{EBSep.pdf}
\includepdf[pages=17,scale=1.,pagecommand={},offset=-60 -100]{EBSep.pdf}
\includepdf[pages=18,scale=1.,pagecommand={},offset=85 -100]{EBSep.pdf}
\includepdf[pages=19,scale=1.,pagecommand={},offset=-60 -100]{EBSep.pdf}
\includepdf[pages=20,scale=1.,pagecommand={},offset=85 -100]{EBSep.pdf}
\includepdf[pages=21,scale=1.,pagecommand={},offset=-60 -100]{EBSep.pdf}
\includepdf[pages=22,scale=1.,pagecommand={},offset=85 -100]{EBSep.pdf}
\includepdf[pages=23,scale=1.,pagecommand={},offset=-60 -100]{EBSep.pdf}

\part{Forecasts on the Physics of the Primordial Universe}
\label{part3}

\chapter{Forecasts on $r$ Detection using $B$ modes} 

\label{Chapter6} 

\lhead{Chapter 6. \textit{Forecasts on $r$ detection}} 
\noindent \hrulefill \\
\textit{The gravitational waves are predicted by general relativity. Their observations would thus test the theory of gravitation but they weakly interact with matter making their detection a current major challenge. An indirect detection has been however made by noticing a diminution of the Hulse Taylor pulsar time period (see \cite{Weisberg_2010}). This progress in gravitational physics has been rewarded by a Nobel Prize in 1993. Since then various experiments aiming at detecting gravitational waves from astrophysical sources such as the VIRGO interferometer (see \cite{virgo_2012}) are acquiring data. A space-based experiment called eLISA (described in \cite{eLISA_2013}) has been recently chosen by ESA as  the L3 mission of their \textit{Cosmic Vision} program. In parallel an effort is being made in the gravitational waves background detection through the CMB polarisation as it would be a smoking gun for cosmic inflation. Until now, two satellite missions aiming at its detection have been proposed in vain to ESA but meanwhile numerous balloon-borne and ground-based experiments are currently giving substantial results.} 
\noindent \hrulefill \\

The recent detections of the CMB $B$ modes has revived the challenge of primordial $B$ modes detection. In particular, the current and nearly forthcoming CMB experiments are intending to measure the tensor-to-scalar ratio $r$ which scales the tensor perturbations. In this scope, the sky coverage along with the noise and the beam have to be carefully selected. Forecasts are helpful for the choice of experimental design and are in this case driven by the accessible values of the tensor-to-scalar ratio $r$. Usually, the Fisher matrix is used together with a naive mode-counting estimation of the uncertainties on the CMB power spectra to derive the expected attainable $r$ for a given experiment. However, such an analysis results in overestimated signal-to-noise ratios on $r$. In the previous Chapter \ref{Chapter5}, the efficiency of the pure method has been stated. It is therefore a method of choice to reconstruct the $B$ modes and thus can be used to forecast $r$ detection. After introducing the Fisher matrix formalism, the impact of the estimation of the variance on the experimental specifications is explored in the scope of optimising the sky coverage of idealised small scale surveys operating from ground or balloon. Furthermore, the expected tensor-to-scalar ratios $r$ detected by a realistic on-going ground-based experiment along with potential telescope array covering half of the sky (half sky survey) and a satellite mission covering the whole celestial sphere (full sky survey) are investigated. 

\section{The Fisher Matrix Formalism}

The Fisher matrix introduced by \cite{Fisher_1935} is widely used in parameter estimation as it quantifies the information contains in a given observable on a parameter. This section qualitatively introduces the Fisher matrix in order to clarify the main ideas and does not pretend to be complete. 

In order to extract the relevant information, a set of data ${d}$ provided by an experiment are translated in observables depending on a set of parameters $\lambda_i$ of a theory. The key issue is to derive the values of the parameters $\lambda_i$ and their uncertainties $\Delta\lambda_i$ corresponding to the data set ${d}$ in the frame of the considered theory. For this purpose, the \textit{likelihood function} $\mathcal{L}$ is introduced. It is the probability of having the data set ${d}$ for the given theory. This probability function is therefore at its maximum when the parameter $\lambda_i$ is the closest from its true value $\bar{\lambda}_i$. It can be intuitively understood that if the likelihood function is highly peaked at its maximum, the estimation of $\bar{\lambda}_i$ is expected to be precise. Conversely, if the curvature of the likelihood function is small, the data are not very constrained. The curvature of the likelihood function thus gives the error on the estimated parameter $\lambda_i$. By Taylor expanding the likelihood function $\mathcal{L}$ around the set of parameters which maximise $\mathcal{L}$, the quadratic terms indeed give the behaviour of $\mathcal{L}$ around its maximum. The useful \textit{Fisher information matrix} $F_{ij}$ is therefore defined as: 
\begin{equation}
F_{ij} = - \left< \frac{\partial^2\ln{\mathcal{L}}}{\partial\lambda_i \partial\lambda_j} \right> \bigg|_{\bar{\lambda_i}\bar{\lambda_j}} .
\end{equation}

It can be shown that any unbiased estimator $\hat{\lambda}_i$ of the parameter $\lambda_i$ has a variance such as: 
\begin{equation}
\left< \Delta\hat{\lambda}_i\right> \geqslant \sqrt{(F_{ii})^{-1}}. 
\end{equation}
This inequality is known as the \textit{Cram\'er-Rao inequality} and indicates that the lowest uncertainties on $\hat{\lambda}_i$ are given by the curvature of the likelihood function \textit{i.e.} by the inverse of the Fisher matrix. If different parameters are jointly estimated, the lowest attainable uncertainties by $\hat{\lambda}_i$ is given by $\sqrt{(F^{-1})_{ii}}$. In the case of a Gaussian likelihood function $\mathcal{L}$, the Fisher matrix is explicitly given by: 
\begin{equation}
F_{ij} = \frac{1}{2}\mathrm{Tr}\left[ \frac{\partial \mathrm{C}}{\partial\lambda_i}\bigg|_{\bar{\lambda_i}} \mathrm{C}^{-1} \frac{\partial \mathrm{C}}{\partial\lambda_j}\bigg|_{\bar{\lambda_i}} \mathrm{C}^{-1} \right].
\label{eq:fisher}
\end{equation}
with C the covariance matrix of the data. This explicit equation is helpful as it enables to compute the curvature of the likelihood function without performing its numerical sampling. Moreover, if the likelihood is Gaussian, the Fisher matrix verifies:
\begin{equation} 
F_{ij} = \left[\mathrm{Cov}(\lambda_i,\lambda_j)\right]^{-1}.
\end{equation}

As it quantifies the shape and the width of $\mathcal{L}$ around its maximum, the Fisher matrix gives an estimate of the constraints set on model parameters by a data set. It is therefore a quantity of choice to forecast parameter detectability depending on experimental specifications.


\section{From $B$ modes detection to $r$ detection}

Applied to the CMB and more specifically to the $B$ modes, the Fisher formalism offers a fast and easy way to forecast the constraints set on $r$ by an experiment dedicated to its detection. As shown in Eq.~(\ref{eq:fisher}), the covariance matrix is however the key quantity of the information Fisher matrix. If not properly done, its estimation can lead to an incorrect estimation of the signal-to-noise ratio (S/N)$_{r}$ on $r$. The conclusions on the required experimental specifications may then not be optimal.

First of all, in the case of $r$ detection through the CMB $B$ modes, it can be shown starting from Eq.~(\ref{eq:fisher}) with $r$ the only parameter and $C_{\ell}^{BB}$ the observable linearly related to $r$, that the sub-part of the Fisher matrix corresponding to $r$ is written as: 
\begin{equation}
F_{rr} = \sum_{\ell \ell'} \left( \frac{\partial C_{\ell}^{BB}}{\partial r} \right) \left({\Sigma}^{-1}\right)_{\ell \ell'} \left( \frac{\partial C_{\ell'}^{BB}}{\partial r} \right),
\end{equation}
with $\left(\Sigma^{-1}\right)_{\ell \ell'} = \mathrm{Cov}(C_{\ell}^{BB},C_{\ell'}^{BB})$ the inverse of the covariance matrix of the $B$ modes power spectrum.
Therefore the uncertainties $\sigma_r^2$ on $r$ are directly the inverse of the Fisher sub-matrix $F_{rr}$: 
\begin{equation}
\sigma_r^2 = F_{rr}^{-1}.
\end{equation}

The $B$ modes power spectrum, being the sum of a primordial and a lensing contribution, is modelled as: 
\begin{equation}
C_{\ell}^{BB}(r) = r~\mathcal{T}_{\ell}^{BB, prim} ~+~ \mathcal{T}^{EE \rightarrow BB, lens}_{\ell},
\end{equation}
with $\mathcal{T}_{\ell}^{BB,prim}$ and $\mathcal{T}_{\ell}^{EE \rightarrow BB,lens}$ two fiducial power spectra not depending on $r$. The former one is the primordial contribution to the $B$ modes power spectrum sourced by inflationary gravitational waves for a fiducial tensor-to-scalar ratio $r_{fid} = 1$. The latter stands for the lensing contribution from the $E$ modes to the $B$ modes power spectrum. Although it gives an important piece of information on the late time universe, this contribution acts as an additional noise to the primordial $B$ modes as it does not depend on $r$. Both fiducial power spectra are easily computed by the use of the Boltzmann code CLASS (\cite{CLASS_2011}). 

The covariance matrix can be derived in several ways. A first approach is the ideal case of naive mode-counting variance as exposed in Chapter~\ref{Chapter4} which ignores the experimental issues such as the $E$-to-$B$ leakage. We recall the expression of the covariance matrix in the present case, for an experiment with a noise power spectrum $N_{\ell}$, a beam $b_{\ell}$ and an observed sky fraction $f_{\mathrm{\mathrm{sky}}}$: 
\begin{equation}
\Sigma_{\ell \ell'} = \frac{2\delta_{\ell \ell'}}{(2\ell+1)f_{\mathrm{\mathrm{sky}}}}\left(C_{\ell}^{BB} + \frac{N_{\ell}(f_{\mathrm{sky}})}{b_{\ell}^2} \right)^2
\end{equation}
Such an approach is valuable for it sets a benchmark as it underestimates the uncertainties on $B$ modes regarding any numerical methods used to derive the covariance matrix.

A second approach consists in the reconstruction of the $B$ modes power spectrum thanks to the pure method. Its efficiency in correcting for the $E$-to-$B$ leakage and giving the lowest error bars on the $B$ modes has been stated in the previous Chapter~\ref{Chapter5}. The covariance matrix is then the covariance obtained from Monte Carlo simulations.

Both approaches can therefore be used to compute the power spectrum covariance matrix involved in the Fisher matrix. In the end, the signal-to-noise ratio on $r$ is given by: 
\begin{equation}
\mathrm{(S/N)}_r = \frac{r}{\sigma_r} = r\times\sqrt{\frac{1}{F_{rr}}}.
\end{equation}
They thus will potentially give different results on the obtained (S/N)$_r$, the naive mode-counting approach being optimistic and the pure method realistic. 

Considerations on the optimisation of experiments dedicated to CMB polarisation in the scope of fundamental physics constraints (such as the neutrino mass or the energy scale of inflation) has already been investigated as in \cite{Verde_2006} and references herein, \cite{Wu_2014} or \cite{Caliguri_2014}. However, the optimistic approach of the mode-counting estimation of the variance is usually adopted. In the frame of the foregrounds contamination, $B$ modes detection has been explored in \cite{Stivoli_2010} in a realistic estimation of the covariance matrix including statistical errors from the pure estimation and from the foregrounds subtraction, although only in the case of a small scale survey with homogeneous noise. The here proposed investigation thus fits into the frame of realistic forecast for a set of current, being deployed or potential experimental set-ups corresponding to Stage II, III (current and upgraded) and IV (forthcoming) CMB experiments (see \cite{Abazajian_2014} for a description of the envisioned CMB experimental evolution). In this Chapter, peculiar attention is paid to the statistical uncertainties coming from the pure estimation of the $B$ modes power spectrum, ensuring realistic forecasts. 

\section{Optimising the scanning strategy of small scale experiments}\label{sec:opt}

A wide range of experiments are aiming at detecting $r$ which accessible values depend on the experimental design. If the tensor-to-scalar ratio $r$ is thought to be as high as claimed by the BICEP2 team in \cite{BICEP2_2014}, thus small scale surveys might be sufficient for a relevant measurement. Their expected performances regarding the width of the survey has been already deeply explored in \cite{Jaffe_2000}. 

The forecasts were however performed using the mode-counting variances which could potentially lead to misleading results as they are idealised. The following analysis thus ensues from two main questions. First of all, for a given sensitivity and time of observation, the instrumental noise per pixel scales with the observed sky fraction. In the scope of deep survey, the sky fraction is chosen to be low therefore ensuring a low noise level. However the large angular scales, crucial to set constraints on $r$, are inaccessible therefore damaging $r$ detection. On the contrary, a shallow survey favours a large sky coverage thus implying a predominant noise level which can prevent from a $r$ detection. Between this two extreme cases lie a sky coverage optimal for $r$ detection. The first enquiry is consequently the dependence of the forecasts on the optimal sky fraction regarding the choice of variances computation (realistic or idealised). The second investigation deals with the obtained signal-to-noise ratio on $r$ for the considered sky fraction in both approaches.

\subsection{Experimental set-ups}

\begin{figure}[!h]
\begin{center}
	\includegraphics[scale=0.27]{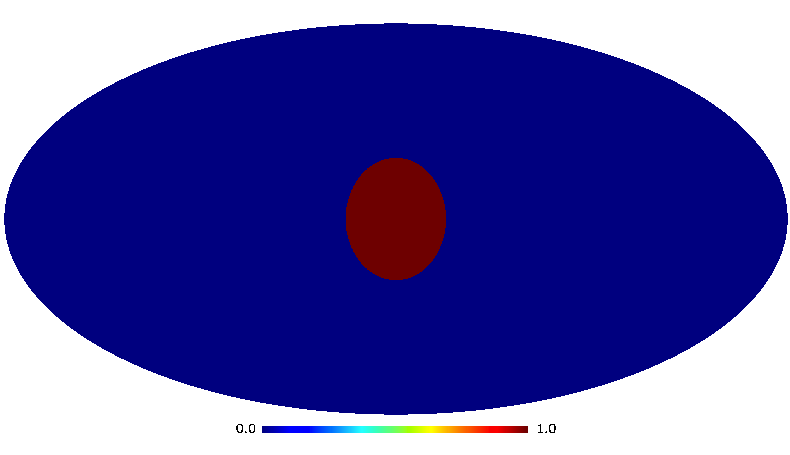}\includegraphics[scale=0.27]{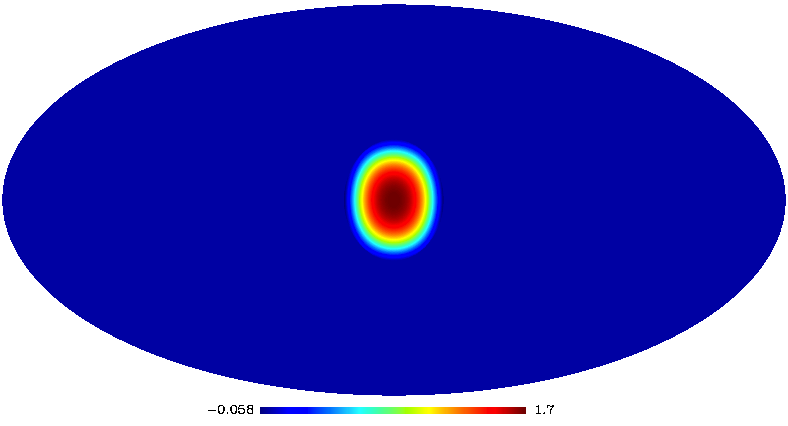}
	\caption{In the left panel figures the binary mask for an observed sky fraction of $4\%$. The corresponding harmonic window function for $r = 2$ and $\ell \in [2;20]$ is displayed in the right panel.}
	\label{fig:small}
\end{center}
\end{figure}

A set of small scale experiments with a beam of $8$ arcmin, a sky fraction ranging from $0.5\%$ to $10\%$ of the sky is considered in this analysis. For a given sensitivity, the noise level is scaling with the observed sky fraction as (\cite{Jaffe_2000}): 
\begin{equation}
\sigma(f_{\mathrm{sky}}) = \sqrt{\frac{f_{\mathrm{sky}}}{f_{\mathrm{sky}}^{0}}} \sigma(f_{\mathrm{sky}}^{0}),
\end{equation}
with $f_{\mathrm{sky}}^{0}$ a fiducial sky fraction. The chosen benchmark is $\sigma(f_{\mathrm{sky}}^{0}) = 5.75 \mu K$-arcmin for $f_{\mathrm{sky}}^{0} = 1\%$ which is typical of balloon-borne experiments such as {\sc EBEX} (see \cite{ebex}) and similar for {\sc ACTPOL} (\cite{Niemack_2010}). For convenience, the sky coverage is chosen to be a simple spherical cap with sky fraction $f_{\mathrm{sky}}$. As an example, the binary mask corresponding to a sky fraction of $f_{\mathrm{sky}} = 4\%$ is displayed on the left panel of Fig.~\ref{fig:small}. 

The pure $B$ modes estimation requires a binning process as explained at the beginning of the previous chapter. The used binning is the same as in Chapter~\ref{Chapter5} \textit{i.e.} $N_{bin} = 26$ bins of band power $\Delta\ell = 40$ except for the first bin $\ell \in [2;20]$. In the Chapter~\ref{Chapter5}, the pure method efficiency has been shown to depend on the choice of the sky apodisation optimisation. The simple contour of the patches however allows for the use of variance-optimised window function in the harmonic domain. A set of harmonic window function for each bin, each $f_{\mathrm{sky}}$ and each $r$ has therefore been computed, the obtained spin-0 window function for $f_{\mathrm{sky}} = 4\%$ and $r = 0.2$ in the multipole range $\ell \in [2;20]$ is shown in the right panel of Fig.~\ref{fig:small}.

\subsection{Numerical results}

The performance of the small scale surveys on $r$ detection is expected to depend on $r$. The BICEP2 team claims for a detection of a signal consistent with a tensor-to-scalar ratio of $r = 0.2$. As this result has to be considered with cautious, an exploration of the detectability for $r = 0.07,0.1,0.15$ and $0.2$ is relevant. The signal-to-noise ratio on $r$ for the different experimental set-ups is therefore derived following the aforementioned Fisher matrix formalism with the covariance matrix computed in the mode-counting and the pure approaches.

\begin{figure}[!h]
\begin{center}
	\includegraphics[scale=0.5]{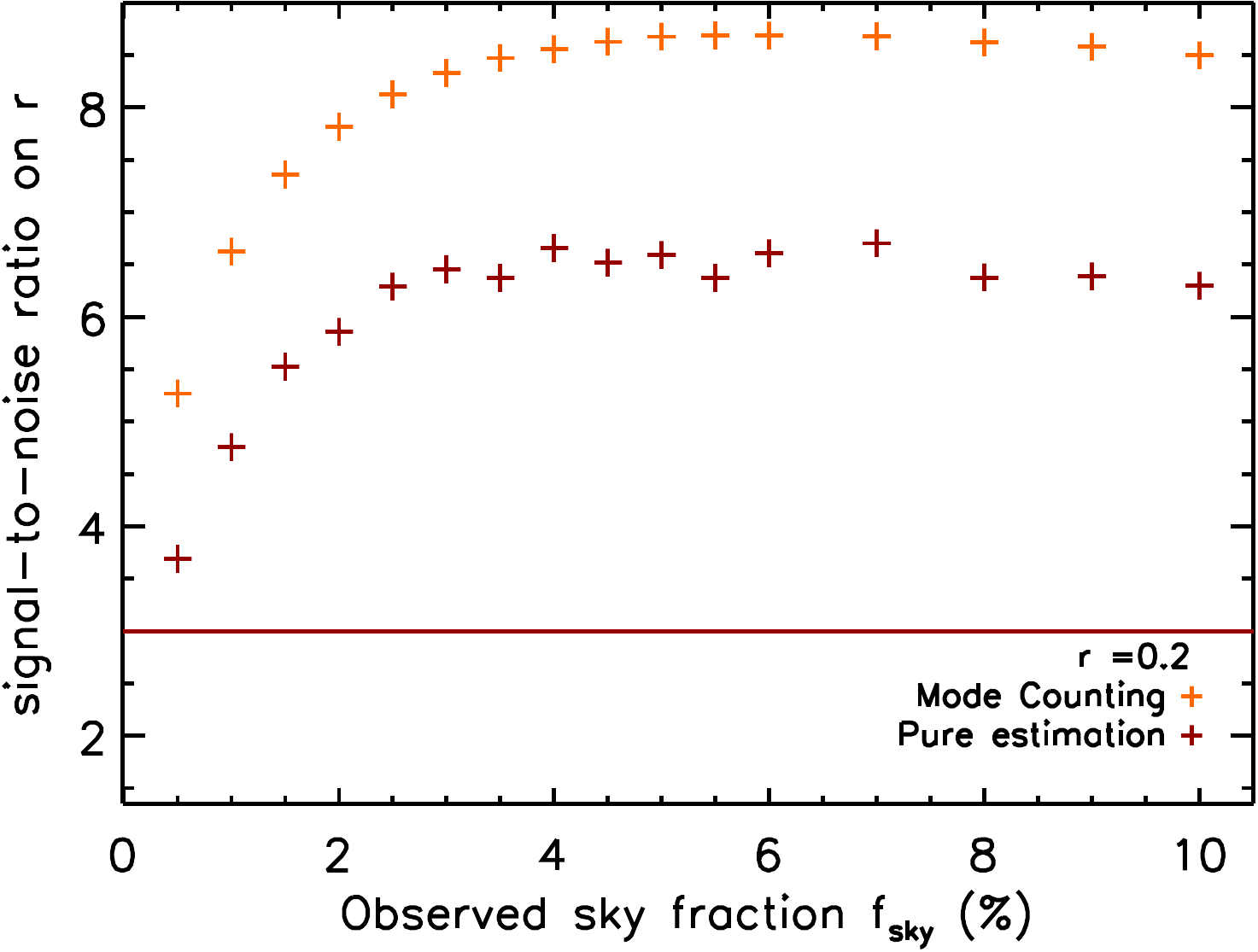}\includegraphics[scale=0.5]{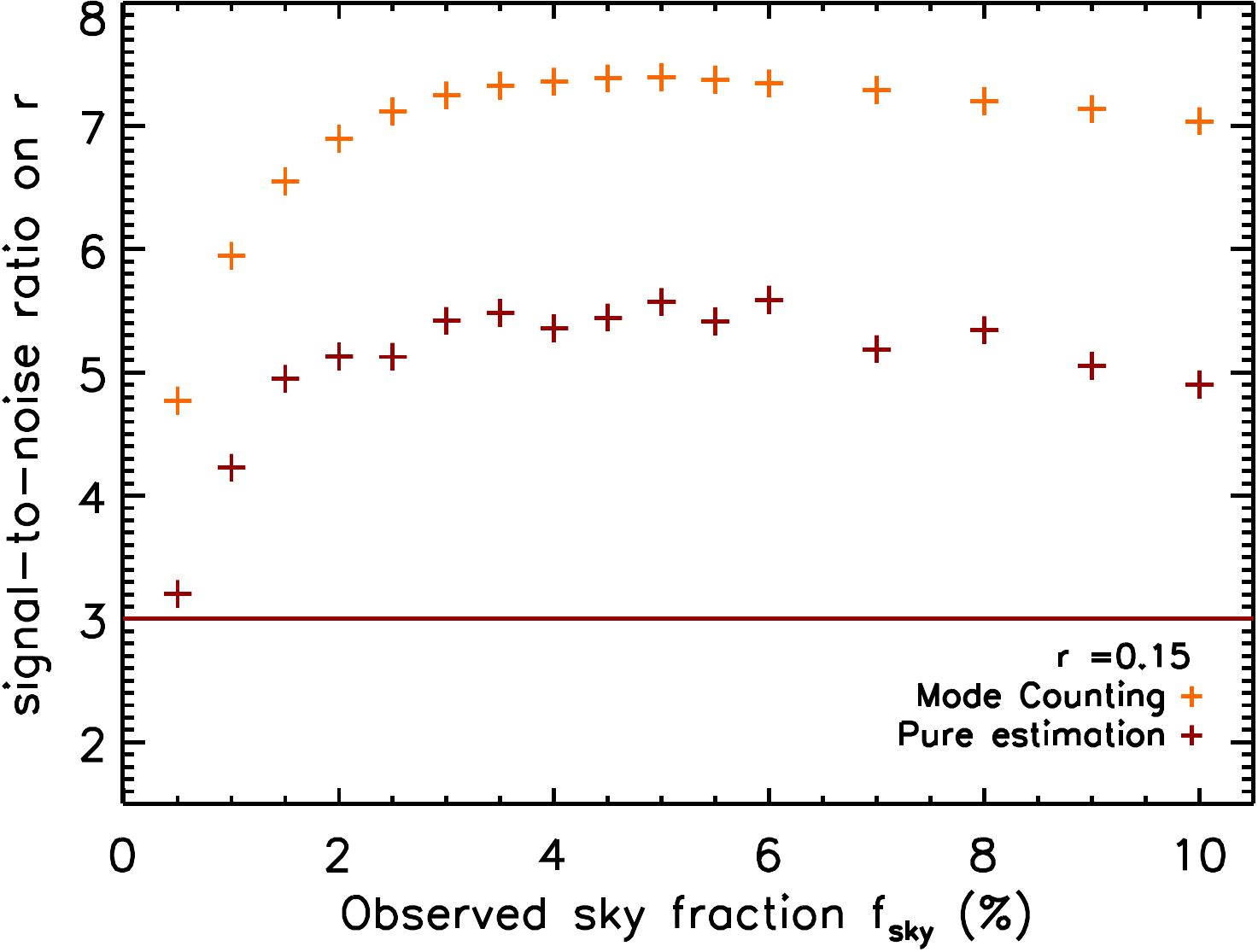}\\
	\includegraphics[scale=0.5]{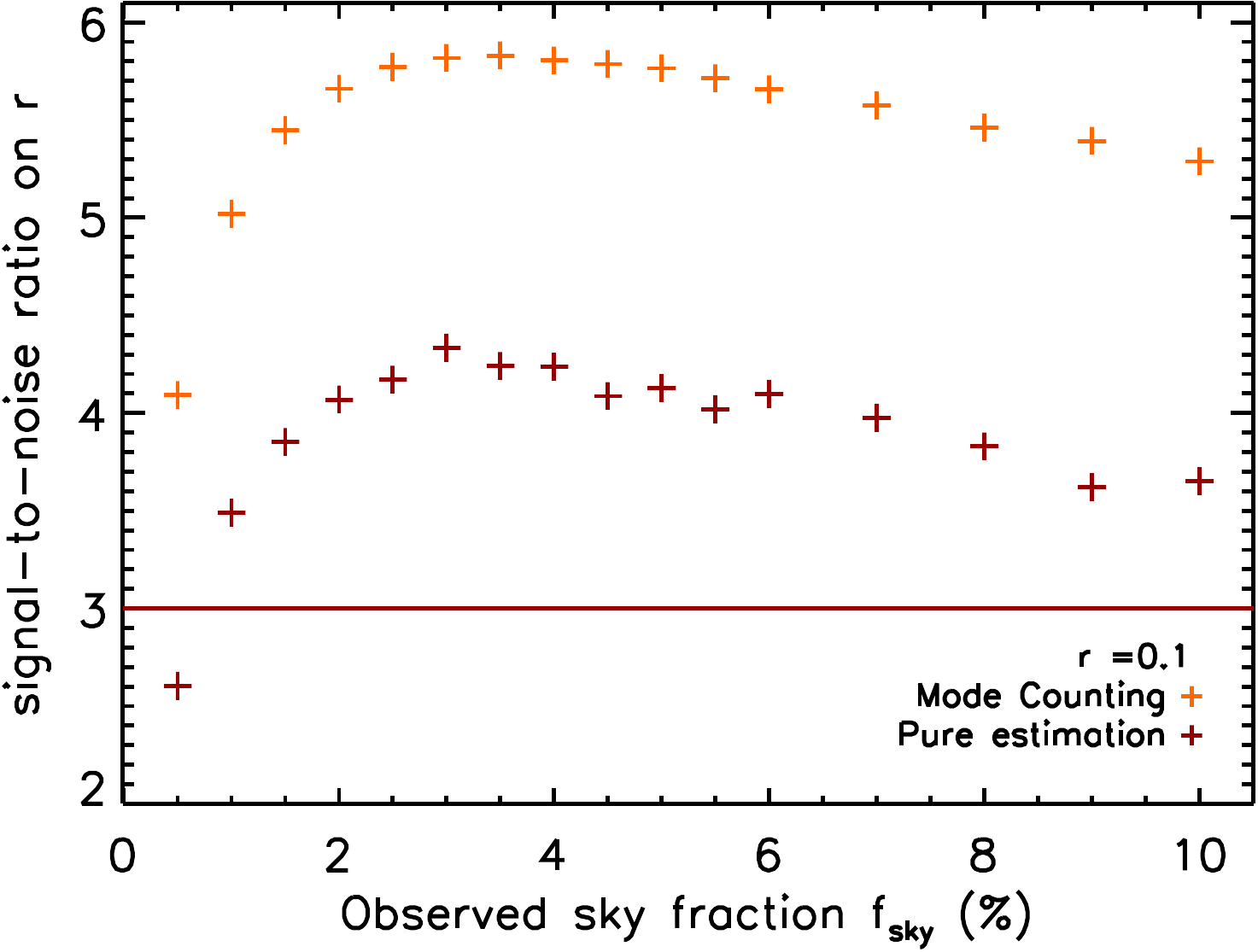}\includegraphics[scale=0.5]{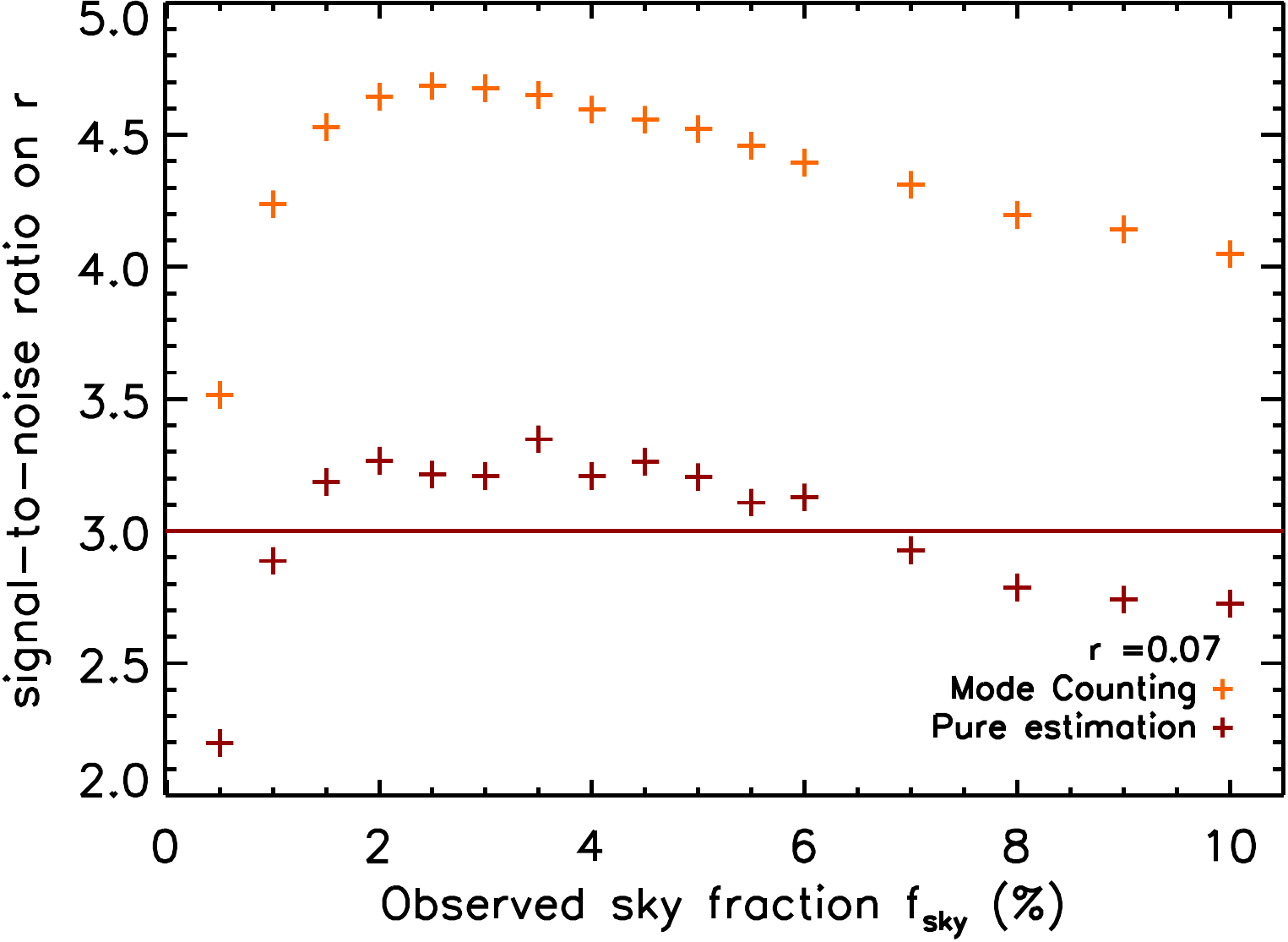}
	\caption{The signal-to-noise ratio for $r = 0.2, 0.15, 0.1, 0.07,$ from top left to bottom right, using a naive mode-counting variance (in yellow), the pure estimation of the $B$ modes (in burgundy). The horizontal red line set the benchmark of $3\sigma$.}
	\label{fig:rfsky}
\end{center}
\end{figure}

The (S/N)$_r$ obtained in the optimistic and realistic approaches are displayed as functions of the observed sky fraction for each $r$ in Fig.~\ref{fig:rfsky}. The yellow crosses stand for the mode-counting estimation of the $B$ modes variance. For all $r$, the increase of (S/N)$_r$ with $f_{\mathrm{sky}}$ for the low values of the observed sky fraction is understood as the increase of the statistics. Being sampling variance dominated, the (S/N)$_r$ indeed roughly behave as $\sqrt(f_{\mathrm{sky}})$. However, for larger $f_{\mathrm{sky}}$, the (S/N)$_r$ decreases with respect to the observed sky fraction with a slope increasing when $r$ decreases. The slope is particularly significant for $r = 0.7$, the bottom right panel of Fig.~\ref{fig:rfsky}. For $r = 0.2$, the slope is so low that (S/N)$_r$ seems to reach a plateau. This behaviour is caused by the increase of the instrumental noise projected on a larger fraction of the sky, which is no longer compensated by the increase of the statistics. 

Besides, the (S/N)$_r$ obtained using the pure $B$ modes estimation (red crosses) have a similar behaviour. Indeed, it shows a steep increase at low $f_{\mathrm{sky}}$ before reaching a maximum followed by a decrease with a slope inversely proportional to $r$. Nonetheless, the (S/N)$_r$ computed thanks to the pure estimation exhibits some fluctuations in the medium range of $f_{\mathrm{sky}}$, being less smooth as the case of the mode-counting approach. It can be understood at the level of the power spectrum reconstruction. Indeed, the behaviour of the variance in the medium bins, in particular around $\ell = 100$, with respect to the observed sky fraction is not evident as the noise competes with the sampling variance. 

Furthermore, as expected, the (S/N)$_r$ using the mode-counting approach over-estimates the signal-to-noise. At its maximum value, the (S/N)$_r$ in the pure estimation is indeed about a factor 1.4 lower than the one derived in the mode-counting estimation. The use of a realistic estimation of the covariance matrix is therefore mandatory for an exact estimation of the (S/N)$_r$. Nevertheless, the observed sky fraction at which the maximum is reached by the signal-to-noise ratio is similar in both cases. For a tensor-to-scalar ratio $r = 0.2$, the optimal observed fraction is reached around $f_{\mathrm{sky}} = 6\%$ while for $r = 0.15$, the optimal sky fraction is $f_{\mathrm{sky}} \sim 5\%$. The optimal sky fraction for $r = 0.1$ is $3\%$ in the mode-counting and the pure approaches. In the case of $r = 0.07$, the conclusions are different due to the aforementioned fluctuations of the (S/N)$_r$ computed with the pure method. Indeed, the optimal sky fraction seems to be $2.5\%$ in the mode-counting approach while it reaches $3.5\%$ for the pure method. The fluctuations are thought to be due to the binning which is suitable for $1\%$ of the sky and might be adapted for higher fraction of the sky. It would however not imply major changes in the results.  

\begin{figure}[!h]
\begin{center}
	\includegraphics[scale=0.5]{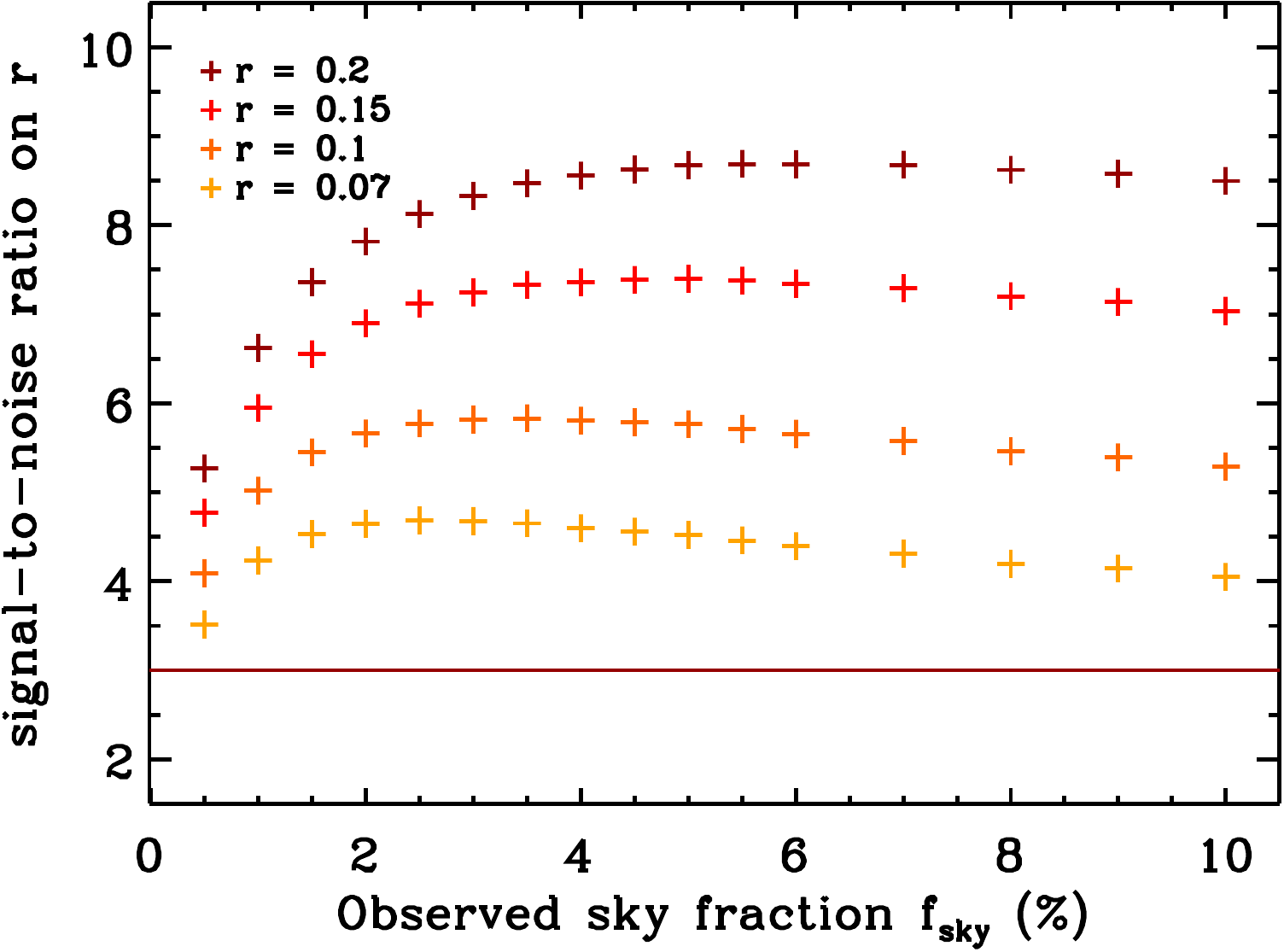}\includegraphics[scale=0.5]{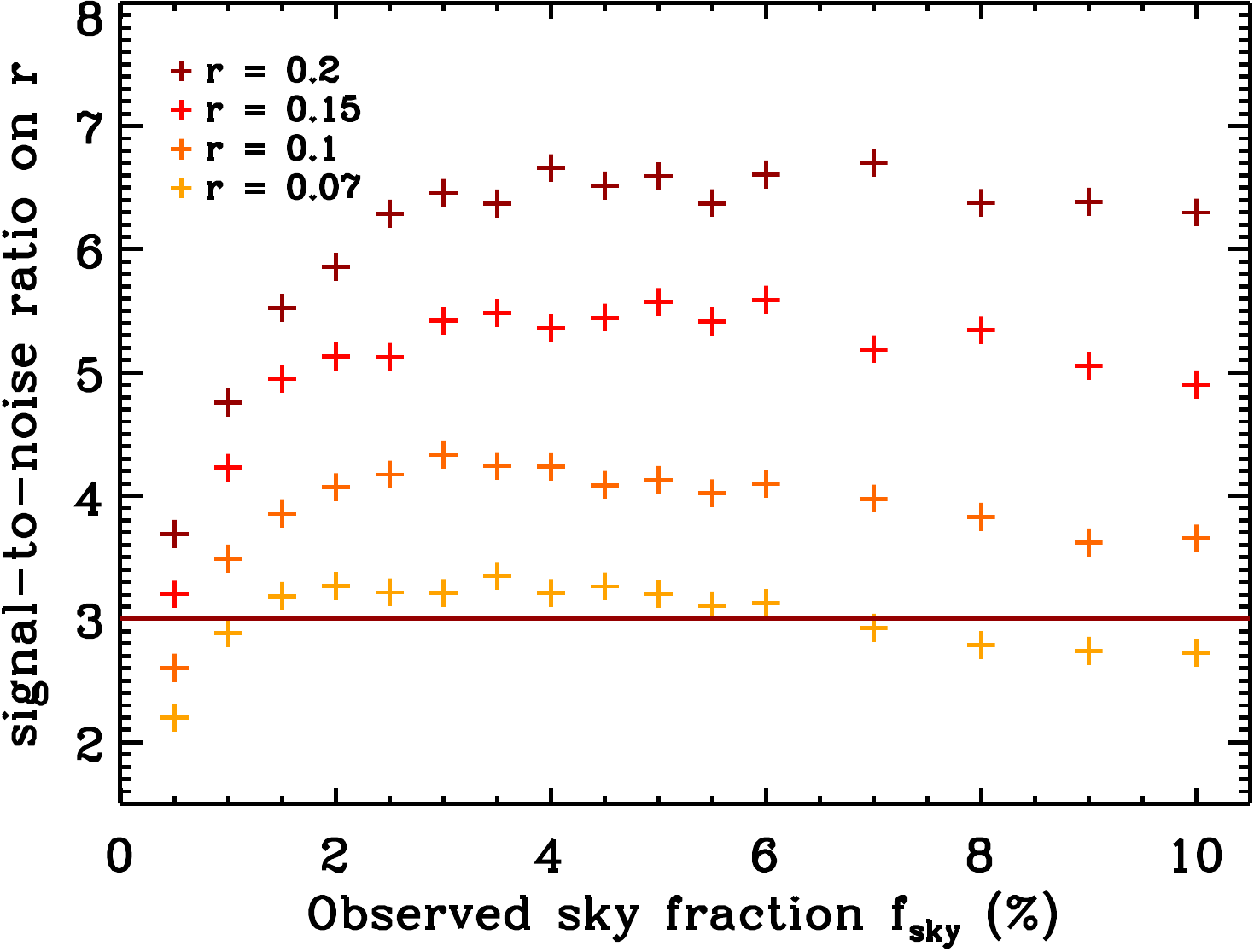}
	\caption{Same results as Fig.~\ref{fig:rfsky} with the signal-to-noise ratio displayed in each panel for all $r$. From left to right, (S/N)$_r$ is computed using the mode-counting approach, the minimal variance quadratic estimator and the pure $B$ modes estimation. The horizontal red line set the benchmark of $3\sigma$.}
	\label{fig:rfsky_app}
\end{center}
\end{figure}

The Fig.~\ref{fig:rfsky_app} shows the same results with all (S/N)$_r$ displayed in each panel corresponding to the mode-counting approach and the pure estimation from left to right respectively. The signal-to-noise ratio on the tensor-to-scalar ratio scales with $r$. For a mode-counting estimation, all $r$ varying from 0.07 to 0.2 are accessible for $f_{\mathrm{sky}} \in [0.5;10\%]$ as the (S/N)$_r$ always exceed $3\sigma$ (displayed as horizontal red line). However, the results are less evident in the pure method. If $r = 0.07$, only an observed sky fraction included between 1.5\% and 6\% would ensure a detection. Also, a coverage of $f_{\mathrm{sky}} = 0.5\%$ would not be enough to constrain $r = 0.1$ while it can reach $4.3\sigma$ for $f_{\mathrm{sky}} = 3\%$. For higher $r$, significant constraints can be set on $r$ for all the considered observed sky fraction. However, the medium $f_{\mathrm{sky}}$ range is favoured as for high sky coverage, the (S/N)$_r$ decreases.  

As a result, while the signal-to-noise ratio is overestimated in the mode-counting approach, the optimal sky fraction is roughly the same in both methods. This analysis validates the use of the mode-counting estimation of the covariance matrix in the perspective of sky optimisation, in the case of small scale surveys with a simple contour. In general, the optimal observed sky fraction is included between $f_{\mathrm{sky}} = 2\%$ (for lower $r$) and $f_{\mathrm{sky}} = 6\%$ (for higher $r$). It is worth noticing that even for $2\%$ of the sky, a high (S/N)$_r$ (about 6) is expected for $r = 0.2$. As the experiment performance seems to depend on $r$, the detectability of $r$ of a given experimental set-up should now be investigated.


\section{Detecting the tensor-to-scalar ratio}

In the scope of current status on primordial $B$ modes and the competitive answers to space agency calls for a satellite mission, the crucial question to be answered to is: what is possible from the ground ? It is known that a space-based experiment is mandatory for the detection of large scales polarised anisotropies and for high quality foregrounds understanding, thanks to electromagnetic spectral coverage. However, regarding the time scale (around twenty years) necessary for the design and construction of such an experiment, the ground based and balloon borne experiments will meanwhile bring substantial constraints on $B$ modes. Intermediate scale survey will also intend to measure $r$. The following study thus consists in forecasting the performances on $r$ detection of three typical experimental set-ups. The first one corresponds to a current suborbital experiment and will be referred to as a \textit{small scale survey}. A satellite-like experiment (\textit{full sky survey}), typical of present proposals to space agency, is also considered. The performances of an intermediate scale experiment, characteristic of an array of telescopes, covering only one hemisphere is under scrutiny as well. The signal-to-noise ratio on $r$ is computed using a naive mode-counting estimation of the uncertainties for a first investigation and finally via the pure estimation of $B$ modes for a realistic forecast. 

\subsection{Experimental set-ups}

\underline{Experimental specifications} 

\begin{figure}[!h]
\begin{center}
	\includegraphics[scale=0.1]{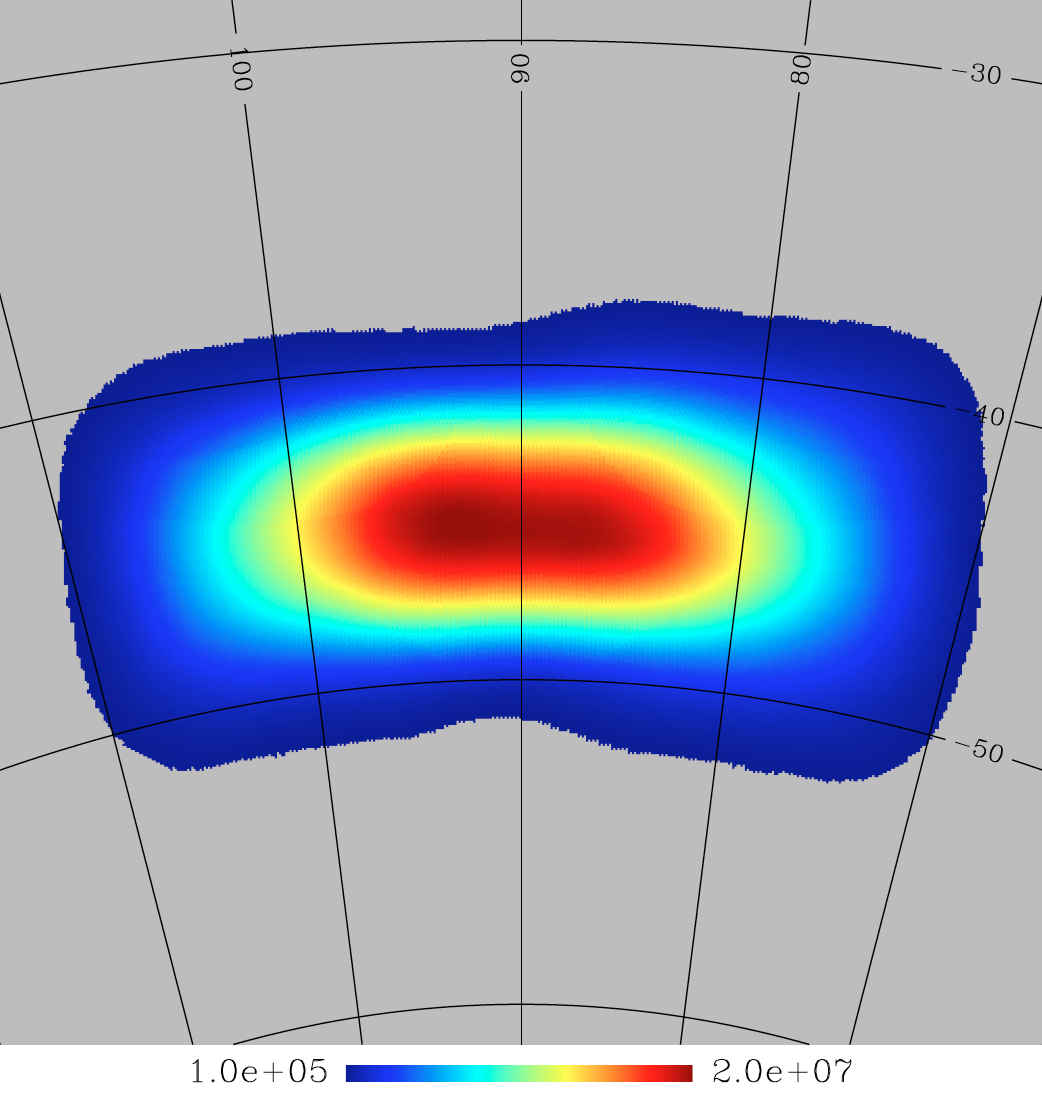}\includegraphics[scale=0.25]{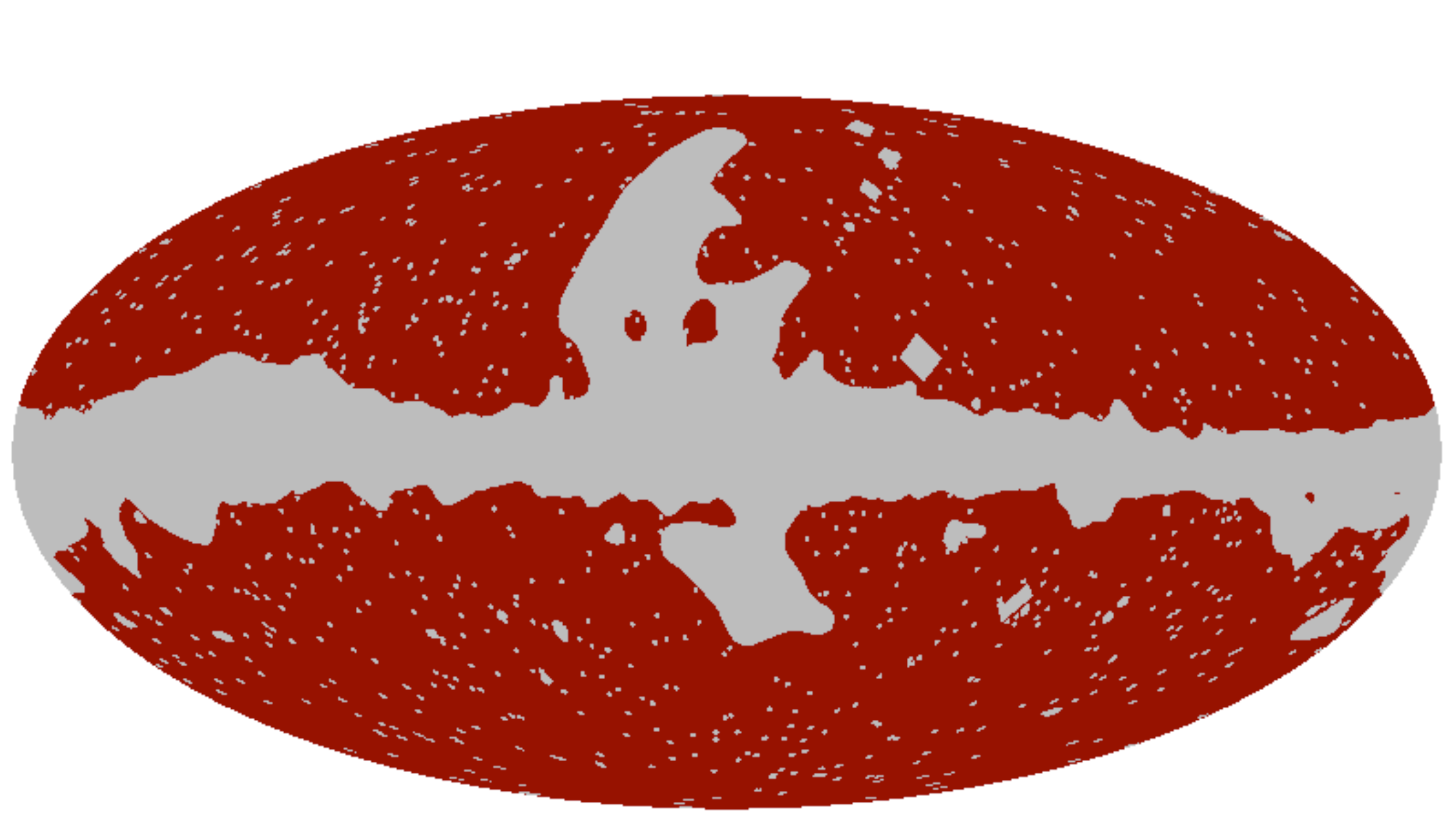}\includegraphics[scale=0.25]{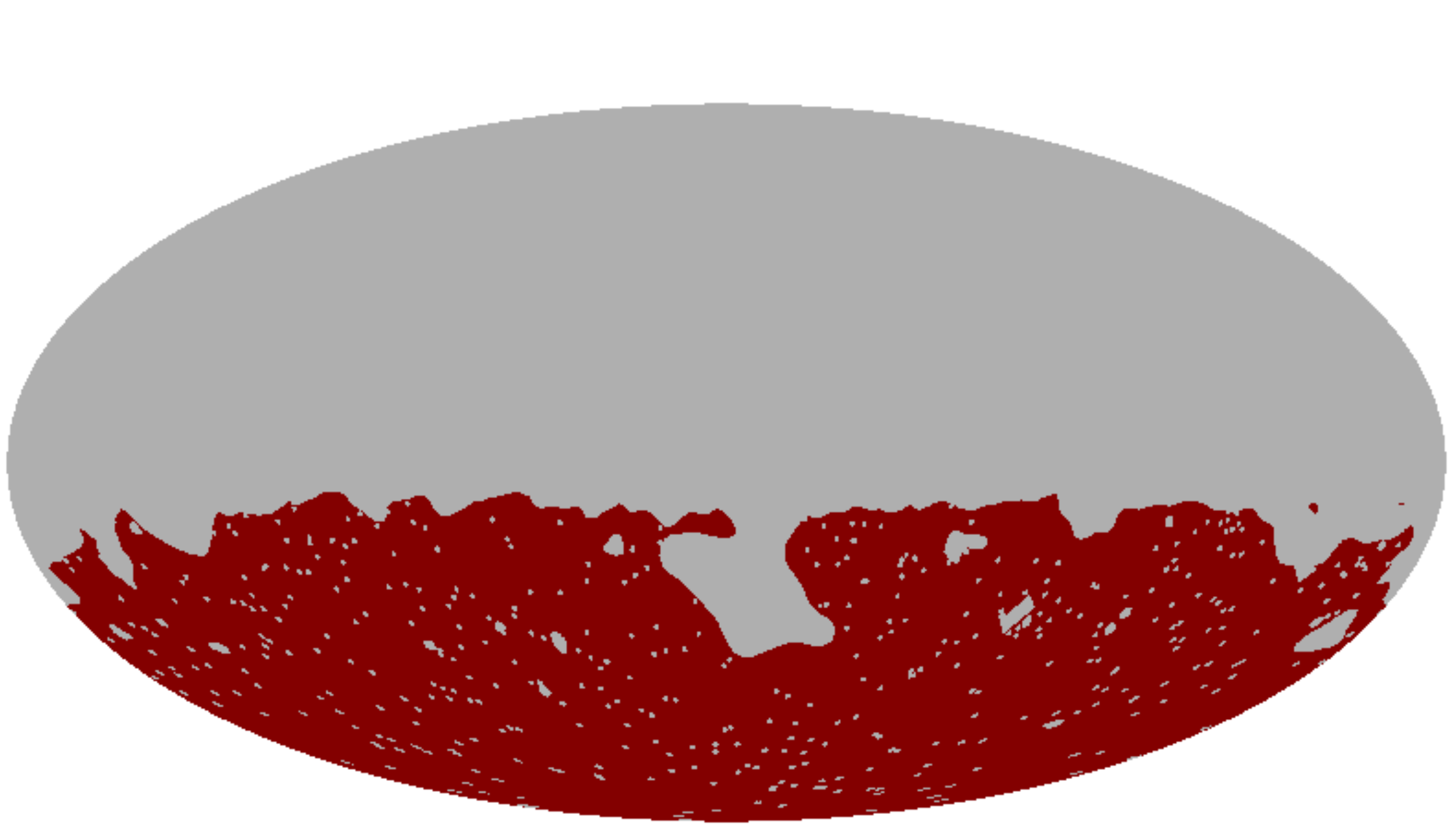}
	\caption{The left panel shows the inhomogeneous noise distribution for an experiment covering $1\%$ of the sky. The binary masks corresponding to an half sky and a full sky survey are displayed in the middle and right panel respectively.}
	\label{fig:mask}
\end{center}
\end{figure}

The previously studied fiducial experiments covering a spherical cap were idealised regarding the mask shape. A more realistic observed patch of the sky is considered in this investigation. The small scale experiment is the one chosen in Chapter \ref{Chapter5}. The observed sky fraction is $1\%$ with a beam with a width of $8$ arcmin. However the noise is chosen here to be inhomogeneous with an average noise level of $\sim 5.75 \mu K$-arcmin. Its distribution is taken from a simulation of the {\sc EBEX} scanning strategy. The noise distribution is displayed in the left panel of Fig.~\ref{fig:mask}. 

The full sky experiment is the fiducial one used in Chapter \ref{Chapter5} covering $71 \%$ of the celestial sphere with a beam of $8$ arcmin and an homogeneous noise of $2.2 \mu K$-arcmin. This experiment is typical for proposed satellite mission dedicated to primordial $B$ modes detection. The corresponding binary mask is shown in the middle panel of Fig.~\ref{fig:mask}. 

The performance of an intermediate scale experiment is also studied in this analysis. This would be the case of an array of ground-based telescopes covering a whole hemisphere. The noise level is of $6.8 \mu K$-arcmin and the beam width of 3 arcmin (it corresponds to a Stage-III experiment). For such a large survey, the galactic emission has to be masked reducing the observed sky fraction to $\sim 36 \%$. The used binary mask is simply the southern hemisphere of the previous full sky binary mask and is depicted in the right panel of Fig.~\ref{fig:mask}

\underline{Window functions} 

\begin{figure}[!h]
\begin{center}
	\includegraphics[scale=0.25]{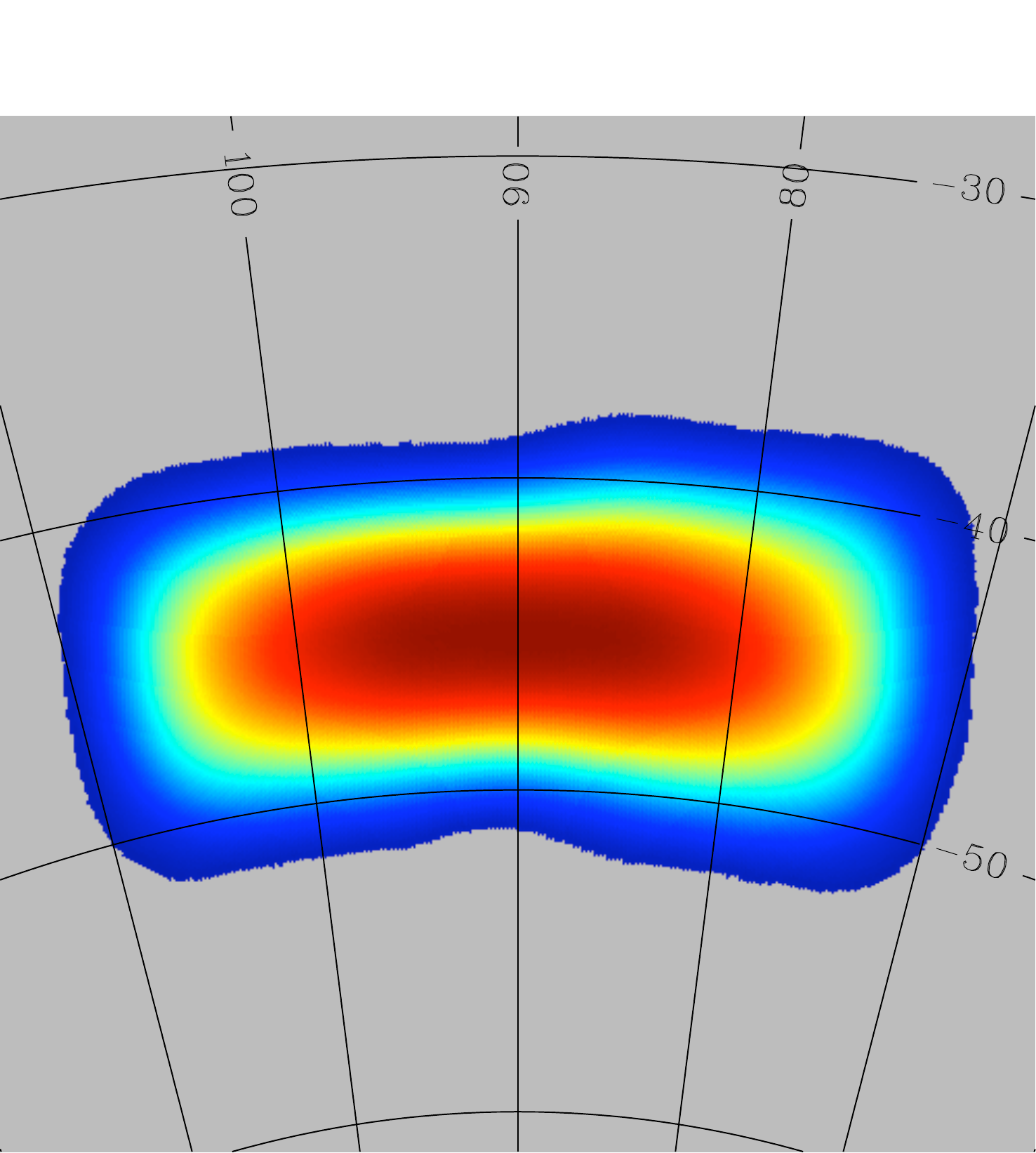}\includegraphics[scale=0.25]{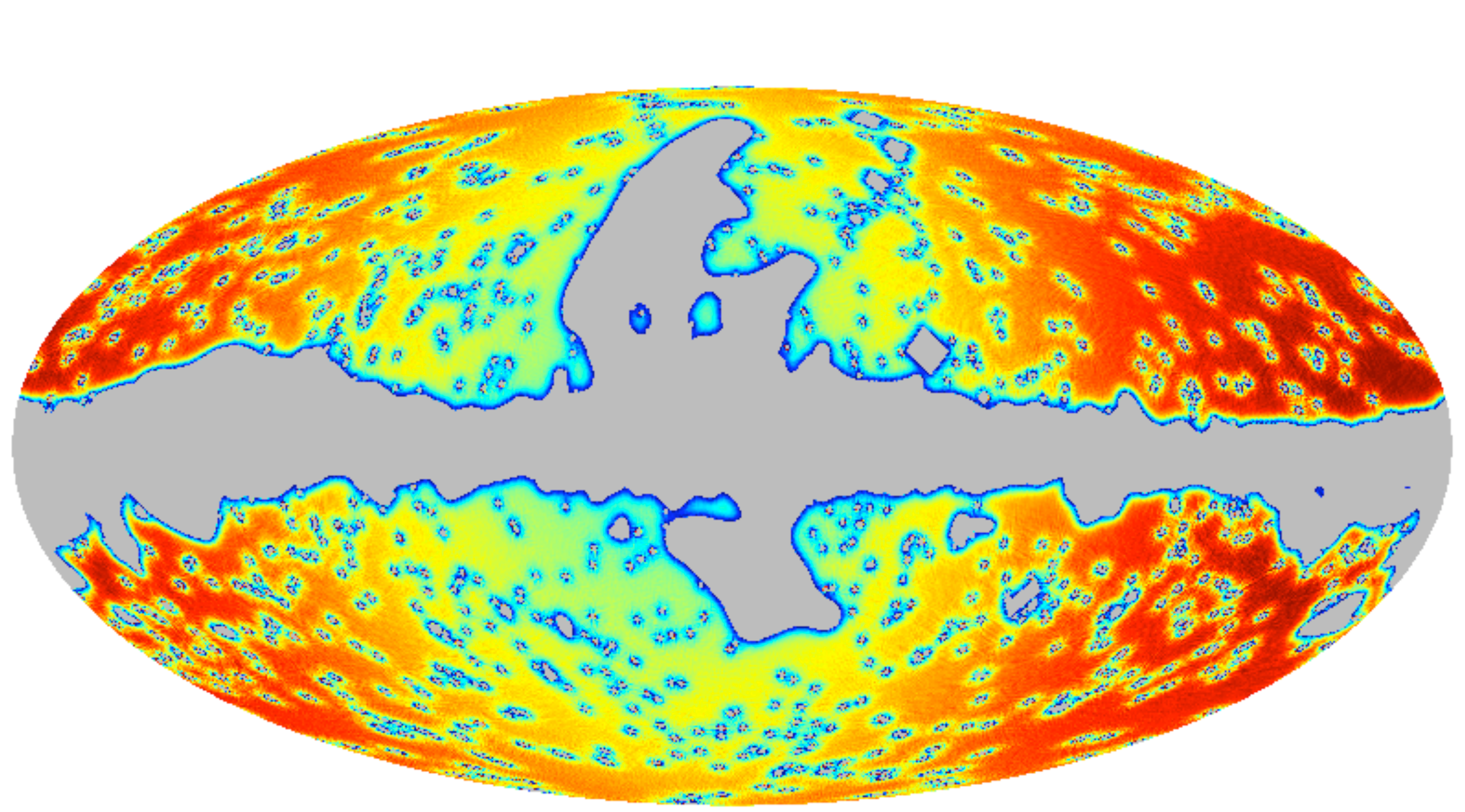}\includegraphics[scale=0.25]{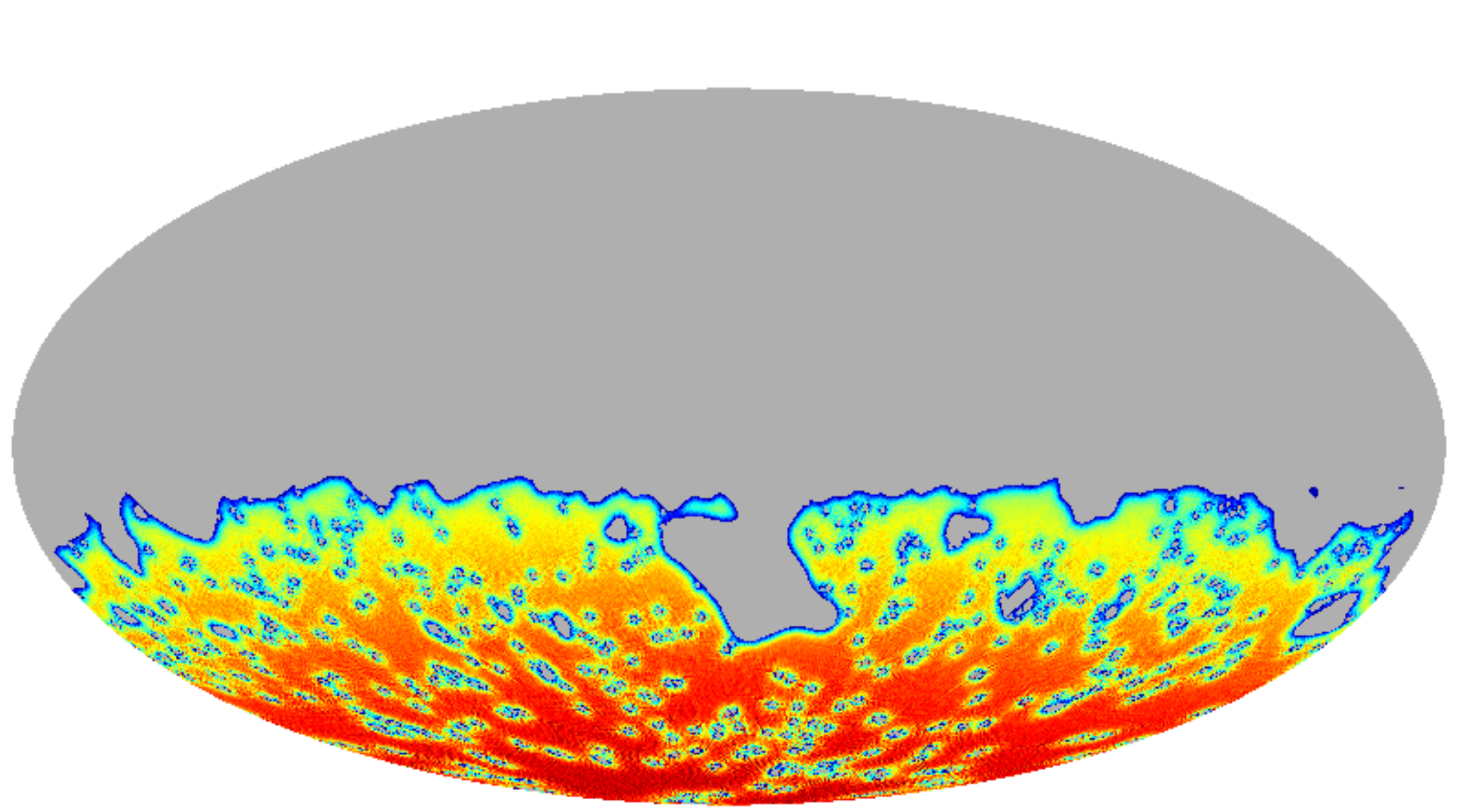}
	\caption{The spin-0 PCG window function optimised in the bin $\ell \in [60;100]$ for small, intermediate and large scale experiments in the left, middle and right panel respectively.}
	\label{fig:opt}
\end{center}
\end{figure}

In the scope of pure estimation of the $B$ modes, the window function applied to the CMB map have to be optimised in order to have the best $B$ modes reconstruction. The need for a pixel-based variance (PCG) optimised window function for small scale experiments with inhomogeneous noise have been already shown in \cite{Grain_2009}. As stated in the previous Chapter \ref{Chapter5}, PCG optimised window functions are also required in order to get the lowest uncertainties on $B$ modes in the case of large scale surveys, due to intricate contours. PCG optimised window functions have thus been used in the present analysis for the three considered set-ups. The obtained spin-0 PCG window function optimised in the third bin $\ell \in [60;100]$ is displayed in Fig.~\ref{fig:opt}, the binning being the same than previously. 

The computation of the PCG optimised window function is numerically heavy and scales as the inverse of the noise level. The required number of iterations to compute the optimised window functions of a full sky survey is around $n_{iter} \sim 100$ while $n_{iter} \sim 10$ are necessary for the small scale survey. In addition, the window function should be optimised per bin ($N_{bin} = 26$) and per $r$. However, we took advantage of the fact that the PCG optimised window function only slightly depend on the $B$ modes signal prior as shown in \cite{Grain_2009}. Therefore, the PCG window functions optimised for $r = 0.05$ are used for the different $r$ values making them slightly suboptimal for $r \neq 0.05$.

\subsection{Numerical results}

The signal-to-noise ratio (S/N)$_r$ on $r$ is obtained using the same reasoning as Sec.~\ref{sec:opt}. The pure estimation of the $B$ modes power spectrum ensure realistic forecasts on $r$ detection. 

\begin{figure}[!h]
\begin{center}
	\includegraphics[scale=0.35]{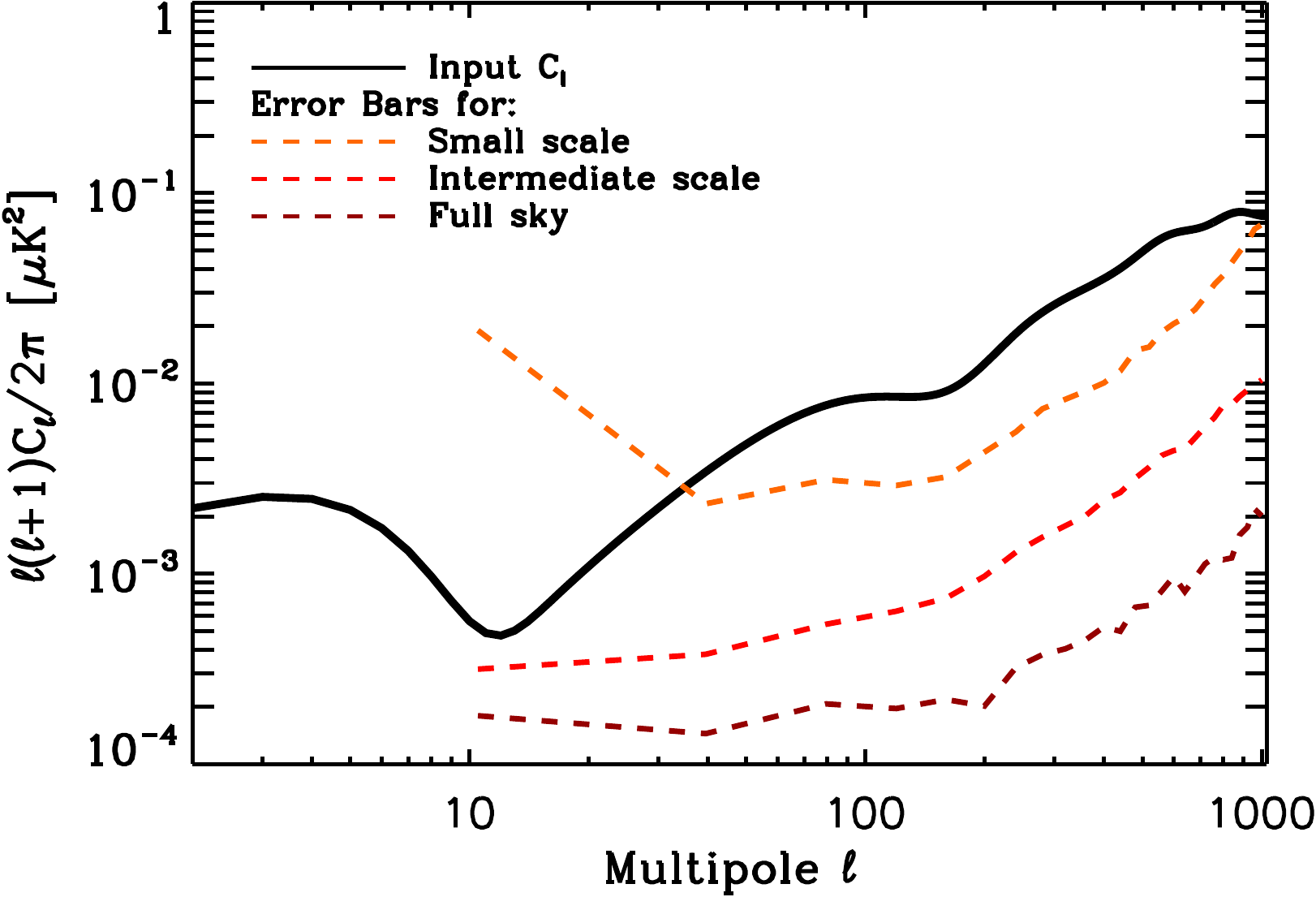}
	\caption{Variances reconstructed with the pure estimation on a theoretical CMB power spectrum with $r = 0.1$ (in solid black line) in the case of a small scale survey (orange dashed curve), an intermediate scale survey (red dashed curve) and a full sky survey (burgundy dashed curve). }
	\label{fig:var_allscale}
\end{center}
\end{figure}

By way of illustration, the variances obtained on $B$ modes power (with $r = 0.1$) estimated with the pure estimation are shown in dashed lines in Fig.~\ref{fig:var_allscale}. The highest ones correspond to the one obtained with a small scale survey while the lowest are obtained for a large scale survey. The middle one results from the estimation on an intermediate sky survey. For each experimental set-up, the forecast on the signal-to-noise ratio (S/N)$_r$ on $r$ is investigated for 6 values of $r$: $r = 0.001, 0.01, 0.05, 0.1, 0.15$ and $0.2$.

\begin{figure}[!h]
\begin{center}
	\includegraphics[scale=0.37]{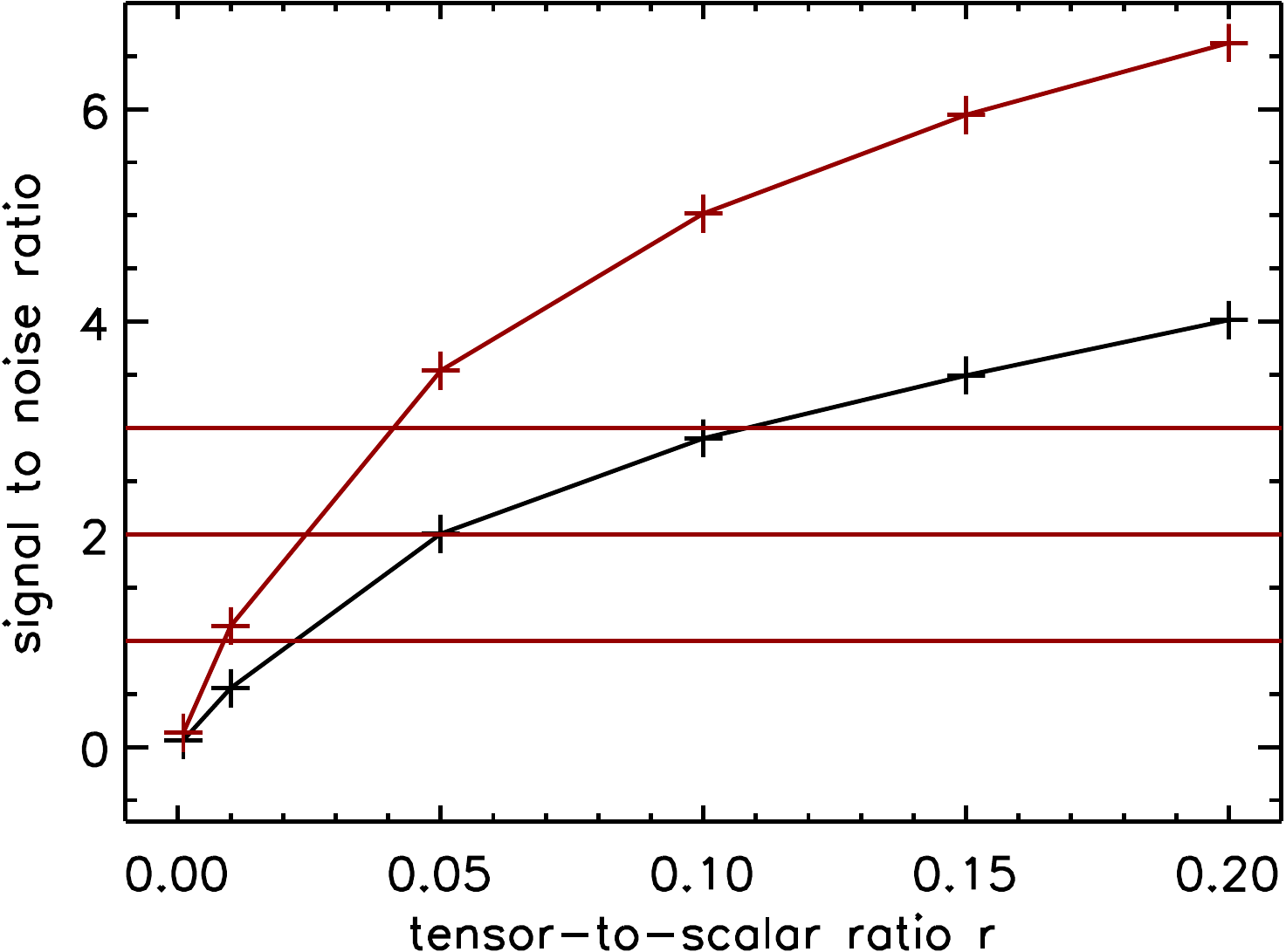}
	\caption{Signal-to-noise ratio on $r$ with respect to $r$ obtained for the fiducial small scale survey. The red (black) crosses connected by red (black) line are the obtained (S/N)$_r$ using the mode-counting (pure) estimation of the covariance matrix in the Fisher matrix. The horizontal red lines stand for $1\sigma$, $2\sigma$ and $3\sigma$ thresholds.}
	\label{fig:ebex}
\end{center}
\end{figure}

The forecast for the fiducial ground-based experiment is shown in Fig.\ref{fig:ebex} where the black (red) crosses stand for the (S/N)$_r$ for a pure estimation (mode-counting) variance. The horizontal red lines delineate a signal-to-noise ratio of $1 \sigma$, $2 \sigma$ and $3 \sigma$, the latter being the threshold for an unambiguous detection. The Fisher matrix computed with mode-counting variance give high (S/N)$_r$ so that a $r$ detection at $3\sigma$ is expected for $r \gtrsim 0.04$. Nonetheless, the realistic pure estimation of the variance ends in lower values for (S/N)$_r$. The discrepancy between both estimations of the variances increases for lower $r$. As an example, the signal-to-noise ratio on $r = 0.05$ obtained using the naive estimation is a factor of 1.75 higher than the pure estimation. In this way, a realistic forecast results in different conclusion: only the $r$ values such as $r \gtrsim 0.1$ are achievable at a $3\sigma$ level. An underestimation of the $B$ modes uncertainties could thus deteriorate a forecast, especially for $r = 0.05$ for which a clear detection is forecast in the former case and not for the other case (lowered from $3\sigma$ to $2\sigma$).

The results for the large scale survey are similar to the previous ones although the discrepancy between the mode-counting and pure predictions is less marked. For $r = 0.05$, the signal-to-noise ratio in the mode-counting approach is 1.5 greater than the one using the pure estimation. Moreover, according to the latter estimation, a $3\sigma$ detection of $r = 0.001$ is impossible while it seems achievable at $3\sigma$ using a mode-counting variance estimation. For $r = 0.1$, (S/N)$_r$ is realistically predicted to be of $45\sigma$.

\begin{figure}[!h]
\begin{center}
	\includegraphics[scale=0.37]{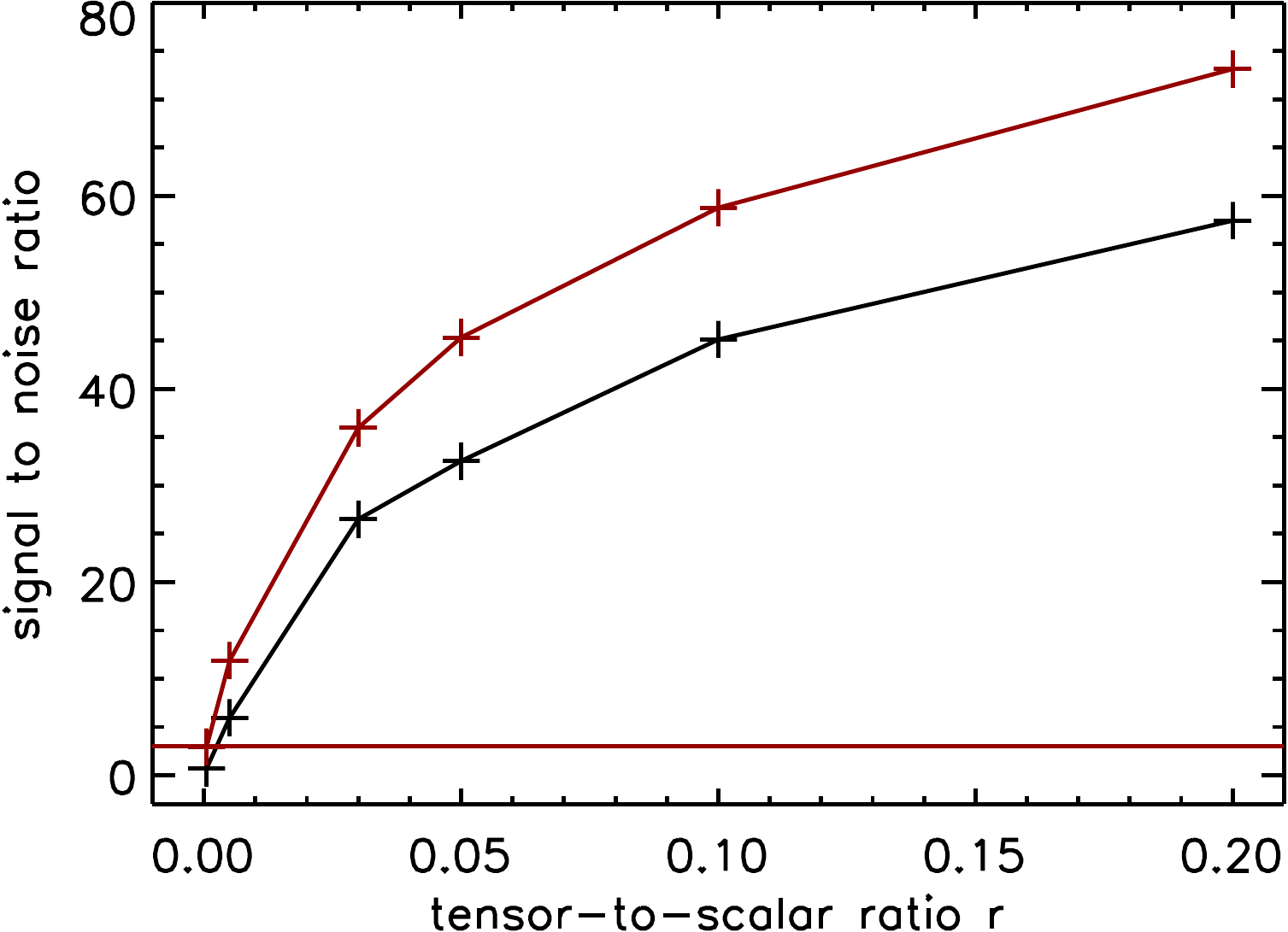}
	\caption{Signal-to-noise ratio on $r$ with respect to $r$ obtained for the fiducial large scale survey. The red (black) crosses connected by red (black) line are the obtained (S/N)$_r$ using the mode-counting (pure) estimation of the covariance matrix in the Fisher matrix. The horizontal red lines stand for $3\sigma$.}
	\label{fig:fullsky}
\end{center}
\end{figure}

\begin{figure}[!h]
\begin{center}
	\includegraphics[scale=0.37]{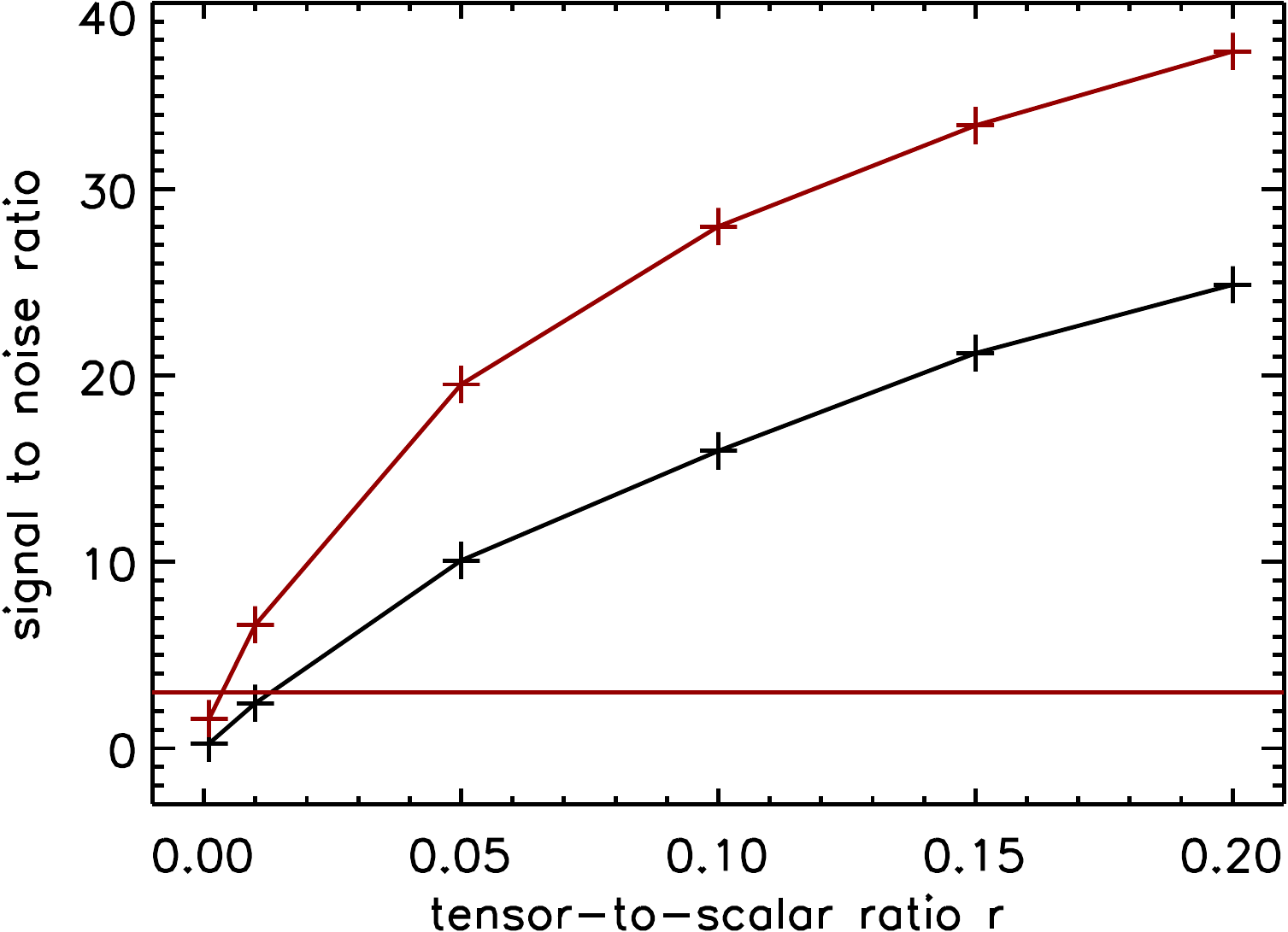}
	\caption{Signal-to-noise ratio on $r$ with respect to $r$ obtained for the fiducial intermediate scale survey. The red (black) crosses connected by red (black) line are the obtained (S/N)$_r$ using the mode-counting (pure) estimation of the covariance matrix in the Fisher matrix. The horizontal red line stand for the $3\sigma$ threshold.}
	\label{fig:halfsky}
\end{center}
\end{figure}

The forecasts on $r$ detection for an intermediate scale experiment are displayed in Fig.~\ref{fig:halfsky}. The main results are similar to the last ones. Owing to the use of a non optimal window function, the disparity between the mode-counting variances and the pure estimation is more marked than previously. The mode-counting estimation of the variance gives (S/N)$_r \sim 20\sigma$ for $r = 0.05$ when the pure estimation forecasts a signal-to-noise ratio on $r$ of $10 \sigma$. In this case also, the discrepancy is higher for lower $r$. This issue is specially damaging for $r = 0.01$: an optimistic forecast gives (S/N)$_r \sim 6\sigma$ ensuring a detection. On the contrary, for a realistic estimation, the obtained (S/N)$_r$ does not exceed $3\sigma$ preventing setting any tight constraints on $r$. In the end, the pure estimation of the variance predicts an unambiguous detection for $r > 0.01$. 

The array~\ref{tab:snr} summarises the forecast performances of the three experimental set-ups. The range of $r$ for which a net detection is ensured is shown in the case of the optimistic and realistic estimation of the variance on $B$ modes power spectrum. The mode-counting estimation is proved to be unreliable to forecast $r$ detectability as the realistic estimation forecasts access to a smallest range of $r$. 

\begin{table}[h!]
\begin{center}
\begin{tabular}{|c||c|c|c|}
	\hline
		&  Suborbital Experiment & Telescopes Array	& Satellite-like Experiment	\\
	\hline\hline
	mode-counting & $r \gtrsim 0.04$ & $r \gtrsim 0.0035$ & $r \gtrsim 0.0005$ \\
	\hline
	Pure estimation  & $r \gtrsim 0.11$	& $r \gtrsim 0.013$ & $r \gtrsim 0.0024$ \\
	\hline
\end{tabular}
	\caption{The minimal accessible $r$ at $3\sigma$ regarding the experimental set-up and the estimation of the variance on the $B$ modes reconstruction by linearly interpolating between the computed (S/N)$_r$.} 
	\label{tab:snr}
\end{center}
\end{table}

The different experimental set-ups give access to different order of magnitude of $r$. The small scale survey is predicted to detect at least $r \sim 10^{-1}$ at $3\sigma$ while a full sky survey would have access to at least $r \sim 10^{-3}$ at $3\sigma$. An intermediate scale experiment could detect $r \sim 10^{-2}$ at $3\sigma$. It would allow to discriminate between large and small field inflationary models. The Lyth bound (\cite{Lyth_1997}) indeed tells that a large field models are required to produce $r \gtrsim 0.01$. Therefore, only satellite mission (stage IV) could ensure a detection of $B$ modes in small field while telescope array (stage III) is sufficient to discriminate between large and small field.

\section*{Conclusion}

In the current context of experimental strategy for $B$ modes detection, the performance forecasts on $r$ detection are crucial. The Fisher matrix allows for a simple translation of the uncertainties on the $B$ modes angular power spectrum on $r$ error bars. However, the involved covariance matrix can be estimated in different ways. The mode-counting variance estimation has the benefit to allow for an utter and fast exploration of the performances. Nonetheless, it underestimates the $B$ modes uncertainties leading to overestimated signal-to-noise ratios on $r$. The pure pseudospectrum $B$ modes estimation, although numerically heavier, remains fast enough to explore realistic forecasts on $r$ detection. In the scope of small scale surveys, the optimal sky coverage has first been investigated in the two approaches. The optimal observed sky fraction is similar regarding the chosen methods and scales with $r$. It is however noticeable that the optimal sky coverage using the pure method is not as pronounced as in the case of mode-counting variance estimation partly due to the binning process. The performance on $r$ detection of three fiducial experiments have then been explored. In this case, a realistic $B$ modes estimation has to be done in order to accurately predict the achievable $r$. With a proper estimation of (S/N)$_r$, the small scale experiment is expected to detect at least $r \sim 10^{-1}$ at $3\sigma$. The intermediate sky survey gives promising results and give access to a $3\sigma$ detection of $r \gtrsim 10^{-2}$. A full sky survey would ensure a clear detection of $r$ in the order of $10^{-3}$. We have thus investigated the information on the inflation that one would extract from data provided by current and forthcoming experiments dedicated to $B$ modes detection. Nonetheless, these results should be compared to maximum likelihood methods. These methods are nevertheless more numerically costly than the approaches used in this analysis. As a starting point, the assumption of azimuthal symmetry of the patch and the noise can be done, simplifying the computation as shown in appendix of \cite{Smith_2006}. In this ideal case, the results are expected to be similar to the ones using the mode-counting variances.

More generally, the CMB polarisation is a mine of information on the primordial physics. For instance, a parity violation at the linear level of gravitation or due to a primordial magnetic field could be constrained in particular through the CMB $TB$ and $EB$ correlations. Both possibilities are investigating in the two following chapters.


\chapter{Primordial Physics through the CMB Polarisation: Chiral Gravity} 

\label{Chapter7} 

\lhead{Chapter 7. \textit{Chiral Gravity}} 
\noindent \hrulefill \\
\textit{In the standard model, the four fundamental interactions are the strong, weak interactions, the electromagnetism and the gravitation. The first three are well described within the quantum mechanism framework by the quantum chromodynamics and the electroweak theory respectively. The quantum formulation of gravitation is however still under scrutiny and suffers from difficulties such as renormalisation issues. Various approaches have been proposed and are now intensively studied, the main ones being the string theory and the loop quantum gravity (LQG). The main difference between both theories is that contrary to string theory, the LQG does not aim at unifying the four fundamental forces as it treats the quantification of gravity independently of the other forces. The seminal paper \cite{Ashtekar_1987} reformulating general relativity sets the founding principles of LQG which now offers physical applications such as the computation of black hole entropy (\cite{Rovelli_1996}). In parallel, the principles of LQG have been applied to cosmology resulting in the Loop Quantum Cosmology (LQC) (a review of which can be found in \cite{Ashtekar_2011}). In this framework, the Universe would have known a Big Bounce followed by a natural inflationary period.} 
\noindent \hrulefill \\

According to some formulation of gravity (\cite{Kibble_1961}, \cite{Ashtekar_1986}), a primordial parity violation could occur at the linear level of general relativity: 

primordial gravitational waves are then chiral which is manifested by left- and right-handed helicities with different power spectrum. Such a parity breaking would lead to non-vanishing CMB $TB$ and $EB$ correlations (refered to odd correlations hereafter). The left-(right-)handed primordial gravitational waves are scaled by $r_{L(R)}$, the tensor-to-scalar ratio for each helicity state. The level of parity breaking $\delta$ is quantified by:
\begin{equation}
\delta = \frac{r_R - r_L}{r_R + r_L} = \frac{r_{(-)}}{r_{(+)}} \leqslant 1.
\end{equation} 
The goal of the present analysis is to realistically forecast the constrains that can be put on $\delta$ from forthcoming measurements of CMB polarised correlations. 

\begin{figure}[!h]
\begin{center}
	\includegraphics[scale=0.45]{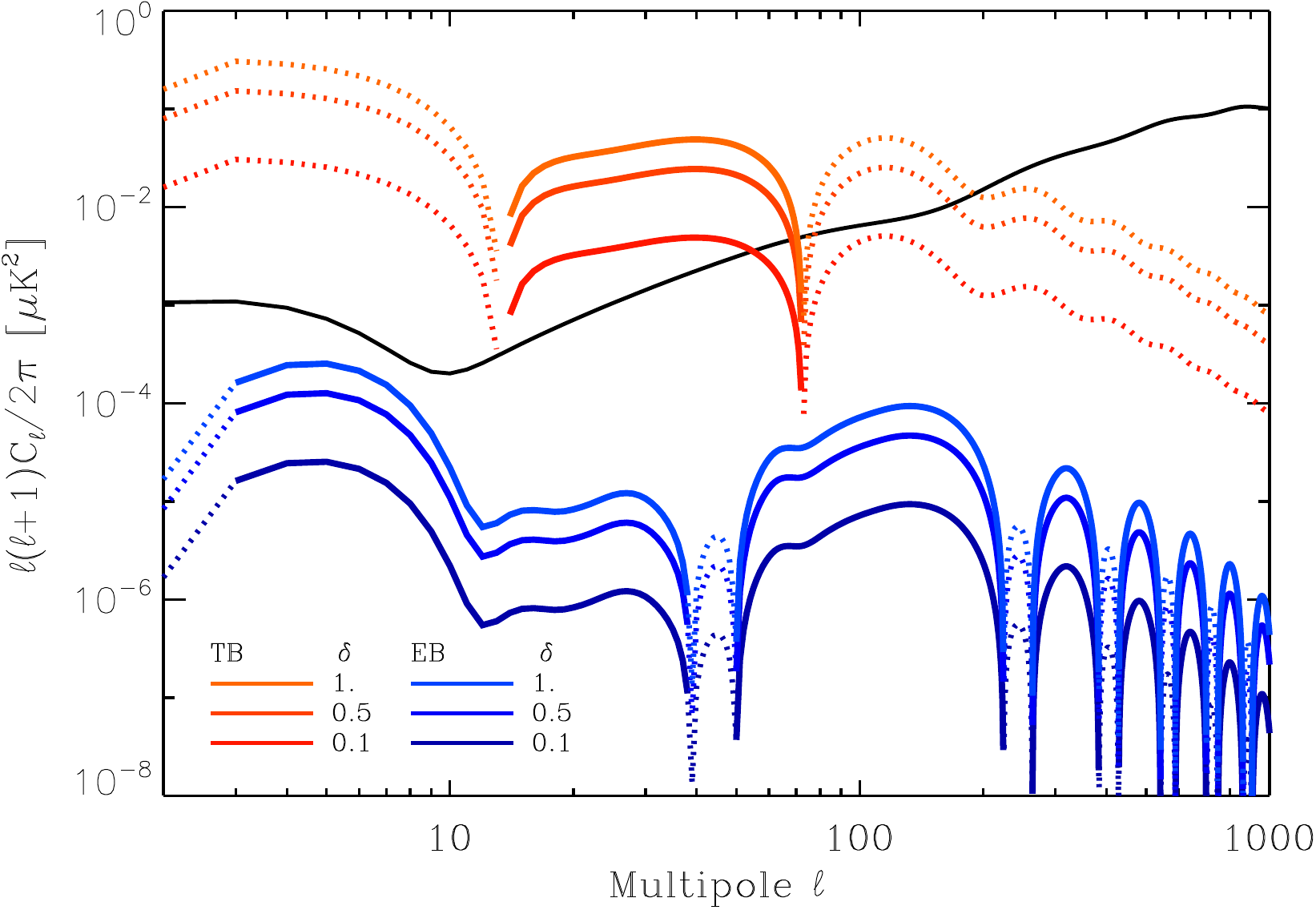}
	\caption{The CMB $B$ modes (in black) power spectrum along with the $TB$ (in red shades) and $EB$ (in blue shades) correlations for $r_{(+)} = 0.05$, including the primordial and lensing contributions. The three red and blue shaded curve correspond to different values of $\delta$: $\delta = 1$ for a maximum parity breaking, $\delta = 0.5$ and $\delta = 0.1$. The dotted lines stand for negative values of the odd correlations.} 
	\label{fig:cltbeb}
\end{center}
\end{figure}

The CMB odd-correlations are sourced by the primordial tensor power spectra modulated by the appropriate transfer function:
\begin{eqnarray}
	C^{BB}_\ell&=&\displaystyle\int dk \left[\Delta^{B}_{\ell,\te}(k,\eta_0)\right]^2\pk^{(+)}_\te(k), \\
	C^{TB}_\ell&=&\displaystyle\int dk \Delta^{T}_{\ell,\te}(k,\eta_0)\Delta^{B}_{\ell,\te}(k,\eta_0)\pk^{(-)}_\te(k), \\
	C^{EB}_\ell&=&\displaystyle\int dk \Delta^{E}_{\ell,\te}(k,\eta_0)\Delta^{B}_{\ell,\te}(k,\eta_0)\pk^{(-)}_\te(k),
\end{eqnarray}
with 
\begin{equation}
	\pk^{(\pm)}_\te(k)=\pk^R_\te(k)\pm\pk^L_\te(k).
\end{equation}

We have implemented the computation of the non-vanishing $TB$ and $EB$ correlations for a given $\delta$ in the Boltzmann code CLASS. Also, we have computed and implemented the impact of the lensing on the odd-correlations due to gravitational potentials of large scale structures. In this way, we have checked that its impact is small as it was expected, although only presumed in the literature (as in \cite{Saito_2007} or \cite{Xia_2012}). The figure~\ref{fig:cltbeb} depicts the total $BB$, $TB$ and $EB$ correlations for three values of $\delta$ ranging from $1\%$ to $100\%$, with $r_{(+)} = 0.05$.

\begin{table}[!h]
\begin{center}
\begin{tabular}{cc||c|c|c|c|c|c}
	& $r_{(+)}$ & $0.2$ & $0.1$ & $0.07$ & $0.05$ & $ 0.03$ & $0.007$ \\
	$r_{(-)}$ & & & & & & & \\
	\hline\hline
	$0.2$    & & 1.22   &         &             &               &               &                \\
	\hline
	$0.1$    & & 0.43   & 0.64 &             &               &               &                \\
	\hline
	$0.07$  & &  0.29  & 0.4  & 0.487    &               &               &              \\
	\hline
	$0.05$  & &  0.2   &0.28 & 0.326    & 0.38     & 		    &  	     \\
	\hline
	$0.03$  & &  0.12  &0.16  & 0.188  & 0.216 & 0.27 &               \\
	\hline
	$0.007$ & & 0.03  &0.037& 0.043 & 0.049 & 0.06  &  0.1   \\
	\hline
\end{tabular}
	\caption{Signal-to-noise on $r_{(-)}$ for different values of $r_{(+)}$ in the case of small-scale (ballon-borne or ground-based) experiments, and using a mode-counting expression for the error bars on the angular power spectra reconstruction.} 
	\label{tab:snrrmsmall}
\end{center}
\end{table}

The forecasts are performed in the frame of two fiducial experiments, typical of a small scale survey and a satellite-like experiment, which are both fully described in Chapter~\ref{Chapter5}. In order to rapidly explore the expected constraints on $\delta$, the power spectra uncertainties are firstly estimated using the naive mode-counting approach. The conclusion is irrevocable for balloon-borne or ground-based experiments: the obtained signal-to-noise ratios are smaller than $2$ even by underestimating the uncertainties. The signal-to-noise on $r_{(-)}$ are summarised in table~\ref{tab:snrrmsmall} where $r_{(-)}$ and $r_{(+)}$ are ranging from 0.07 to 0.2. Therefore, a satellite-like experiment is required for the detection of chiral gravity. 

The case of the nearly-full sky survey is more intricate. As a preliminary work, the signal-to-noise ratio on $r_{(-)}$ goes up to $10\sigma$ for the most optimistic configuration with a naive mode-counting estimation of the power spectra uncertainties. The pure power spectrum estimation of the CMB polarised power spectra extended to $EB$ and $TB$ correlations (\cite{Grain_2012}) is then used to realistically estimate the uncertainties on their detection. The table~\ref{tab:snr} shows the resulting realistic signal-to-noise ratio on $r_{(-)}$ $\mathrm{(S/N)}_{r_{(-)}}$ for $r_{(+)} = 0.2, 0.1$ and $0.05$ with $\delta = 100\%$ and $50\%$ (reminding that $r_{(-)} = \delta \times r_{(+)}$). A range of model is therefore accessible for such a nearly-full sky experiment for high values of $r_{(+)}$ and $\delta$. Moreover, $\mathrm{(S/N)}_{r_{(-)}}$ decreases from $10\sigma$ for a naive estimation of the variance to about $5.5\sigma$. A careful estimation of the CMB power spectra uncertainties is consequently crucial to perform realistic forecasts, in the scope of chiral gravity detection. 

Nonetheless, the odd-correlations are usually set equal to zero to calibrate polarisation detectors as these correlations do vanish in the standard model of cosmology. In particular, it enables the estimation of a possible miscalibration angle $\Delta\Psi$ of the global orientation of the detectors. We have therefore studied the impact of a joint reconstruction of $\delta$ and $\Delta\Psi$ on $\mathrm{(S/N)}_{r_{(-)}}$. The $r_{(-)}$ estimation is consequently biased with a level growing with $\Delta\Psi$. Furthermore, we have shown that $\mathrm{(S/N)}_{r_{(-)}}$ is not degraded with respect to $\Delta\Psi$ if the variances are estimated via the naive mode-counting approach: $\Delta\Psi$ and $r_{+/-}$ are not degenerate. However, the pure estimation of the odd-correlations breaks this non degeneracy and the signal-to-noise ratio on $r_{(-)}$ is reduced by a factor of $\sim 2.4$ for $r_{(-)} = r_{(+)} = 0.2$ and $\Delta\Psi = 1^o$. A proper estimation of the miscalibration angle is subsequently crucial for constraints on chiral gravity.  

\begin{table}[!h]
\begin{center}
\begin{tabular}{cc||ccc|ccc}
	& & & & & & & \\
	& & & $\delta=1$ & & & $\delta=0.5$ & \\ \hline\hline
	& $r_{(+)}=0.2$ & & 5.46 & & & 2.5 & \\ \hline
	& $r_{(+)}=0.1$ & & 3.67 & & & 1.51 & \\ \hline
	& $r_{(+)}=0.05$ & & 2.35 & & & 1.11 & \\
	\hline
\end{tabular}
	\caption{Signal-to-noise ratio on $r_{(-)}$, (S/N)$_{r_{(-)}}$, as derived from a pure estimation of the angular power spectra. (For a given value of $r_{(+)}$ and $\delta$, the value of $r_{(-)}$ is $r_{(-)}=\delta\times r_{(+)}$.)}
	\label{tab:snr}
\end{center}
\end{table}

If $TB$ and $EB$ correlations detections are consistent with zero, upper bound can be put on $\delta$ owing to the noise and sampling variance. It can be translated in exclusion range for theoretical parameters. In particular, a non detection of odd correlations for a satellite-like survey would translate in values of the Barbero-Immirzi parameter $\gamma$ (\cite{Magueijo_2011}, \cite{Bethke_2012}) such as the range: $0.2 \leqslant |\gamma| \leqslant 4.9$ is excluded at $3\sigma$ for $r_{(+)} = 0.2$. The complete analysis and the consequences on the theoretically relevant parameters are detailed in the following article \cite{Ferte_2014}. 

The CMB odd correlations can thus be exploited to explore the primordial universe physics. The presence of a primordial magnetic field can also be constrained through the $B$ modes and CMB $EB$ and $TB$ correlations as shown in the next Chapter~\ref{Chapter8}.

\includepdf[pages=1,scale=1.,pagecommand={},offset=85 -100]{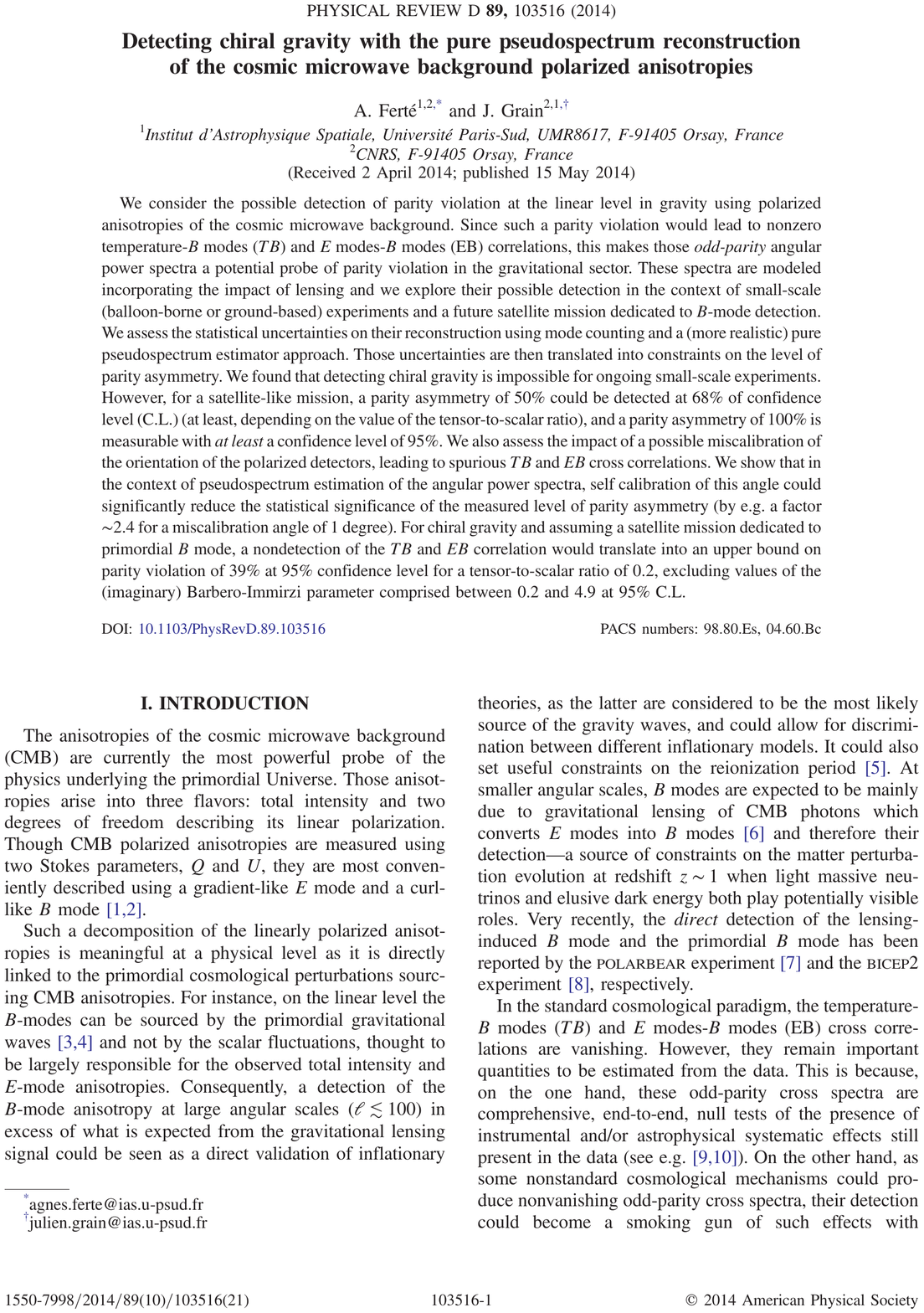}
\includepdf[pages=2,scale=1.,pagecommand={},offset=-60 -100]{ChiralGrav.pdf}
\includepdf[pages=3,scale=1.,pagecommand={},offset=85 -100]{ChiralGrav.pdf}
\includepdf[pages=4,scale=1.,pagecommand={},offset=-60 -100]{ChiralGrav.pdf}
\includepdf[pages=5,scale=1.,pagecommand={},offset=85 -100]{ChiralGrav.pdf}
\includepdf[pages=6,scale=1.,pagecommand={},offset=-60 -100]{ChiralGrav.pdf}
\includepdf[pages=7,scale=1.,pagecommand={},offset=85 -100]{ChiralGrav.pdf}
\includepdf[pages=8,scale=1.,pagecommand={},offset=-60 -100]{ChiralGrav.pdf}
\includepdf[pages=9,scale=1.,pagecommand={},offset=85 -100]{ChiralGrav.pdf}
\includepdf[pages=10,scale=1.,pagecommand={},offset=-60 -100]{ChiralGrav.pdf}
\includepdf[pages=11,scale=1.,pagecommand={},offset=85 -100]{ChiralGrav.pdf}
\includepdf[pages=12,scale=1.,pagecommand={},offset=-60 -100]{ChiralGrav.pdf}
\includepdf[pages=13,scale=1.,pagecommand={},offset=85 -100]{ChiralGrav.pdf}
\includepdf[pages=14,scale=1.,pagecommand={},offset=-60 -100]{ChiralGrav.pdf}
\includepdf[pages=15,scale=1.,pagecommand={},offset=85 -100]{ChiralGrav.pdf}
\includepdf[pages=16,scale=1.,pagecommand={},offset=-60 -100]{ChiralGrav.pdf}
\includepdf[pages=17,scale=1.,pagecommand={},offset=85 -100]{ChiralGrav.pdf}
\includepdf[pages=18,scale=1.,pagecommand={},offset=-60 -100]{ChiralGrav.pdf}
\includepdf[pages=19,scale=1.,pagecommand={},offset=85 -100]{ChiralGrav.pdf}
\includepdf[pages=20,scale=1.,pagecommand={},offset=-60 -100]{ChiralGrav.pdf}
\includepdf[pages=21,scale=1.,pagecommand={},offset=85 -100]{ChiralGrav.pdf}


\chapter{Primordial Physics through the CMB Polarisation: Primordial Magnetic Field} 

\label{Chapter8} 

\lhead{Chapter 8. \emph{Primordial Magnetic Field}} 
\noindent \hrulefill \\
\textit{Lodestone is a mineral which is naturally magnetised. Its ability to attract iron has been noticed since thousands of years and was used to built compasses for instance. The origin of this phenomenon was however a mystery that, during Antiquity, Thales of Miletus or Lucretius had attempted to elucidate. In the 13th Century, Pierre de Maricourt gave a first description of the magnet properties in \cite{Maricourt_1269}, introducing the magnetic poles. 
The magnetic field was explicitly introduced much later by Faraday in the 19th century, setting the foundation for the Maxwell's theory of electromagnetism in \cite{Maxwell_1865}. Since then magnetic fields are observed at different scales and strengths. The Earth indeed possesses a magnetic field of $\sim 50 \mu T$ while the highest ones are produced by a type of neutron stars with an intensity of $10^{11} T$ (\cite{Spruit_2008}).
While the origin of such fields is more and more understood, the origin of the ones present on the largest scales of the Universe remains a mystery that the study of the primordial Universe might help solving.} 
\noindent \hrulefill \\

The origin of the cosmic magnetic field is unsettled. One possibility is that it has been produced in the early
Universe. Though potentially requiring exotic physics, such a primordial generation
would explain the coherence of the magnetic field on cosmological scales. However, an inverse-cascade phenomenon transferring power from small scales to large scales is required to reach the observed magnetic field intensity at large scales. This is possible if the magnetic field has an helical component, which is predicted according to some mechanisms that could occur during the primordial universe. The produced magnetic field would therefore be a source of temperature and polarised CMB anisotropies and in particular would give $TB$ and $EB$ correlations. The CMB polarisation is thus a probe of choice to set constraints on a potential helical primordial magnetic field. We propose a preliminary study, which is still in progress, to estimate the constraints set by current experiments.


\section{Primordial Magnetic Field}

Magnetic fields are observed at large scales in the Universe with an amplitude of the microGauss in galaxies clusters (\cite{Clarke_2001},\cite{Bonafede_2010}) and of at least the femtoGauss in intergalactic medium (\cite{Tavecchio_2010},\cite{Neronov_2010}). Their origin remains an open issue although various scenarios for their creation and evolution have been proposed taking place either in the late time or in the primordial Universe. In the latter case, the production of the magnetic field precedes the structure formation, occurring during the inflationary period (\cite{Turner_1988},\cite{Ratra_1992}) or during a phase transition (\cite{Vachaspati_1991}). Such fields generated during the early universe are called \textit{primordial magnetic field} (PMF). 

Usually, a magnetic field $\vec{B}$ is said to be \textit{helical} if the magnetic helicity does not vanish: 
\begin{equation} 
\vec{B} . \left( \vec{\nabla} \times \vec{B} \right)  \ne 0.
\end{equation}
 
In particular, it has been shown that a PMF can be generated with helicity (\cite{Cornwall_1997},\cite{Vachaspati_2001} or very recently in \cite{Caprini_2014}). A non vanishing helicity ensures the amplification of the magnetic field via inverse cascade: the small scales power the largest scales. The helical PMF therefore constitutes a potential origin for the cosmic magnetic field. 

In the present study, the generation of a PMF taking place before the radiation-matter equality by a given process is assumed. We consider the produced PMF to be a Gaussian random field with an helical component. Thus all its statistical information is contained in its two point correlation function which, following \cite{Caprini_2004}, can be written as: 
\begin{equation}
\left<B_i(\vec{k})B_j(\vec{k'})\right> = \frac{(2\pi)^3}{2} \delta(\vec{k}-\vec{k'}) \left[ \left(\delta_{ij} - \frac{k_ik_j}{|k|^2} \right) \mathrm{S}(k)  + 
i\epsilon_{ijk}  \frac{k^k}{|\vec{k}|} \he(k) \right],
\end{equation}
with $\epsilon_{ijk}$ is the totally antisymmetric tensor and $B_i(k)$ the Fourier transform of the magnetic field. $\mathrm{S}$ and $\he$ are the symmetric and helical term respectively of the PMF power spectrum and are assumed to follow a power law:  
\begin{equation}
\mathrm{S}(k) = \left\{ \begin{array}{lc} \mathrm{S}_0 k^{\mathrm{n_\mathrm{S}}} & \mathrm{if} k<k_D, \\ 0 & \mathrm{otherwise}, \end{array} \right.
\end{equation}
and,
\begin{equation}
\mathrm{H}(k) = \left\{ \begin{array}{lc} \mathrm{H}_0 k^{\mathrm{n_\mathrm{H}}} & \mathrm{if} k<k_D, \\ 0 & \mathrm{otherwise}, \end{array} \right.
\end{equation}
with $\mathrm{S}_0(\he_0)$ and $\mathrm{n_{\sa(\he)}}$ the amplitude and spectral index of the (anti)symmetric term of the magnetic field power spectrum respectively. The wavenumber $k_D$ stands for the damping of the magnetic field at small scales (\cite{Durrer_2000}) and is assumed here to equal to $10^4 Gpc^{-1}$. 

As shown in \cite{Caprini_2004}, the spectral indices $\mathrm{n}_{\he/\mathrm{S}}$ have to verify some conditions. Firstly, in order to avoid divergence of the magnetic field amplitude at large scales, they have to be such as:  $\mathrm{n}_{\mathrm{S}} > -3$ and $\mathrm{n}_{\he} > -4$. Secondly, as the symmetric term has to be larger than the absolute value of the helical term, the condition: $\mathrm{n}_{\he} \geqslant \mathrm{n}_{\mathrm{S}}$. Thirdly, if the considered PMF is generated by a mechanism taking place after the inflationary period, this mechanism has to be \textit{causal} boiling down to a suppression of the power spectrum on very large scales.
 This leads to the following conditions on the spectral indices: 
\begin{equation}
\left\{ \begin{array}{c} \mathrm{n}_{\mathrm{S}} \geqslant 2, \mathrm{and~even~integer}, \\ \mathrm{n}_{\he} \geqslant 3, \mathrm{and~odd~integer}.  \end{array} \right.
\end{equation}

An helical PMF would therefore be a promising source of the magnetic fields observed on cosmological scales today. Being primordial, it could have left its imprints in the CMB allowing us to set constraints on such a field.


\section{Impact on the CMB and Forecasts}

The presence of a stochastic magnetic field in the early universe would have perturbed the scalar, vector and tensor parts of the metric (see reference within \cite{Durrer_2013}). A PMF would have thus contribute to the CMB temperature and polarisation power spectra. In particular, an helical PMF induces a symmetry breaking which is translated in non-vanishing $TB$ and $EB$ correlations. In the following, we recap all the resulting computations regarding the kind of perturbations and the reader is referred to the mentioned articles for the details. As in \cite{Caprini_2004}, the CMB $XY$ correlations $C_{\ell}^{XY}$ (with $X,Y$ the $T$, $E$ and $B$ modes) are split in contributions from the symmetric $C_{(\sa) \ell}^{XY}$ and helical $C_{(\he) \ell}^{XY}$ parts:
\begin{equation}
C_{\ell}^{XY} = C_{(\sa) \ell}^{XY} - C_{(\he) \ell}^{XY}.
\end{equation}
Also, we have introduced an exponential damping of the CMB power spectra above a certain multipole $\ell_{cut}$:
\begin{equation}
C_{\ell}^{XY,cut} = C_{\ell}^{XY} \times \mathrm{exp}\left(-\frac{\ell}{\ell_{cut}}\right).
\end{equation}

Furthermore, different quantities and notations are introduces. The helicity density parameter $\Omega_\he$ is defined as: 
\begin{equation}
\Omega_\he = \frac{\mathcal{B}_{\lambda}}{8\pi\rho_c}\left(k_D \lambda \right)^{n_\he+3}
\end{equation}
and its analogue quantity for the magnetic field energy density: 
\begin{equation}
\Omega_\sa = \frac{B_{\lambda}}{8\pi\rho_c}\left(k_D \lambda \right)^{n_\sa+3}
\end{equation}
with $B_{\lambda}$ the averaged magnetic field energy density smoothed over a sphere of comoving radius $\lambda$ and $\mathcal{B}_{\lambda}$ is the same for the helicity of the magnetic field.
Besides, we introduce the following quantities in order to simplify the expressions of the CMB power spectra:
\begin{eqnarray}
C_{(\sa) 0} & = & \frac{\mathcal{A}_{(\sa) 0}}{(2 n_\sa + 3) \Gamma^2(\frac{n_\sa+3}{2})}, \\
C_{(\he) 0} & = & \frac{\mathcal{A}_{(\he) 0}}{(2 n_\he + 3) \Gamma^2(\frac{n_\he+4}{2})},
\end{eqnarray}
with: 
\begin{eqnarray}
\mathcal{A}_{(\sa) 0} & = & \left[\frac{\Omega_\sa}{\Omega_r}\mathrm{ln}(\frac{z_{in}}{z_{eq}}) \right]^2, \\
\mathcal{A}_{(\he) 0} & = & \left[\frac{\Omega_\he}{\Omega_r}\mathrm{ln}(\frac{z_{in}}{z_{eq}}) \right]^2, \\
\mathcal{A}_{(\times) 0} & = & \frac{\Omega_\he\Omega_\sa}{\Omega_r^2(n_\he+n_\sa+2)\Gamma\left(\frac{n_\he+4}{2}\right)\Gamma\left(\frac{n_\sa+3}{2}\right)}\left[\mathrm{ln}(\frac{z_{in}}{z_{eq}}) \right]^2, 
\end{eqnarray}
where $\Omega_r$ is radiation density parameter today, $z_{in/eq}$ is the redshift of the production of the magnetic field and the matter radiation equality respectively. We also denote $\eta_0$ the distance to the last scattering surface.

\underline{Scalar contribution}

The scalar perturbations generated by an helical PMF are smaller than the tensor and vector ones. They are therefore not considered in our analysis (\cite{Mack_2002}).

\underline{Vector contribution}

An helical PMF acts as a source of vector perturbations giving a non negligible vectorial contribution to the CMB power spectra, explicitly computed in \cite{Kahniashvili_2005}. We recap their expression in the following. For convenience, the CMB power spectra can be expressed in the form: 
\begin{equation}
C_{\ell}^{XY} = C_{(\sa) \ell}^{XY} \left( 1 - \frac{2\left( 2n_\sa +3 \right)}{3\left( 2n_\he +3 \right)}  \frac{H_{\lambda}^2\Gamma\left(\frac{n_\sa}{2}+\frac{3}{2}\right) k_D^{n_\he-n_\sa} }{B_{\lambda}\Gamma\left( \frac{n_\sa}{2} +2\right)} \right) \mathcal{R}^{XY}
\end{equation}
with $X,Y$, the $T$, $E$ or $B$ modes.
Besides, we introduce: 
\begin{eqnarray}
C_{(\sa) 0}^{(V)} & = & \alpha_{dec} v_{A\lambda}^4 \\
C_{(\times) 0}^{(V)} & = & \beta_{dec} \frac{v_{A\lambda}^2v_{H\lambda}^2}{27(n_\he+n_\sa+2)\Gamma\left(\frac{n_\sa}{2}+\frac{3}{2}\right)\Gamma\left(\frac{n_\he}{2}+2\right)} 
\end{eqnarray}
with $\alpha_{dec}$ and $\beta_{dec}$ pre-factors depending on the cosmology, explicitly expressed in \cite{Kaniashvili_2005}, $v_{A \lambda}$ and $v_{H \lambda}$ the Alvén and helicity velocity respectively.

The symmetric and $\mathcal{R}^{BB}$ contributions to the $BB$, and equivalently to $EE$, power spectre are written as:
\begin{eqnarray}
\ell^2C_{\ell}^{BB} & = & \left\{ \begin{array}{ll} \frac{(2\pi)^{2n_\sa+10}}{54} C_{\sa 0}^{(V)} \frac{\ell^4}{\Gamma\left( \frac{n_\sa+3}{2}\right)(2n_\sa+3)} \frac{(k_D\eta0)^{2n_\sa+3}}{(k_{\lambda}\eta_0)^{2n_\sa+6}} \left[ (k_\sa\eta_0)^3-\ell^3 \right] & \mathrm{~for~} n_\sa > \frac{-3}{2}, \\
\frac{(2\pi)^{2n_\sa+10}}{36} C_{\sa 0}^{(V)} \frac{n_\sa \ell^4}{\Gamma\left( \frac{n_\sa+3}{2}\right)(2n_\sa+3)(n_\sa+3)^2(k_\lambda\eta_0)^{2n_\sa+6}}  \left[ (k_\sa\eta_0)^{2n_\sa+6}-\ell^{2n_\sa++6} \right] & \mathrm{~for~} -3 < n_\sa < \frac{-3}{2},
\end{array} \right.
\end{eqnarray}
and, for $n_\he > -3/2$:
\begin{eqnarray}
\mathcal{R}^{BB} & \simeq & \left\{ \begin{array}{ll} 1 & \mathrm{~for~} n_\he > \frac{-3}{2}, \\
\frac{2(n_\he+3)^2}{n_\sa} \left(\frac{k_D}{k_\sa}\right)^{2n_\sa+3} & \mathrm{~for~} -3 < n_\he < \frac{-3}{2}, 
\end{array} \right .
\end{eqnarray}
and, for $-3 < n_\sa \leqslant n_\he < -\frac{3}{2}$:
\begin{eqnarray}
\mathcal{R}^{EE} & \simeq & \frac{(n_\he-1)(n_\sa+3)^2}{n_\sa(n_\he+4)(n_\he+3)} \left( \frac{k_\sa}{k_D} \right)^{2(n_\he-n_\sa)},
\end{eqnarray}
with $k_\sa = \frac{2\pi}{L_\sa}$ where $L_\sa$ is Silk scale.

The $TB$ correlation is written as: 
\begin{eqnarray}
\hspace{-1cm}C_{\ell}^{TB} = \left\{ \begin{array}{ll}  -\ell^2 \frac{(2\pi)^{n_\sa+n_\he+8}2^{n_\sa+n_\he+4}}{27} C_{(\times) 0}^{(V)} \frac{(k_D\eta_0)^{n_\sa+n_\he+2}}{(k_\lambda\eta_0)^{n_\sa+n_\he+6}} (k_\sa\eta_0)^3 & \mathrm{~for~} n_\sa + n_\he > -2, \\
-\ell^{n_\sa+n_\he+7} \frac{(2\pi)^{n_\sa+n_\he+7}2^{n_\sa+n_\he+4}}{9} C_{(\times) 0}^{(V)} \frac{1}{(k_\lambda\eta_0)^{n_\sa+n_\he+6}} \frac{n_\he-1}{n_\sa+3}  \frac{\Gamma\left(-n_\sa -n_\he-5 \right)}{\Gamma\left( -\frac{n_\sa}{2} - \frac{n_\he}{2} - 2\right)^2} & \mathrm{~for~} -6 < n_\sa + n_\he < -5, \\
-\ell^2 \frac{(2\pi)^{n_\sa+n_\he+7}}{9(n_\sa+n_\he+5)} C_{(\times) 0}^{(V)} \frac{(k_\sa\eta_0)^{n_\sa+n_\he+5}}{(k_\lambda\eta_0)^{n_\sa+n_\he+6}} \frac{n_\he-1}{n_\sa+3}  & \mathrm{~for~} -5 < n_\sa + n_\he < -2. \\
\end{array} \right .
\end{eqnarray}

The vector contribution to the $EB$ correlation is negligible with respect to the $TB$ correlation and is therefore set to zero in our analysis.

\underline{Tensor contribution}

In the case of the tensor contributions, \cite{Caprini_2004}, extending the work presented in \cite{Pogosian_2002}, has derived all the CMB power spectra for a stochastic helical PMF. The computations of the symmetric part of the power spectra are given in \cite{Mack_2002}.

The symmetric and helical contributions to the $TT$ power spectrum are written as:
\begin{eqnarray}
\ell^2C_{(\sa) \ell}^{TT} & \simeq & \left\{ \begin{array}{ll} \frac{32(4\pi)^2}{27} C_{(\sa) 0} (\frac{\ell}{k_D\eta_0})^3 & \mathrm{~for~} n_{\sa} > -3/2, \\
\frac{2(4\pi)^4}{9\sqrt{\pi}} C_{(\sa) 0} \frac{\Gamma(\frac{1}{2} -n_\sa)}{\Gamma(1-n_\sa)} \frac{n_\sa}{n_\sa + 3}(\frac{\ell}{k_D\eta_0})^{2n_\sa+6}   & \mathrm{~for} -3 < n_{\sa} < -3/2,
 \end{array} \right . \\
\ell^2C_{(\he) \ell}^{TT} & \simeq & \left\{ \begin{array}{ll} \frac{32(4\pi)^2}{27} C_{(\he) 0} (\frac{\ell}{k_D\eta_0})^3 & \mathrm{~for~} n_{\he} > -3/2, \\
 \frac{(16\sqrt{\pi})^3}{9} C_{(\he) 0}  \frac{\Gamma(\frac{1}{2} -n_\he)}{\Gamma(1-n_\he)} \frac{n_\he-1}{n_\he+4} (\frac{\ell}{k_D\eta_0})^{2n_\he+6} & \mathrm{~for~} -3 < n_{\he} < -3/2.
 \end{array} \right .
\end{eqnarray} 

The symmetric and helical contributions to the $EE$ power spectrum are written as:
\begin{eqnarray}
\ell^2C_{(\sa) \ell}^{EE} & \simeq & \left\{ \begin{array}{ll} \frac{(4\pi)^3}{9} C_{(\sa) 0} (\frac{\ell}{k_D\eta_0})^2 & \mathrm{~for~} n_{\sa} > -3/2, \\
 \frac{(4\pi)^3}{9} C_{(\sa) 0} \frac{n_\sa}{(n_\sa+3)(2n_\sa+4)} (\frac{\ell}{k_D\eta_0})^2 & \mathrm{~for~} -2 < n_{\sa} < -3/2, \\
- \frac{(4\pi)^3}{9} C_{(\sa) 0} \frac{3}{2} \mathrm{ln}(\frac{k_D\eta_0}{\ell^2}) (\frac{\ell}{k_D\eta_0})^2 & \mathrm{~for~} n_{\sa} = -2, \\
  \frac{2(2\pi)^4}{9\sqrt{\pi}} C_{(\sa) 0}  \frac{\Gamma(-n_\sa-2)}{\Gamma(-n_\sa-\frac{3}{2})} \frac{n_\sa-1}{n_\he+4} (\frac{\ell}{k_D\eta_0})^{2n_\sa+6} & \mathrm{~for~} -3 < n_{\sa} < -2,
 \end{array} \right . \\
\ell^2C_{(\he) \ell}^{EE} & \simeq & \left\{ \begin{array}{ll} \frac{(4\pi)^3}{9} C_{(\he) 0} (\frac{\ell}{k_D\eta_0})^2 & \mathrm{~for~} n_{\he} > -3/2, \\
 \frac{(4\pi)^3}{9} C_{(\he) 0} \frac{n_\he-1}{(n_\he+4)(2n_\he+4)} (\frac{\ell}{k_D\eta_0})^2 & \mathrm{~for~} -2 < n_{\he} < -3/2, \\
- \frac{(4\pi)^3}{9} C_{(\he) 0} \frac{3}{2} \mathrm{ln}(\frac{k_D\eta_0}{\ell^2}) (\frac{\ell}{k_D\eta_0})^2 & \mathrm{~for~} n_{\he} = -2, \\
 \frac{2(2\pi)^4}{9\sqrt{\pi}} C_{(\he) 0}  \frac{\Gamma(-n_\he-2)}{\Gamma(-n_\he-\frac{3}{2})} \frac{n_\he-1}{n_\he+4} (\frac{\ell}{k_D\eta_0})^{2n_\he+6} & \mathrm{~for~} -3 < n_{\he} < -2.
 \end{array} \right .
\end{eqnarray} 

The symmetric and helical contributions to the $BB$ power spectrum are similar to the ones contributing to the $EE$ power spectrum:
\begin{eqnarray}
\ell^2C_{(\sa) \ell}^{BB} & \simeq & \left\{ \begin{array}{ll} \frac{(4\pi)^3}{9} C_{(\sa) 0} (\frac{\ell}{k_D\eta_0})^2 & \mathrm{~for~} n_{\sa} > -3/2, \\
 \frac{(4\pi)^3}{9} C_{(\sa) 0} \frac{n_\sa}{(n_\sa+3)(2n_\sa+4)} (\frac{\ell}{k_D\eta_0})^2 & \mathrm{~for~} -2 < n_{\sa} < -3/2, \\
- \frac{(4\pi)^3}{9} C_{(\sa) 0} \frac{3}{2} \mathrm{ln}(\frac{k_D\eta_0}{\ell^2}) (\frac{\ell}{k_D\eta_0})^2 & \mathrm{~for~} n_{\sa} = -2, \\
  \frac{2(2\pi)^4}{9\sqrt{\pi}} C_{(\sa) 0}  \frac{\Gamma(-n_\sa-2)}{\Gamma(-n_\sa-\frac{3}{2})} \frac{n_\sa-1}{n_\he+4} (\frac{\ell}{k_D\eta_0})^{2n_\sa+6} & \mathrm{~for~} n_{\sa} < -2,
 \end{array} \right . \\
\ell^2C_{(\he) \ell}^{BB} & \simeq & \left\{ \begin{array}{ll} \frac{(4\pi)^3}{9} C_{(\he) 0} (\frac{\ell}{k_D\eta_0})^2 & \mathrm{~for~} n_{\he} > -3/2, \\
 \frac{(4\pi)^3}{9} C_{(\he) 0} \frac{n_\he-1}{(n_\he+4)(2n_\he+4)} (\frac{\ell}{k_D\eta_0})^2 & \mathrm{~for~} -2 < n_{\he} < -3/2, \\
- \frac{(4\pi)^3}{9} C_{(\he) 0} \frac{3}{2} \mathrm{ln}(\frac{k_D\eta_0}{\ell^2}) (\frac{\ell}{k_D\eta_0})^2 & \mathrm{~for~} n_{\he} = -2, \\
 \frac{2(2\pi)^4}{9\sqrt{\pi}} C_{(\he) 0}  \frac{\Gamma(-n_\he-2)}{\Gamma(-n_\he-\frac{3}{2})} \frac{n_\he-1}{n_\he+4} (\frac{\ell}{k_D\eta_0})^{2n_\he+6} & \mathrm{~for~} n_{\he} < -2.
 \end{array} \right .
\end{eqnarray}

The symmetric and helical contributions to the $TE$ power spectrum are written as:
\begin{eqnarray}
\ell^2C_{(\sa) \ell}^{TE} & \simeq & \left\{ \begin{array}{ll} \frac{2(2\pi)^4}{9\sqrt{\pi}} C_{(\sa) 0} \frac{\Gamma(\frac{3}{4})}{\Gamma(\frac{5}{4})} (\frac{\ell}{k_D\eta_0})^3 & \mathrm{~for~} n_{\sa} > -3/2, \\
 \frac{2(2\pi)^4}{9\sqrt{\pi}} C_{(\sa) 0} \frac{\Gamma(-\frac{3}{4}-n_\sa)}{-\Gamma(\frac{1}{4}-n_\sa)} \frac{n_\sa-1}{(n_\sa+4)} (\frac{\ell}{k_D\eta_0})^{2n_\he+6} & \mathrm{~for~} -3 < n_{\sa} < -3/2. \\
 \end{array} \right . \\
\ell^2C_{(\he) \ell}^{TE} & \simeq & \left\{ \begin{array}{ll} \frac{2(2\pi)^4}{9\sqrt{\pi}} C_{(\he) 0} \frac{\Gamma(\frac{3}{4})}{\Gamma(\frac{5}{4})} (\frac{\ell}{k_D\eta_0})^3 & \mathrm{~for~} n_{\he} > -3/2, \\
 \frac{2(2\pi)^4}{9\sqrt{\pi}} C_{(\he) 0} \frac{\Gamma(-\frac{3}{4}-n_\he)}{-\Gamma\left(\frac{1}{4}-n_\he\right)} \frac{n_\he-1}{(n_\he+4)} (\frac{\ell}{k_D\eta_0})^{2n_\he+6} & \mathrm{~for~} -3 < n_{\he} < -3/2. \\
 \end{array} \right .
\end{eqnarray}

The $TB$ and $EB$ correlations do not vanish in the presence of an helical PMF and their complete expressions are: 
\begin{eqnarray}
\ell^2C_{\ell}^{TB} & \simeq & \left\{ \begin{array}{ll} - 4 \sqrt{\frac{\pi}{2}} (2\pi)^4 C_{(\times) 0} (\frac{\ell}{k_D\eta_0})^4 & \mathrm{~for~} n_{\sa}+n_\he > -2, \\
- \frac{4(4\pi)^4}{9\sqrt{\pi}} C_{(\times) 0} \frac{\Gamma\left(-\frac{n_\he}{2} -\frac{n_\sa}{2} -\frac{3}{4} \right)}{\Gamma\left(-\frac{n_\he}{2} -\frac{n_\sa}{2} -\frac{1}{4} \right)} \frac{n_\he-1}{n_\sa+3} (\frac{\ell}{k_D\eta_0})^{n_\he+n_\sa+6} & \mathrm{~for~} -6 < n_{\sa}+n_\he < -2, \\
 \end{array}   \right.
\end{eqnarray}
and
\begin{eqnarray}
\ell^2C_{\ell}^{EB} & \simeq & \left\{ \begin{array}{ll} \frac{4(4\pi)^3}{9} C_{(\times) 0} \frac{n_\he-1}{n_\sa+3} \frac{(-1)^\ell}{k_D\eta_0} \mathrm{sin}(2\eta_0k_D) (\frac{\ell}{k_D\eta_0})^2 & \mathrm{~for~} -3 < n_{\sa}+n_\he < -2, \\
\frac{4(4\pi)^3}{9} C_{(\times) 0} \frac{n_\he-1}{n_\sa+3} \frac{(-1)^{\ell+1}}{(k_D\eta_0)^2} \mathrm{sin}(2\ell^2) (\frac{\ell}{k_D\eta_0})^{n_\he+n_\sa+4} & \mathrm{~for~} -4 < n_{\sa}+n_\he < -3, \\
- \frac{(4\pi)^4}{9\sqrt{\pi}} C_{(\times) 0} \frac{\Gamma\left( -\frac{n_\he}{2} -\frac{n_\sa}{2} -\frac{3}{2} \right)}{\Gamma\left( -\frac{n_\he}{2} -\frac{n_\sa}{2} - 1 \right)}  \frac{n_\he-1}{n_\sa+3} (\frac{\ell}{k_D\eta_0})^{n_\he+n_\sa+6} & \mathrm{~for~} -6 < n_{\sa}+n_\he < -4.
 \end{array}   \right.
\end{eqnarray}

\begin{figure}[!h]
\begin{center}
	\includegraphics[scale=0.5]{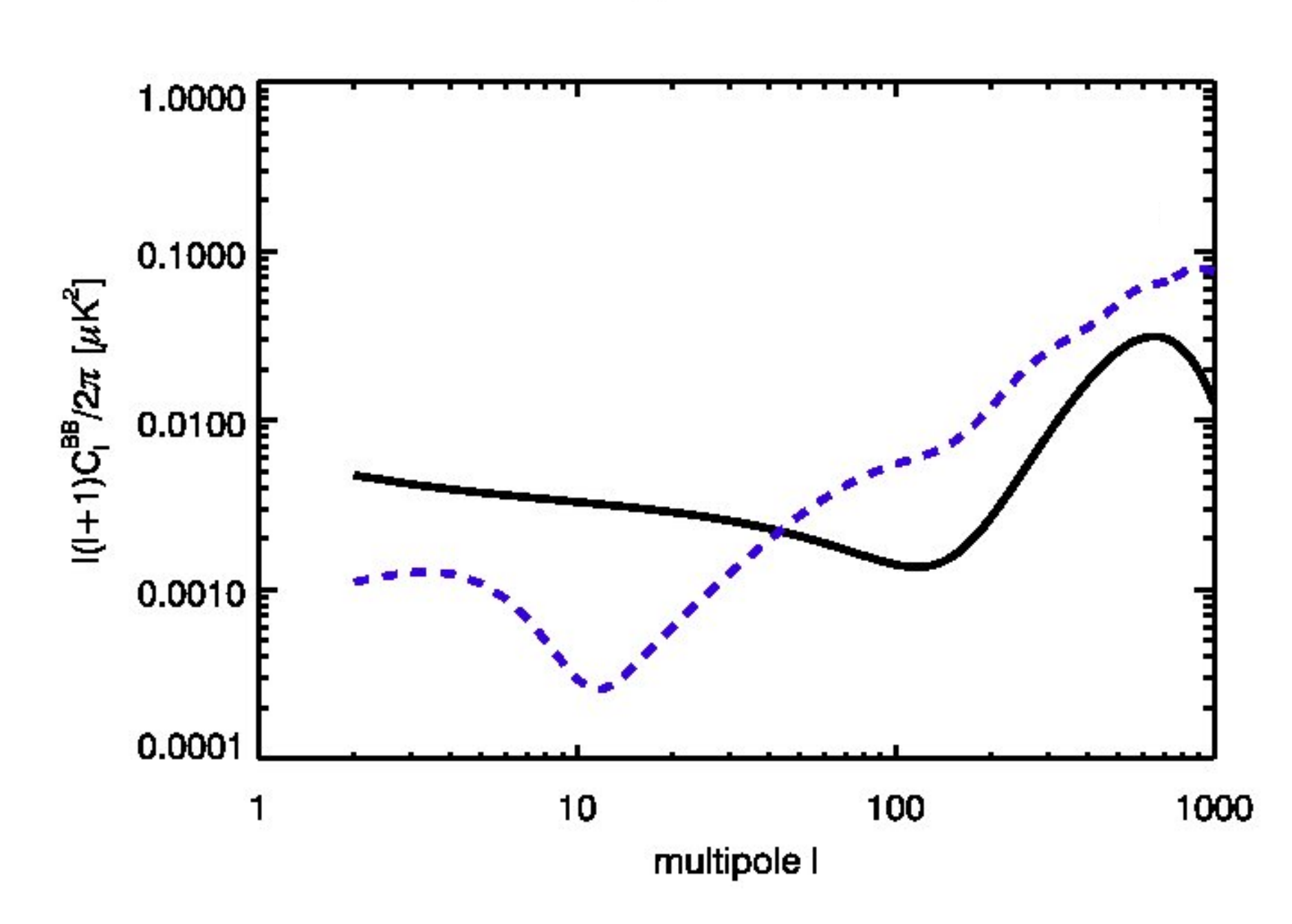}
	\caption{The CMB $BB$ power spectrum expected in the standard model with $r = 0.05$ in black and the contribution from an helical PMF in dashed blue with $n_\sa = 2.99$ and $n_\he = 2.5$.}
	\label{fig:bb_helicalB}
\end{center}
\end{figure}

\begin{figure}[!h]
\begin{center}
	\includegraphics[scale=0.5]{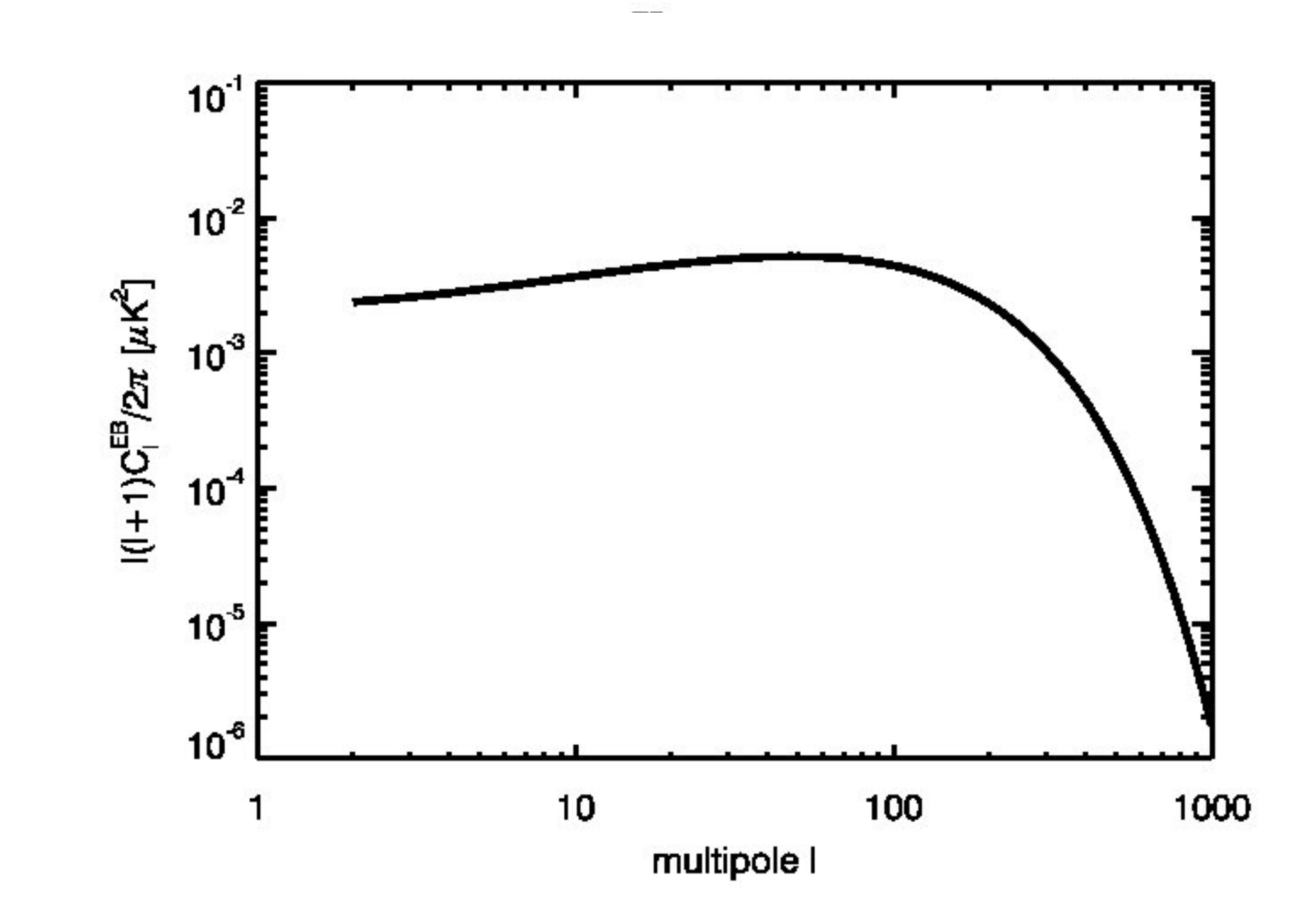}
	\caption{The CMB $EB$ correlation from an helical PMF with $n_\sa = 2.99$ and $n_\he = 2.5$.}
	\label{fig:eb_helicalB}
\end{center}
\end{figure}

In order to produce potential CMB power spectra which include the effect of an helical PMF, we have implemented the above analytic formulas with realistic values of the different parameters. As an example, the figures~\ref{fig:bb_helicalB},\ref{fig:eb_helicalB} show the resulting $BB$ and $EB$ power spectra respectively for the spectral indices: $n_\sa = 2.99$ and $n_\he = 2.5$. The contribution to the $BB$ power spectrum by an helical PMF is shown along with the one from the standard model with $r = 0.05$. Up to $\ell = 40$, it exceeds the signal from the standard $BB$ power spectrum: the largest angular scales are thus of interest. Similarly, the contribution the helical PMF to the $EB$ correlation is dominant at low $\ell$ and then damped by the introduced cut-off ($\ell_{cut} = 100$ in this case). 

The polarised power spectra and in particular the $TB$ and $EB$ correlations offer a promising probe of primordial magnetic field. They can indeed help to set constraints on the amplitudes $\sa_0$, $\he_0$ and the spectral indices $n_\sa$, $n_\he$ of the PMF. In the same way as in the previous chapter, we have thus performed a Fisher analysis to state on the detection of an helical PMF with current experimental set up. We have studied the case of a CMB polarisation detection by a small scale experiment such as the one described in Chapter~\ref{Chapter5}. Also, as the PMF affects the $BB$, $TB$ and $EB$ correlations, we have considered the Fisher matrix given by these three correlators. Furthermore, we have considered the estimation of the PMF parameters jointly with the cosmological parameter $r$. Our preliminary results were obtained using a mode counting estimation of the uncertainties on the power spectra. The results show that, in this optimistic case, the magnetic field parameters (apart from the amplitude of the helical part) can be detectable at $3\sigma$ with a suborbital experiments. This study needs now to be continued in the light of the results from the previous chapters and especially to be performed using the pure estimation of the CMB power spectra.


\section*{Conclusion}

The origin of the magnetic field on cosmological scales remains a mystery. In the recent years, various mechanisms giving an explanation for its creation have been proposed. A promising scenario is the generation of an helical magnetic field during the primordial universe, before the radiation-matter equality. Such a magnetic field would have left its imprints in the CMB, mainly leading to non vanishing $TB$ and $EB$ correlators due to the symmetry breaking induced by the helical component of the PMF. We have forecast the constraints that one could set on an helical PMF in the case of current ground based or balloon borne experiments thanks to analytic expressions of the CMB power spectra induced by an helical PMF, found in the literature. The results show that valuable constraints might be set on such a PMF. This work has been made as part of my master 2 internship at IAS in 2011 and will be updated and continued in the coming months, taking lessons from the results obtained during my PhD. In particular, the uncertainties on the CMB power spectra must be realistically estimated to state on PMF detectability. As shown in Chapter~\ref{Chapter5}, the pure method would have therefore to be used along with variance optimised window function. Also a satellite-like survey might be necessary to give constraints on the magnetic field parameters as its contribution in the CMB anisotropies mainly dominates at large angular scales. Furthermore, an additional source of vector and tensor perturbations could be add in the CLASS code and thus could be used to simulate CMB power spectra in the presence of a PMF.

\part{Conclusion and perspectives}
\label{part4}

\chapter{Conclusion \& Perspectives} 

\label{Conclusion} 

\lhead{\textit{Conclusion \& Perspectives}} 



The CMB $B$ modes at large angular scales, when correctly estimated, are a unique probe to explore the physics of the primordial Universe. They indeed are a smoking gun of the tensor perturbations which are thought to be generated during the cosmic inflation. Owing to their low amplitude, the CMB $B$ modes detection is an instrumental and data analysis challenge. This year 2014 has however started a new era for the CMB $B$ modes exploration. Thanks to instrumental improvements, two experiments have recently announced having succeed in detecting the CMB $B$ modes. Firstly, the {\sc POLARBEAR} experiment has directly detected the lensed $B$ modes, thus corroborating the standard model of cosmology. Later, the {\sc BICEP2} experiment claimed that they had achieved a direct detection of the primordial $B$ modes. Although controversial, this discovery augurs outstanding performances of forthcoming experiments to set constraints on the primordial universe. 

Unfortunately, both space-based or suborbital experiments dedicated to CMB polarisation detection have only access to an incomplete part of the sky. A statistical issue arises when constructing the $B$ modes power spectrum on a masked sky: the \textit{$E$-to-$B$ leakage}. The $E$ modes signal indeed pollutes the amplitude and the variance of the $B$ modes. This effect can compromise $B$ modes detection due to the high level of the $E$ modes with respect to the $B$ modes. It could thus damage the instrumental and data analysis efforts. Pseudospectrum methods, the so-called pure, \zb~and \kn~methods, constructing $B$ modes free from any leakage have therefore been proposed. By construction, they \textit{theoretically} exactly correct for the $E$-to-$B$ leakage. The reconstructed pseudospectra however show that, \textit{in practice}, the leakage does not vanish when applied to pixelised CMB maps. Moreover, although the three considered pseudospectrum approaches rely on the same concept -- the reconstruction of the masked $\chi^B$ field --, they have distinct numerical implementations. It thus leads to different efficiency on $B$ modes reconstruction in terms of statistical uncertainties. I have consequently performed an analysis, at the power spectrum level, to state on their respective performances. 

In the case of a small scale survey typical of current suborbital experiments, the sampling variance does not allow a detection of the reionisation bump at $\ell \sim 2-10$. A pure estimation of the $B$ modes power spectrum is however the most efficient approach as it, at least, gives access to the recombination bump at $\ell \sim 90$ while the \kn~method fails at reconstructing the primordial $B$ modes. The \zb~approach provides a moderate detection of the recombination bump with a lower signal-to-noise ratio than using a pure estimation. 

The case of a large scale survey characteristics of potential satellite experiment dedicated to $B$ modes detection is more striking. The \zb~and \kn~approaches only ensure a detection of the recombination bump. The pure method provides remarkable results as the $B$ modes power spectrum is accurately reconstructed in all bins including the reionisation bump, the obtained variances closely following the ideal ones. This performance originates from the use of the pixel-based variance-optimised (PCG) window functions which are adapted to the pure method. Although the sky coverage is high ($\sim 70\%$), an observed region with intricate shape (high ratio between the perimeter and the area covered by the mask) indeed leads to an amount of leakage comparable to the case of a small scale survey. I have therefore stated that in the case of a large scale survey the use of PCG window functions offers the flexibility required to optimise such intricate mask thus allowing for an optimal $B$ modes reconstruction. This statement is a key issue that arises during my PhD work for it was not expected. 

\begin{figure}[!h]
\begin{center}
	\includegraphics[scale=0.45]{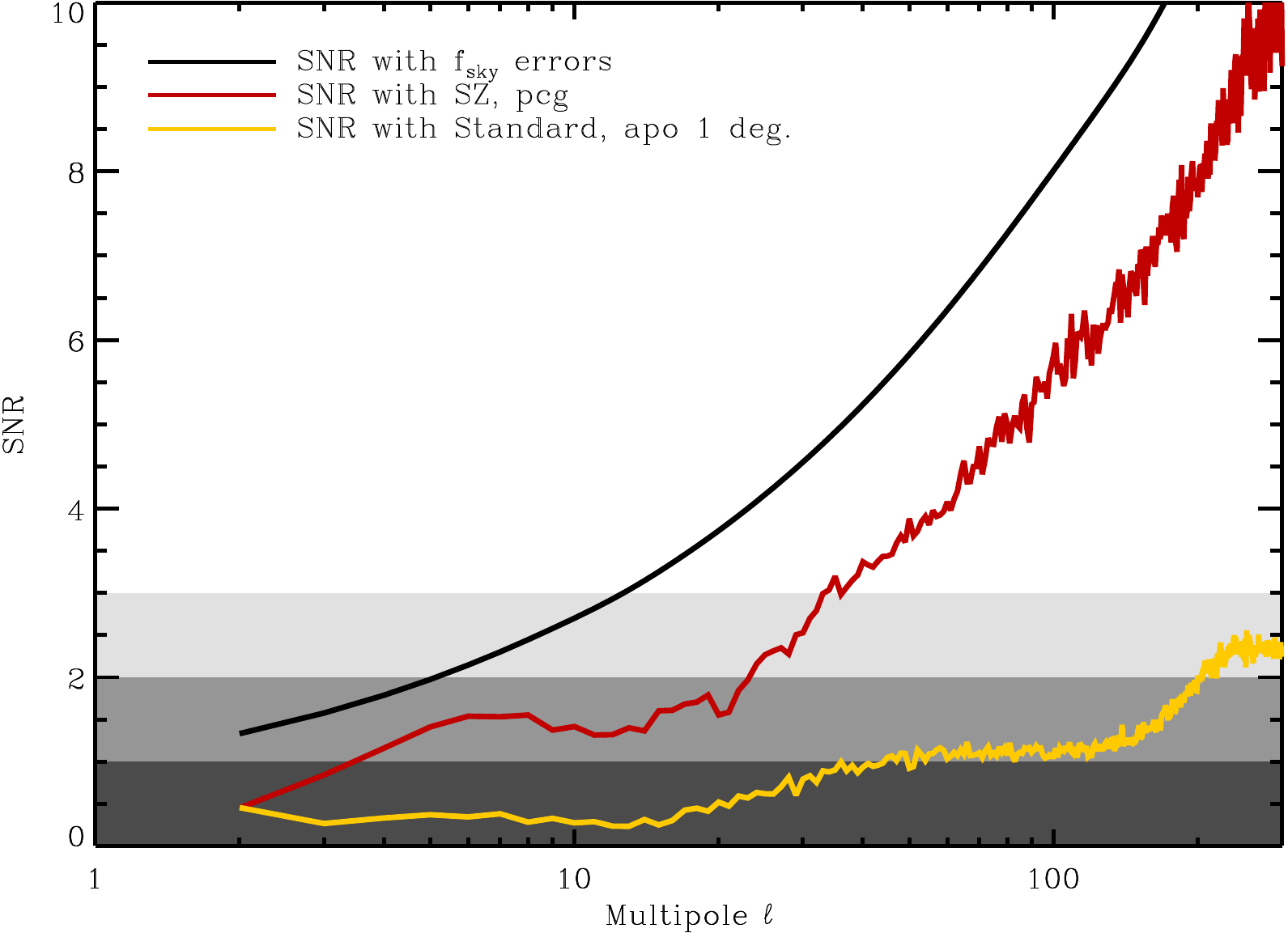}
	\caption{The signal-to-noise ratio of $B$ modes detection for a satellite-like experiment. The solid black line stands for the one using the ideal mode-counting estimation of the variance. The red solid line is the one obtained from a pure $B$ modes estimation along with pixel-based variance optimised window function. The yellow line displays the signal-to-noise ratio using the standard method to etimate $B$ modes. The grey shaded areas set the $1\sigma$, $2\sigma$ and $3\sigma$ limits. The $B$ modes have to be carefully estimated to ensure at least a $3\sigma$ detection. (Taken from \cite{Ferte_2013}). } 
	\label{fig:snr_wmap}
\end{center}
\end{figure}

As a consequence, the pure estimation using PCG optimised window functions results in an efficient $B$ modes detection. As an illustration, the figure~\ref{fig:snr_wmap} displays the signal-to-noise ratio $C_{\ell}^{BB}/\sqrt{\Sigma_{\ell \ell}}$ on the $\ell$-by-$\ell$ $B$ modes reconstruction using the ideal mode-counting variances (in black), the pure estimation (in red) and the standard method (in yellow).  A standard pseudospectrum reconstruction of the CMB polarised power spectra, thus not correcting for the $E$-to-$B$ leakage, prevent from a detection of the primordial $B$ modes as the signal-to-noise ratio is below 1 for $\ell \lesssim 100$. It is therefore necessary to use the pure estimation with PCG window functions for a primordial $B$ modes detection. It results in a $3\sigma$ detection for $\ell > 40$. As a result, the estimation of CMB $B$ modes power has to be carefully done for an optimal primordial $B$ mode detection, the window function computation is consequently crucial particularly in the case of large scale survey.    

The physics of the primordial Universe -- such as the energy scale of inflation or the presence of a magnetic field -- can be probed by the CMB polarisation and in particular the $B$ modes. I was interested in the constraints that will be set by forthcoming or under development experiments dedicated to the $B$ modes detection. The adopted strategy is to estimate the CMB polarised power spectra along with their statistical uncertainties and propagate them on relevant theoretical parameters thanks to the Fisher information matrix. The mode-counting estimation of the uncertainties enable a first exploration of the parameters to be measured and is commonly used to make forecast. It however underestimates the power spectrum uncertainties implying the production of misleading forecasts. I therefore proposed to use the pure method for $B$ modes reconstruction, which efficiency has been previously stated, to perform realistic forecasts.

I have first focused on the detection of the tensor-to-scalar ratio $r$, providing the amplitude of the primordial gravitational waves and thus giving the energy scale of inflation. In the scope of small scale surveys with a given sensitivity, an optimal observed sky fraction will provide the highest signal-to-noise ratio on a given $r$. I have shown that the mode counting estimation of the variances on $B$ modes is reliable to find the optimal observed sky fraction. However, the resulting signal-to-noise ratio on $r$ is overestimated and the pure estimation is mandatory to obtain realistic forecasts. 

Moreover, I have performed forecast on $r$ detection by three fiducial experiments corresponding to a current suborbital experiment, a forthcoming telescopes array and a potential satellite mission. The mode-counting results are spurious as it overestimates the forecasts by a factor of 3 to 5. A pure estimation of the $B$ modes predicts that a large scale survey would have access to $r \sim 10^{-3}$, a half sky survey to $r \sim 10^{-2}$ and a small scale survey to $r \sim 10^{-1}$, at $3\sigma$. A half sky survey would therefore be tolerable to discriminate between large and small field inflation while a satellite experiment would be able detect small field inflation. 

A parity violation in the primordial Universe would leave its imprints in the CMB $TB$ and $EB$ correlations, vanishing in the standard model of cosmology. Following the same strategy as previously, I have investigated the detectability of a parity breaking amount by the parameter $\delta$ in the case of a small scale and a full sky survey. The results are unequivocal: no constraints can be set on $\delta$ using a small scale experiment alone. Nevertheless, as it gives access to the largest angular scales, a satellite experiment would provide a detection of a range of $\delta$. A $2\sigma$ detection is expected for a maximum of parity violation $\delta = 100\%$ for at least $r = 0.05$. Also, I have shown that the miscalibration angle of the detectors has to be well estimated for an accurate estimation of $\delta$ in the case of a pseudospectrum approach. Such a parity breaking can be due to a parity violation at the linear level of gravitation: $\delta$ is thus related to the Barbero-Immirzi parameter $\gamma$. A detection of $TB$ and $EB$ correlations consistent with zero would exclude: $0.2 \leqslant |\gamma| \leqslant 4.9$ at $3\sigma$ for $r = 0.2$. 
The presence of a magnetic field with an helical component in the primordial universe would also induce $TB$ and $EB$ correlations. The amplitude and the spectral index of its power spectrum are proved to be poorly constrained, even in the scope of a satellite experiment. 
 
\vspace{1cm}
\hspace{1cm} All this work has been done focusing on the pseudospectrum approaches to reconstruct the $B$ modes in order to set constraints on the primordial universe. This study can be extended to data analysis aspect or to the establishment of a model for the CMB polarised anisotropies. 

The pseudospectrum methods are a method of choice for they are fast and reliable. Although they are close to be optimal, we have shown that they nonetheless are not optimal specially for the lowest multipoles which are precisely of interest to constrain the primordial Universe. A maximum likelihood (ML) approach or a minimum variance quadratic estimator would ensure an optimal reconstruction of the CMB polarised power spectra. A comparison with these optimal approaches would therefore validate the use of pseudospectrum method. Also, the strategy chosen to constrain the relevant parameters such as $r$ or $\delta$ involves the Fisher information matrix. This implies to assume that the likelihood is Gaussian which is thought to be a fair approximation for $\ell >$ few dozens. The constraints set on the parameters could therefore be non optimal and potentially slightly overestimated. A ML or minimum variance quadratic estimator approaches to estimate the parameters would assert the results of our analysis. As such optimal approaches are numerically costly, the analysis can first be simplified under the assumption of azimuthal symmetry of the noise and observed sky fraction. It has been implemented by J. Peloton and we are now comparing the performances on $r$ detection in order to validate the use of the $B$ modes pure estimation. In the end, this assumption is obviously not verified but the implementation of optimal methods is numerically heavy avoiding to perform simulations. If validate, the pure method would thus be a valuable tool for data analysis.  

Moreover, the current implementation of the pure method only allows an estimation of the variance using Monte Carlo simulations. In the scope of the data analysis, it can be misleading as it implicitly assumes that the power spectrum is already well estimated or sufficiently \textit{a priori} known. An analytical expression has therefore to be found to allow an estimation of the variance directly from the data and not from the simulations. This would potentially be done using the same approach as the {\sc Xspect} method in \cite{Tristram_2005}.  

More generally, the {\sc x2pure} code, along with the window functions computation, are currently used daily for data analysis, tests or forecasts. In the scope of proposals for satellite mission and of all the current or forthcoming experiments dedicated to the CMB $B$ modes, a robust and reliable algorithm for CMB power spectra estimation is necessary. Several possible extents to the {\sc x2pure} code are relevant as it would allow its application to different experimental contexts. Firstly, the current version of the {\sc x2pure} code allows for a cross-spectrum estimation only on the common region of the cross-correlated maps. Most current experiments are however based on array of detectors, each of them reconstructing different maps which are thus not exactly completely recovering. Also, the estimation of the CMB power spectra over two disjoint regions of the sky would be of interest. The {\sc x2pure} code could therefore be adapted for a cross-spectrum estimation over not completely covering or totally disjoint sky patches. Secondly, future experiments (foreseen for the years $\sim 2020$) are expected to lower their instrumental noise down to $1 \mu K$-arcmin. In the low noise limit, the computation of the PCG window functions appear to be very long: about $1000$ iterations are required versus only dozens for typical current experiments. The algorithm of the PCG window function computation have thus to be optimised in the limit of low noise. 

The auto- or cross-correlations of the CMB $B$ modes with the temperature or $E$ modes are used to calibrate the polarisation detectors but above all are an important probe of the primordial universe. As shown in the present study, it can indeed test a parity violation at the level of the gravitational waves or the presence of a magnetic field in the primordial universe. The $TB$ and $EB$ correlations are however also a precious probe of the late time Universe. Induced by Faraday rotation in clusters, they could indeed probe the large scales magnetic field structure. Besides, these correlations can be used to test extensions of the standard model such as the cosmic birefringence.

\vspace{1cm}
\hspace{1cm} In summary, I have validated the use of efficient statistical tools to estimate the CMB polarised power spectra, a crucial preliminary step to constrain fundamental physics in the primordial or late time Universe. The CMB polarised anisotropies containing a lot of information, I have studied various cosmological effects in order to distinguish their different features in the CMB.


 \addtocontents{toc}{\vspace{2em}} 

 \appendix 



\chapter{The Mixing kernels $K_{\ell\ell'}$} 

\label{AppendixA} 

\lhead{Appendix A. \emph{Mixing kernels}} 

This appendix is dedicated to the expression of the mixing kernels in each considered pseudospectrum approaches. Their exact calculations can be found in the relevant articles. 

The  linear system relating the temperature and polarised pseudospectra to the true angular power spectra is:
\begin{eqnarray}
\left( \begin{array}{c} \tilde{C}_{\ell}^{TT} \\ \tilde{C}_{\ell}^{EE}  \\  \tilde{C}_{\ell}^{BB} \end{array} \right)
= \left( \begin{array}{ccc} K_{\ell \ell'}^{\mathrm{method},TT} & 0 & 0 \\ 0 & K_{\ell \ell'}^{\mathrm{method},+} & K_{\ell \ell'}^{\mathrm{method},-} \\ 0 & K_{\ell \ell'}^{\mathrm{method},-} & K_{\ell \ell'}^{\mathrm{method},+} \end{array} \right) \left( \begin{array}{c} C_{\ell}^{TT} \\  C_{\ell}^{EE} \\  C_{\ell}^{BB} \end{array} \right).
\end{eqnarray}

We introduce the following notation: 
\begin{equation}
W_{\ell}^{XY} = \sum\limits_m w_{\ell m}^X w_{\ell m}^{Y*},
\end{equation}
with $w_{\ell m}^{X/Y}$ the multipoles of the window function $W$ corresponding to the $X/Y$ polarisation modes: $T$, $E$ or $B$. We denote:
\begin{equation}
\wjjj{\ell}{\ell'}{\ell''}{s}{s'}{s''},
\end{equation}
the Wigner $3-j$ symbol, $s$ being the spin.
For convenience, we also define: 
\begin{displaymath}
	J^{\pm}_{s}(\ell,\ell',\ell'')=\left(\begin{array}{ccc}
		\ell & \ell' & \ell'' \\
		-2+s & 2 & -s
		\end{array}\right)\pm\left(\begin{array}{ccc}
		\ell & \ell' & \ell'' \\
		2-s & -2 & s
		\end{array}\right).
\end{displaymath}

\section*{Standard Method}

The calculations of the mixing kernels in the standard approach can be found in Appendix of \cite{Tristram_2005}. Their expressions are such as:

\begin{equation}
K_{\ell \ell'}^{\mathrm{std},TT} = \frac{2\ell'+1}{4\pi}\sum_{\ell''} W_{\ell''}^{TT} \wjjj{\ell}{\ell'}{\ell''}{0}{0}{0}^2 ,
	\label{eq:kernel_final}
\end{equation}
with $\wjjj{\ell}{\ell'}{\ell''}{0}{0}{0}$ the Wigner $3-j$ symbol.

\begin{equation}
K_{\ell \ell'}^{\mathrm{std},\pm} = \frac{2\ell'+1}{16\pi}\sum_{\ell''} W_{\ell''}^{XY} \left[  \wjjj{\ell}{\ell'}{\ell''}{-2}{2}{0} \pm  \wjjj{\ell}{\ell'}{\ell''}{-2}{2}{0} \right]^2,
	\label{eq:kernel_final}
\end{equation}
with $X/Y$ standing for $E$ or $B$ modes. 

\section*{Pure Method}

The explicit calculations of the mixing kernels in the pure method can be found in \cite{Grain_2009}. 

If the spin window function are not independent, they write:
\begin{eqnarray}
	K^{\mathrm{pure},+}_{\ell\ell'}&=&\frac{2\ell'+1}{16\pi}\displaystyle\sum_{\ell''}(2\ell''+1) W_{\ell''} \left[J^{+}_0(\ell,\ell',\ell'')+2\sqrt{\frac{(\ell+1)!(\ell-2)!(\ell''+1)!}{(\ell-1)!(\ell+2)!(\ell''-1)!}}J^{+}_1(\ell,\ell',\ell'')\right.
	\label{eqn:mlldiaganal}
	\\
	&&\left.+ \sqrt{\frac{(\ell-2)!(\ell''+2)!}{(\ell+2)!(\ell''-2)!}}J^{+}_2(\ell,\ell',\ell'')\right]^2, \nonumber \\
	K^{\mathrm{pure},-}_{\ell\ell'}&=&\frac{2\ell'+1}{16\pi}\displaystyle\sum_{\ell''}(2\ell''+1) W_{\ell''}\left[J^{-}_0(\ell,\ell',\ell'')+2\sqrt{\frac{(\ell+1)!(\ell-2)!(\ell''+1)!}{(\ell-1)!(\ell+2)!(\ell''-1)!}}J^{-}_1(\ell,\ell',\ell'')\right.
	\label{eqn:mlloffanal}
	\\
	&&\left.+\sqrt{\frac{(\ell-2)!(\ell''+2)!}{(\ell+2)!(\ell''-2)!}}J^{-}_2(\ell,\ell',\ell'')\right]^2. \nonumber
\end{eqnarray}

If the spin-weighted window functions are independent, they write: 
\begin{eqnarray}
	K^{\mathrm{pure},+}_{\ell\ell'}&=&\frac{2\ell'+1}{16\pi}\displaystyle\sum_{\ell''m''}\left|w^{(E)}_{0,\ell''m''}J^+_{0}+2\sqrt{\frac{(\ell+1)!(\ell-2)!}{(\ell-1)!(\ell+2)!}}w^{(E)}_{1,\ell''m''}J^+_{1}+\sqrt{\frac{(\ell-2)!}{(\ell+2)!}}w^{(E)}_{2,\ell''m''}J^+_{2}\right|^2
	\label{eqn:mlldiaganalEB}
	 \\
	&&+\left|2\sqrt{\frac{(\ell+1)!(\ell-2)!}{(\ell-1)!(\ell+2)!}}w^{(B)}_{1,\ell''m''}J^-_{1}+\sqrt{\frac{(\ell-2)!}{(\ell+2)!}}w^{(B)}_{2,\ell''m''}J^-_{2}\right|^2, \nonumber \\
	K^{\mathrm{pure},-}_{\ell\ell'}&=&\frac{2\ell'+1}{16\pi}\displaystyle\sum_{\ell''m''}\left|w^{(E)}_{0,\ell''m''}J^-_{0}+2\sqrt{\frac{(\ell+1)!(\ell-2)!}{(\ell-1)!(\ell+2)!}}w^{(E)}_{1,\ell''m''}J^-_{1}+\sqrt{\frac{(\ell-2)!}{(\ell+2)!}}w^{(E)}_{2,\ell''m''}J^-_{2}\right|^2	
	\label{eqn:mlloffanalEB}
	 \\
	&&+\left|2\sqrt{\frac{(\ell+1)!(\ell-2)!}{(\ell-1)!(\ell+2)!}}w^{(B)}_{1,\ell''m''}J^+_{1}+\sqrt{\frac{(\ell-2)!}{(\ell+2)!}}w^{(B)}_{2,\ell''m''}J^+_{2}\right|^2. \nonumber
\end{eqnarray}
where $w_{s,\ell m}^{(X)}$ are the multipoles of the spin-weighted window functions using the $E$ and $B$ decomposition: 
\begin{eqnarray}
w_{0,\ell m}^{(E)} & = & -w_{0,\ell m}, \\
w_{s,\ell m}^{(E)} & = & -\frac{1}{2} \left( w_{s,\ell m} + (-1)^sw_{-s,\ell m}\right), \\
w_{s,\ell m}^{(B)} & = & \frac{i}{2}\left( w_{s,\ell m} - (-1)^sw_{-s,\ell m}\right).
\end{eqnarray}


\section*{\zb~and \kn~Method}

The mixing kernels in the scope of the \zb~and \kn~method are written taking advantage of the fact that $\chi^B$ is a scalar field. The mixing kernel therefore writes:
\begin{equation}
\label{eq:kernelScalar}
	K^{\mathrm{\kn/\zb},+}_{\ell\ell'}= \frac{(2\ell'+1)\alpha^2_{\ell'}}{4\pi \alpha^2_{\ell}} \sum_{\ell''} W_{\ell''}^{XX} \left(\begin{array}{ccc}
		\ell & \ell' & \ell'' \\
		0 & 0 & 0
	\end{array}\right)^2.
\end{equation}


The $K_{\ell \ell'}^{\mathrm{\kn/\zb},-}$ mixing kernels are set equal to zero by construction.



 \addtocontents{toc}{\vspace{2em}} 

 \backmatter


\label{Bibliography}

\lhead{\emph{Bibliography}} 

\bibliographystyle{plainnat85} 

\bibliography{Bibliography} 

\end{document}